\documentclass[a4paper,reqno,phd,abbrevs,11pt]{msthesis}

\usepackage{latexsym}
\usepackage{amsthm,amsmath,amsfonts,amscd}

\usepackage[dvips]{epsfig}

\usepackage{epic}
\usepackage{rotating}

\newtheorem{theorem}[equation]{Theorem}
\newtheorem{proposition}[equation]{Proposition}
\newtheorem{lemma}[equation]{Lemma}
\newtheorem{corollary}[equation]{Corollary}
\newtheorem{claim}[equation]{Claim}
\newtheorem*{quotedtheorem}{Theorem}
\newtheorem*{quotedproposition}{Proposition}
\newtheorem{defn+}[equation]{Definition}
\newtheorem*{rmks+}{Remarks}
\newtheorem{conjecture}{Conjecture}
\newenvironment{definition}{\begin{defn+}\normalfont}{\end{defn+}}

\def \qed{\hfill{\hbox{$\Box$}}}
\def \proof{\noindent{\bf Proof:  }}
\def \eproof{\qed\vskip .3cm}

\def \A{{\mathbb A}}
\def \Aq{{\mathbb A}^q}
\def \C{{\mathbb C}}
\def \Z{{\mathbb Z}}
\def \E{{\mathbb E}}
\def \Eq{{\mathbb E}^q}
\def \R{{\mathbb R}}
\def \F{{\mathbb F}}

\def \B{{\mathbb B}}

\def \SUn{SU(n)}
\def \Un{U(n)}
\def \sun{{\mathfrak su}(n)}
\def \un{{\mathfrak u}(n)}
\def \sutwo{{\mathfrak su}(2)}

\def \Lh{\hat{L}}
\def \curlyg{{\mathfrak g}}

\def \sphinf{S^{2}_{\infty}}
\def \Phiinf{\Phi_{\infty}}
\def \Einf{E_\infty}
\def \Ainf{A_\infty}
\def \threeball{\overline{B}^3}

\def \muo{\mu_{0}}
\def \ko{k_{0}}
\def \xo{x_{0}}
\def \dxo{dx_{0}}

\def \per{\frac{2\pi}{\muo}}
\def \perflat{2\pi / \muo}
\def \Ieps{I_{\epsilon}}
\def \Iper{[0,\perflat]}

\def \rfour{{\mathbb R}^{4}}
\def \rthree{{\mathbb R}^{3}}

\def \Xo{X^{o}}

\def \dX{\partial X}
\def \So{S^{1}_{\perflat}}
\def \Sdual{S^{1}_{\muo}}
\def \cyl{\So\times\threeball}
\def \cylo{\So\times\rthree}

\def \Splus{S^{+}}
\def \Sminus{S^{-}}
\def \spinthree{S_{(3)}}
\def \DAxiplus{D_{\A,\xi}^+}
\def \DAximinus{D_{\A,\xi}^-}

\def \trace{{\textrm {tr\ }}}
\def \Spin{{\textrm{Spin}}}
\def \deg{{\textrm {deg~}}}

\def \ch{{\textrm{ch}}}
\def \chtwo{{\textrm{ch}_{2}}}
\def \End{{\textrm{End}}}
\def \dim{{\textrm{dim\ }}}
\def \ind{{\textrm{ind\ }}}
\def \supp{{\textrm{Supp}}}
\def \Map{\textrm{Map}}
\def \ker{{\textrm{ker\ }}}
\def \im{{\textrm{im\ }}}
\def \coker{{\textrm{coker\ }}}
\def \diag{{\textrm {diag}}}

\def \Auto{\textrm{Aut}_{0}~}
\def \SAuto{\textrm{SAut}_{0}~}

\def \rotNahm{\rho_{{\mathcal {N}}}}
\def \rotbdary{\rho_{\partial}}
\def \rotcal{\rho_{{\mathcal {C}}}}

\def \bdarydata{(\ko,\vec k,\muo,\vec\mu)}
\def \monbdarydata{(\vec k,\vec\mu)}

\def \curlyL{{\mathcal{L}}}

\newcommand{\spc}[2]{{\mathcal{#1}}{#2}}
\newcommand{\spcSD}[2]{{\mathcal{#1}}^{\ast}{#2}}
\newcommand{\mon}[2]{{\mathcal{#1}}_{\textrm{Mon}}{#2}}
\newcommand{\monSD}[2]{{\mathcal{#1}}^{\ast}_{\textrm{Mon}}{#2}}
\def \monopoles{{\mathcal{M}}on\monbdarydata}

\newcommand{\Lbar}[1]{\bar{L}^2_{#1}}
\newcommand{\Ldot}[1]{\dot{L}^2_{#1}}

\def \COKER{{\mathsf{Coker}}}

\def \Wo{\overset{\circ}{W}}
\newcommand{\Woone}[1]{{\Wo\raisebox{0.85ex}{\scriptsize{1}}\negmedspace}_{#1}}
\def \Zeros{{\mathcal{J}}}
\def \Nzer{N_{\textrm{zero}}}
\def \Srep{{\mathcal{S}}}
\def \modelop{\tilde{\Delta}}
\def \Fred{\mathcal{F}_{-n}}
\def \Fredi{\mathcal{F}_{-n}^i}

\def \Ph{\hat{P}}
\def \Xising{\xi_{\textrm{sing}}}
\def \IopN{I^\circ_{p,N}}
\def \Iop{I^\circ_{p}}
\def \PhiDO{$\Psi$DO}

\def \Lker{\textrm{$L^2$-ker~}}

\newcommand{\rest}{|_}

\begin{document}

\title{The Geometry of Calorons}
\author{Thomas M. W. Nye}
\submityear{2001} 
\maketitle

\begin{abstract}
Calorons, or periodic instantons, are anti-self-dual (ASD) connections on $S^1\times\rthree$, and form an intermediate case between instantons (ASD connections on $\rfour$) and monopoles (translation invariant instantons). 
Complete constructions of instantons and monopoles have been found: 
there is a complete construction of instantons from algebraic data, the ADHM construction due to Atiyah and others, 
while Nahm gave a construction of monopoles from solutions to a system of ODEs known as Nahm's equation. 
Both these constructions can be thought of as generalizations of a correspondence between ASD connections on the $4$-torus, and ASD connections over the dual $4$-torus, originally due to Mukai and Braam-van Baal. 
This correspondence, often called the `Nahm transform', is invertible and the inverse of the transform is the transform itself up to sign. 
Given an ASD connection on the $4$-torus it is defined in terms of the kernel of a family of Dirac operators parameterized by the dual torus. 
The aim of this thesis is to generalize the Nahm transform to the caloron case. 
In particular, our approach is via analysis of these families of Dirac operators rather than via twistor theory. 

We start by exploring topological aspects of calorons and boundary conditions. 
These are needed to ensure that the Dirac operators that define the Nahm transform are Fredholm. 
Our main innovation is to regard $\rthree$ as the interior of the closed $3$-ball $\threeball$, and to stipulate fixed behaviour on the boundary, rather than imposing asymptotic boundary conditions. 
The boundary conditions for calorons can be stated as follows: given a bundle on $S^1\times\threeball$ we fix some gauge $f$ on the boundary, and we require that in the gauge $f$, a $U(n)$ caloron must resemble the pull-back of a $U(n)$ monopole. 
There is a topological obstruction to extending $f$ to the interior of $S^1\times\threeball$, which we call the `instanton charge' of the caloron. 

The Nahm transform of a caloron consists of a solution to Nahm's equation on $S^1$, which we refer to as Nahm data. 
Many aspects of the $4$-torus transform generalize readily to the caloron case, and the construction of calorons from Nahm data is very similar to the construction of monopoles. 
The main difficulty in the construction lies with recovering the boundary conditions for the caloron and calculating its instanton charge. 
The caloron constructed from a set of Nahm data is defined using a family of Dirac operators $\Delta(x)$ parameterized by $x\in S^1\times\rthree$. 
Our approach is to deform $\Delta(x)$ to some model $\tilde{\Delta}(x)$ for which we can recover the boundary conditions and calculate the instanton charge. 
We then show that this deformation does not affect the behaviour on the boundary. 
Thus we prove that every set of Nahm data on $S^1$ gives rise to a caloron via the Nahm transform. 

Going the other way, from the caloron to the Nahm data, we encounter two main problems: first, we must calculate the rank of the Nahm data, which can jump at isolated points on $S^1$; and secondly, we must show that the Nahm data has certain prescribed singularities at these points. 
The transform is defined in terms of a family of Dirac operators parameterized by $S^1$. 
We show that the caloron boundary conditions ensure this family of Dirac operators is Fredholm away from the prescribed points on $S^1$. 
We also prove an index theorem for Dirac operators coupled to connections on $S^1\times\rthree$ that allows us to calculate the rank of the Nahm data. 
We obtain partial results concerning the behaviour of the Nahm data at singularities. 
These are based on Nakajima's method for recovering the singularities in Nahm data constructed from SU(2) monopoles. 
\end{abstract}

\dedication{To my parents.}

\begin{acknowledgements}
I would like to thank my supervisor, Michael Singer, for all his help with this thesis, and for being so generous with his time. 
In particular, our joint publication would not have been possible without his work. 
Others in the Department I would like to thank are Toby Bailey, my second supervisor; Elmer Rees, for helping with some homotopy; and Allan Sinclair, for reading an early draft of the thesis. 
My contemporaries Ann and Des have been a huge source of support, Des providing wise advice at crucial moments, and Ann just putting up with me day after day. 
They have played a large part in seeing this thesis finished. 
I must also thank the EPSRC for financial support. 

Outside the Department my thanks go to Ingrid for all our adventures together, and to all the other friends I have been out on the hill with. 
Thanks also go to Braemar Mountain Rescue Team for the helicopter ride. 
My friends Ross and Jo helped by taking me out for a few shandies once in a while. 
From further back I would like to thank Dan Scott, for his influence on my mathematical career, and Rob Rule, for years of friendship and good advice. 

Above all, however, I would like to thank my parents for their constant support and faith in me, throughout all the ups and downs. 
I cannot thank them enough. 
\end{acknowledgements}

\standarddeclaration

\tableofcontents

\chapter{Introduction}

Anti-self-dual (ASD) connections have been studied intensively by mathematicians over the last three decades, and have many remarkable properties and applications.
One of the original problems in the area, the construction of finite-action ASD connections on $\rfour$, was solved in 1978 by Atiyah and others \cite{adhm78,ati79}. 
They gave a complete construction of all such connections---the celebrated ADHM construction of instantons---by means of twistor theory and algebraic geometry. 
For the purposes of this introduction we will call any anti-self-dual connection on $\rfour$ an \emph{instanton}. 
Instantons are of great interest to physicists, since they are solutions to the Yang-Mills equation in a gauge field theory, and so represent minimum-energy configurations. 
They have had an important impact on quantum chromodynamics (QCD), with applications to symmetry breaking, tunnelling, and confinement. 

In fact the ADHM construction can be regarded as a special case of a correspondence we will refer to as the \emph{Nahm transform}. 
If $\Lambda\subset\rfour$ is a sub-group of translations,\label{pag:lattice} and $\Lambda^\ast\subset(\rfour)^\ast$ is the dual of $\Lambda$ (so that $\Lambda^\ast$ consists of elements taking integer values on $\Lambda$), then the Nahm transform is a correspondence between ASD connections on $\rfour / \Lambda$ and ASD connections on $(\rfour)^\ast / 2\pi\Lambda^\ast$. 
The details of the transform vary depending on the nature of $\Lambda$, and the existence and invertibility of the transform have been proved rigorously in several cases. 
For example, the case $\Lambda=\{ 0 \}$ corresponds to the ADHM construction, as presented by Donaldson and Kronheimer \cite{don90}.

Another area of interest to mathematicians and physicists alike is that of monopoles, which are solutions to the Bogomolny equation on $\rthree$. 
The Bogomolny equation is a translation reduction of the ASD equation, and so monopoles can be thought of as translation-invariant instantons, or equivalently, ASD connections on $\rfour / \Lambda$ where $\Lambda=\R$. 
In this case, the Nahm transform is a correspondence between monopoles and objects defined on the `dual torus' $(\rfour)^\ast / 2\pi\Lambda^\ast = \R$, which we will refer to as \emph{Nahm data}. 
The construction of monopoles from Nahm data and the inverse correspondence has been described by Hitchin \cite{hit83} (for $SU(2)$ monopoles) and Hurtubise--Murray \cite{hur89} (for arbitrary gauge group).

Periodic connections on $\rfour$ correspond to $\Lambda=\Z$, and form an intermediate case between instantons and monopoles. 
A periodic anti-self-dual connection over $\rfour$ is called a periodic instanton, or \emph{caloron}. 
The aim of this thesis is to study the geometry of calorons and the Nahm transform. 
Since $\Lambda=\Z$, the transform will be a correspondence between calorons lying on $\rfour / \Lambda=S^1\times\rthree$ and Nahm data on $(\rfour)^\ast / 2\pi\Lambda^\ast = S^1$. 
In some sense a caloron can be thought of as a hybrid between an instanton and a monopole. 
For example, a caloron has an `instanton charge' (an invariant that depends on the $4$-dimensional topology of the caloron) together with `monopole charges' (which characterize the $3$-dimensional topology). 
We will see this reflected in the Nahm transform for calorons, as it shares features with both the ADHM construction of instantons and the construction of monopoles. 
Originally introduced by Nahm in \cite{nah83}, the transform for calorons has been studied recently for calorons with unit instanton charge and zero monopole charges, in a series of papers \cite{kra98a,kra98b,kra00} and \cite{lee98}. 

Completeness of the ADHM construction of instantons and construction of monopoles was originally proved using twistor theory. 
In the case of monopoles, there is a correspondence between monopoles and certain algebraic curves in twistor space, called spectral curves, and this was used in~\cite{hit83} and~\cite{hur89} to go from monopoles to their Nahm data. 
The twistor picture for calorons was studied in~\cite{gar88}, and there is a similar correspondence between calorons and their spectral curves. 
While the twistor picture could be used to prove the existence of the Nahm transform for calorons, the approach in this thesis is via analysis of families of Dirac operators, regarding the caloron case as a generalization of the Nahm transform on the $4$-torus.  

With the scene set, we go on to introduce anti-self-duality and the Nahm transform more formally in Section~\ref{sec:ASDandNahm}. 
In Section~\ref{sec:review} we review existing work on calorons, before giving an overview of the aims and results of this thesis in Section~\ref{sec:overview}. 
From the outset, we draw the reader's attention to the Glossary of Notation on page~\pageref{pag:glo} in the hope it will make the thesis easier to read. 

\section{Anti-self-dual connections and the Nahm transform}\label{sec:ASDandNahm}

\subsection{The anti-self-duality equation on $\rfour$}\label{sec:asdrfour}

Let $x_0,x_1,x_2,x_3$\label{glo:rfourcoords} be the standard coordinates on $\rfour$, and equip $\rfour$ with the standard Euclidean inner product. 
Fix an orientation by decreeing that the ordered basis of $1$-forms $dx_0,dx_1,dx_2,dx_3$ be positive. 
The space of $p$-forms on $\rfour$ is denoted $\Lambda^p\rfour$, and a $p$-form $\alpha$ will be represented by its skew-symmetric covariant tensor $\alpha_{a_1,a_2,\ldots,a_p}$ of components, defined by
\begin{equation*}
\alpha=\frac{1}{p!}\sum_{a_1,\ldots,a_p}\alpha_{a_1,a_2,\ldots,a_p}dx_{a_1}\wedge\ldots\wedge dx_{a_p}. 
\end{equation*}
In general, the Hodge star\label{glo:HodgeStar} operator is defined on any $n$-dimensional Riemannian manifold $M$ with volume form $\eta$. 
It is the linear map\label{glo:ast} $\ast:\Lambda^p M \rightarrow \Lambda^{n-p} M$ defined pointwise by
\begin{equation*}
\langle\alpha,\beta\rangle \eta = \ast\alpha\wedge\beta
\end{equation*}
where $\alpha,\beta\in\Lambda^p M$ and $\langle\alpha,\beta\rangle$ is the inner product between $p$-forms defined by the metric. 
Integrating over the manifold $M$ gives an inner product on forms defined globally: 
\begin{equation}\label{eq:ipandstar}
\langle \alpha,\beta\rangle = \int_M\langle \alpha,\beta\rangle \eta = 
\int_M \ast\alpha\wedge\beta.
\end{equation}
On $\rfour$ the Hodge star becomes 
\begin{gather*}
  \ast : \Lambda^{2}\rfour \longrightarrow \Lambda^{2}\rfour \\
  \left( \ast \alpha \right)_{ab} = \frac{1}{2}\sum_{c,d} \epsilon_{abcd}\alpha_{cd}
\end{gather*}
for any $2$-form $\alpha$, where $\epsilon_{abcd}$ is the $4$-dimensional alternating tensor with $\epsilon_{0123}=1$. 
Since $\ast\ast=1$, $\ast$ has eigenspaces with eigenvalues $\pm 1$. 
A $2$-form $\alpha$ is self-dual\label{glo:SD} (SD) if $\ast \alpha=\alpha$ and anti-self-dual\label{glo:ASD} (ASD) if $\ast \alpha=-\alpha$. 
In terms of components, $\alpha$ is anti-self-dual if 
\begin{equation}\label{eq:ASDcmpts}
 \alpha_{01}+ \alpha_{23}=0, \quad \alpha_{02}+\alpha_{31}=0, 
  \quad \textrm{and}\quad \alpha_{03}+\alpha_{12}=0.
\end{equation}
Given a connection $\A$ on a bundle $\E$ on $\rfour$, $\A$ is anti-self-dual, or \emph{satisfies the anti-self-duality equation}, if its curvature $F_{\A}$ is anti-self-dual as an endomorphism-valued $2$-form \ie if
\begin{equation}\label{eq:anti-self-duality}
\ast F_\A = -F_\A.
\end{equation}
If we consider the action of a discrete sub-group $\Lambda\subset\rfour$ of translations on $\rfour$, so that $\rfour / \Lambda$ is $4$-dimensional, the anti-self-duality condition makes sense on the quotient manifold $\rfour / \Lambda$. 
The ASD $2$-forms are those that satisfy~$\eqref{eq:ASDcmpts}$ where $x_0,x_1,x_2,x_3$ are the coordinates on $\rfour / \Lambda$ corresponding to the standard coordinates on $\rfour$.
Note that throughout we will use the symbols $\E,\A$ \etc to refer to vector bundles and connections over $4$-manifolds and the corresponding symbols $E,A$ \etc to refer to vector bundles and connections over $3$-manifolds. 

An \emph{instanton}\label{glo:instanton} is a unitary ASD connection $\A$ on $\rfour$ whose action
\begin{equation*}
\| F_\A \|^2=\int_{\rfour}\trace \ast F_\A\wedge F_\A
\end{equation*}
is finite. 
Uhlenbeck \cite{uhl82} showed that any such connection extends smoothly to the compactification $S^4$ of $\rfour$, and that every ASD connection on $S^4$ arises in this way. 
Thus instantons really live on $S^4$ and are characterized by their second Chern class, which is often called the instanton number, or charge. 

\subsection{Translation reduction: the Bogomolny and Nahm equations}\label{sec:transredASD}

Next, we define the Bogomolny equation and Nahm's equation, and show how these can be thought of as translation reductions of the anti-self-duality equation. 
The Bogomolny equation applies to connections over $\rthree$, and we need to fix the following conventions. 
Throughout, we will consider $\rthree$ as a slice of $\rfour$ of the form $\xo =$ constant, oriented so that the ordered basis $dx_1,dx_2,dx_3$ is positive. 
The Hodge star operator is given in components by 
\begin{gather*}
  \ast : \Lambda^{2}\rthree \longrightarrow \Lambda^{1}\rthree \\
  ( \ast \alpha )_{a} = \frac{1}{2}\sum_{b,c} \epsilon_{abc}\alpha_{bc}
\end{gather*}
where $\epsilon_{abc}$ is the $3$-dimensional alternating tensor with $\epsilon_{123}=1$.
To distinguish between the star operators on $\rthree$ and $\rfour$ we will sometimes write them as $\ast_3$\label{glo:ast34} and $\ast_4$. 
 
Consider what happens to the anti-self-duality equation~$\eqref{eq:anti-self-duality}$ if we decree that a connection $\A$ is translation invariant.
Of course, this is equivalent to looking at connections on the `generalized torus' $\rfour / \Lambda$ with $\Lambda=\R$, that satisfy a symmetry reduction of the ASD condition.  
Let $\A_0,\A_1,\A_2,\A_3$ be the matrices representing $\A$ in some global trivialisation of $\E$, and suppose that the matrices are independent of $\xo$. 
Then, using~$\eqref{eq:ASDcmpts}$ and  $(F_\A)_{ij}=\partial_i\A_j-\partial_j\A_i+[\A_i,\A_j]$, we see that $F_\A$ is ASD iff
\begin{equation}\label{eq:reducetoBog} 
\partial_2\A_3-\partial_3\A_2+[\A_2,\A_3]=\partial_1\A_0+[\A_1,\A_0] 
\end{equation} 
holds, together with the two equations obtained by cyclic permutations in $\{ 1,2,3 \}$. 
Let $E$ be a unitary bundle over $\rthree$ with some fixed trivialisation. 
From this point on we will restrict to unitary bundles for the rest of the thesis. 
Let $A$ be the connection on $E$ with components $\A_1,\A_2,\A_3$ in this trivialisation and let $\Phi$ be the endomorphism of $E$ represented by $\A_0$. 
Then~$\eqref{eq:reducetoBog}$ and its cyclic permutations can be written as
\begin{equation}\label{eq:Bogomolny}
 \ast F_{A}=\nabla_A \Phi.
\end{equation}
This is the Bogomolny equation\label{glo:Bogomolny}. 
Note that conversely, 
any solution to the Bogomolny equation can be used to construct a translation invariant ASD connection on $\rfour$ via the same argument. 
The endomorphism $\Phi$ is called a \emph{Higgs field}\label{glo:Higgs}. 

A \emph{monopole}\label{glo:monopole} is a unitary solution $(A,\Phi)$\label{glo:monconfig} to the Bogomolny equation whose energy
\begin{equation*}
\| F_A \|^2 +\| \nabla_A \Phi \|^2 =
\int_{\rthree}\trace\{ \ast F_A\wedge F_A +\ast\nabla_A\Phi\wedge\nabla_A\Phi\}
\end{equation*}
is finite. 
The finite energy condition can be re-expressed in terms of the asymptotic behaviour of $A$ and $\Phi$. 
For $SU(2)$ monopoles the asymptotic condition is that $\| \Phi \|\rightarrow\mu$ as $r\rightarrow\infty$ where $r$ is the standard polar coordinate on $\rthree$. (In fact there are additional conditions that will not concern us till later.) 
On the $2$-sphere at infinity $\Phi$ therefore has eigenvalues $\pm i\mu$, defining two eigenbundles. 
We let $\pm k$ be the first Chern classes of these eigenbundles, and say that the monopole $(A,\Phi)$ has charge $k$. 
Note that instantons cannot be translation invariant, since this would contradict the finite action condition. 

We can perform further translation reduction: consider an ASD connection $\A$ on $\rfour$ whose components $\A_0,\A_1,\A_2,\A_3$ are independent of $x_1,x_2,x_3$ in some trivialisation. 
The anti-self-duality equation becomes
\begin{equation}\label{eq:reducetoNahm}
\partial_0\A_1 +[\A_0,\A_1] + 
[\A_2,\A_3]=0
\end{equation}
plus cyclic permutations in $\{ 1,2,3 \}$. 
In a similar way to the reduction to the monopole case above, $\A$ determines a connection $\nabla$ and three endomorphisms $T_1,T_2,T_3$ on a vector bundle over $\R$; from~$\eqref{eq:reducetoNahm}$ we see that these satisfy
\begin{equation}\label{eq:Nahm}
\nabla T_i + \frac{1}{2}\sum_{j,k}\epsilon_{ijk}[T_j,T_k]=0.
\end{equation}
This is Nahm's equation\label{glo:Nahmequn}. 
Following the sketch of the Nahm transform at the start of the Chapter, we expect solutions to Nahm's equation to correspond to solutions of the Bogomolny equation under the transform.

\subsection{Gauge transformations}

Monopoles and instantons are studied modulo the action of the group of bundle automorphisms preserving the base manifold. 
These automorphisms are referred to as \emph{gauge transformations}. 
In particular, the Nahm transform is defined modulo this gauge action. 
If local trivialisations are fixed, the action of a gauge transformation is the same as a change of trivialisation. 
We therefore sometimes refer to fixing a local trivialisation as `fixing a gauge'.

For an instanton $\A$ on a $U(n)$ bundle $\E\rightarrow\rfour$, a gauge transformation $g$ acts on sections of $\E$ by $s\mapsto gs$ and on $\A$ by $\nabla_\A\mapsto g\nabla g^{-1}$. 
In a local trivialisation over some open region $U\subset\rfour$, $g$ becomes a map $g:U\rightarrow U(n)$ and $\nabla_\A$ is represented by matrix-valued functions $\A_0,\A_1,\A_2,\A_3:U\rightarrow\un$. 
The action of $g$ is given by 
\begin{equation*}
\A_a\mapsto g \A_a g^{-1}-dgg^{-1}.
\end{equation*}
Similarly, for a monopole $(A,\Phi)$ on a $U(n)$ bundle $E\rightarrow\rthree$, a gauge transformation $g$ acts on sections of $\E$ by $s\mapsto gs$, on $A$ by $\nabla_A\mapsto g\nabla_A g^{-1}$, and on $\Phi$ by $\Phi\mapsto g\Phi g^{-1}$. 
In a local trivialisation over some open region $U\subset\rthree$, $g$ becomes a map $g:U\rightarrow U(n)$, $\Phi$ becomes a map $\Phi:U\rightarrow\un$, and $\nabla_A$ is represented by $A_1,A_2,A_3:U\rightarrow\un$.
The action of $g$ on $\Phi$ is given by 
\begin{equation*}
\Phi\mapsto g \Phi g^{-1}
\end{equation*}
and on $A$ by
\begin{equation*}
A_a\mapsto g A_a g^{-1}-dgg^{-1}.
\end{equation*}
Gauge transformations for bundles over the $4$-torus and over $S^1\times\rthree$ are defined in an analogous way.

\subsection{Spin structures and Dirac operators}

The Nahm transform is defined using Dirac operators on $\rfour / \Lambda$.
In this Section we fix our notation and conventions for such operators. 
We start, however, by recalling the definition of a spin structure on an arbitrary manifold. 
References \cite{don90}, \cite{roe88} and \cite{gro83} provide more background material, and proofs of all the statements. 

Suppose $M$ is an oriented smooth $n$-dimensional Riemannian manifold. 
Let $F$ be the principal $SO(n)$ bundle of oriented orthogonal frames for the tangent bundle.  
A spin structure for $M$ is a pair $(\tilde{F},\sigma)$ where $\tilde{F}$ is a principal $\Spin(n)$-bundle over $M$, and $\sigma:\tilde{F}\rightarrow F$ is a two-to-one covering such that the restriction to each fibre is the double covering $\Spin(n)\rightarrow SO(n)$. 
%\begin{equation*}
%\begin{CD}
%Spin(n) @>>> \tilde{F} @>>> M \\
%@VVV @VV{\sigma}V @VV{\textrm{id}}V \\
%SO(n)@>>>F@>>>M
%\end{CD}
%\end{equation*}
%is a commutative diagram. 
The obstruction to the existence of a spin structure is the second Stiefel-Whitney class, which is contained in $H^2(M,\Z_2)$ (see \cite[Section $1.1.4$]{don90}), and the number of spin structures is counted by $H^1(M,\Z_2)$. 
If a spin structure exists then $M$ is called a spin manifold. 
The spin bundle\label{glo:spinbdl} $S\rightarrow M$ is defined to be the vector bundle associated to the $\Spin(n)$ principal bundle $\tilde{F}$ via the spin representation. 
It comes equipped with a representation $\gamma$ of the Clifford algebra of the tangent space $T_x M$ on the fibre $S_x$ for each $x\in M$, and this is used to define the Dirac operator. 

When $M=\rfour$, it is easy to check that there is a unique spin structure. 
Since $\Spin(4)=SU(2)\times SU(2)$ it follows that the spin bundle $S$ decomposes into two $SU(2)$ bundles $S^+,S^-$\label{glo:spin4} with $S=S^+\oplus S^-$ (see \cite[Section $3.1.1$]{don90}). 
The spin representation $\gamma$ respects this decomposition in the following way. 
Given the orthonormal frame $(\partial_{x_0},\partial_{x_1},\partial_{x_2},\partial_{x_3})$ the endomorphisms $\gamma(\partial_{x_a}):S\rightarrow S$ decompose as 
\begin{equation*}
\gamma(\partial_{x_a})=
\begin{pmatrix}
0 & \gamma_a^\ast \\ \gamma_a & 0
\end{pmatrix} : \Splus\oplus\Sminus\rightarrow\Splus\oplus\Sminus,
\end{equation*}
for $a=0,1,2,3$. 
Since $\gamma$ is a representation of the Clifford algebra these endomorphisms satisfy
\begin{equation}\label{eq:cliffordalg}
\gamma^\ast_a \gamma_b +\gamma^\ast_b\gamma_a = 2\delta_{ab}.
\end{equation}
In particular, we can choose bases for $\Splus$ and $\Sminus$ in which the endomorphisms are given by matrices\label{glo:gammaj}
\begin{equation}\label{eq:defgammas}
\gamma_0 =
\begin{pmatrix}
1 & 0 \\ 0 & 1
\end{pmatrix},\quad
\gamma_1 =
\begin{pmatrix}
i & 0 \\ 0 & -i
\end{pmatrix},\quad
\gamma_2 =
\begin{pmatrix}
0 & -1 \\ 1 & 0
\end{pmatrix},\quad
\gamma_3 =
\begin{pmatrix}
0 & i \\ i & 0
\end{pmatrix}.
\end{equation}
Note that these satisfy 
%\begin{multline}\label{eq:gammarelns}
%\gamma_1^\ast=-\gamma_1, ~\gamma_1\gamma_2=-\gamma_3, \quad
%\textrm{and the corresponding relations}\\
%\textrm{from cyclic permutations of $\{ 1,2,3 \}$}.
%\end{multline}  
\begin{equation}\label{eq:gammarelns}
\gamma_i^\ast=-\gamma_i, ~\textrm{and\ }\gamma_i\gamma_j=-\gamma_k,
\end{equation}
where $\{i,j,k\}$ is a cyclic permutation of $\{1,2,3\}$. 
Let $\nabla$ be the covariant differential operator on some bundle $\E\rightarrow\rfour$ associated to some unitary connection $\A$, and let $\nabla_0,\nabla_1,\nabla_2,\nabla_3$ be its components in the frame $(\partial_{x_0},\partial_{x_1},\partial_{x_2},\partial_{x_3})$. 
The Dirac operators\label{glo:Diracop}\label{glo:Gamma}\label{glo:DA}
\begin{equation}\label{eq:diracopspaces}
D_\A^+ : \Gamma(\Splus\otimes\E)\rightarrow \Gamma(\Sminus\otimes\E)
\quad\textrm{and}\quad
D_\A^- : \Gamma(\Sminus\otimes\E)\rightarrow \Gamma(\Splus\otimes\E)
\end{equation}
are defined by 
\begin{equation}\label{eq:defDAplus}
D_\A^+ s =\sum_{a=0}^{3}\gamma_a\nabla_a s 
\end{equation}
and
\begin{equation}\label{eq:defDAminus}
D_\A^- s =-\sum_{a=0}^{3}\gamma_a^\ast\nabla_a s,
\end{equation}
and are called the Dirac operators \emph{coupled to the bundle $\E$ via the connection $\A$}. 

Next consider the two cases $M=\rfour / \Lambda$ where $\Lambda=\Z$ or $\Z^4$ (\ie $M=S^1\times\rthree$ or $M=T^4$). 
The spin bundle and spin representation $\gamma$ are invariant under the action of $\Lambda$, and so descend to the quotient. 
Thus in both cases $M$ has a spin bundle $S=\Splus\oplus\Sminus$, together with Dirac operators defined by equations~$\eqref{eq:diracopspaces}$--$\eqref{eq:defDAminus}$. 

The following Weitzenb\"ock\label{glo:weitz} formula holds on $M=\rfour$, $S^1\times\rthree$, and $T^4$:

\begin{lemma}\label{lem:weitzenbock}
Given any section $s$ of $\Splus\otimes\E$ we have
\begin{equation*}
D_\A^- D_\A^+ s = -\sum_a \nabla_a \nabla_a s -\sum_{a<b}\gamma_a^\ast\gamma_b 
(F^+_{\A})_{ab}
\end{equation*}
where $F^+_\A$ is the self-dual part of the curvature $F_\A$ of $\A$. 
\end{lemma}

\proof
Using the relations~$\eqref{eq:cliffordalg}$ and definitions~$\eqref{eq:defDAplus}$--$\eqref{eq:defDAminus}$ we have 
\begin{align*}
D_\A^- D_\A^+ 
& = -\sum_{a,b=0}^3\gamma^\ast_a\gamma_b\nabla_a\nabla_b \\
&=-\big( \sum_{a} \gamma_a^\ast\gamma_a\nabla_a\nabla_a \big)
-\big( \sum_{a\neq b} \gamma_a^\ast\gamma_b\nabla_a\nabla_b \big) \\
&= -\sum_{a} \nabla_a\nabla_a 
-\sum_{a<b}\gamma_a^\ast\gamma_b(\nabla_a\nabla_b-\nabla_b\nabla_a).
\end{align*}
The last term can be rewritten as
\begin{equation}\label{eq:SDcurvfromdirac}
\sum_{a<b}\gamma^\ast_a\gamma_b(F_\A)_{ab}.
\end{equation}
A basis of anti-self-dual $2$-forms is given by 
\begin{equation*}
dx_0\wedge dx_1 - dx_2\wedge dx_3,\ 
dx_0\wedge dx_2 - dx_3\wedge dx_1,\ 
dx_0\wedge dx_3 - dx_1\wedge dx_2. 
\end{equation*}
Substituting these forms into~$\eqref{eq:SDcurvfromdirac}$ and using~$\eqref{eq:gammarelns}$, we see that the final term vanishes on the  anti-self-dual part of $F_\A$, establishing the lemma.
\eproof

Finally, consider $M=\rthree$. 
In this case the spin bundle is a $\Spin(3)=SU(2)$ bundle which we denote $\spinthree$\label{glo:spinthree}. 
We can find a trivialisation of $\spinthree$ in which the spin representation $\gamma(\partial_j)$ for $j=1,2,3$ is given by $\gamma_j$, where the $\gamma_j$ are defined by~$\eqref{eq:defgammas}$. 
Given a unitary bundle $E$ over $\rthree$ and a unitary connection $A$ on $E$, the Dirac operator coupled to $A$ is defined by 
\begin{gather*}
D_A: \Gamma(\spinthree\otimes E)\rightarrow \Gamma(\spinthree\otimes E) \\
D_A = \sum_{j=1}^{3}\gamma_j \nabla_j.
\end{gather*}
Note that we will often want to use identifications
\begin{equation}\label{eq:idspinthreefour}
S^+\cong \pi^\ast \spinthree,\quad S^-\cong \pi^\ast \spinthree,
\end{equation}
where $\pi$ is the projection $\pi:\rfour\rightarrow\rthree$ or $\pi:S^1\times\rthree\rightarrow\rthree$. 

\subsection{The Nahm transform on the $4$-torus}\label{sec:nahmtorus}

As described on page~\pageref{pag:lattice}, the Nahm transform is a correspondence between ASD connections on $\rfour / \Lambda$ and ASD connections on $(\rfour)^\ast / 2\pi\Lambda^\ast$, where $\Lambda\subset\rfour$ is a group of translations.  
The Nahm transform on the $4$-torus (when $\Lambda=\Z^4$) is in some ways the most natural version, and the other cases of the transform can be regarded as generalizations of the $4$-torus version. 
In this Section we present the Nahm transform on $T^4$ following Braam-van Baal \cite{bra89} and Donaldson-Kronheimer \cite[Section $3.2$]{don90}; later, we will use this as the framework into which the other cases fit, in particular the transform for calorons. 

Let $\Lambda$ be a maximal lattice in $\rfour$ and let \label{glo:TTast}$T=\rfour / \Lambda$. 
As before, the dual lattice $\Lambda^\ast$ consists of elements of $(\rfour)^\ast$ taking integer values on $\Lambda$, and we define the dual torus to be $T^\ast = (\rfour)^\ast / 2\pi\Lambda^\ast$. 
We equip $T^\ast$ with the flat Riemannian metric induced from $(\rfour)^\ast$, and denote the spin bundles $\hat{S}^\pm$\label{glo:spindual}.  
Note that if $\Lambda$ is generated by $\{\mu_0 e_0,\ldots, \mu_3 e_3 \}$ where $\mu_a\in\R$ and $e_0,\ldots,e_3$ is the standard basis of $\rfour$ then $T^\ast$ has periods $2\pi / \mu_a$ for $a=0,1,2,3$.  

The dual torus\label{glo:dualtorus} $T^\ast$ parameterizes flat $U(1)$ connections on $T$ in the following way. 
Any \label{glo:xi}$\xi\in(\rfour)^\ast$ can be regarded as a $1$-form with constant coefficients on $T$ via $\xi\mapsto\sum\xi_a dx_a$. 
The connection $d-i\xi$ on the trivial line bundle $\C\times T$ is flat, and we denote the line bundle with this connection by $L_\xi$. 
Two points $\xi_1,\xi_2\in(\rfour)^\ast$ determine gauge equivalent connections iff there is a well-defined gauge transformation $g$ satisfying 
\begin{align}
i(\xi_1-\xi_2) &= dgg^{-1}\notag \\
\Leftrightarrow\qquad\quad  g(x)&=\exp i(\xi_1-\xi_2)\cdot x \label{eq:torusGT}
\end{align}
where $x$ is the coordinate on $\rfour$. 
The map $g$ is a well-defined gauge transformation $g:T\rightarrow U(1)$ iff $\xi_1-\xi_2\in 2\pi\Lambda^\ast$. 
Thus $L_{\xi_1}$ and $L_{\xi_2}$ are gauge equivalent iff $\xi_1-\xi_2\in 2\pi\Lambda^\ast$, and $T^\ast$ parameterizes gauge equivalence classes of flat $U(1)$ connections on $T$. 

Next, fix a unitary vector bundle $\E$ over $T$ with a unitary ASD connection $\A$. 
We restrict to connections that are `irreducible' in the following sense (the definition is taken directly from \cite{don90}):

\begin{definition}
The connection $\A$ is WFF \label{glo:WFF}(without flat factors) if there is no splitting $\E=\E' \oplus L_\xi$ compatible with $\A$ for any flat line bundle $L_\xi$. 
\end{definition}

For each $\xi\in(\rfour)^\ast$ we can consider the bundle $\E_\xi = \E\otimes L_\xi$ equipped with the induced connection $\A_\xi$. 
Using some fixed trivialisation of $\E$, $\A_\xi$ is represented by $\A\otimes 1 - 1\otimes i\xi$ where (by abuse of notation) $\A$ is a matrix of $1$-forms. 
It is easy to check that $\A_\xi$ is ASD for all $\xi$. 
We can write down the Dirac operators on $T$ coupled to $\E_\xi$ via $\A_\xi$, following definitions~$\eqref{eq:diracopspaces}$--$\eqref{eq:defDAminus}$:\label{glo:Dxi} 
\begin{equation*}
D_\xi^\pm : \Gamma(S^\pm\otimes\E\otimes L_\xi)\rightarrow \Gamma(S^\mp\otimes\E\otimes L_\xi)
\end{equation*}
where 
\begin{equation}\label{eq:defDxi}
D_\xi^+  =D_\A^+ -i\gamma(\xi),\quad D_\xi^-  =D_\A^- +i\gamma^\ast(\xi),
\end{equation}
and $\gamma$ is the spin representation for $\rfour$. 
Since $\A_\xi$ is ASD, applying Lemma~\ref{lem:weitzenbock} we have
\begin{equation*}
D^-_\xi D^+_\xi=\nabla^\ast_{\A_\xi}\nabla_{\A_\xi}.
\end{equation*}
Thus $s\in \ker D^+_\xi$ iff $\| \nabla_{\A_\xi}s \|=0$, in which case $s$ is a covariant constant section of $\E\otimes L_\xi$, and, if non-trivial, yields a splitting $\E=\E'\oplus L_\xi^\ast$. 
Since we are assuming that $\A$ is WFF it follows that $D_\xi^+$ is injective. 

Via standard results on elliptic operators, $D_\xi^+$ is Fredholm for all $\xi\in T^\ast$ and has constant $L^2$-index. 
Since $D_\xi^+$ is injective, using the Fredholm alternative it follows that $\dim\ker D^-_\xi = - \textrm{index}~D_\xi^+$ and this is independent of $\xi$. 
Moreover, the fibres $\hat{\E}_\xi = \coker D_\xi^+=\ker D_\xi^-$ define a vector bundle $\hat{\E}$ over $(\rfour)^\ast$ which inherits an hermitian metric from the $L^2$ hermitian metric on $\Gamma(S^-\otimes\E)$. 
Let $\hat{\F}$ be the bundle over $(\rfour)^*$ whose fibre at $\xi$ consists of $L^2$ sections of $S^-\otimes\E\otimes L_\xi$ so that $\hat{\E}$ is a sub-bundle of $\hat{\F}$, and let $\hat{P}_\xi$\label{glo:Phat} be the $L^2$ orthogonal projection onto $\ker D_\xi^-$. 
Let $\hat{P}$ be the map on $\hat{\F}$ given fibrewise by $\hat{P}_\xi$. 
Using the standard covariant derivative $d$ on the trivial bundle $\hat{\F}$, we can define a unitary connection $\hat{\A}$ on $\hat{\E}$ by
\begin{equation}\label{eq:projconnection}
\nabla_{\hat{\A}} = \hat{P}\cdot d.
\end{equation}
Finally, note that 
\begin{equation*}
D^-_{\xi+\eta}=(\exp i\eta\cdot x)D_\xi^-(\exp -i\eta\cdot x)
\end{equation*}
for all $\eta\in 2\pi\Lambda^\ast$ so that $(\exp i\eta\cdot x) : T\rightarrow U(1)$ is a gauge transformation identifying $\ker D^-_{\xi}$ with $\ker D^-_{\xi+\eta}$. 
Hence $2\pi\Lambda^\ast$ acts on $\hat{\E}\rightarrow(\rfour)^\ast$, and the quotient is a bundle over $T^\ast$, which we also denote $\hat{\E}$. 
The connection $\hat{\A}$ respects this action and descends to a connection over $T^\ast$ in the same way. 

\begin{definition}
If $\A$ is a unitary WFF ASD connection on a unitary bundle $\E\rightarrow T$ then $(\hat{\E},\hat{\A})$\label{glo:nahmtrans} is called the \emph{Nahm transform} of $(\E,\A)$, where $\hat{\E}\rightarrow T^\ast$ is given fibrewise by $\hat{\E}_\xi=\coker D^+_\xi$ and $\hat{\A}$ is defined by~$\eqref{eq:projconnection}$. 
\end{definition}

The transform is non-trivial, in that, when $c_1(\E)=0$, the rank and Chern classes of $\hat{\E}$ are given by:
\begin{align*}
\textrm{rank}(\hat{\E}) &= c_2(\E) \\
c_1(\hat{\E}) &= 0 \\
c_2(\hat{\E}) &=\textrm{rank}(\E).
\end{align*}
(When $c_1(\E)\neq 0$ there is a very similar formula.) 
The relations are a direct result of the index theorem for families due to Atiyah-Singer, and a proof is given in \cite[Section $3.2.2$]{don90}. 

The key point is the following:

\begin{proposition}\label{prop:NahmgivesASD}
The Nahm transform $(\hat{\E},\hat{\A})$ of $(\E,\A)$ is ASD.
\end{proposition}

\proof
The curvature $F_{\hat{\A}}$ of $\hat{\A}$ is given by
\begin{equation*}
F_{\hat{\A}} = \hat{P}d\hat{P}d\hat{P}\hat{P},
\end{equation*}
where $\hat{P}$ is the $L^2$-projection onto $\ker D_\xi^-$ at each point $\xi\in T^\ast$. 
Now 
\begin{equation*}
\hat{P}_\xi = 1-D_\xi^+ G_\xi D_\xi^-
\end{equation*}
where 
\begin{equation*}
G_\xi = ( D_\xi^- D_\xi^+)^{-1}.
\end{equation*}
Substituting this into the expression for $F_{\hat{\A}}$ we obtain
\begin{equation*}
F_{\hat{\A}} = \hat{P} (dD_\xi^+) G (dD_\xi^-) \hat{P}.
\end{equation*}
From~$\eqref{eq:defDxi}$, we have
\begin{equation*}
dD^+_\xi = -i\sum_a \gamma_a d\xi_a\quad\textrm{and\ }
dD^-_\xi = i\sum_b \gamma_b^\ast d\xi_b.
\end{equation*}
Since
\begin{equation*}
(\sum_a \gamma_a d\xi_a)\wedge (\sum_b \gamma_b^\ast d\xi_b)
\end{equation*}
is ASD (which can be checked by substituting in the matrices $\gamma_a$ from~$\eqref{eq:defgammas}$ and comparing with the anti-self-duality equation), it follows that $\hat{\A}$ is ASD provided $G_\xi$ commutes with Clifford multiplication for all $\xi$. 
But Lemma~\ref{lem:weitzenbock} shows that $G_\xi^{-1}=D_\xi^-D_\xi^+$ commutes with Clifford multiplication, so the Proposition follows. 
\eproof

Another remarkable property of the Nahm transform is that---like the Fourier transform---it is invertible, and the inverse is the transform itself (up to factors of $2\pi$ and sign). 
It turns out that given an ASD WFF connection $(\E,\A)$, not only is $\hat{\A}$ ASD, but it is also WFF, so the transform can be applied again to $(\hat{\E},\hat{\A})$. 
The inverse transform is defined in the following way.

Just as $T^\ast$ parameterizes the flat line bundles on $T$, the torus $T$ parameterizes flat line bundles over $T^\ast$ via the identification $x\mapsto d - i\sum x_a d\xi_a$.  
We write $\hat{L}_x$ to denote the line bundle and flat connection corresponding to $x\in T$, and, given a unitary bundle $\F$ and unitary connection $\B$ over $T^\ast$, let $D^\pm_x$ be the Dirac operators coupled to $\F\otimes\hat{L}_x$ via $\B$ twisted by $\hat{L}_x$:\label{glo:Dx} 
\begin{equation*}
D^\pm_x: \Gamma(\hat{S}^\pm\otimes\F\otimes\hat{L}_x)\rightarrow \Gamma(\hat{S}^\mp\otimes\F\otimes\hat{L}_x).
\end{equation*}
If $\B$ is WFF and ASD then the inverse Nahm transform, $(\check{\F},\check{\B})$\label{glo:invnahmtrans}, is defined entirely analogously to the original transform, so that $\check{\F}$ has fibre $\ker D_x^-$ for each $x\in T$. 

\begin{theorem}\label{thm:inversion}
If $\A$ is a WFF ASD unitary connection on $\E\rightarrow T$ then $\hat{\A}$ is WFF. 
Hence $(\check{\raisebox{1.8ex}{~}}\hat{\E},\check{\raisebox{1.8ex}{~}}\hat{\A})$ is well defined and there is a natural isomorphism $\omega:\check{\raisebox{1.8ex}{~}}\hat{\E}\rightarrow\E$ such that $\omega^\ast(\A)=\check{\raisebox{1.8ex}{~}}\hat{\A}$. 
\end{theorem}

The analogy with the Fourier transform\label{glo:FT} for functions on $\R^n$ is obvious---indeed Donaldson-Kronheimer call the Nahm transform the `Fourier transform for ASD connections'. 

There are two approaches to proving the Theorem, as described by Braam-van Baal \cite{bra89} and Donaldson-Kronheimer \cite{don90} respectively. 
The first approaches uses relations between harmonic spinors on $T$ and harmonic spinors on $T^\ast$. 
Given an ASD connection ${\A}$ and its transform $\hat{\A}$, there is an elegant relation between the Greens function $\hat{G}_x$ of $D^-_x D^+_x$ on $T^\ast$ and solutions $\psi_\xi(x)$ to $D^-_\xi$ on $T$. 
The Greens function $\hat{G}_x$ is given very explicitly, and it follows quite readily that $\hat{\A}$ is WFF. 
There is also a formula expressing the solutions $\hat{\psi}_x(\xi)$ to $D^-_x$ on $T^\ast$ in terms of the solutions $\psi_\xi(x)$ to $D_\xi^-$. 
Using these two relations one constructs the desired isometry $\omega:\check{\raisebox{1.8ex}{~}}\hat{\E}\rightarrow\E$. 
The second approach to the Theorem is to convert the problem into an equivalent one in holomorphic geometry that can be solved using $\bar{\partial}$-cohomology and spectral sequences. 
ASD connections on $T$ are characterized by the following: a unitary connection $\A$ on a unitary bundle $\E$ is ASD if and only if it defines a holomorphic structure on $\E$ for each complex structure on $T$. 
There is also an identification between the spaces of forms $\Omega^{0,0}\oplus\Omega^{0,2}$, and $\Omega^{0,1}$ over $T$ with the spin bundles $\Splus,\Sminus$. 
This characterization allows one to move between ASD connections on unitary bundles and $\bar{\partial}$-operators on holomorphic bundles. 
The holomorphic version of the transform was first given by Mukai \cite{muk81}\label{pag:mukai}; Donaldson-Kronheimer also give a proof of the holomorphic version together with the details of how to move between ASD connections and holomorphic bundles. 
\vspace{1.5ex}
\newline
\noindent
\textbf{Remarks}
\vspace{1.0ex}
\newline
%\begin{remarks}
%\item 
($1$) In algebraic geometry the transform of Mukai between holomorphic bundles over complex tori has been generalized to give the `Fourier-Mukai' transform\label{glo:FMT}---a transform between bundles over algebraic varieties (\eg elliptic surfaces, K$3$ surfaces). 
This has been very successful in the study of moduli spaces of bundles over these surfaces. 
\vspace{1.0ex}
\newline
%\item 
($2$) The Nahm transform is a hyperK\"ahler\label{glo:HK} isometry between the moduli spaces ${\mathcal{M}}(\E)$ and ${\mathcal{M}}(\hat{\E})$ of WFF ASD connections on $\E$ and $\hat{\E}$. 
The moduli spaces are smooth manifolds away from the connections with flat factors and are equipped with natural metrics, given by the $L^2$ metrics on $1$-forms over $T$ and $T^\ast$. 
The ASD equation implies that these metrics are hyperK\"ahler. 
Braam-van Baal \cite{bra89} prove the Nahm transform on the $4$-torus is a hyperK\"ahler isometry. 
It is conjectured that this holds for the transform for other cases of $\Lambda$, and this has been proved in certain cases. 
%\end{remarks}
\vspace{1.5ex}
\newline
This concludes our description of the Nahm transform on $T^4$. 
Next we consider the transform when $\Lambda$ is a general group of translations, reviewing the cases that have been studied in the literature. 

\subsection{The Nahm transform on the generalized torus}\label{sec:reviewNahm}

We want to consider the `generalized torus' $T=\rfour / \Lambda$ where $\Lambda=\lambda_0\times\lambda_1\times\lambda_2\times\lambda_3$ and $\lambda_a=\{0\}$, $\Z$, or $\R$ for each $a=0,\ldots,3$. 
It can be thought of as the limit of the $4$-torus with generators $\{ \mu_0 e_0,\ldots,\mu_3 e_3 \}\subset\rfour$ where some of the $\mu_a$ tend to zero or infinity. 
Recall that the dual torus has periods $2\pi / \mu_a$, so `shrinking' $\mu_a$ to zero (\ie taking $\lambda_a=\R$) corresponds to `stretching' a period on the dual torus (\ie taking $\lambda_a^\ast =\{0\}$ where $\Lambda^\ast=\lambda_0^\ast\times\ldots\times\lambda^\ast_3$) and vice versa. 
Given some generalized torus, one can attempt to carry across all the theory in Section~\ref{sec:nahmtorus} to give a version of the Nahm transform in this new setting. 
Many aspects carry across readily, but two main problems are encountered:
\begin{enumerate}
\item In all cases other than the $4$-torus we can assume $T$ is non-compact (since if $T$ is compact then $T^\ast$ must be non-compact and we can swap the two around). 
It becomes necessary to impose boundary conditions at infinity on the connection $\A$ to ensure that $D_\xi^+$ is Fredholm---and even with these conditions $D_\xi^+$ may fail to be Fredholm for certain values of $\xi$. 
The Nahm data will contain singularities at the points where $D_\xi^+$ is not Fredholm. 
In fact $\hat{\E}$ will not in general be a single vector bundle, but may change rank at these singular points. 
This introduces a new element into the problem: one must give boundary conditions for $\A$ and $\hat{\A}$ at infinity and singular points, and then show that these can be recovered from each direction of the transform. 
\item A related problem is that the index theorem for families relating the topology of $\E$ and $\hat{\E}$ does not hold on non-compact manifolds or for singular connections. 
Furthermore, while index theorems on non-compact manifolds do exist in the literature, they depend heavily on the precise nature of the geometry at infinity. 
It may therefore be necessary to prove a new index theorem depending on $\Lambda$ and the nature of the boundary conditions being imposed, for each different case of the transform.  
\end{enumerate}

Various cases of the transform on the generalized torus exist in the literature, and we review these next. 
We describe the constructions of instantons and monopoles in some detail, as these cases are directly relevant to the caloron case, but delay commenting on the literature until Section~\ref{sec:litcom}. 
We let $T^n$\label{glo:Tn} denote the $n$-torus $S^1\times\cdots\times S^1$.
\vspace{1.0ex}
\newline
\noindent
\textbf{The case $T=\rfour$, $T^\ast=\{0\}$: the ADHM construction of instantons.} 
A description of the ADHM construction\label{glo:ADHM} as a generalization of the Nahm transform on the $4$-torus is given in \cite[Chapter $3$]{don90}. 
We give a brief sketch of the construction, comparing with what you might expect na\"\i vely from the $4$-torus transform. 

Let $(\E,\A)$ be an instanton with charge $k$ (recall the definition in Section~\ref{sec:asdrfour}). 
There is a unique flat line bundle over $\rfour$ and the dual torus $T^\ast$ is a single point, so to perform the Nahm transform we consider only the Dirac operator $D_\A^+$ rather than the family of operators $D_\xi^+$. 
The boundary condition on $(\E,\A)$ (\ie the assumption that the instanton extends to the compactification $S^4$) allows the Atiyah-Singer index theorem to be applied. This shows that $D_\A^+$ is Fredholm with index $-k$, so $\hat{\E}$ is just the $k$-dimensional vector space $\coker D_\A^+$. 
In analogy with the transform for $T^4$, we expect $\hat{\A}$ to be an ASD connection over $\rfour$ invariant under any translation. 
This would be represented by skew-hermitian endomorphisms 
\begin{equation}\label{eq:badTmatrix}
T_a s= i\hat{P}(x_a s)
\end{equation}
for $s\in\coker D_\A^+$, satisfying 
\begin{equation}\label{eq:badasdequn}
[T_0,T_1]+[T_2,T_3]=0
\end{equation}
and cyclic permutations in $\{ 1,2,3 \}$ (the fully translation invariant ASD equation). 
In fact when we perform the transform, the RHS of equation~$\eqref{eq:badasdequn}$ turns out to be non-zero. 
This `surprise' arises as a direct result of the non-compactness of $\rfour$. 

The correct dual picture is the following. 
Let
\begin{equation}\label{eq:ADHMoperator}
\Delta(x) = \sum_a 
\begin{pmatrix}
\lambda_a \\ T_a - ix_a
\end{pmatrix}
\otimes\gamma_a :
\C^k\otimes \hat{S}^+\rightarrow \C^{k+1}\otimes \hat{S}^-
\end{equation}
where $T_a$ is a $k\times k$ skew-hermitian matrix and $\lambda_a$ a rank $k$ row vector, for $a=0,\ldots,3$. 
The map $\Delta(x)$ is the analogue of the Dirac operator $D_x^+$. 
Under the assumptions
\begin{gather}
\Delta(x)\textrm{~is injective for all $x$, and} \notag \\
\Delta^\ast(x)\Delta(x)\textrm{~commutes with the $\gamma$ matrices for all $x$},
\label{eq:adhmasd}
\end{gather}
$\coker\Delta(x)$ defines an $SU(2)$ bundle $\E$ over $\rfour$ as $x$ varies. 
This has $c_2(\E)=k$, and an analogue of Proposition~\ref{prop:NahmgivesASD} shows that the induced connection $\A$ is ASD. 
Atiyah and others \cite{adhm78,ati79} proved that every $SU(2)$ instanton can be constructed in this way---this is the ADHM construction. 
Expanding~$\eqref{eq:adhmasd}$ gives
\begin{equation*}
[T_0,T_1]+[T_2,T_3]=(\lambda_0^\ast\lambda_1 - \lambda_1^\ast\lambda_0)+
(\lambda_2^\ast\lambda_3 - \lambda_3^\ast\lambda_2)
\end{equation*}
and cyclic permutations in $\{ 1,2,3 \}$; this is the correct version of equation~$\eqref{eq:badasdequn}$. 
To obtain the ADHM data $\{ T_a,\lambda_a : a=0,\ldots,3 \}$ from a given instanton $(\E,\A)$ one must consider the Nahm transform less na\"\i vely than we did above. 
Careful analysis of the asymptotic behaviour of the solutions in $\coker D_\A^+$ can be used to obtain the $\lambda_a$. 
\vspace{1.0ex}
\newline
\noindent
\textbf{The case $T=\rthree$, $T^\ast=\R$: the construction of monopoles.} 
The following sketch of the Nahm transform for $SU(2)$ monopoles as a generalization of the transform on the $4$-torus follows Nakajima~\cite{nak90}. 
We comment on other approaches in the literature in Section~\ref{sec:litcom}. 

We have already seen how a translation invariant anti-self-dual connection $\A$ is equivalent to a solution $(A,\Phi)$ of the Bogomolny equation. 
We can perform a similar reduction on the Dirac operator $D_\xi^+$.
Using the identifications~$\ref{eq:idspinthreefour}$, $D_\xi^+$ reduces to
\begin{equation}\label{eq:diracopformono}
D_\xi=D_A+\Phi-i\xi : 
\Gamma(\spinthree\otimes E)\rightarrow\Gamma(\spinthree\otimes E)
\end{equation}
where $\xi\in\R$ is the coordinate on the dual torus. 
The rank of the Nahm transform is given by the dimension of the cokernel of $D_\xi$. 

An important r\^ole will be played in this thesis by Callias' index theorem\label{glo:cal1} \cite{cal78}, which gives a formula for the index of operators like $D_\xi$ on odd dimensional manifolds. \label{pag:callias}
The original version of the theorem applied to operators on bundles over $M=\R^m$ for odd $m$, but was generalized by Anghel \cite{ang93b} and R{\aa}de \cite{rad94} to apply more widely. 
The generalized version applies to operators of the form
\begin{equation*}
D_{A,\Phi}=D_A+1\otimes\Phi:C^\infty(M,S\otimes E)\rightarrow C^\infty(M,S\otimes E)
\end{equation*}
where $M$ is a complete open odd-dimensional spin manifold with spin bundle $S$; E is a unitary vector bundle on $M$; $D_A$ is the Dirac-operator coupled to $E$ via a unitary connection $A$; and $\Phi$ is a skew-adjoint endomorphism of $E$. 
The main condition required is that $\Phi$ should be invertible outside some compact set $M_0\subset M$ and on the boundary $\partial M_0$ (some mild additional conditions are also required). 
If this is the case, then $\Phi$ decomposes $E |_{\partial M_0}$ as a direct sum $E |_{\partial M_0}= E^+\oplus E^-$ where $E^+$ consists of eigenvectors of $-i\Phi$ that have positive eigenvalue, and $E^-$ consists of eigenvectors with negative eigenvalue. 

\begin{quotedtheorem}[Callias-Anghel-R\aa de]
Under the assumptions above, $D_{A,\Phi}$ is Fredholm with $L^2$-index given by
\begin{equation}\label{eq:AhatCallias}
\ind D_{A,\Phi} = -\int_{\partial M_0}\hat{A}(\partial M_0)\wedge \ch(E^{+})
\end{equation}
where $\ch$ denotes the Chern character of a bundle and $\hat{A}$ is the $\hat{A}$-genus. 
(See \cite[Chapter 2]{roe88} for background on characteristic classes and genera.)
\end{quotedtheorem}

Returning to the operator $D_\xi$ defined by~$\eqref{eq:diracopformono}$, we want to apply Callias' theorem to compute its index. \label{pag:Calliascomp}
Suppose that $(A,\Phi)$ satisfies the $SU(2)$ monopole boundary condition, so that at infinity $\Phi$ has eigenvalues $\pm i\mu$ defining eigenbundles with Chern classes $\pm k$.
Since the $\hat{A}$-genus of the two sphere is trivial, $\eqref{eq:AhatCallias}$ gives 
\begin{equation*}
\ind D_\xi = -c_1(E^+)[S^2_R]
\end{equation*}
provided $\Phi-i\xi$ is invertible on $S^2_R$ for all sufficiently large $R$. 
Here $E^+$ is the eigenbundle over $S^2_R$ on which $-i(\Phi-i\xi)$ is positive, $c_1$ denotes the first Chern class, and $S^2_R$ is the $2$-sphere with radius $R$. 
Since this result is independent of $R$ we can take the limit as $R\rightarrow\infty$, and write $\ind D_\xi = -c_1(E^+)[\sphinf]$.
When $\xi > \mu$, $E^+$ is trivial so the index is zero; when $\xi < -\mu$, $E^+$ is the whole vector bundle over $S^2_R$, so the index is zero again; and when $\mu >\xi >-\mu$, $E^+$ is the eigenbundle with Chern class $k$ so the index is $-k$.  
A Weitzenb\"ock formula like Lemma~\ref{lem:weitzenbock} shows that $D_\xi$ is injective, so the Nahm transform of $(A,\Phi)$ consists of a bundle $\hat{E}$ over the interval $(-\mu,\mu)\subset\R$ with rank $k$. 

The analogue of $\hat{\A}$ is a connection $\nabla$ and skew-adjoint endomorphisms $T_1,T_2,T_3$ on $\hat{E}$ defined by
\begin{align}
\nabla s &= \hat{P}(\partial_\xi s) \label{eq:nahmmoncon}\\
T_j s &= i\hat{P}( x_j s),\quad j=1,2,3 \label{eq:nahmmonmtx}
\end{align}
for a family $s(\xi)\in\coker D_\xi$. 
These satisfy Nahm's equation~$\eqref{eq:Nahm}$. 
Near the singularities $\xi=\pm\mu$ there is a parallel gauge in which the $T_j$ have a simple pole:
\begin{equation*}
T_j(\xi)=\frac{R_j^\pm}{\xi\pm\mu} +\textrm{analytic function,}
\end{equation*}
and at each singularity the residues $R^\pm_j$ define an irreducible representation\label{glo:irrep1} of $\sutwo$ with dimension $k$. 

Conversely, given such a connection $\nabla$ and endomorphisms $T_j$ on a rank $k$ bundle $\hat{E}\rightarrow (-\mu,\mu)$, the analogue of $D_x^+$ is the operator
\begin{equation}\label{eq:firstNahmop}
\Delta(x)=\nabla + \sum_{j=1}^3 T_j\otimes\gamma_j - i\sum_{j=1}^3 x_j\otimes\gamma_j : 
\Gamma(\hat{E}\otimes\hat{S}^+)\rightarrow\Gamma(\hat{E}\otimes\hat{S}^-)
\end{equation}
for $x\in\rthree$. 
Nahm's equation implies $\Delta$ is injective, and the singularity condition implies that $\Delta$ has index $-2$, so $\coker \Delta(x)$ is a rank $2$ bundle over $\rthree$. 
We then define
 \begin{align*}
\nabla_A s &= {P}(d s) \\
\Phi s &= i{P}( \xi s)
\end{align*}
where $s(x)\in\coker\Delta(x)$ for each $x$. 
An analogue of Proposition~\ref{prop:NahmgivesASD} shows $(A,\Phi)$ satisfies the Bogomolny equation~$\eqref{eq:Bogomolny}$, and the monopole also satisfies the $SU(2)$ monopole boundary condition with eigenvalues $\pm i\mu$ and charge $k$. 
\vspace{1.0ex}
\newline
\noindent
\textbf{The case $T=T^1\times\rthree$, $T^\ast=T^1$: calorons.}
Existing work on the transform for calorons is reviewed in Section~\ref{sec:review}. 
\vspace{1.0ex}
\newline
\noindent
\textbf{The case $T=T^2\times\R^2$, $T^\ast=T^2$.}
Jardim \cite{jar01} has recently proved the existence and invertibility of the transform for this case. 
The transform takes instantons on $T^2\times\R^2$ satisfying certain decay conditions to solutions of the so-called Hitchin equations on $T^2$ with point singularities. 
The proof has a strong algebro-geometric flavour by regarding $T$ as a complex manifold $T^2\times\C$, and in broad terms follows the proof of the transform on $T^4$ in \cite[Chapter $3$]{don90}. 
\vspace{1.0ex}
\newline
\noindent
\textbf{The case $T=T^3\times\R$, $T^\ast=T^3$.}
Van Baal \cite{baa96,baa99} has studied instantons on $T^3\times\R$ and the Nahm transform, in particular with `twisted boundary conditions' (conditions on every half period, as well as full periodicity). 

\subsection{Notes on the literature for the Nahm transform}\label{sec:litcom}

In Section~\ref{sec:reviewNahm} we presented the constructions of instantons and monopoles as generalizations of the Nahm transform on the $4$-torus. 
Historically, this is not how these constructions arose---in this Section we describe approaches to the Nahm transform for instantons and monopoles as they occurred in the literature. 

The ADHM construction of instantons~\cite{adhm78} arose from developments in twistor theory\label{glo:twistorthy} in the late 1970's. 
Ward~\cite{war77} proved a correspondence between instantons on $S^4$ and certain holomorphic bundles on the twistor space ${\mathbf{P}}_3(\C)$ of $S^4$.
In turn, the ADHM construction~\cite{adhm78} was a complete construction of these bundles, thereby giving a complete construction of instantons. 

In the early 1980's Nahm~\cite{nah82,nah83} sketched the Nahm transform, showing how it encompassed the ADHM construction, and describing the transform for monopoles and calorons. 
It was apparent from this work, but not explicitly stated, that the transform would work on $T^4$ and generalized tori. 
At roughly the same time Mukai~\cite{muk81} proved the existence and invertibility of the transform between holomorphic bundles over complex tori which we mentioned on page~\pageref{pag:mukai}. 
Later, Corrigan-Goddard~\cite{cor84} provided some of the details missing from Nahm's original work on the transform for instantons and monopoles. 

%%%%%%%%%%%%%%%%%%%%%%%%%%%%%%%%%%%%%%%%%%%%%%%%%%%%%%%%%%%%%%%%%%%%%%%%%%%%%%%
% FIGURE: Hitchin's circle of ideas
%%%%%%%%%%%%%%%%%%%%%%%%%%%%%%%%%%%%%%%%%%%%%%%%%%%%%%%%%%%%%%%%%%%%%%%%%%%%%%%%
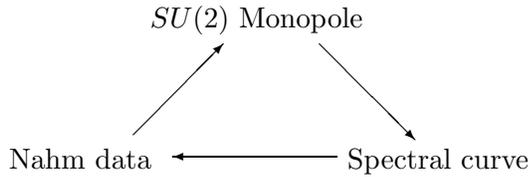
\begin{figure}
\setlength{\unitlength}{0.75cm}
\begin{center}
\begin{picture}(6,4)(-3,0)
\put(-2.0,2.5){$SU(2)$ Monopole}
\put(1,2.2){\vector(1,-1){1.7}}
\put(1.5,0){Spectral curve}
\put(1.3,0.2){\vector(-1,0){2.9}}
\put(-4.5,0){Nahm data}
\put(-2.3,0.6){\vector(1,1){1.6}}
\end{picture}
\end{center}
\caption{Hitchin's circle of ideas}
\label{fi:Hitchinscircle}
\end{figure}
%%%%%%%%%%%%%%%%%%%%%%%%%%%%%%%%%%%%%%%%%%%%%%%%%%%%%%%%%%%%%%%%%%%%%%%%%%%%%%
% END FIGURE: Hitchin's circle of ideas
%%%%%%%%%%%%%%%%%%%%%%%%%%%%%%%%%%%%%%%%%%%%%%%%%%%%%%%%%%%%%%%%%%%%%%%%%%%%%%%%

Up to this point there was no rigorous proof of the transform for monopoles including the singularities in the Nahm data. 
Hitchin \cite{hit83} proved the correspondence between $SU(2)$ monopoles and Nahm data via spectral curves\label{glo:spectral1} of monopoles. 
Using twistor theory he proved a correspondence between $SU(2)$ monopoles and certain algebraic curves in the twistor space $T{\mathbf{P}}_1(\C)$ of $\rthree$ \cite{hit82}. 
Instead of constructing the Nahm data directly from the cokernel of a Dirac operator coupled to a monopole, Hitchin \cite{hit83} considered the spectral curve of the monopole and constructed a set of Nahm data from this.
By going round the circle shown in Figure~\ref{fi:Hitchinscircle} and proving the monopole obtained is isomorphic to the monopole started from, Hitchin showed that the construction of $SU(2)$ monopoles from Nahm data is complete. 
Hurtubise-Murray \cite{hur89} adopted a similar approach to prove completeness of the Nahm transform for $SU(n)$ monopoles. 

In 1989 Braam-van Baal \cite{bra89} described the Nahm transform on $T^4$ as we presented it in Section~\ref{sec:nahmtorus}, in terms of ASD connections and Dirac operators, rather than the holomorphic approach of Mukai~\cite{muk81}. 
At a similar time, Nakajima~\cite{nak90} proved the existence and invertibility of the Nahm transform for $SU(2)$ monopoles by `direct' means (\ie via analysis of the relevant Dirac operators obtained by dimensional reduction of the transform on $T^4$), rather than by the spectral curve method. 
A direct proof for $SU(n)$ monopoles is yet to be given. 

\subsection{Twistor theory and spectral curves for monopoles}\label{sec:scformonopoles}

In this Section we expand on the twistor theory used to prove the existence of the Nahm transform for monopoles. 
The twistor space for $\rthree$ is the set of oriented geodesics in $\rthree$. 
This can be identified with the tangent space to the $2$-sphere, $TS^2\cong T{\mathbf{P}}_1(\C)$. 
Hitchin \cite{hit82} proved a correspondence between certain rank $2$ holomorphic bundles over twistor space\label{glo:twistorspace} $T{\mathbf{P}}_1(\C)$, and $SU(2)$ solutions $(A,\Phi)$ to the Bogomolny equation on a bundle $E\rightarrow\rthree$. 
Given a solution $(A,\Phi)$, the fibre $\tilde{E}_z$ at $z\in T{\mathbf{P}}_1(\C)$ of the corresponding bundle is given by 
\begin{equation*}
\tilde{E}_z = \{ s\in\Gamma(E |_{\gamma_z}) : (\nabla_U -i\Phi)s=0 \}
\end{equation*}
where $\gamma_z$ is the oriented geodesic in $\rthree$ corresponding to $z$, and $U$ is the unit vector along $\gamma_z$. 
Imposing the $SU(2)$ boundary conditions on the monopole, some analysis reveals that there is a holomorphic rank $1$ sub-bundle $L^{+}$ of $\tilde{E}$, whose fibre at $z$ is given by sections of $E$ over $\gamma_{z}$ solving
\begin{equation} \label{eq:twistortrans}
(\nabla_U -i\Phi)s=0 
\end{equation}
that decay in the direction of the oriented geodesic $\gamma_{z}$.
Similarly there is a holomorphic rank $1$ sub-bundle $L^{-}$ whose fibre at $z$ is given by solutions of~$\eqref{eq:twistortrans}$ over $\gamma_{z}$ that decay in the  opposite direction.  
The spectral curve\label{glo:spectral2} $S\subset T{\mathbf{P}}_1(\C)$ is then defined by
\begin{equation*}
  S=\{ z\in T{\mathbf{P}}_1(\C) : L^{+}_{z}=L^{-}_{z} \}
\end{equation*}
so $S$ consists of geodesics over which there exist solutions to~$\eqref{eq:twistortrans}$ which decay at both ends of the geodesic. 
The spectral curve is a compact algebraic curve, and from it one can reconstruct $\tilde{E}$ and hence the original monopole.
In other words, monopoles are uniquely determined by their spectral curves. 
Proofs for all these statements are given in \cite{hit82}. 
As described in Section~\ref{sec:litcom}, the spectral curve is used to prove completeness of the construction of $SU(2)$ monopoles from Nahm data in \cite{hit83}.
The Nahm data is given in terms of a flow on the Jacobian of the spectral curve.

As we mentioned above, twistor theory and spectral curves have also been used to prove completeness of the Nahm transform for $SU(n)$ monopoles. 
In \cite{hur89} Hurtubise-Murray describe spectral curves and Nahm data for $SU(n)$ monopoles, and give a chain of constructions equivalent to those in Figure~\ref{fi:Hitchinscircle}. 
The spectral curve of an $SU(n)$ monopole has $(n-1)$ components in $T{\mathbf{P}}_1(\C)$ with some prescribed intersection relations. 
More relevant to us, however, is the Nahm picture for $SU(n)$ monopoles. 

%%%%%%%%%%%%%%%%%%%%%%%%%%%%%%%%%%%%%%%%%%%%%%%%%%%%%%%%%%%%%%%%%%%%%%%%%%%%%%
% FIGURES: boundary conditions in Nahm data
%%%%%%%%%%%%%%%%%%%%%%%%%%%%%%%%%%%%%%%%%%%%%%%%%%%%%%%%%%%%%%%%%%%%%%%%%%%%%%%
\begin{figure}
\setlength{\unitlength}{1.0cm}
\begin{center}
\begin{picture}(14,6.5)(-1,-0.5)
\footnotesize
%axes
\put(0,0){\vector(1,0){10}}
\put(0,0){\vector(0,1){4.5}}
\put(-0.5,4.7){Rank}
\put(10.0,-0.4){$\xi$}
\put(1.8,-0.4){$\mu_4$}
\put(3.8,-0.4){$\mu_3$}
\put(5.8,-0.4){$\mu_2$}
\put(7.8,-0.4){$\mu_1$}

%block 1
\put(2.0,0){\line(0,1){2.5}}
\put(2.0,2.5){\line(1,0){2}}
\put(4,2.5){\line(0,-1){2.5}}
\put(0.7,2.7){Rank$=k_1+k_2+k_3$}

%block 2
\put(4,0){\line(0,1){4}}
\put(4,4){\line(1,0){2}}
\put(6,4){\line(0,-1){4}}
\put(3.75,4.2){Rank$=k_1+k_2$}

%block 3
\put(6,2){\line(1,0){2}}
\put(8,2.0){\line(0,-1){2.0}}
\put(8.25,1.1){Rank$=k_1$}

%\put(6.1,3){\oval(0.25,1.8)[r]}
%\put(6.1,1){\oval(0.25,1.8)[r]}
%\put(6.225,3){\line(2,1){2.0}}
%\put(6.225,1){\line(2,1){2.0}}
\put(6.1,2.85){\LARGE$\Bigg\}$}
\put(6.1,0.85){\LARGE$\Bigg\}$}

\put(8.5,5.5){
\parbox[t]{4.0cm}{\sloppy
Terminating ~~~component: this piece of the Nahm data has a singularity exactly like that for $SU(2)$ data.}}
\put(8.4,4.75){\vector(-1,-1){1.7}}

\put(8.5,3.5){
\parbox[t]{4.0cm}{\sloppy
Continuing component: this piece of the Nahm data is continuous across the join.}}
\put(8.4,2.9){\vector(-1,-1){1.7}}

\end{picture}
\end{center}
\caption{Typical $U(4)$ monopole Nahm data illustrating gluing conditions at the singularities}
\label{fi:nahmbcsa}
\end{figure}
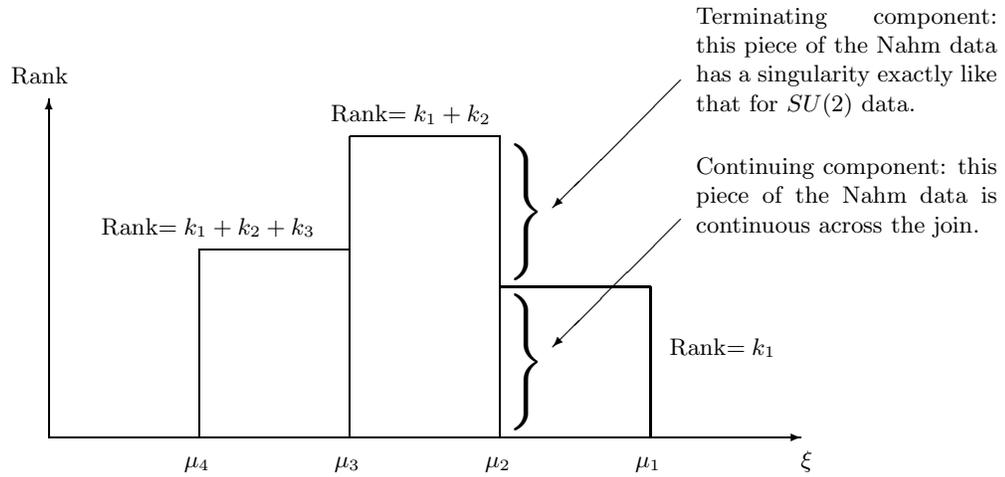

\begin{figure}
\setlength{\unitlength}{1.0cm}
\begin{center}
\begin{picture}(13,6.0)(-1,-0.5)
\footnotesize
%axes
\put(0,0){\vector(1,0){10}}
\put(0,0){\vector(0,1){4.5}}
\put(-0.5,4.7){Rank}
\put(10.0,-0.4){$\xi$}
\put(1.8,-0.4){$\mu_4$}
\put(3.8,-0.4){$\mu_3$}
\put(5.8,-0.4){$\mu_2$}
\put(7.8,-0.4){$\mu_1$}

%block 1
\put(2.0,0){\line(0,1){4}}
\put(2.0,4){\line(1,0){2}}
\put(4,4){\line(0,-1){4}}

%block 2 & 3
\put(4,2){\line(1,0){4}}
\put(6,2){\line(0,-1){2}}
\put(8,2.0){\line(0,-1){2.0}}

\put(5.5,2.2){$k_2=0$}

\put(6.1,0.85){\LARGE$\Bigg\}$}

\put(7.0,4.0){
\parbox[t]{4.0cm}{\sloppy
Zero jump: the Nahm data has a prescribed discontinuity across the join.}}
\put(8.4,2.9){\vector(-1,-1){1.7}}

\end{picture}
\end{center}
\caption{Typical $U(4)$ monopole Nahm data with a zero jump}
\label{fi:nahmbcsb}
\end{figure}
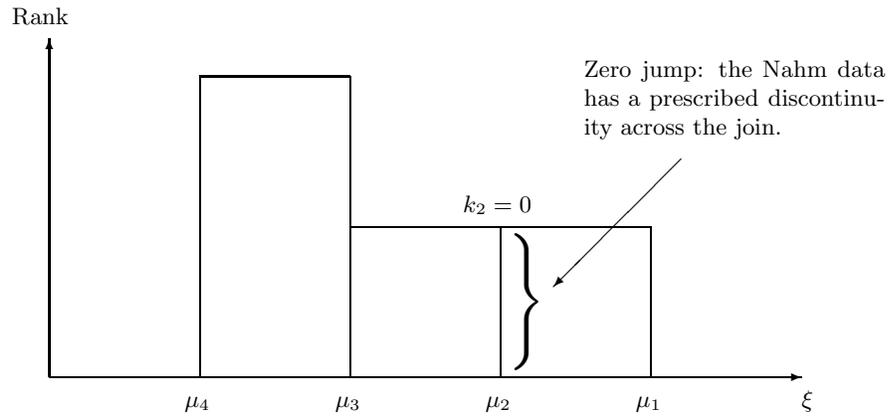
%%%%%%%%%%%%%%%%%%%%%%%%%%%%%%%%%%%%%%%%%%%%%%%%%%%%%%%%%%%%%%%%%%%%%%%%%%%%%%%
% END FIGURES: boundary conditions in Nahm data
%%%%%%%%%%%%%%%%%%%%%%%%%%%%%%%%%%%%%%%%%%%%%%%%%%%%%%%%%%%%%%%%%%%%%%%%%%%%%%%
  
Given an $SU(n)$ monopole $(A,\Phi)$ we assume that at infinity $\Phi$ has eigenvalues $i\mu_1,\ldots, i\mu_n$ which define eigenbundles with Chern classes $k_1,\ldots,k_n$. 
(There are other boundary conditions which we ignore for the moment.) 
The $\mu_j$ are ordered so that $\mu_n\leq\ldots\leq\mu_1$. 
Hurtubise-Murray \cite{hur89} show that the Nahm picture for $(A,\Phi)$ consists of $(n-1)$ bundles $X_p$\label{glo:Xp1} over the intervals $[\mu_{p+1},\mu_p]$, $p=1,\ldots,n-1$, such that
\begin{equation}\label{eq:ranksformono}
\textrm{rank}~X_p=k_1+\cdots +k_p.
\end{equation} 
Each bundle is equipped with a connection $\nabla_p$\label{glo:nabp1} and endomorphisms $T_p^j$\label{glo:Tjp1}, $j=1,2,3$, satisfying Nahm's equation. 
At each $\mu_p$ there is a gluing condition between the bundles $X_p$ and $X_{p-1}$, which depends on the ranks of $X_p$ and $X_{p-1}$: see Figures~\ref{fi:nahmbcsa} and~\ref{fi:nahmbcsb}. 
When $k_p=0$, rank $X_p$ $=$ rank $X_{p-1}$ and we call this a `zero jump'\label{glo:zerojump1}. 

This form for the Nahm data is precisely what you expect if you consider the cokernel of $D_\xi$, the Dirac operator defined by equation~$\eqref{eq:diracopformono}$. 
Applying Callias' index theorem in a similar way as that on page~\pageref{pag:Calliascomp}, $D_\xi$ is Fredholm except when $\xi\in\{ \mu_1,\ldots,\mu_n \}$, and the dimension of the cokernel is given by 
\begin{equation*}
\dim\coker D_\xi = 
\begin{cases}
k_1+\cdots +k_p~ & \textrm{when}~\xi\in(\mu_{p+1},\mu_p), \\
0~ & \textrm{when}~\xi<\mu_n ~\textrm{or}~ \xi>\mu_1.
\end{cases}
\end{equation*}
This agrees with equation~$\eqref{eq:ranksformono}$.

\section{Review of existing work on calorons}\label{sec:review}Recent work on calorons consists of two main strands. 
First there is the work of Garland-Murray in which $SU(n)$ calorons are regarded as monopoles whose structure group is the loop-group of $SU(n)$. 
Secondly, Kraan and others have proved the existence of a version of the Nahm transform for a special kind of caloron, namely those with unit instanton charge and vanishing monopole charges. 
Before reviewing this work, however, we make some remarks on other places calorons appear in the literature. 

Calorons and their applications to QCD at finite temperature were first studied in the late 1970's: \cite{gro81} contains a review of this work. 
In \cite{har78}, Harrington-Shepard constructed an explicit $SU(2)$ caloron by arranging a periodic array of charge-$1$ instantons in $\rfour$. 
Later, others showed how to take monopole and instanton limits of the Harrington-Shepard\label{glo:Harrington} caloron (by letting the period tend to zero or infinity respectively). 
Other explicit solutions have been constructed by Chakrabarti using various ans\"atze, including calorons with non-trivial monopole charges: see \cite{cha99} and the references therein. 
Calorons and the Nahm transform have an interpretation in string and brane theory\label{glo:stringthy}: 
a caloron can be regarded as a periodic arrangement of strings and D-branes; the Nahm transform is a correspondence between different brane configurations, called `T-duality'. 
In this thesis, however, we steer clear of these stringy issues. 

\subsection{The loop group point of view}\label{sec:loopgroup}

Garland and Murray \cite{gar88} make the remarkable observation that $SU(n)$ periodic instantons are the same as monopoles with the group $\Lh SU(n)$ as their structure group, where $\Lh SU(n)$ is the semi-direct product\label{glo:semidirect} of the loop group\label{glo:loopgroup} $LSU(n)$ and $U(1)$.
In other words, they show that the anti-self-duality equation~$\eqref{eq:anti-self-duality}$ for a $SU(n)$ connection on $\rfour$ is equivalent to the Bogomolny equation~$\eqref{eq:Bogomolny}$ for a $\Lh SU(n)$ monopole.  
Garland and Murray go on to develop twistor theory and the spectral curve picture for periodic instantons, by regarding them as loop-valued monopoles, and extending known results for regular monopoles. 
In a similar vein, Norbury \cite{nor00} has extended the rational map construction of $SU(n)$ monopoles \cite{don84} to a construction of $\Lh SU(n)$ monopoles based on holomorphic maps from $S^2$ to a certain flag manifold. 

In general we will not make much use of the loop group approach in this thesis; on the other hand it often gives clues as to what one might expect for calorons, by extending existing results for monopoles. 
In this Section we prove the correspondence between calorons and loop-group monopoles, before sketching Garland and Murray's work \cite{gar88} on the spectral curve of a caloron. 
Finally we describe the form we expect caloron Nahm data to take, as sketched by Garland and Murray \cite[Section 8]{gar88}. 
First, however, we make the following definitions. 

Let $G$ be a Lie group and $\curlyg$ its Lie algebra. 
The loop group $LG$\label{glo:LG} of $G$ is the group of smooth maps from $S^{1}$ to $G$ with pointwise composition.
Let $L\curlyg$\label{glo:Lalg} denote the Lie algebra of $LG$, that is $L\curlyg = \textrm{Map}(S^{1},\curlyg)$.
Define the Lie group $\Lh G$\label{glo:LhG} to be the semi-direct product of $LG$ and $U(1)$: 
as a set
\begin{equation*}
  \Lh G = LG \times U(1)
\end{equation*}
and the composition is
\begin{equation*}
  (g_{1}(\theta ), e^{i\alpha_{1}})\circ
  (g_{2}(\theta ), e^{i\alpha_{2}}) = 
  (g_{1}(\theta - \alpha_{2})\circ g_{2}(\theta ), 
  e^{i\left( \alpha_{1}+\alpha_{2}\right)}).
\end{equation*}
(Note that there is a choice as to how we let $U(1)$ act on $LG$. We have taken the choice as above since, as we shall see, it gives an adjoint action corresponding to ordinary gauge transformation on $S^1\times\rthree$.)
The Lie algebra $\Lh \curlyg$ is then
\begin{equation*}
  \Lh\curlyg = L\curlyg\oplus i\R.
\end{equation*}
The adjoint action of $\Lh G$ on $\Lh \curlyg$ is given by
\begin{equation} \label{eq:AdLGLg}
  Ad_{(g(\theta ), e^{i\alpha})}(\xi(\theta),i\lambda) =
  ( g(\Theta)\xi(\Theta)g^{-1}(\Theta) - 
  \lambda\frac{\partial g}{\partial\theta}(\Theta)g^{-1}(\Theta), i\lambda )
\end{equation}
where $\Theta = \theta + \alpha$,
and the Lie bracket is 
\begin{equation} \label{eq:loopbracket}
  [(\xi_{1}(\theta ), i\lambda_{1}),(\xi_{2}(\theta ), i\lambda_{2})]
= ([\xi_{1}(\theta ),\xi_{2}(\theta )] +
  \lambda_{1}\frac{\partial\xi_{2}}{\partial\theta} - 
  \lambda_{2}\frac{\partial\xi_{1}}{\partial\theta},0).
\end{equation} 

Next we want to describe the correspondence between $SU(n)$ calorons and $\Lh SU(n)$ monopoles. 
We will consider calorons with period $\perflat$ for some \label{glo:mu01}$\mu_0\in\R$. 
These can be thought of as connections on bundles over $\cylo$ where \label{glo:So}$\So = \R / (\per\Z)$. 
Given a connection $\A$ on a bundle 
$\E\rightarrow \cylo$ fix a global trivialisation of $\E$, and define $A$ and $\Phi$ by 
\begin{equation}\label{eq:calasloop1}
\nabla_\A = \nabla_A +dx_0(\partial_{x_0} +\Phi)
\end{equation}
so that $A$ is a loop of connections on a bundle $E\rightarrow\rthree$ and $\Phi$ a loop of endomorphisms on $E$, parameterized by $\xo\in\So$. 
We then define\label{glo:loopmon} 
\begin{equation}\label{eq:calasloop2}
\hat{A}=(A(\theta),0), \quad\textrm{and\ }\hat{\Phi}=(\Phi(\theta),i\muo)
\end{equation}
to be the $\Lh SU(n)$ monopole configuration on $E$ corresponding to $\A$, where $\theta=\muo\xo$. 
Conversely, given an $\Lh SU(n)$ monopole configuration of the form~$\eqref{eq:calasloop2}$ on a bundle $E$, let $\E=p^\ast E$ where $p$ is the projection $p:\cylo\rightarrow\rthree$, and define $\A$ using~$\eqref{eq:calasloop1}$. 
(Garland and Murray show that by gauge transforming, every finite energy $\Lh SU(n)$ monopole can be written in the form~$\eqref{eq:calasloop2}$.) 
We want to show that this correspondence is gauge invariant. 
Let $g:\cylo\rightarrow SU(n)$ be a gauge transformation on $\E$ and let $g(\theta)$ be the corresponding map $\rthree\rightarrow LSU(n)$. 
The $\Lh SU(n)$ gauge transformation corresponding to $g$ is then 
\begin{equation*}
\hat{g}=(g(\theta),0):\rthree\rightarrow \Lh SU(n).
\end{equation*}
Under $\hat{g}$, $\hat{A}$ and $\hat{\Phi}$ transform as 
\begin{align*}
  \hat{A} & \longmapsto \textrm{Ad}_{\hat{g}}\hat{A} -d \hat{g} \hat{g}^{-1} = 
  (gAg^{-1} -dgg^{-1},0) \\
  \hat{\Phi} & \longmapsto \textrm{Ad}_{\hat{g}}\hat{\Phi}  =  (g\Phi g^{-1}-\muo\frac{\partial g}{\partial\theta}g^{-1}, i\muo)
\end{align*}
using the adjoint action~$\eqref{eq:AdLGLg}$. 
Since $\muo \partial_\theta = \partial_{x_0}$ this is exactly the same as the action of $g$ on $\A$: 
\begin{equation*}
\A_a\mapsto g\A_a g^{-1}-\partial_a g g^{-1}.
\end{equation*}

\begin{quotedproposition}[Garland and Murray \cite{gar88}]
Under the correspondence given above, $\A$ is ASD if and only if $\hat{A},\hat{\Phi}$ satisfy the Bogomolny equation. 
\end{quotedproposition}

\proof
Fixing a global trivialisation of $\E$, $\A$ is represented by a matrix of $1$-forms
\begin{equation*}
\Phi~dx_0 + \sum_{i=1}^{3}A_i dx_i.
\end{equation*}
The coordinate form of the anti-self-duality equation~$\eqref{eq:ASDcmpts}$ then implies
\begin{align*}
& ~(F_\A)_{23}+(F_\A)_{01}=0 \\
\Leftrightarrow & ~\partial_2 A_3 -\partial_3 A_2 +[A_2,A_3] +
\partial_0 A_1 -\partial_1 \Phi +[\Phi, A_1] =0\\
\Leftrightarrow & ~\partial_2 A_3 -\partial_3 A_2 +[A_2,A_3] +
 =\partial_1\Phi +[ A_1,\Phi]-\muo\frac{\partial A_1}{\partial\theta}
\end{align*}
and cyclic permutations in $\{ 1,2,3 \}$. 
These equations can be written invariantly as
\begin{equation}\label{eq:GMSD}
\ast F_A = \nabla_A\Phi -\muo\frac{\partial A}{\partial\theta}.
\end{equation}
On the other hand, the $\Lh SU(n)$ Bogomolny equation is 
\begin{equation}\label{eq:loopbogomolny}
\ast \hat{F}_{\hat{A}} =\nabla_{\hat{A}}\hat{\Phi}
\end{equation}
where $\hat{F}_{\hat{A}}$ is the curvature of $\hat{A}$. 
However $\hat{F}_{\hat{A}}=(F_A(\theta),0)$, while from the bracket~$\eqref{eq:loopbracket}$ we have
\begin{equation*}
\nabla_{\hat{A}}\hat{\Phi} = ( (\nabla_A \Phi)(\theta)-(\muo\partial_\theta A)(\theta) ,0).
\end{equation*}
Hence~$\eqref{eq:GMSD}$ and~$\eqref{eq:loopbogomolny}$ are equivalent. 
\eproof

With the correspondence between $SU(n)$ calorons and $\Lh SU(n)$ monopoles established, Garland and Murray go on to consider the twistor theory and spectral curves of calorons. 
The twistor space of $\cylo$ can be obtained by lifting the translation $\xo\mapsto\xo+\perflat$ on $\rfour$ to an action on the twistor space ${\mathbf{P}}_3(\C)\setminus{\mathbf{P}}_1(\C)$ of $\rfour$, and quotienting by this action. 
(The twistor space $T{\mathbf{P}}_1(\C)$ of $\rthree$ can be obtained in a similar fashion.) 
The quotient is a bundle ${\mathcal{T}}^\circ$ over $T{\mathbf{P}}_1(\C)$ with fibre $\C^\times = \C \setminus \{ 0 \}$, and it can be embedded in a fibre bundle ${\mathcal{T}}\rightarrow T{\mathbf{P}}_1(\C)$ with fibre ${\mathbf{P}}_1(\C)$. 
Just like the monopole case, standard twistor theory methods show that a caloron $(\E,\A)$ determines a holomorphic bundle $\tilde{\E}^\circ$ over twistor space ${\mathcal{T}}^\circ$. 
Garland and Murray show that if the caloron satisfies certain boundary conditions, then $\tilde{\E}^\circ$ extends to the compactification to define a holomorphic bundle $\tilde{\E}\rightarrow{\mathcal{T}}$. 
The bundle $\tilde{\E}$ can then be used to define $n$ spectral curves in $T{\mathbf{P}}_1(\C)$: a point $z\in T{\mathbf{P}}_1(\C)$ lies in a given curve depending on the existence of certain sections of $\tilde{\E}$ over the fibre of ${\mathcal{T}}\rightarrow T{\mathbf{P}}_1(\C)$ above $z$. 
In particular, sections over a fibre ${\mathbf{P}}_1(\C)$ of ${\mathcal{T}}$ are characterized by how they extend from $\C^\times$ to zero and infinity. 
Garland and Murray then show how a caloron is determined by its spectral curves, constructing a caloron from a set of spectral data.  

Given the twistor picture for calorons, and arguing by analogy with the monopole case, one can predict what the Nahm picture for calorons should look like --- this is sketched in \cite[Section 8]{gar88}. 
First, however, we need to describe Garland and Murray's boundary conditions for calorons. 
They require that there is a gauge at infinity in which a $\Lh SU(n)$ monopole $(\hat{A},\hat{\Phi})$ of the form~$\eqref{eq:calasloop2}$ agrees with a static ($\theta$-independent) $SU(n)$ monopole configuration. 
Thus a caloron can be characterized asymptotically by the eigenvalues $i\mu_1,\ldots,i\mu_n$ of the Higgs field in this gauge, and the Chern classes $k_1,\ldots,k_n$ (the `monopole charges'\label{glo:moncharge}). 
There is an additional topological characteristic, denoted $k_0$ and called the `instanton charge'\label{glo:instantoncharge}, that is the obstruction up to deformation on the interior of $\rthree$ to the entire configuration being $\theta$-independent. 
In terms of an $SU(n)$ periodic instanton $(\E,\A)$ the boundary condition requires that there is a gauge at infinity in which $\A_0=\Phi$ has eigenvalues $i\mu_1,\ldots,i\mu_n$ etc. 

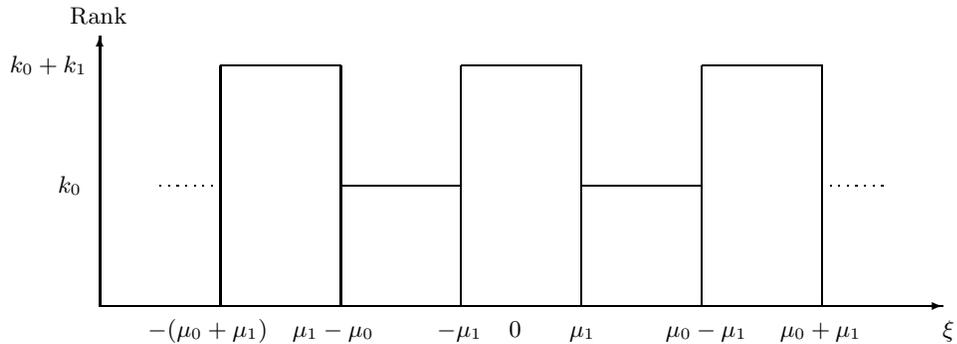
\begin{figure}
\setlength{\unitlength}{0.8cm}
\begin{center}
\begin{picture}(16,6)(-2,-0.5)
\footnotesize
%axes
\put(0,0){\vector(1,0){14}}
\put(0,0){\vector(0,1){4.5}}
\put(-0.5,4.7){Rank}
\put(14.0,-0.5){$\xi$}
\put(0.8,-0.5){$-(\mu_0+\mu_1)$}
\put(3.2,-0.5){$\mu_1-\mu_0$}
\put(5.6,-0.5){$-\mu_1$}
\put(7.8,-0.5){$\mu_1$}
\put(9.4,-0.5){$\mu_0-\mu_1$}
\put(11.3,-0.5){$\mu_0+\mu_1$}
\put(6.8,-0.5){$0$}

\put(2.0,0){\line(0,1){4}}
\put(4.0,0){\line(0,1){4}}
\put(6.0,0){\line(0,1){4}}
\put(8.0,0){\line(0,1){4}}
\put(10.0,0){\line(0,1){4}}
\put(12.0,0){\line(0,1){4}}

\put(2.0,4){\line(1,0){2}}
\put(6.0,4){\line(1,0){2}}
\put(10.0,4){\line(1,0){2}}

\put(4.0,2){\line(1,0){2}}
\put(8.0,2){\line(1,0){2}}

\dottedline{0.15}(1.0,2.0)(2.0,2.0)
\dottedline{0.15}(12.0,2.0)(13.0,2.0)

\put(-0.7,1.9){$k_0$}
\put(-1.5,3.9){$k_0+k_1$}

\end{picture}
\end{center}
\caption{Typical $SU(2)$ caloron Nahm data}
\label{fi:SU2calNahmdata}
\end{figure}

We expect the Nahm data corresponding to such a caloron to consist of $n$ bundles $X_p$\label{glo:Xp2}, $p=1,\ldots,n$, over the\label{glo:Ip1} intervals 
\begin{equation}\label{eq:bdlintervals}
I_p:=[\mu_{p+1},\mu_p]\subset\R / \mu_0\Z, ~p=1,\ldots,n-1, \quad\textrm{and}\quad I_n:=[\mu_1-\mu_0,\mu_n]\subset\R / \mu_0\Z. 
\end{equation}
Each bundle $X_p$ is equipped with a connection and endomorphisms satisfying Nahm's equation. 
Furthermore, we anticipate the rank of the data to be given by
\begin{equation}\label{eq:bdlranks}
{\textrm{rank}}~X_p = k_0+k_1+\cdots+k_p.
\end{equation}
(Garland-Murray do not give this formula, but it is implicit from their work.)
At each point $\xi=\mu_p$ we expect the Nahm data to satisfy conditions entirely similar to those for $SU(n)$ monopoles (recall Figures~\ref{fi:nahmbcsa} and~\ref{fi:nahmbcsb}).  
A typical set of $SU(2)$ caloron Nahm data is illustrated in Figure~\ref{fi:SU2calNahmdata}. 

%%%%%%%%%%%%%%%%%%%%%%%%%%%%%%%%%%%%%%%%%%%%%%%%%%%%%%%%%%%%%%%%%%%%%%%%%%%%%%
% FIGURE: caloron Nahm data
%%%%%%%%%%%%%%%%%%%%%%%%%%%%%%%%%%%%%%%%%%%%%%%%%%%%%%%%%%%%%%%%%%%%%%%%%%%%%%%
%%%%%%%%%%%%%%%%%%%%%%%%%%%%%%%%%%%%%%%%%%%%%%%%%%%%%%%%%%%%%%%%%%%%%%%%%%%%%%
% END FIGURE: caloron Nahm data
%%%%%%%%%%%%%%%%%%%%%%%%%%%%%%%%%%%%%%%%%%%%%%%%%%%%%%%%%%%%%%%%%%%%%%%%%%%%%%

\subsection{Calorons with vanishing monopole charges}\label{sec:kraan}

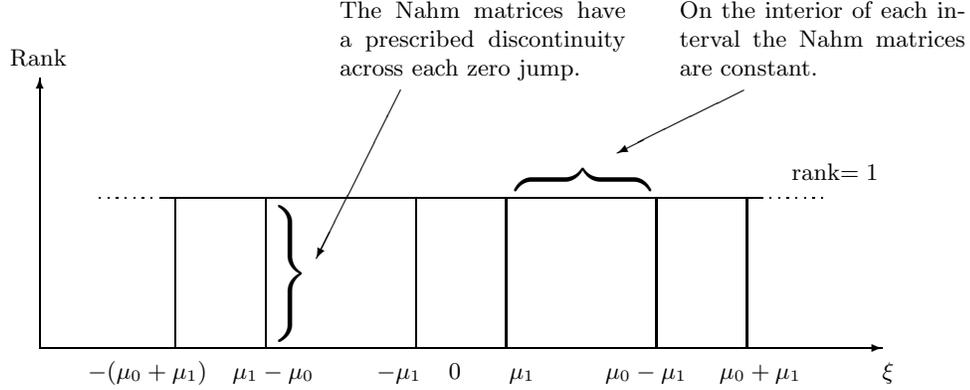
\begin{figure}
\setlength{\unitlength}{0.8cm}
\begin{center}
\begin{picture}(16.5,6)(-1,-0.5)
\footnotesize
%axes
\put(0,0){\vector(1,0){14}}
\put(0,0){\vector(0,1){4.5}}
\put(-0.5,4.7){Rank}
\put(14,-0.5){$\xi$}
\put(0.8,-0.5){$-(\mu_0+\mu_1)$}
\put(3.2,-0.5){$\mu_1-\mu_0$}
\put(5.6,-0.5){$-\mu_1$}
\put(7.8,-0.5){$\mu_1$}
\put(9.4,-0.5){$\mu_0-\mu_1$}
\put(11.3,-0.5){$\mu_0+\mu_1$}
\put(6.8,-0.5){$0$}

\put(2.0,2.5){\line(1,0){10}}
\put(2.25,0){\line(0,1){2.5}}
\put(3.75,0){\line(0,1){2.5}}
\put(6.25,0){\line(0,1){2.5}}
\put(7.75,0){\line(0,1){2.5}}
\put(10.25,0){\line(0,1){2.5}}
\put(11.75,0){\line(0,1){2.5}}

\dottedline{0.15}(1.0,2.5)(2.0,2.5)
\dottedline{0.15}(12.0,2.5)(13.0,2.5)

\put(12.5,2.8){rank$=1$}

\put(3.85,1.05){\LARGE$\Bigg\}$}

\put(4.85,5.5){
\parbox[t]{3.8cm}{\sloppy
The Nahm matrices have a prescribed discontinuity across each zero jump.}}

\put(10.5,5.5){
\parbox[t]{3.8cm}{\sloppy
On the interior of each interval the Nahm matrices are constant.}}

\put(6.0,4.3){\vector(-1,-2){1.4}}
%\put(12.0,4.0){\vector(-1,-1){2.8}}

\put(7.75,2.5){
\begin{sideways}\LARGE$\Bigg\}$\end{sideways}}
\put(11.7,4.3){\vector(-2,-1){2.1}}

\end{picture}
\end{center}
\caption{Typical $SU(2)$ caloron Nahm data with vanishing monopole charges and unit instanton charge}
\label{fi:physicistsNahmdata}
\end{figure}

The Nahm transform for calorons has been studied recently for $SU(n)$ calorons with unit instanton charge and vanishing monopole charges\label{glo:VMC1} in a series of papers \cite{kra98a,kra98b,kra00} and \cite{lee98}. 
In terms of the boundary conditions described in Section~\ref{sec:loopgroup}, these calorons have $k_0=1$ and $k_j=0$ for $j=1,\ldots,n$. 
In \cite{kra98a}, \cite{kra98b}, and \cite{kra00}, Kraan and others construct such calorons from infinite arrays of ordinary ADHM data, corresponding to an arrangement of instantons in $\rfour$ repeated periodically, possibly with some topological `twist'. 
(This is, of course, a similar approach to the Harrington-Shepard construction, but more general.)
The caloron is given by the cokernel of an infinite-dimensional matrix operator $\Delta(x)$, similar to the operator~$\eqref{eq:ADHMoperator}$ for the regular ADHM construction, but constructed from this array of data. 
By regarding the algebraic equation $\Delta^\ast(x)v=0$ as a equation for the Fourier coefficients of a function on $S^1$, Kraan obtains the Nahm picture for calorons. 
This agrees exactly with the Nahm picture conjectured at the end of Section~\ref{sec:loopgroup} and consists of bundles $X_p$, $p=1,\ldots,n$ over the intervals~$\eqref{eq:bdlintervals}$, each equipped with a connection and skew-adjoint endomorphisms $T_1,T_2,T_3$. 
The bundles all have rank $1$, and so Nahm's equation reduces to
\begin{equation*}
\nabla T_j = 0,\quad j=1,2,3,
\end{equation*}
since the $T_j$ commute. 
Working in a parallel gauge on each bundle, the matrices $T_1,T_2,T_3$ are constant. 
At each point $\xi=\mu_p$ there is a `zero jump' (\ie rank $X_{p+1}=$ rank $X_p$), and the matrices have some prescribed discontinuity there. 
Kraan and van Baal \cite{kra98b} show that each discontinuity determines (and is determined by) a vector $y_p\in\rthree$. 
Note that each block of data $X_p$ is exactly the same as the Nahm data for a charge-$1$ $SU(2)$ monopole, and so the caloron can be said to consist of $n$ `constituent monopoles'. 
Kraan and van Baal show that the constituent monopoles are located at the points $y_p\in\rthree$. 
This interpretation does not hold for higher charge calorons: for example, consider a caloron with $k_0 = 2$ and vanishing monopole charges. 
The Nahm data for the caloron should be discontinuous, but not singular, at each $\xi=\mu_p$, while the data for a charge-$2$ $SU(2)$ monopole has singularities at its endpoints. 
The caloron Nahm data therefore cannot be assembled from $n$ sets of monopole Nahm data. 
Figure~\ref{fi:physicistsNahmdata} illustrates Nahm data for Kraan's calorons.

%%%%%%%%%%%%%%%%%%%%%%%%%%%%%%%%%%%%%%%%%%%%%%%%%%%%%%%%%%%%%%%%%%%%%%%%%%%%%%
% FIGURE: physicists Nahm data
%%%%%%%%%%%%%%%%%%%%%%%%%%%%%%%%%%%%%%%%%%%%%%%%%%%%%%%%%%%%%%%%%%%%%%%%%%%%%%%
%%%%%%%%%%%%%%%%%%%%%%%%%%%%%%%%%%%%%%%%%%%%%%%%%%%%%%%%%%%%%%%%%%%%%%%%%%%%%%
% END FIGURE: physicists Nahm data
%%%%%%%%%%%%%%%%%%%%%%%%%%%%%%%%%%%%%%%%%%%%%%%%%%%%%%%%%%%%%%%%%%%%%%%%%%%%%%

Since Nahm's equation can be solved completely in the case of unit instanton charge and vanishing monopole charges, the construction of calorons can be given very explicitly. 
Roughly speaking, the locations of the `constituent monopoles' $y_1,\ldots,y_n$ and the values $\mu_1,\ldots,\mu_n$ determine the Nahm data completely: expressions for the corresponding caloron are given in terms of these data in \cite{kra98b}. 
Kraan and van Baal go on to consider the moduli space of calorons, and argue completeness of their construction by counting parameters. 
They also take various limits for the $SU(n)$ caloron, obtaining monopole ($\mu_0\rightarrow\infty$) and instanton ($\mu_0\rightarrow 0$) limits on the moduli space. 
Shrinking one interval of the Nahm data gives a caloron with a `massless' constituent monopole---the Harrington-Shepard caloron can be obtained in this way. 
Independently of Kraan, Lee \cite{lee98} has given a very similar construction of calorons from this kind of Nahm data.

\section{Overview of results}\label{sec:overview}

The aim of this thesis is to prove the existence of a version of the Nahm transform for calorons as a generalization of the transform on the $4$-torus. 
As indicated in Section~\ref{sec:reviewNahm}, many aspects of the Nahm transform on the $4$-torus carry across directly to the caloron case, but we encounter the following difficulties:
\begin{enumerate}
\item \textbf{Boundary conditions.} 
We need to specify boundary conditions on the caloron that are sufficiently strong for the transform from the caloron to the Nahm data to be possible, but sufficiently weak to be recovered for a caloron constructed from some set of Nahm data. 
The `dual' problem---specifying conditions for the Nahm data at singularities---is much more straightforward as we take these to be exactly the same as the conditions for $SU(n)$ monopoles. 
The main difficulty in the construction of calorons from Nahm data is proving the caloron obtained satisfies the boundary conditions; conversely, obtaining the singularity conditions for the Nahm data constructed from a caloron is also a difficult problem.
\item \textbf{The index formula.}
Given a caloron $\A$ we need a formula for the $L^2$-index of the Dirac operator $\DAxiplus = D^+_{\A} -i\xi$ in order to calculate the rank of the Nahm data obtained from the transform. 
\end{enumerate} 

Our approach to these problems and our main results can be summarized in the following way:
\vspace{2.0ex}
\newline
\noindent
\textbf{Chapter 2: the topology of calorons.}
\vspace{1.0ex}
\newline
We define boundary conditions for calorons and explore topological aspects such as deformation of caloron configurations. 
The main innovation is that we work on closed manifolds with boundary rather than open manifolds with asymptotic boundary conditions, and thus consider calorons on $S^1\times\threeball$ where $\rthree$ is the interior of the closed ball $\threeball$\label{glo:threeball}. 
This allows the boundary conditions to be stated very succinctly, but the drawback is that we have to do more work to recover them in the Nahm transform. 
A $U(n)$ bundle $\E\rightarrow S^1\times\threeball$ is \emph{framed}\label{glo:framedbdl1} if it is equipped with a trivialisation $f$ at infinity. 
There is a topological obstruction, denoted $c_2(\E,f)$\label{glo:c21}, to extending this to a global trivialisation of $\E$. 
Our boundary condition for a caloron $\A$ on $\E$ is that it should resemble the pull-back of a $U(n)$ monopole configuration in the trivialisation $f$. 
Thus calorons are characterized by the eigenvalues $i\mu_1,\ldots,i\mu_n$ of the Higgs field at infinity, the Chern classes $k_1,\ldots,k_n$ of the corresponding eigenbundles, the period $\perflat$, and the invariant $\ko=c_2(\E,f)$ of the framed bundle. 
We go on to define a map between calorons in different topological classes which corresponds to a rotation of the Nahm data round $S^1$. 
This `rotation map' has been considered previously by Lee~\cite{lee98,lee98b} for calorons with vanishing monopole charges, but its relation with the Nahm transform has not been fully explored. 
It plays an important r\^ole in our construction of calorons from Nahm data.
\vspace{2.0ex}
\newline
\noindent
\textbf{Chapter 3: the transform from Nahm data to calorons.}
\vspace{1.0ex}
\newline
Nahm data for calorons was described at the end of Section~\ref{sec:loopgroup}. 
Let $\spcSD{N}{\bdarydata}$ denote the collection (modulo gauge transformations) of Nahm data characterized by $\bdarydata$, so that the data is singular at $\xi=\mu_1,\ldots,\mu_n$ and has rank given by~$\eqref{eq:bdlranks}$. 
Here $\vec k =(k_1,\ldots,k_n)$ and similarly for $\vec \mu$, and we call $\bdarydata$ a set of caloron boundary data. 
Let $\spcSD{C}{\bdarydata}$ be the collection of calorons (modulo gauge transformations) with boundary conditions defined by $\bdarydata$. 
We prove the following in Chapter $3$:

\begin{quotedtheorem}[Nahm data $\rightarrow$ caloron]
For each set of boundary data $\bdarydata$, the Nahm transform is a well-defined map from $\spcSD{N}{\bdarydata}$ to $\spcSD{C}{\bdarydata}$. 
\end{quotedtheorem}

Showing that the connection constructed from the Nahm data is ASD, periodic, and $SU(n)$, is relatively easy, and the main difficulty lies in recovering the boundary conditions. 
In the notation of Section~\ref{sec:nahmtorus}, $D^+_x$ is the Dirac operator whose cokernel gives the caloron. 
We define a model operator $\tilde{D}^+_x$---a deformation of $D^+_x$---and prove that the cokernel of $\tilde{D}^+_x$ gives a periodic connection satisfying the desired boundary conditions, but not the ASD condition. 
We then prove that the boundary conditions are not affected by the deformation, and conclude that the cokernel of $D^+_x$ therefore satisfies the desired boundary conditions. 
This approach was used by Hitchin \cite{hit83} to recover the boundary conditions for an $SU(2)$ monopole. 
We have extended it in three ways: firstly to deal with zero jumps (which cannot occur for an $SU(2)$ monopole); secondly to work right up to the boundary of $\threeball$; and finally to model $D^+_x$ on the interior as well as at infinity. 
This last point is necessary to recover $\ko$, which is the obstruction to extending the framing of the caloron to the interior, and so requires understanding of $D^+_x$ (at least up to deformation) when $x$ is in the interior of $S^1\times\threeball$. 
Defining the model operator $\tilde{D}_x^+$ requires some ingenuity, and recovering the boundary conditions occupies the majority of Chapter 3. 
\vspace{2.0ex}
\newline
\noindent
\textbf{Chapter 4: the transform from calorons to Nahm data.}
\vspace{1.0ex}
\newline
We aim to prove:

\begin{quotedtheorem}[Caloron $\rightarrow$ Nahm data]
For each set of boundary data $\bdarydata$, the Nahm transform is a well-defined map from $\spcSD{C}{\bdarydata}$ to $\spcSD{N}{\bdarydata}$. 
\end{quotedtheorem}

In particular, we have to prove that the rank of the Nahm data obtained is given by~$\eqref{eq:bdlranks}$, and that the Nahm data obtained satisfies the desired singularity conditions. 
The rank condition follows from the following theorem:

\begin{quotedtheorem}[The index theorem]
Given a caloron $\A$ on a framed bundle $(\E,f)$ which satisfies the boundary conditions specified by $\bdarydata$, $\DAxiplus=D^+_\A-i\xi$ is Fredholm with $L^2$-index
\begin{equation*}
\ind\DAxiplus = -(k_0+k_1+\ldots+k_p)
\end{equation*}
when $\xi\in\textrm{interior}~I_p$ and $I_p$ is defined by~$\eqref{eq:bdlintervals}$ for $p=1,\ldots,n$. 
\end{quotedtheorem}

The proof of the index theorem involves two main steps. 
The first is a calculation of the index when $k_0=c_2(\E,f)=0$, in which case the caloron can be deformed through the space of calorons satisfying the boundary conditions, until it is independent of $\xo$. 
Callias' index theorem \cite{cal78} is used to compute the index in this case. 
The second step uses an excision theorem of Gromov-Lawson \cite{gro83} to reduce the problem to the case $k_0=0$. 
The theorem has been published, together with some material from Chapter 2, in \cite{nye00}, and a copy of this paper is attached to the thesis. 

The singularity conditions are recovered, in part at least, by generalizing Nakajima's analysis \cite{nak90} of the singularities for $SU(2)$ monopoles. 
We successfully recover the singularity conditions at points $\xi=\mu_p$ where $k_p\neq 0$ under the assumption of two analytic conjectures given in Section~\ref{sec:decompkpneq0}. 
Note that we do not give a complete proof of the Theorem (Caloron $\rightarrow$ Nahm data): we do not recover the gluing conditions at zero jumps in the Nahm data, and we do not show that the Nahm data obtained gives rise to an injective operator $D^+_x$---a necessary condition for the Nahm data to lie inside $\spcSD{N}{\bdarydata}$.

\section{Open problems}\label{sec:further}

\textbf{Behaviour at the singular points.}
There are a few problems concerning the behaviour of the Nahm data constructed from a monopole or caloron which we do not resolve. 
These are explained in Section~\ref{sec:decompkpneq0}. 
\vspace{1.0ex}
\newline
\noindent
\textbf{Injectivity of the Nahm operator and invertibility of the transform.} 
An obvious `next step' following this thesis is to prove that the transform from Nahm data to calorons and the transform from calorons to Nahm data described in Chapters $3$ and $4$ are mutually inverse. 
One approach to proving invertibility is to adapt Donaldson and Kronheimer's method \cite[Chapter $3$]{don90} that uses holomorphic geometry and $\bar{\partial}$-cohomology. 
Another approach, more in keeping with the rest of this thesis, is to adapt Nakajima's analytic proof that the Nahm transform for $SU(2)$ monopoles is invertible \cite[Sections $4$ and $5$]{nak90}. 
We remarked above that our proof of the Theorem (Caloron $\rightarrow$ Nahm data) is incomplete because we do not prove injectivity of the `Nahm operator' $D_x^+=\Delta(x)$ constructed from the Nahm data in a similar way to~$\eqref{eq:firstNahmop}$. 
Without this established, it may not be possible to apply the inverse transform to go back to the caloron. 
This is not a serious problem: if we perform the transform on a caloron, and then form the Nahm operator $\Delta$ from the Nahm data obtained, it is easy to prove that $\Delta(x)$ is injective away from a finite collection of points, so the inverse transform is defined almost everywhere. 
A first step towards proving invertibility is to express the Greens function of $\Delta^\ast(x)\Delta(x)$ in terms of smooth solutions $\psi$ of $D_\xi^-\psi=(D^-_\A+i\xi)\psi=0$ (see \cite[Section $2.3$]{bra89} for the $4$-torus calculation). 
This Greens function should have some canonical form in terms of the spinors $\psi$. 
A proof by contradiction shows that $\Delta(x)$ is injective: if it is not then the spinors $\psi$ cannot be smooth. 
Braam-van Baal \cite[Proposition $2.5$]{bra89} and Corrigan-Goddard \cite[Section $3$]{cor84} use an argument like this for the $4$-torus and ADHM construction respectively. 
With this in place, adapting Nakajima's proof of invertibility should be quite straight-forward. 
\vspace{1.0ex}
\newline
\noindent
\textbf{The moduli space of calorons.}
There are many open problems concerning the moduli space of calorons, including fundamental problems such as proving existence and smoothness, calculating the dimension of the space, and proving the existence of a hyperK\"ahler metric. 
These fundamental problems also apply to the space of Nahm data. 
With these problems solved, it might be possible to prove that the Nahm transform is a hyperK\"ahler isometry between the two spaces. 
One could then explore the moduli space of calorons by working on the space of Nahm data. 
Many of the problems investigated for monopoles, for example calculation of the metric for widely separated constituents or scattering for symmetric configurations, could be carried over to calorons. 
\vspace{1.0ex}
\newline
\noindent
\textbf{Monopole and instanton limits of calorons.}
One of the reasons for studying calorons is that they form an interpolating case between monopoles and instantons. 
An obvious question to ask is whether we can find families of calorons with a monopole or instanton limit. 
In terms of moduli spaces it might be possible to prove that monopoles and instantons form the boundary of the caloron moduli space in some sense. 
Kraan and van Baal \cite{kra00,kra98b} have constructed such families for calorons with vanishing monopole charges and unit instanton charge. 
However, these involve taking the limit as one interval of the Nahm data is contracted to zero (taking the `massless monopole limit' in physics language). 
This is the same as looking at monopoles and calorons with non-maximal symmetry breaking---opening up a wide range of problems. 
Throughout this thesis we will only consider calorons with maximal symmetry breaking. 

\chapter{The Topology of Calorons}

This Chapter is concerned with topological aspects of calorons, especially boundary conditions. 
Based on the boundary conditions used by Garland-Murray \cite[Section $3$]{gar88}, in Section~\ref{sec:boundaryconds} we define a set of boundary conditions which are sufficiently strong for the Nahm transform from calorons to Nahm data to be possible, but which can be recovered for a caloron constructed from some set of Nahm data. 
We draw the reader's attention to definitions~\ref{def:framedbdl}, \ref{def:c2bdl}, and \ref{def:caloron}---these are central to the rest of the thesis.
In Section~\ref{sec:rotationbdary} we study `large' gauge transformations (non-periodic gauge transformations that leave the caloron periodic) and a map between different calorons given by such a gauge transformation, that in some sense corresponds to rotation of the Nahm data round $S^1$. 
Section~\ref{sec:loopbcs} contains a slight digression in which the caloron boundary conditions are derived by regarding a $SU(n)$ caloron as a $\Lh SU(n)$ monopole and imposing the monopole boundary conditions.

\section{Boundary conditions}\label{sec:boundaryconds}
  
Boundary conditions for objects on open manifolds can be regarded from two points of view:
\begin{enumerate}
\item one can specify a set of asymptotic conditions that control behaviour as `infinity' is approached;
\item alternatively, one can glue on a boundary to obtain a compact manifold with boundary, and demand that objects extend to the boundary and have fixed behaviour there.
\end{enumerate}
The second approach is motivated by the work of Melrose (see \cite{maz99,mel95}). 
It allows the boundary conditions to be stated more concisely, and is the approach we adopt to define the boundary conditions in Sections~\ref{sec:monopolebcs} and~\ref{sec:caloronbcs}.
I acknowledge the help of my supervisor, Michael Singer, with the definitions in this Section, many of which appeared in our joint publication \cite{nye00}.

We compactify $\rthree$ by identifying it with the interior of the closed $3$-ball $\threeball$. 
As previously, we consider calorons with period $\perflat$\label{glo:mu02}, and let $\So=\R / (\per\Z)$\label{glo:So2}. 
Let $X=\So\times\threeball$\label{glo:X}, denote the boundary by $\dX=\So\times\sphinf$\label{glo:partialX}\label{glo:sphinf}, and let $p:X\rightarrow\threeball$\label{glo:projp} be the projection on to $\threeball$.
The interior $\Xo=X \setminus \dX$\label{glo:Xo} can be identified with $\So\times\rthree$. 
Let $x_0,\ldots,x_3$ be the coordinates on $\Xo$ corresponding to the standard coordinates on $\rfour$ under projection  $\rfour\rightarrow\Xo$, and orient $\Xo$ so that $dx_0,dx_1,dx_2,dx_3$ is positive. 
Let the metric $g$ on $\So\times\rthree$ be the standard flat product metric that gives the circle length $2\pi / \muo$.
Next we write down coordinates near the boundary of $X$ and derive the form of the metric $g$ in these coordinates.
Let $r,y_1,y_2$\label{glo:ry1y2} be polar coordinates\label{glo:polarcoords} on $\rthree$, so that $r$ is the distance from the origin in $\rthree$ and
$y_1$, $y_2$ are some local angular coordinates on $\sphinf$. 
We suppose $y_1$ and $y_2$ are chosen so that $g$ takes the form
\begin{equation*}
g = dr^2 + r^2(h_{1}dy_1^2 + h_{2}dy_2^2) + \dxo^2,
\end{equation*}
for some positive locally-defined functions $h_1,h_2$. 
Local coordinates near the boundary of $X$ will be $\chi = r^{-1}$\label{glo:chi}, $y_1,y_2$
and $\xo$, so that $\chi$ becomes a boundary defining function: 
$\chi\geq 0$ on $X$, with equality only at $\dX$, and $d\chi\neq 0$ on $\partial X$. 
Writing $g$ in terms of $\chi$, 
\begin{equation}\label{eq:metricinchi}
g = \frac{d\chi^2}{\chi^4} + h_1\frac{dy_1^2}{\chi^2} + h_2\frac{dy_2^2}{\chi^2} + \dxo^2,
\end{equation}
so $g$ is singular at the boundary. 

\subsection{Boundary conditions for monopoles}\label{sec:monopolebcs}

Let $E\rightarrow\threeball$\label{glo:E} be the trivial $\Un$ vector bundle and let $\Einf=E|_{\sphinf}$\label{glo:Einf}.
Suppose $\Ainf$\label{glo:Ainf} is a $\Un$ connection on $\Einf$, $\Phiinf$ is a skew-adjoint endomorphism on $\Einf$, and that the following conditions are satisfied:
\begin{equation}\label{eq:Phiframed}
\Phiinf\ \textrm{has $n$ distinct constant eigenvalues, and}
\end{equation}
\begin{equation}\label{eq:Aframed}
\Ainf = \bigoplus_{j=1}^{n}P_j \cdot d
\end{equation}
where $P_j$ is projection onto the $j$-th eigenbundle, and $d$ is the covariant derivative on $\Einf$.
It follows that $\nabla_{A_\infty}\Phiinf=0$. 
The condition that the eigenvalues are distinct is that of maximal symmetry breaking\label{glo:maxSB}: although this is not required for many of our results, we only prove the existence of the Nahm transform for calorons with maximal symmetry breaking.
We therefore assume maximal symmetry breaking from the outset. 
Note that we also assume $n\geq 2$ (since we are ultimately interested in the gauge group $SU(n)$). 

\begin{definition}\label{def:monopole}
A \emph{$\Un$ monopole configuration framed\label{glo:framedmon} by $(\Ainf,\Phiinf)$} is a unitary connection $A$ on $E$ and a skew-adjoint endomorphism $\Phi$ on $E$ that satisfy
\begin{equation*}
A|_{\sphinf} = \Ainf\ \textrm{and}\ \Phi|_{\sphinf} = \Phiinf.
\end{equation*}
\end{definition}
We will also work with $\SUn$ monopole configurations, in which case we require $\Phi$ and $\Phiinf$ to be trace-free and $A$ to be compatible with the volume form. 
Note that a monopole configuration is not required to satisfy the Bogomolny equation. 
The gauge transformations are the unitary bundle automorphisms of $E$ that are the identity at infinity. 

Let $i\mu_1,\ldots,i\mu_n$\label{glo:muj} be the eigenvalues of $\Phiinf$, and order them so that $\mu_n<\mu_{n-1}<\cdots <\mu_1$.
Let $k_j$\label{glo:kj} be the Chern class of the eigenbundle with eigenvalue $i\mu_j$.
Thus we have $\vec k =(k_1,\ldots,k_n)\in\Z^n$ and $\vec \mu =(\mu_1,\ldots,\mu_n)\in\R^n$ satisfying:
\begin{enumerate}
\item $\sum_1^n k_j = 0$,
\item $\mu_n<\mu_{n-1}<\cdots <\mu_1$.
\end{enumerate}

This choice of notation matches that in \cite{gar88} and \cite{hur89}. 

\begin{definition}\label{def:monbdarydata}
A pair $\monbdarydata$\label{glo:veckvecmu} is a set of \emph{$\Un$ monopole boundary data}\label{glo:monbdarydata} if it satisfies  these two conditions. 
It is a set of $\SUn$ monopole boundary data if in addition it satisfies  $\sum_1^n \mu_j =0$.
\end{definition}

We have shown that on one hand a pair $\Ainf,\Phiinf$ satisfying the boundary conditions determines a set of boundary data; on the other hand
note that a set of monopole boundary data $\monbdarydata$ determines $\Ainf,\Phiinf$ uniquely up to isomorphism.

\subsection{Framed bundles}\label{sec:framedbundles}

\begin{definition}\label{def:framedbdl}
A $\Un$ \emph{framed bundle}\label{glo:framedbdl2} over $X$ consists of a pair $(\E,f)$ where $\E\rightarrow X$ is a $\Un$ vector bundle and $f:\E\rest{\dX}\rightarrow p^{\ast}\Einf$ is a unitary bundle-isomorphism.
\end{definition}

There exists a topological obstruction to extending $f$ to a global identification of $\E$ with $p^{\ast}E$, and we use this to define an invariant $c_2(\E,f)$ of a framed bundle.
Consider what happens when we try to extend $f$:
for each $s$ we can find an identification $F_{(s)}:\E\rest{\xo=s}\rightarrow E$ that agrees with $f_{(s)}=f\rest{\xo=s}$ on $\sphinf$.
We can do this continuously, and obtain a path of maps $F_{(s)}$ for $s$ in some interval $\Ieps = (-\epsilon,\perflat+\epsilon)$\label{glo:Ieps}.
Define 
\begin{equation}\label{eq:cintermsofF}
c(s) = F_{(s+\perflat)}F^{-1}_{(s)}\in\Auto E
\end{equation}
where $\Auto E$\label{glo:AutoE} is the group of unitary automorphisms of $E$ that are the identity on $\sphinf$.
Then $c$ represents an element of 
\begin{equation*}
\pi_{0}\Auto E = \pi_0 \Map(S^3,\Un)= \pi_3 \Un = \Z
\end{equation*}
which we call $\deg c$\label{glo:deg}.
Note that $\deg c$ is independent of the choice of $F$, that the choice of $F$ corresponds to a bundle automorphism of $\E$, and that $F$ can be chosen so that $c$ is independent of $s$.
If $c$ is sufficiently smooth then there is the following integral formula for $\deg c$:
\begin{equation}\label{eq:degc}
\deg c =\frac{1}{24\pi^2}\int_{\threeball}\trace (dcc^{-1})^3.
\end{equation}
To establish this formula it is easier to work with a map $c:S^3\rightarrow U(n)$ and prove that $\deg c$ is given by the same integral over $S^3$.
It is easy to check that the formula holds when $c$ is the standard map with $\deg c=1$:
\begin{equation*}
c(x)=\sum_0^3 \gamma_a x_a\quad\textrm{where~}\|x\|=1.
\end{equation*} 
To prove that the integral depends only on the homotopy class of $c$, consider a family of maps $c_t$ defined for all $t$ in some open interval in $\R$. 
The form
\begin{equation*}
\trace \partial_t(dc_t c_t^{-1})^3
\end{equation*}
is exact, so
\begin{equation*}
\partial_t \int_{S^3}\trace (dc_t c_t^{-1})^3 =0
\end{equation*}
and it follows that the integral is homotopy invariant. 
Finally, given some $c:S^3\rightarrow U(n)$ and some $k\in\Z$, a short calculation shows that 
\begin{equation*}
\trace [d(c^k)c^{-k}]=k\trace (dcc^{-1})+\textrm{exact terms.}
\end{equation*} 
This proves the integral formula for $\deg c$ for $c\in\Map(S^3,\Un)$. 
Identifying $\Auto E$ with $\Map(S^3,\Un)$ then gives~$\eqref{eq:degc}$. 

\begin{definition}\label{def:c2bdl}
Given a framed bundle $(\E,f)$ let $c_2(\E,f)=\deg c$\label{glo:c22}. 
We may also write $c_2(\E,f)[X]$. 
\end{definition}

We can equally well work with gauge group $\SUn$, in which case the clutching map $c$ takes values in $\SAuto E$\label{glo:SAutoE}, the group of special unitary automorphisms of $E$ that are the identity on the boundary. 
The same formula holds for $\deg c$.

\subsection{Boundary conditions for calorons}\label{sec:caloronbcs}

Let $\Ainf,\Phiinf$ be a $\Un$ connection and endomorphism on $\Einf$ satisfying~$\eqref{eq:Phiframed}$ and~$\eqref{eq:Aframed}$.

\begin{definition}\label{def:caloron}
Let $\A$ be a unitary connection on a framed bundle $(\E,f)$.
Then $\A$ is a \emph{$\Un$ caloron configuration\label{glo:framedcal} framed by $\Ainf,\Phiinf$} if
\begin{equation*}
\A\rest{\dX} = p^{\ast}\Ainf + p^{\ast}\Phiinf\dxo
\end{equation*}
where the framing $f$ is being used to identify $\E\rest{\dX}$ with $p^{\ast}\Einf$.
\end{definition}
We can define $\SUn$ caloron configurations in a similar way by equipping $\E$ with a parallel volume form. 
Note that the gauge transformations are (strictly periodic) unitary bundle automorphisms of $\E$. 
A gauge transformation $g$ acts on the framing $f$ by $f\mapsto f g^{-1}$, and acts on the connection in the usual way. 

Just as for monopoles, the boundary conditions for a caloron configuration determine (and are determined by) a set of boundary data $\bdarydata$.
In particular the Higgs field at infinity, $\Phiinf$, determines a set of monopole boundary data $\monbdarydata$ and we assume that $\mu_n<\ldots<\mu_1$.
In addition the caloron configuration is characterized by $\muo$ and by\label{glo:k0} 
\begin{equation*}
\ko=c_2(\E,f).
\end{equation*}
We require two further conditions:
\begin{equation}\label{eq:posweylchamber}
\mu_0 - (\mu_1 - \mu_n) > 0.
\end{equation}
and
\begin{equation}\label{eq:mppos}
\sum_{j=0}^p k_j \geq 0\ \textrm{for\ }p=0,1,\ldots,n
\end{equation}
The first condition is equivalent to saying that, regarding the caloron as a loop-group monopole as in Section~\ref{sec:loopgroup}, the Higgs field at infinity lies in the positive Weyl chamber of the Lie algebra. 
Garland and Murray~\cite{gar88} discuss this condition in greater detail. 
The second condition ensures that each block of Nahm data has positive rank. 
It can be derived from the spectral curve picture---see \cite[Section $4$]{gar88}. 

\begin{definition}\label{def:bdarydata}
A set of \emph{caloron boundary data} is a set $\bdarydata$\label{glo:bdarydata} where $\monbdarydata$ is a set of monopole boundary data, and the two conditions~$\eqref{eq:posweylchamber}$ and~$\eqref{eq:mppos}$ are satisfied. The boundary data is said to be \emph{principal}\label{glo:principal1} if
\begin{equation*}
\sum_1^p k_j \geq 0\ \textrm{for\ }p=1,\ldots,n.
\end{equation*}
\end{definition}
The condition of being principal is important in the context of the rotation map which we discuss in Section~\ref{sec:rotationbdary}, 
and is equivalent to saying that the lowest rank of any block of the Nahm data is $\ko$.

For each set of caloron boundary data we need the following quantities.
Let\label{glo:mp}
\begin{equation}\label{eq:constituentcharges}
m_p = \sum_0^p k_j
\end{equation}
for $p=1,\ldots,n$
and 
\begin{equation}\label{eq:constituentmasses}
\lambda_p = \mu_{p}-\mu_{p+1}
\end{equation}
for $p=1,\ldots,n-1$ and take $\lambda_n=(\mu_0+\mu_n)-\mu_1$.
In the physics literature $m_p$ and $\lambda_p$ are called the charges and masses of the `constituent monopoles' of the caloron respectively. 
Note that $\bdarydata$ is principal iff $m_n=\textrm{min}\{ m_1,\ldots,m_n \}$. 

\subsection{Framed quasi-periodic connections}\label{sec:quasi-per}

Let $\Ieps=(-\epsilon,\perflat+\epsilon)$ be some open neighbourhood of the interval $\Iper$ with coordinate $s$.
Let $q$\label{glo:q} be the projection $q:\Ieps\times\threeball \rightarrow \threeball$ and let $\Eq=q^\ast E$\label{glo:Eq}.
There is an obvious correspondence between caloron configurations and connections on $\Eq$---we spell out the details in this Section.
Let $\Ainf,\Phiinf$ satisfy~$\eqref{eq:Phiframed}$ and~$\eqref{eq:Aframed}$.

\begin{definition}\label{def:quasiper}
A $U(n)$ connection $\Aq$\label{glo:Aq} on $\Eq$ is \emph{quasi-periodic with clutching map $c$}\label{glo:quasiper} if 
\begin{equation*}
\Aq(\perflat+s) = (c^{-1})^\ast\Aq(s)
\end{equation*}
for some map
\begin{gather*}
c:(-\epsilon,\epsilon) \rightarrow \Auto E \\
c(s) : \Eq |_{s} \rightarrow \Eq |_{s+\perflat}.
\end{gather*}
We say that $\Aq$ \emph{clutches} with clutching function $c$. 
The map $c$ has a degree since it represents an element of $\pi_0(\Auto E)$.
\end{definition}

$\SUn$ quasi-periodic connections are defined in exactly the same way, except the clutching function $c$ takes values in $\SAuto E$.

\begin{definition}\label{def:framedquasiper}\label{def:framedquas}
A connection $\Aq$ on $\Eq$ is framed by $\Ainf,\Phiinf$ if 
\begin{equation*}
\Aq |_{\sphinf} = q^\ast\Ainf + q^\ast\Phiinf ds.
\end{equation*}
\end{definition}

There is a $1-1$ correspondence between caloron configurations and  quasi-periodic connections framed by $\Ainf,\Phiinf$ (up to bundle isomorphism).
Given a caloron configuration $\A$ on $(\E,f)$, extend $f$ by $F$ (as in Section \ref{sec:framedbundles}).
Let $\Aq=(F^{-1})^{\ast}\A$. 
Then $\Aq$ is a framed quasi-periodic connection with clutching function given by~$\eqref{eq:cintermsofF}$. 
Conversely, given a framed quasi-periodic connection $\Aq$ with clutching function $c$, quotienting by the action of $c$ gives a framed caloron configuration $\A$ on a framed bundle $(\E,f)$ with $c_2(\E,f)=\deg c$.
The framing $f$ is properly periodic on $\E$ because $c=1$ on $\sphinf$.
The correspondence is determined up to bundle automorphisms on $\Eq$ that are the identity on the boundary $\sphinf$, and periodic bundle automorphisms of $\E$.
Given a caloron configuration $\A$, we call the corresponding framed quasi-periodic connection $\Aq$ the \emph{quasi-periodic pull-back} of $\A$.

\subsection{Calorons as loops of monopoles}\label{sec:calaspath}

There is a correspondence between loops of monopoles `with a twist' and caloron configurations. 
Fix a set of monopole boundary data $\monbdarydata$ and let $A_\infty,\Phi_\infty$ be the connection and Higgs field on $\sphinf$ this determines.
Let $\monopoles$\label{glo:spcmon} denote the set of all $\Un$-monopole 
configurations which are framed by $A_\infty,\Phi_\infty$ 
(as defined by Definition~\ref{def:monopole}).  
$\monopoles$ is equipped with gauge group $\Auto E$, 
and $c\in\Auto E$ acts according to
\begin{equation*}
c(A) = cAc^{-1} - dcc^{-1},\quad c(\Phi)=c\Phi c^{-1}.
\end{equation*}
Now
$\pi_0(\Auto E) = \pi_3(\Un)= \Z$. 
Moreover, $\monopoles$ is an affine 
space, hence contractible, so
$\pi_1(\monopoles /\Auto E) = \pi_0(\Auto E) = \Z$.
Let $\curlyL\bdarydata$\label{glo:loopsofmon} denote the smooth (free) loops in $\monopoles /\Auto E$ with degree $\ko$ and parameterized by $s\in\Iper$.
Some care is needed to ensure loops are smooth across the ends of paths, so we note the following characterization of smooth loops.
Suppose $A(s),\Phi(s)$ is some path in $\monopoles$ such that
\begin{equation*}
A(\perflat)=c(A(0)),\ \Phi(\perflat)=c(\Phi(0))
\end{equation*}
for some $c\in\Auto E$. 
The path defines a smooth loop in $\monopoles /\Auto E$ if and only if it can be extended to a path defined for $s\in(-\epsilon,\perflat+\epsilon)$, for some small $\epsilon$, such that
\begin{equation}\label{eq:smoothclutch}
A(\perflat+s)=c(A(s)),\ \Phi(\perflat+s)=c(\Phi(s))
\end{equation}
for $s\in(-\epsilon,\epsilon)$. 
Note that $c$ is independent of $s$. 

There is a correspondence up to isomorphism between caloron configurations with boundary data $\bdarydata$ and elements of $\curlyL\bdarydata$, which follows immediately using the quasi-periodic pull-back of a caloron configuration.
A loop in $\monopoles$ whose ends are related by~$\eqref{eq:smoothclutch}$ determines a framed quasi-periodic connection via 
\begin{equation}\label{eq:univsplitting}
\Aq=A+\Phi ds
\end{equation}
and hence a caloron configuration with boundary data $\bdarydata$.
On the other hand, given a framed caloron configuration consider its framed quasi-periodic pull-back $\Aq$ with clutching map $c$.
As we have already seen, $\Aq$ can be chosen so that $c$ is independent of $s$. 
The splitting~$\eqref{eq:univsplitting}$ then determines an element of $\curlyL\bdarydata$. 
Note that the correspondence does not restrict to a correspondence between loops of monopoles satisfying the Bogomolny equation and anti-self-dual calorons (compare the Bogomolny equation~$\eqref{eq:Bogomolny}$ with the equation for loop-group monopoles~$\eqref{eq:GMSD}$). 
Also note the difference between the picture of a caloron as a twisted loop of framed monopole configurations (where the twisting occurs on the interior of $\threeball$) and the loop-group picture (where the twisting is at infinity). 

In the light of this correspondence, it is clear that a monopole configuration can be pulled-back from $\threeball$ to $\cyl$ to give a caloron configuration with $k_0=0$. 
We will need the following converse: 

\begin{lemma}\label{lem:caltomon}
Let $\A$ be a framed $U(n)$ caloron on a framed bundle with $c_2(\E,f)[X]=0$. 
Then there is a deformation $\B$ of $\A$ 
(through framed $U(n)$ caloron configurations), such that $\B$ is the 
pull-back of a monopole.
\end{lemma}

\proof
Let $\Aq$ be a quasi-periodic pull-back of $\A$. 
Then $\Aq$ has clutching map $c$ with $\deg c=0$. 
Let $c_{\textrm{ext}}$ be any smooth unitary automorphism of $\Eq$ satisfying 
\begin{equation*}
c_{\textrm{ext}}=
\begin{cases}
c & \textrm{on\ }(\per-\epsilon,\per+\epsilon)\times\threeball, \\
1 & \textrm{on\ }\Ieps\times\sphinf,\\
1 & \textrm{on\ }(-\epsilon,\epsilon)\times\threeball.
\end{cases}
\end{equation*}
Such an extension exists if and only if $\deg c =0$. 
Acting on $\Aq$ by $c_{\textrm{ext}}$, we reduce to the case $c\equiv 1$. If we define a connection and endomorphism $(A,\Phi)$ on $E$ by 
\begin{equation*}
A+\Phi~ds=\Aq(s=0) 
\end{equation*}
then 
\begin{equation*}
\B^q(s) = A+\Phi ds
\end{equation*}
is a framed quasi-periodic connection which is the pull-back of the monopole configuration $(A,\Phi)$. 
Moreover, $\A^q$ can be deformed to $\B^q$ through framed quasi-periodic connections with $c\equiv 1$ via the obvious linear path. 
\eproof

\subsection{Smoothness at the boundary}\label{sec:smooth}

Up to this point we have been deliberately vague about the precise degree of smoothness up to the boundary that we are assuming (\ie whether connections are continuous or smooth up to the boundary)---Definitions~\ref{def:monopole}, \ref{def:framedbdl}, \ref{def:c2bdl}, \ref{def:caloron}, and~\ref{def:framedquas} all make sense if we assume only continuity up to the boundary. 
In this Section we specify precise smoothness conditions for our objects. 
We also give a brief comparison of our boundary conditions with the asymptotic boundary conditions for calorons used by Garland and Murray \cite{gar88} and others. 

We will need the following spaces of functions:

\begin{definition}
$C^k_\chi(X)$\label{glo:Ckchi} is the space of functions $f$ on $X$ such that $f$ is smooth on $\Xo$, and for all $\alpha,\beta,\gamma$ and all $l\leq k$, $\partial_\chi^l \partial_{y_1}^\alpha \partial_{y_2}^\beta 
\partial_{x_0}^\gamma f$ is continuous up to the boundary. 
\end{definition}

\begin{definition}
A $1$-form $\alpha$ on $X$ is $C_\chi^{0,1}$\label{glo:C01chi} if the $d\chi$ component is $C_\chi^0$ and the other components are $C_\chi^1$.  
\end{definition}

Let $\A$ be a caloron configuration on a framed bundle $(\E,f)$ that is continuous up to the boundary, and framed by $\Ainf,\Phiinf$. 
Then there exist local gauges on $\E$ defined for sufficiently small $\chi\geq 0$ and all $\xo$, in which
\begin{equation}\label{eq:smo1}
\A_{\xo} = \diag( i\mu_1, \ldots, i\mu_n ),
\end{equation}
and 
\begin{equation}\label{eq:smo2}
\A_{{y_j}} = \diag( \langle \partial_{y_j} e_1, e_1 \rangle, \ldots, \langle \partial_{y_j} e_n, e_n \rangle ),\quad j=1,2,
\end{equation}
on $\sphinf$, where $e_1, \ldots, e_n$ is a local trivialisation of $\Einf$ respecting the decomposition into eigenbundles, and such that $\A_{x_0},\A_{y_1},\A_{y_2},\A_\chi$ are continuous up to the boundary. 
Here $\A_{\xo},\A_{y_j},\A_\chi$ are the matrices representing $\A$ in the fixed gauge. 
Conversely, if such gauges exist for some connection $\A$ on $\E$, then $\A$ is framed by $\Ainf,\Phiinf$ and some map $f:\E|_{\partial X}\rightarrow p^\ast\Einf$. 
We restrict attention to caloron configurations for which there exist local gauges satisfying~$\eqref{eq:smo1}$ and~$\eqref{eq:smo2}$ in which $\A$ is $C_\chi^{0,1}$. 
We call these \emph{$C_\chi^{0,1}$ caloron configurations}.

In addition, we also require that, on the boundary, $\A_\chi$ is diagonal and independent of $\xo$ in these gauges. 
We impose this condition to ensure that $\A_{\xo}$ has the following asymptotic behaviour: 
\begin{equation}\label{eq:asympphi}
\A_{\xo} = \diag(i\mu_1,\ldots,i\mu_n) -\frac{\chi}{2}\diag(ik_1,\ldots,ik_n)+
\textrm{higher order terms}
\end{equation}
when $\A$ is anti-self-dual. 
To obtain this, extend the framing $f$ to a neighbourhood of the boundary, perform the ``$3+1$'' decomposition~$\eqref{eq:calasloop1}$, and
consider the $d\chi$ component of the anti-self-duality equation~$\eqref{eq:GMSD}$:
\begin{equation}\label{eq:loopbog}
\ast_3 F_A = \nabla_A\Phi-\partial_{\xo}A.
\end{equation} 
Working in the gauges described above, on the boundary $\partial X$ the $d\chi$ component of the LHS of~$\eqref{eq:loopbog}$ is $-\frac{1}{2}\diag(ik_1,\ldots,ik_n)$, because $\Ainf$ is the standard connection on each constituent line bundle of $\Einf$. 
On the RHS of~$\eqref{eq:loopbog}$ the second term vanishes on the boundary, while the $d\chi$ component of the first is $\partial_\chi \A_{\xo}$. 
Equating the two sides of~$\eqref{eq:loopbog}$ shows that the $O(\chi)$ term of $\A_{\xo}$ is $-\frac{1}{2}\diag(ik_1,\ldots,ik_n)$, and so we obtain~$\eqref{eq:asympphi}$. 
We will need the expansion~$\eqref{eq:asympphi}$ in Chapter $4$ when we construct Nahm data from a given caloron.

\begin{definition}
Given a set of $\Un$ caloron boundary data, let $\spc{C}{\bdarydata}$\label{glo:spccal} be the set of gauge equivalence classes of $U(n)$ $C_\chi^{0,1}$ caloron configurations $(\E,\A)$ satisfying these smoothness conditions, whose boundary conditions are determined by $\bdarydata$.
Let $\spcSD{C}{\bdarydata}$\label{glo:spcSDcal} denote the subset of anti-self-dual caloron configurations modulo gauge. 
When the boundary data is $SU(n)$ we restrict $\mathcal{C}$ and ${\mathcal{C}}^\ast$ to $SU(n)$ configurations. 
\end{definition}
 
It is convenient at this point to compare our boundary conditions for calorons with the `BPS' decay conditions used in \cite{gar88}. 
Given a framed bundle $(\E,f)$, we can extend $f$ to a neighbourhood of the boundary $\partial X$, and perform the ``$3+1$'' decomposition~$\eqref{eq:calasloop1}$ to define
a loop of connections $A$ on $E$ over this neighbourhood, and a loop of endomorphisms $\Phi$. 
Garland and Murray \cite{gar88} impose the condition that $A$ and $\Phi$ satisfy the `BPS' boundary conditions for monopoles uniformly in $\xo$. 
(Various versions of the `BPS' monopole boundary conditions exist: see \cite{mur84} and \cite{hit82} for example.)
For a $C_\chi^{0,1}$ framed caloron, $\nabla_{y_j}\Phi$ and $\partial_{x_0}A_j$ are $O(\chi)$ as $\chi\rightarrow 0$, while $\nabla_\chi\Phi$ and $\partial_{\xo}A_\chi$ are $O(1)$. 
Using the form of the metric in equation~$\eqref{eq:metricinchi}$ it follows that
\begin{equation*}
\| \nabla_A \Phi \|=O(\chi^2) \qquad\textrm{and\ }
\| \partial_{x_0}A \|=O(\chi^2)
\end{equation*}
so
\begin{equation}\label{eq:decayforfred}
\| \nabla_A\Phi -\partial_{x_0}A \|=O(\chi^2).
\end{equation}
This is a gauge invariant quantity---it does not depend on the choice of framing $f$. 
Thus our boundary conditions imply the following:
\begin{itemize}
\item there are local gauges in which 
$\Phi = \diag(i\mu_1,\ldots,i\mu_n) -\frac{\chi}{2}\diag(ik_1,\ldots,ik_n)+$ higher order terms, 
\item $\| \nabla_A \Phi \|=O(\chi^2)$,
\item $\frac{\partial \| \Phi \|}{\partial y_j}=O(\chi^2)$ for $j=1,2$,
\end{itemize}
and these estimates are uniform in $\xo$. 
But these conditions are just the BPS monopole boundary conditions described in \cite{hit82}.

\subsection{Chern-Weil theory}\label{sec:chernweil}

Given a caloron configuration $\A\in\spc{C}{\bdarydata}$ on a framed bundle $\E$, we will need the `Pontryagin integral'\label{glo:Pontryagin}
\begin{equation}\label{chargeintegral}
\int_{\So\times\threeball} \chtwo (\E,\A ) = -\frac{1}{8\pi^{2}}\int_{\cyl} \trace 
F_{\A}\wedge F_{\A},
\end{equation}
which we calculate in this section. 
Here $\chtwo(\E,\A)$\label{glo:ch} denotes the second order term of the Chern character of $\E$ (which also depends on the connection $\A$ since we are working on a manifold with boundary). 
The integral has been evaluated by different means in \cite{gro81} and \cite{gar88}.
Pulling $\A$ back to a framed quasi-periodic connection $\Aq$ on $\Eq$ gives:
\begin{equation*}
-\frac{1}{8\pi^{2}}\int_{\So\times\threeball} \trace F_{\A}\wedge F_{\A} = 
-\frac{1}{8\pi^{2}}\int_{\Iper\times\threeball} \trace F_{\Aq}\wedge F_{\Aq}
\end{equation*}
Using the familiar trick of writing
\begin{equation*}
\trace F_{\Aq}\wedge F_{\Aq} = d~\trace \{d\Aq\wedge\Aq +\frac{2}{3}\Aq\wedge\Aq\wedge\Aq \}
\end{equation*}
the integral becomes an integral over the boundary of the rectangle $\Iper\times\threeball$:
\begin{multline}\label{eq:chernsimons}
-\frac{1}{8\pi^{2}}\int_{\Iper\times\threeball} \trace F_{\Aq}\wedge F_{\Aq} =
\\ -\frac{1}{8\pi^{2}}\int_{\partial (\Iper\times\threeball)} \trace \{d\Aq\wedge\Aq +\frac{2}{3}\Aq\wedge\Aq\wedge\Aq\}.
\end{multline}
Working on the boundary requires the smoothness assumptions made in Section~\ref{sec:smooth}.
Regarding $\Aq$ as a path in $\monopoles / \Auto E$, we obtain a path $A(s),\Phi(s)$, whose ends are related by~$\eqref{eq:smoothclutch}$.
Evaluating~$\eqref{eq:chernsimons}$ on the component $(\partial \Iper)\times\threeball$ and using the clutching formula gives
\begin{multline*}
-\frac{1}{8\pi^{2}}\int_{(\partial \Iper)\times\threeball} \trace \{d\Aq\wedge\Aq +\frac{2}{3} \Aq\wedge\Aq\wedge\Aq\} = \\
-\frac{1}{24\pi^2}\int_{\threeball}\trace (dcc^{-1})^3 + \frac{1}{8\pi^2}
\int_{\threeball}d~\trace \{A(0)c^{-1}dc\}.
\end{multline*}
The first term is $-\deg c=-c_2(\E,f)[X]$, and the second can be re-expressed as an integral on $\sphinf$ which vanishes because $c=1$ on $\sphinf$.
On the other component of the boundary we obtain
\begin{multline*}
 -\frac{1}{8\pi^{2}}\int_{\Iper\times\sphinf} \trace \{d\Aq\wedge\Aq +\frac{2}{3}\Aq\wedge\Aq\wedge\Aq \}= \\
 -\frac{1}{8\pi^{2}}\int_{\Iper\times\sphinf} \trace \{
2F_{A}\wedge\Phi ds - dA\wedge\Phi ds + A\wedge d\Phi\wedge ds +
\partial_s A\wedge A\wedge ds\}.
\end{multline*}
The final term vanishes because $\partial_s A=0$ on $\sphinf$, and the sum of the  middle two terms is exact, so does not contribute.
Since the connection $A_\infty$ is compatible with $\Phiinf$, the first term is given by\label{glo:c1}
\begin{equation*}
-\frac{1}{8\pi^{2}}\int_{\Iper\times\sphinf} \trace
2F_{A}\wedge\Phi ds = - \frac{1}{\muo}\sum_{1}^{n}\mu_j 
c_{1}(E_{\mu_j})[\sphinf]
\end{equation*} 
where $E_{\mu_j}\subset\Einf$ is the eigenbundle of $\Phiinf$ with 
eigenvalue $i\mu_j$.
Putting the terms together, we arrive at the expression
\begin{align} 
\int_{\So\times\threeball} \chtwo(\E,\A) &= -c_2(\E,f)[X] - \frac{1}{\muo}\sum_{1}^{n}\mu_j  
c_{1}(E_{\mu_j})[\sphinf] \label{eq:upcaloroncharge} \\
& = -k_0 - \frac{1}{\mu_0}(\mu_1 k_1 + \cdots +\mu_n k_n).
\label{eq:caloroncharge}
\end{align}

Just like regular instantons, calorons minimize the action\label{glo:action} within each topological class. 
Expanding $\| F_\A+\ast F_\A \|$ and using~$\eqref{eq:ipandstar}$ gives
\begin{equation*}
\textrm{action}= \| F_\A \|^2 =
\frac{1}{2}\| F_\A +\ast F_\A \|^2 -\int_{\cylo}\trace F_\A\wedge F_\A.
\end{equation*}
The second term is constant within each topological class, so the action is minimized when $F_\A + \ast F_\A=0$ \ie when $\A$ is ASD. 
A similar result is obtained by regarding a $SU(n)$ caloron as a $\Lh SU(n)$ monopole, evaluating the energy $\| F_{\hat{A}} \|^2 + \| \nabla_{\hat{A}}\hat{\Phi} \|^2$, and performing the `Bogomolny trick' (re-arranging in terms of $\| \ast_3 F_{\hat{A}}- \nabla_{\hat{A}}\hat{\Phi}\|$). 
%Given the pullback $\Aq$ of a caloron, and regarding it as a loop of monopole configurations $(A,\Phi)$, we obtain
%\begin{equation*}
%F_{\Aq} = F_A +(\nabla_A \Phi - \partial_{s}A)\wedge dx_0
%\end{equation*}
%and
%\begin{equation*}
%\trace F_{\Aq} \wedge F_{\Aq} = 2\trace F_A\wedge (\nabla_A \Phi - \partial_{s}A)\wedge dx_0.
%\end{equation*}
%We include these equations only for comparison with the integrals in \cite[Section $1$]{gar88}. 

\section{The rotation map}\label{sec:rotationbdary}

Up to this point we have considered two caloron configurations to be equivalent if related by a strictly periodic bundle isomorphism, or, in other words, we have taken gauge transformations to be strictly periodic. 
However, we can consider `large' gauge transformations\label{glo:largeGT}---transformations that are non-periodic but leave the caloron strictly periodic. 
In the quasi-periodic picture, these are equivalent to bundle automorphisms of $\Eq$ that are not necessarily the identity at infinity, and such gauge transformations affect the framing (recall the final paragraph of Section~\ref{sec:quasi-per}). 
We therefore expect a large gauge transformation (if such an object exists) to be a map between calorons with different framings, and possibly with different boundary data. 

On the other hand, consider a caloron constructed from some set of Nahm data. (For the present we assume we have a construction like that conjectured in the Introduction.) 
The choice of origin on the circle $T^\ast = \Sdual=\R / \mu_0\Z$ should have no effect on the caloron constructed from the Nahm data (as a connection over $\rfour$) because the inner product defined on sections of the Nahm data is independent of the origin. 
However, changing the origin \emph{does} change the values $\mu_1,\ldots,\mu_n$ and therefore the framing of the caloron obtained. 
The Nahm construction gives a connection over $\rfour$ which we quotient by some action of $\Z$ to obtain a framed connection on $\cylo$, and there is some freedom as to how we take the quotient and make the framing. 
The choice of origin for the circle $T^\ast = \Sdual$ corresponds in some sense to a choice of the quotient and framing. 
Furthermore, calorons obtained by different quotients are related by large gauge transformations. 

All this will be made rigorous later, but we can draw the following conclusion: given a caloron with boundary data $B$ we expect that by applying a large gauge transformation we can obtain a caloron with boundary data $B'$, where $B$ and $B'$ are related by shifting the origin on $\Sdual=\R / \mu_0\Z$. 
We call this the `rotation map'\label{glo:rotation1} and prove its existence in this Section. 
We will explore its relation to the Nahm transform later. 
Lee \cite{lee98} has described the rotation map for calorons with vanishing monopole charges, and explained it in representation theoretical terms~\cite{lee98b}. 

%%%%%%%%%%%%%%%%%%%%%%%%%%%%%%%%%%%%%%%%%%%%%%%%%%%%%%%%%%%%%%%%%%%%%%%%%%%%%%%
% FIGURES: rotating Nahm data
%%%%%%%%%%%%%%%%%%%%%%%%%%%%%%%%%%%%%%%%%%%%%%%%%%%%%%%%%%%%%%%%%%%%%%%%%%%%%%%
\begin{figure}
\setlength{\unitlength}{0.75cm}
\begin{center}
\begin{picture}(14,6)(-2,-1)
%axis
\put(-2,0){\vector(1,0){14}}
\put(12.2,-0.1){$\xi$}
%\put(0,0){\line(0,-1){0.25}}
%\put(10,0){\line(0,-1){0.25}}

%block 1
\put(-1,0){\line(0,1){4}}
\put(-1,4){\line(1,0){2}}
\put(1,4){\line(0,-1){4}}
\put(-0.75,4.5){$m_{n-1}=k_{0} + \hdots +k_{n-1}$}

%block 2
\put(1,0){\line(0,1){3}}
\put(1,3){\line(1,0){2}}
\put(3,3){\line(0,-1){3}}
\put(1.25,3.5){$m_{n-2}=k_{0} + \hdots +k_{n-2}$}

%block 4
\put(5,0){\line(0,1){2}}
\put(5,2){\line(1,0){2}}
\put(7,2){\line(0,-1){2}}
\put(4.75,2.5){$m_{1}=k_{0}+k_{1}$}

%block 5
\put(7,0){\line(0,1){1}}
\put(7,1){\line(1,0){2}}
\put(9,1){\line(0,-1){1}}
\put(7.25,4.5){$m_{0}=k_{0}$}

%block 6
\put(9,0){\line(0,1){4}}
\put(9,4){\line(1,0){2}}
\put(11,4){\line(0,-1){4}}

%mu labels
\put(-1.5,-1){\begin{picture}(12,1)(-1,0)
              \put(-0.8,0){$\mu_{n}$}
              %\put(0.4,0){$0$}
              \put(1,0){$\mu_{n-1}$}
              \put(3,0){$\mu_{n-2}$}
              \put(4.3,0){$\hdots$}
              \put(5.25,0){$\mu_{2}$}
              \put(7.25,0){$\mu_{1}$}
              \put(8.5,0){$\muo + \mu_{n}$}
              %\put(10.25,0){$\muo$}
              \put(10.75,0){$\muo+\mu_{n-1}$}
              \end{picture}}

\end{picture}
\end{center}
\caption{Typical $U(n)$ Nahm data}
\label{fi:sunNahmdata}
\end{figure}
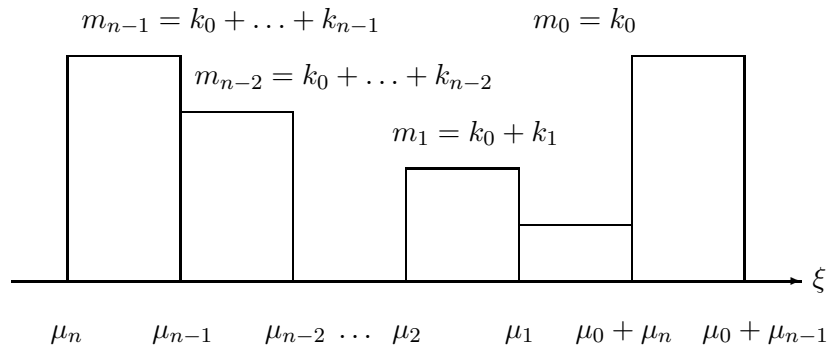

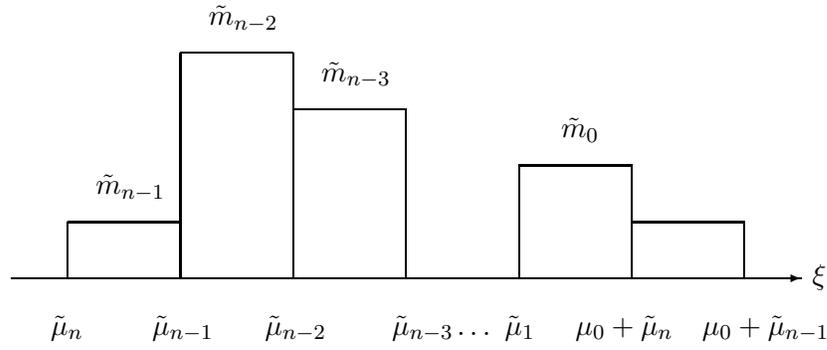
\begin{figure}
\setlength{\unitlength}{0.75cm}
\begin{center}
\begin{picture}(14,6)(-2,-1)
%axis
\put(-2,0){\vector(1,0){14}}
\put(12.2,-0.1){$\xi$}
%\put(0,0){\line(0,-1){0.25}}
%\put(10,0){\line(0,-1){0.25}}

%block 1
\put(-1,0){\line(0,1){1}}
\put(-1,1){\line(1,0){2}}
\put(1,1){\line(0,-1){1}}
\put(-0.5,1.5){$\tilde{m}_{n-1}$}

%block 2
\put(1,0){\line(0,1){4}}
\put(1,4){\line(1,0){2}}
\put(3,4){\line(0,-1){4}}
\put(1.5,4.5){$\tilde{m}_{n-2}$}

%block 3
\put(3,0){\line(0,1){3}}
\put(3,3){\line(1,0){2}}
\put(5,3){\line(0,-1){3}}
\put(3.5,3.5){$\tilde{m}_{n-3}$}

%block 5
\put(7,0){\line(0,1){2}}
\put(7,2){\line(1,0){2}}
\put(9,2){\line(0,-1){2}}
\put(7.75,2.5){$\tilde{m}_{0}$}

%block 6
\put(9,0){\line(0,1){1}}
\put(9,1){\line(1,0){2}}
\put(11,1){\line(0,-1){1}}

%mu labels
\put(-1.5,-1){\begin{picture}(12,1)(-1,0)
              \put(-0.8,0){$\tilde{\mu}_{n}$}
              %\put(0.4,0){$0$}
              \put(1,0){$\tilde{\mu}_{n-1}$}
              \put(3,0){$\tilde{\mu}_{n-2}$}
              \put(5.25,0){$\tilde{\mu}_{n-3}$}
              \put(6.4,0){$\hdots$}
              \put(7.25,0){$\tilde{\mu}_{1}$}
              \put(8.5,0){$\muo + \tilde{\mu}_{n}$}
              %\put(10.25,0){$\muo$}
              \put(10.75,0){$\muo+\tilde{\mu}_{n-1}$}
              \end{picture}}

\end{picture}
\end{center}
\caption{The same $U(n)$ Nahm data after rotation}
\label{fi:rotNahmdata}
\end{figure}
%%%%%%%%%%%%%%%%%%%%%%%%%%%%%%%%%%%%%%%%%%%%%%%%%%%%%%%%%%%%%%%%%%%%%%%%%%%%%%%

Fix a set of boundary data $B=\bdarydata$ and consider the following map:
\begin{align*}
\lambda_p &\mapsto\lambda_{p+1},\ m_p\mapsto m_{p+1},\ 
\textrm{for\ }p=1,\ldots,n-1,\\
\lambda_n &\mapsto\lambda_{1},\ m_n\mapsto m_{1},
\end{align*}
where $m_p$ and $\lambda_p$ are defined by~$\eqref{eq:constituentcharges}$ and~$\eqref{eq:constituentmasses}$.
This map permutes---or rotates---the `constituent monopoles'. 
We have defined it so that it corresponds to a rotation of the Nahm data---the map can be obtained by comparing Figures~\ref{fi:sunNahmdata} and~\ref{fi:rotNahmdata}. 
For example, the highest rank block has width $\lambda_{n-1}=\mu_{n-1}-\mu_n$ before rotation and width $\tilde{\lambda}_{n-2}=\tilde{\mu}_{n-2}-\tilde{\mu}_{n-1}$ after rotation, so $\lambda_{n-1}=\mu_{n-1}-\mu_n = \tilde{\mu}_{n-2}-\tilde{\mu}_{n-1}=\tilde{\lambda}_{n-2}$. 
In terms of $k_p$ and $\mu_p$ the map is given by:
\begin{align}
  k_{0}  & \mapsto \tilde{k}_{0}= k_{0}+k_{1}
  \qquad & & \textrm{$\muo$ is fixed,}  \notag \\
  k_{1}  & \mapsto \tilde{k}_{1}= k_{2}      
  \qquad & \mu_{1} & \mapsto \tilde{\mu}_{1}={\muo} / {n}+\mu_{2}\notag \\
         & \cdots \qquad & & \cdots \notag\\
  k_{n-2}& \mapsto \tilde{k}_{n-2}= k_{n-1}    
  \qquad & \mu_{n-2} & \mapsto\tilde{\mu}_{n-2}= {\muo} / {n}+\mu_{n-1}\notag \\
  k_{n-1} & \mapsto \tilde{k}_{n-1}= k_n
  \qquad & \mu_{n-1} & \mapsto\tilde{\mu}_{n-1}= {\muo} / {n} +\mu_{n}\notag \\
  k_{n} & \mapsto \tilde{k}_{n}= k_{1}    
  \qquad & \mu_{n} & \mapsto\tilde{\mu}_{n}=\mu_{1} -{(n-1)\muo} / {n}.
\label{eq:defnrotbdary}
\end{align}
It is easy to check that the result is a new set of boundary data (\ie it satisfies the conditions of Definition~\ref{def:bdarydata}), 
hence we have a map $\rotbdary$\label{glo:rotbdary} defined on sets of caloron boundary data.
Note that $(\rotbdary)^n$ is the identity and that the quantity  
$\mu_0 k_0 +\ldots+\mu_n k_n$
remains constant under the action of the rotation.
Each orbit under $\rotbdary$ contains at least one set of principal boundary data, corresponding to $m_n$ being the lowest rank (recall Definition~\ref{def:bdarydata} and the remarks following it).

Our aim is to construct a map $\rotcal$\label{glo:rotcal} on caloron configurations that changes the boundary data in the same way as $\rotbdary$.
In other words we want a map
\begin{equation*}
\rotcal : \spc{C}{(B)}\longrightarrow\spc{C}{(\rotbdary B)}.
\end{equation*}
Fix a set of boundary data $B$ and some framed caloron configuration $(\E,\A)\in\spc{C}{(B)}$. 
Let $\Ainf,\Phiinf$ be the connection and Higgs field on $\sphinf$ fixed by $B$.
The map $\rotbdary$ is a large gauge transformation on the quasi-periodic pull-back of the caloron configuration. 
Let $(\Eq,\Aq)$ be the quasi-periodic pull-back of $(\E,\A)$ in the sense of Section~\ref{sec:quasi-per}. 
Let $c:(-\epsilon,\epsilon)\rightarrow\Auto E$ be the clutching map.
Start by defining a family of maps $\rho(s):\Einf\rightarrow\Einf$ for $s\in(-\epsilon,\perflat+\epsilon)$, by
\begin{equation*}
\rho(s)=
\begin{cases}
\exp (i\muo s(n-1) /n) & \textrm{on\ }E_{\mu_1}\ 
\textrm{(the eigenbundle of $\Einf$ with eigenvalue $i\mu_1$),} \\
\exp (-i\muo s/n) & \textrm{on the other eigenbundles.}
\end{cases}
\end{equation*}
Hence $\rho(s)\in SU(n)$, and $\rho(\perflat)=\exp(-2\pi i / n)\textrm{id}=\omega$.
Note that $\omega$ lies in the centre of $\SUn$ and acts trivially as a bundle automorphism. 
The rotation map can be thought of as an action of the centre of $SU(n)$ (which is $\Z_n$) on the space of $SU(n)$ caloron configurations. 
Next we extend $\rho$ arbitrarily (but smoothly) to the interior of $\threeball$ to obtain a family of maps $\rho(s):E\rightarrow E$. 
Now $\rho$ defines a bundle automorphism of $\Eq$---but it does not necessarily define a (periodic) automorphism of $\E$.
 
Consider the action of $\rho$ on $\Aq$.
Our claim is that $\rho(\Aq)$ is the pull-back of an element of $\spc{C}{(\rotbdary B)}$: 
we have to show $\rho(\Aq)$ is framed correctly and that it clutches correctly.
Split $\Aq$ as $\Aq=A+\Phi ds$ using the framing at infinity; we know that on $\sphinf$, $A(s)=\Ainf$ and $\Phi(s)=\Phiinf$.
But $\rho$ acts on $\Phi$ by
\begin{equation*}
\rho(\Phi)=\rho\Phi\rho^{-1}-\frac{\partial\rho}{\partial s}\rho^{-1}.
\end{equation*}
It is easy to check that $\rho(\Phi)|_{\sphinf}$ is independent of $s$ and has eigenvalues $i\tilde{\mu}_1,\ldots,i\tilde{\mu}_n$ defined by~$\eqref{eq:defnrotbdary}$.
Moreover the eigenbundle with eigenvalue $i\tilde{\mu}_j$ has Chern class $\tilde{k}_j$, so $\rho(\Phi)$ is framed in the desired manner.
The map $\rho$ preserves the eigenbundles of $\Phiinf$, and since $\Ainf$ is compatible with the decomposition of $\Einf$ into eigenbundles, $\rho(A)|_{\sphinf}=\Ainf$. 
Hence $\rho(A)$ is also framed in the desired manner.
It remains to show that $\rho(\Aq)$ clutches correctly.
Now
\begin{equation*}
\rho(\Aq)_{(\perflat+s)}=\rho_{(\perflat+s)}c\rho^{-1}_{(s)}\rho(\Aq)_{(s)}
\end{equation*}
so $\rho(\Aq)$ has clutching function
\begin{equation*}
c_\rho = \rho_{(\perflat+s)}c\rho^{-1}_{(s)}.
\end{equation*}
However, $c_\rho |_{\sphinf} = \omega$ so, as it stands, $\rho(\Aq)$ does not clutch correctly (the clutching function should be $1$ on $\sphinf$).
But $\omega$ acts trivially as a gauge transformation since it is in the centre of $\SUn$, so if we redefine
\begin{equation}\label{eq:rotclutch}
c_\rho = \omega^{-1}\rho_{(\perflat+s)}c\rho^{-1}_{(s)}
\end{equation}
it becomes a well-defined clutching function.
We want to show $\deg c_\rho = k_0+k_1$ so that $\rho$ has the correct action on $k_0$. 
We do this indirectly by considering the action of the caloron configurations.
The map $\rho$ is a bundle isomorphism on $\Eq$ so it preserves the quantity
\begin{equation*}
\frac{1}{8\pi^2}\int_{\Iper\times\threeball}\trace F_{\Aq}\wedge F_{\Aq}.
\end{equation*}
Using~$\eqref{eq:caloroncharge}$ this implies that
\begin{equation*}
\muo k_0 + \mu_1 k_1+\cdots+\mu_n k_n =
\muo\deg c_\rho + \tilde{\mu}_1\tilde{k}_1 +\cdots+\tilde{\mu}_n\tilde{k}_n.
\end{equation*}
Hence $\deg c_\rho = k_0 + k_1=\tilde{k}_0$, and this completes the proof of the claim.
Note that the choice of extension of $\rho$ corresponds to a bundle automorphism on $\E$, so that $\rotcal$ is really defined on isomorphism classes of connections.

Having defined $\rotcal$ we will consider its relation to the Nahm transform in subsequent Chapters. 
My thanks go to Michael Murray for a useful exchange of emails about the definition of the rotation map $\rotcal$.

\section{Boundary conditions from the loop group point of view}\label{sec:loopbcs}

The boundary condition for $SU(n)$ monopoles (Definition~\ref{def:monopole}) implies that there is a gauge at infinity in which the Higgs field, $\Phiinf$, lies in some adjoint orbit of $SU(n)$. 
Let $\hat{A},\hat{\Phi}$ be a $\Lh SU(n)$ monopole configuration on $\threeball$. 
The corresponding boundary condition is that
%\begin{quote}
%there exists a gauge at infinity in which $\hat{\Phi}_\infty:=\hat{\Phi}|\sphinf$ lies in an adjoint orbit of $\Lh SU(n)$ on $\Lh \sun$.
%\vspace{-5.0ex}
%\begin{flushright}
%$(\dag)$
%\end{flushright}
%\end{quote}
\begin{multline}\label{eq:loopbc}
\textrm{there exists a gauge at infinity in which $\hat{\Phi}_\infty:=\hat{\Phi}|_{\sphinf}$ lies in an adjoint}\\ 
\textrm{orbit of $\Lh SU(n)$ on $\Lh \sun$.}
\end{multline}
Our aim is to interpret this boundary condition in terms of the caloron $\A$ corresponding to $\hat{A},\hat{\Phi}$ and compare it with Definition~\ref{def:caloron}. 
This attempt to `justify' our boundary conditions is not used at any point later, and is simply intended as a comparison of the two view-points. 

The first problem is to identify the adjoint orbits of $\Lh SU(n)$ on $\Lh \sun$. 
From~$\eqref{eq:AdLGLg}$, the adjoint action of $\Lh SU(n)$ is the action of $LSU(n)$ followed by a rotation in $\theta$. 
The following Proposition determines the orbits under the action of $LSU(n)$. 

\begin{proposition}[Pressley and Segal \cite{pre86}]
Suppose $G$ is a compact simply connected Lie group with Lie algebra $\curlyg$. 
Let $\lambda\in\R$ be some non-zero constant. 
Given $(\xi,i\lambda)\in\Lh\curlyg$, solutions $h:\R\rightarrow G$ to 
\begin{equation*}
\frac{\partial h}{\partial \theta}h^{-1}= -\lambda^{-1}\xi,\quad h(0)=1
\end{equation*}
satisfy $h(\theta+2\pi)=h(\theta)M_\xi$ for some $M_\xi \in G$ called the holonomy of $(\xi,i\lambda)$. 
The map
\begin{equation*}
Ad_{LG}(\xi,i\lambda)\mapsto Ad_{G} M_\xi
\end{equation*} 
is an isomorphism between the adjoint orbits of $LG$ on $\Lh\curlyg$ and conjugacy classes in $G$. 
\end{proposition}

Fix a $\Lh SU(n)$ monopole configuration $\hat{A}=(A(\theta),0),\hat{\Phi}=(\Phi(\theta),i\muo)$ on a bundle $E\rightarrow\rthree$ and let $(\E,\A)$ be the corresponding caloron configuration so that $\E=p^\ast E$ and $\A$ is given by~$\eqref{eq:calasloop1}$. 
The boundary condition $\eqref{eq:loopbc}$ says that there is a trivialisation of $E|_{\sphinf}$ in which $\hat{\Phi}_\infty$ lies in an adjoint orbit of $\Lh SU(n)$. 
Since the adjoint action of $\Lh SU(n)$ is the action of $LSU(n)$ followed by a rotation in $\theta$, the orbits of the two groups are the same, because two elements of $\Lh\curlyg$ related by a rotation in $\theta$ have the same holonomy so lie in the same orbit of $LG$. 
For each $y\in\sphinf$, the holonomy map defined in the Proposition takes $\hat{\Phi}_\infty(y)\in\Lh\sun$ to $M(y)\in SU(n)$ so that $M(y)$ lies in a fixed conjugacy class of $SU(n)$ as $y$ varies. 
Thus
\begin{equation*}
M(y)= \gamma(y) \bar{M}\gamma^{-1}(y)
\end{equation*}
for some fixed $\bar{M}$ which we can assume is diagonal, where $\gamma:\sphinf\rightarrow SU(n) / \textrm{Stab}~\bar{M}$. 
Define $\bar{m}$ by
\begin{equation*}
\bar{M} = \exp \big(-\frac{2\pi\bar{m}}{\mu_0}\big)
\end{equation*}
and let
\begin{equation*}
m(y) = \gamma(y) \bar{m}\gamma^{-1}(y).
\end{equation*}
(Note Stab $\bar{M}=$ Stab $\bar{m}$ because $\bar{M}$ and $\bar{m}$ are diagonal.)
Clearly $m(y)$ lies in a fixed adjoint orbit of $SU(n)$ as $y$ varies. 
By construction, the constant loop $(m(y),i\muo)\in\Lh\sun$ and $\hat{\Phi}_\infty(y)$ have the same holonomy for each $y$, and so lie in the same adjoint orbit of $LSU(n)$. 
In other words, there is a gauge transformation $g:\sphinf\rightarrow LSU(n)$ taking $\hat{\Phi}_\infty$ to $(m,i\muo)$. 
Thus we have constructed a trivialisation of $\E |_{\partial X}$ in which $\Phi=m$ where $m:\sphinf\rightarrow\sun$ lies in a fixed adjoint orbit of $SU(n)$ as $y$ varies. 

Since a trivialisation at infinity is really a framing, the interpretation of $\eqref{eq:loopbc}$ in terms of calorons can therefore be stated as follows: 
\begin{quote}
given a connection $\A$ on $\E$ there is a framing $f:\E|_{\partial X}\rightarrow p^\ast E$ in which 
\begin{equation*}
\A = A(\xo) +p^\ast\Phiinf dx_0
\end{equation*}
where $A(\xo)$ is a loop of connections on $E$ and $\Phiinf$ has eigenvalues independent of $y\in\sphinf$. 
\end{quote}
In other words, the $dx_0$ component of $\A$ is framed at infinity. 
Comparing this with Definition~\ref{def:caloron}, we have successfully derived a weak version of the caloron boundary conditions from~$\eqref{eq:loopbc}$. 

\chapter{From Nahm Data to Calorons}

We present the construction of calorons from Nahm data. 
Section~\ref{sec:nahmformon} contains material from \cite{hur89} on the construction of $SU(n)$ monopoles from Nahm data, including the precise conditions imposed on the Nahm data at singularities, which we use in our definition of Nahm data for calorons. 
In Section~\ref{sec:calnahmdata} we prove that the connection constructed from our caloron Nahm data is periodic and ASD. 
As we explained in Section~\ref{sec:overview}, the main difficulty in the construction of calorons lies in recovering the boundary conditions. 
This occupies Sections~\ref{sec:modelop} to~\ref{sec:deform} and follows the deformation method outlined in Section~\ref{sec:overview}.

\section{The Nahm transform for $SU(n)$ monopoles}\label{sec:nahmformon}

\subsection{Nahm data for $SU(n)$ monopoles.}\label{sec:nahmdataformon}

Nahm data for $SU(n)$ monopoles are defined in \cite{hur89} and \cite{hur89b}, and the following definition is taken almost directly from these papers.
Given a set of boundary data $\monbdarydata$, define $m_p=k_1 +\cdots+k_p$ for $p=1,\ldots,n-1$ and fix the conventions $m_0=m_n=0$.
A set of monopole Nahm data consists of the following:
\begin{description}
\item[Bundles:] hermitian vector bundles $X_{p}$\label{glo:Xp3} of rank $m_{p}$ on each interval $I_{p}=[\mu_{p+1},\mu_{p}]$ for $p=1,\ldots,n-1$.
We fix the conventions $I_0=[\mu_1,\infty), I_n=(-\infty,\mu_n]$ and take $X_0,X_n$ to be rank zero bundles.
Let $\xi$ be a coordinate on $\bigcup I_p = \R$.
\item[Connections and endomorphisms:] an analytic connection $\nabla_{p}$\label{glo:nabp2} on $I_p$ and analytic skew-herm\-it\-ian endomorphisms $T^{j}_{p}\label{glo:Tjp2}, j=1,2,3$, on the interior of $I_{p}$ for each $p=1,\ldots,n-1$. 
Note that the connection $\nabla_p$ is defined on an open neighbourhood of the closed interval $I_p$. 
The connection and endomorphisms satisfy Nahm's equation on the interior of each interval:
\begin{equation}\label{eq:nahm}
\nabla_{p}T^{i}_p + \frac{1}{2}\sum_{j,k}\epsilon_{ijk}[T^{j}_p,T^{k}_p]=0.
\end{equation}
\end{description}
The bundles $X_{p}$ come with a means of gluing them together at each $\mu_p$:
\begin{description}
\item[When $m_{p}>m_{p-1}$:] an injection $X_{p-1}\rest{\mu_p}\hookrightarrow X_{p}\rest{\mu_p}$.
\item[When $m_{p}<m_{p-1}$:] an injection $X_{p}\rest{\mu_p}\hookrightarrow X_{p-1}\rest{\mu_p}$.
\item[When $m_{p}=m_{p-1}$:] an identification $X_{p}\rest{\mu_p}\leftrightarrow X_{p-1}\rest{\mu_p}$. We call this a \emph{zero jump}\label{glo:zerojump2}.
\end{description}
Each map preserves the hermitian structure.
At each boundary point $\mu_p$ the data satisfy the following boundary conditions:
\begin{description}
\item[When $m_{p}>m_{p-1}$:]  
we require that for each $j=1,2,3$,
\begin{equation*}
T_{p-1}^{j,+}=\lim_{\xi\rightarrow\mu_{p}^{+}}T_{p-1}^{j}
\end{equation*}
exists and $T_{p-1}^{j}$ is analytic at $\mu_{p}$.
Fix a parallel unitary basis for $X_{p-1}$ in a neighbourhood of $\mu_p$.
Using the injection into $X_{p}$ this determines a unitary parallel basis of a rank $m_{p-1}$ sub-bundle of $X_p$ in a neighbourhood of $\mu_p$, which we can extend to a unitary parallel basis of $X_{p}$.
In this gauge there is a decomposition:\label{pag:decompTj}
%%%%%%%%%%%%%%%%%%%%%%%%%%%%%%%%%%%%%%%%%%%%%%%%%%%%%%%%%%%%%%%%%%%%%%%%%%%%%%%
% Decomposition of Nahm data at singularities
%%%%%%%%%%%%%%%%%%%%%%%%%%%%%%%%%%%%%%%%%%%%%%%%%%%%%%%%%%%%%%%%%%%%%%%%%%%%%%%
\begin{center}
\setlength{\unitlength}{1.2cm}
\begin{picture}(8,2.5)(-4,-1.0)

\put(-2,0){\line(1,0){4}}
\put(0,-0.6){\line(0,1){1.2}}

\put(-2,0.25){$T^{j,+}_{p-1}+O(t)$}
\put(0.1,0.25){$O(t^{(k_{p}-1)/2})$}
\put(-2,-0.5){$O(t^{(k_{p}-1)/2})$}
\put(0.1,-0.5){$R^{j}_{p}/t+O(1)$}

\put(-3.15,-0.1){$T_{p}^{j} = $}
\put(-2.4,-0.1){$\Bigg($}
\put(2.1,-0.1){$\Bigg)$}

\put(-1.4,0.9){$m_{p-1}$}
\put(3,0.25){$m_{p-1}$}
\put(0.8,0.9){$k_{p}$}
\put(3,-0.5){$k_{p}$}

\put(2.75,0.05){\vector(0,1){0.55}}
\put(2.75,0.6){\vector(0,-1){0.55}}
\put(2.75,-0.05){\vector(0,-1){0.55}}
\put(2.75,-0.6){\vector(0,1){0.55}}

\put(1.4,1.0){\vector(1,0){0.6}}
\put(0.6,1.0){\vector(-1,0){0.6}}
\put(-1.5,1.0){\vector(-1,0){0.55}}
\put(-0.6,1.0){\vector(1,0){0.55}}

%\put(5.0,-0.5){$(\dag)$}

\end{picture}
\end{center}
%%%%%%%%%%%%%%%%%%%%%%%%%%%%%%%%%%%%%%%%%%%%%%%%%%%%%%%%%%%%%%%%%%%%%%%%%%%%%%%
where the upper diagonal block corresponds to the image of $X_{p-1}$ in $X_{p}$.
The upper diagonal block is analytic in $t=\xi-\mu_{p}$; the lower diagonal block is meromorphic in $t$; and the off-diagonal blocks are of the form $t^{(k_{p}-1)/2}\times (\textrm{analytic in\ }t)$.
The residues $R^{j}_{p}$\label{glo:Rjp} define an irreducible representation of $\sutwo$:
in particular the map
\begin{equation}\label{eq:irrepbc}
\rho : \lambda_1 \gamma_1 +\lambda_2 \gamma_2+\lambda_3 \gamma_3\mapsto -2(\lambda_1 R_p^1 +\lambda_2 R_p^2+\lambda_3 R_p^3)
\end{equation}
is the unique irreducible representation\label{glo:irrep2} $\Srep^{k_p-1}$ \label{glo:srep} on homogeneous polynomials in $(z_0,z_1)$ of degree $k_p-1$, where the $\gamma_j$ are defined by~$\eqref{eq:defgammas}$. 
\item[When $m_{p}<m_{p-1}$:] the situation is just the previous case but with $X_{p}$ and $X_{p-1}$ swapped round.
\item[When $m_{p}=m_{p-1}$:] working in a gauge that is parallel either side of the join and continuous across the join,
we require that limits $T^{j,-}_{p}$ of $T^{j}_{p}$ and $T^{j,+}_{p}$ of $T^{j}_{p-1}$ exist. 
Setting
\begin{equation*}
A^{\pm}(\zeta) = (T^{2,\pm}_{p}+iT^{3,\pm}_{p}) + (2iT^{1,\pm}_{p})\zeta +
(T^{2,\pm}_{p}-iT^{3,\pm}_{p})\zeta^{2}
\end{equation*}
we require that for all $\zeta\in\C$
\begin{equation*}
A^{+}(\zeta)-A^{-}(\zeta) = (u-w\zeta)(w^{\ast}+u^{\ast}\zeta)
\end{equation*}
for some $m_{p}$-dimensional column vectors $u,w$.
At the singularity we therefore have\label{glo:Tpplus}
\begin{equation*}
T^{+}_p - T^{-}_p =\sum_j\gamma_j\otimes T^{j,+}_{p} - \sum_j\gamma_j\otimes T^{j,-}_{p} =
\alpha\alpha^{\ast} - \frac{1}{2}\langle\alpha , \alpha\rangle 
\end{equation*}
where $\alpha$ is an element of $\C^2\otimes\C^{m_{p}}$ formed from $u$ and $w$.
\end{description}

The gauge transformations on a set of Nahm data consist of bundle automorphisms on each of the bundles $X_p$, such that the automorphism on $X_p$ is the identity at $\xi_p$ and $\xi_{p+1}$ for all $p$. 
A gauge transformation $g_1,\ldots,g_n$ acts on $\nabla_p$ by $\nabla_p\mapsto g\nabla_p g^{-1}$ and on $T_p^j$ by $T_p^j\mapsto gT_p^j g^{-1}$. 

\begin{definition}
Let $\monSD{N}{\monbdarydata}$\label{glo:nahmmonSD} be the space of gauge equivalence classes of monopole Nahm data with boundary data $\monbdarydata$. 
\end{definition}

\subsection{Definition of the Nahm operator $\Delta$}\label{sec:mononahmop}

The next task is to show how to construct the analogue of the Dirac operator $D_x^+$ from a set of monopole Nahm data. 
Consider adopting a na\"\i ve approach to generalizing the $4$-torus Nahm transform to the monopole case, as we did in Section~\ref{sec:reviewNahm}. 
Dimensional reduction of the Dirac operator $D_x^+$ by $\Lambda^\ast=\rthree$ gives an operator
\begin{equation}\label{eq:naivenahmop}
\nabla +\sum_{j=1}^3 \gamma_j\otimes T_j - ix_0 -i\sum_{j=1}^3 \gamma_j\otimes x_j : C^\infty(\R,\hat{S}^+\otimes X)\rightarrow C^\infty(\R,\hat{S}^-\otimes X)
\end{equation}
where $\nabla$ is a connection on a bundle $X\rightarrow \R$, and $T_1,T_2,T_3$ are skew-adjoint endomorphisms of $X$, satisfying Nahm's equation. 
Of course, this picture is not quite correct, since the rank of the `bundle' may jump, but using~$\eqref{eq:naivenahmop}$ and the definition of the Nahm data it is clear how to define $D^+_x$ on the interior of the intervals $I_p$.  
We call the analogue of $D_x^+$ the \emph{Nahm operator}\label{glo:nahmop}, $\Delta (x)$\label{glo:Deltax}. 
In fact, following conventions in the monopole literature, we introduce a factor of $i$ and define $\Delta(x)$ to be the analogue of $i\times D_x^+$. 
Some care is needed at the points $\xi=\mu_p$ in the definition of $\Delta(x)$. 
In particular, we want to ensure that $\Delta(x)$ is Fredholm with index $-n$ and is injective for all $x\in\rfour$, so that the cokernel of $\Delta(x)$ defines a rank $n$ bundle over $\rfour$. 
If we can construct $\Delta(x)$ with these properties then we can make the following definition:

\begin{definition}\label{def:COKER}
Suppose $\Delta(x):W\rightarrow V$ is a family of bounded linear maps between Hilbert spaces $W,V$, parameterized by $x\in\rfour$, such that $\Delta(x)$ is injective, Fredholm, and has index $-n$ for all $x$. 
Define $\COKER~\Delta$\label{glo:COKER} to be the $U(n)$ bundle over $\rfour$ with fibre $\coker\Delta(x)$, equipped with the connection
\begin{equation*}
P\cdot d
\end{equation*}
where $P$\label{glo:P} is orthogonal projection from $V$ onto $\coker\Delta(x)$ for each $x$, and $d$ is the standard covariant derivative on the trivial bundle $V\times\rfour$. 
If $v_1,\ldots,v_n$ is a local trivialisation of $\COKER~\Delta$, then in this gauge the connection is represented by matrices 
\begin{equation}\label{eq:cokerinlocalgauge}
(\A_a)_{ij} = \langle \partial_a v_i, v_j \rangle
\end{equation}
for $a=0,1,2,3$. 
\end{definition} 

% Prove that COKER is a bundle. 
To see that $\COKER~\Delta$ is a smooth bundle we find a trivialisation on a neighbourhood of the origin in $\rfour$. 
First note that $\coker\Delta(x)=\ker\Delta^\ast(x)$ for all $x$. 
Identifying $\ker\Delta^\ast(0)$ with $\C^n$, there is a decomposition 
\begin{equation*}
\Delta^\ast(0)=(P_0,0):V'\oplus\C^n\rightarrow W
\end{equation*} 
where $V=V'\oplus\C^n$ and $P_0$ is invertible.
Relative to this decomposition of $V$ we can write $\Delta^\ast(x)=(P_x,Q_x)$ where $P_x$ is invertible for sufficiently small $x$. 
The null-space of $\Delta^\ast(x)$ is a graph over $\C^n$:
\begin{equation*}
\Delta^\ast(x)(u,v)=0\quad\Rightarrow\quad u=-P_x^{-1}Q_x v. 
\end{equation*}
Thus the map
\begin{equation*}
v\mapsto (-P_x^{-1}Q_x v,v)
\end{equation*}
is a smooth isomorphism from $\C^n$ to $\ker\Delta^\ast(x)$ for all sufficiently small $x$. 
The same argument gives a trivialisation round an arbitrary point, and the transition between different trivialisations is smooth.

Working with some fixed set of $U(n)$ monopole Nahm data, let $X_p$, $p=1,\ldots,n-1$, be the corresponding vector bundles. 
Using the trivialisations of the spin spaces $\hat{S}^+$ and $\hat{S}^-$ fixed by~$\eqref{eq:defgammas}$ we identify $\hat{S}^+,\hat{S}^-$ with $\C^2$ throughout this Chapter. 
Let $Y_p=\C^2\otimes X_p$\label{glo:Yp}. 
Let $W_p^l$\label{glo:Wlp} be the Sobolev space of sections of $Y_p$ with $l$ derivatives in $L^2$, 
where the $L^2$ inner product (conjugate linear in the second entry) is defined by
\begin{equation*}
\langle v,w \rangle_{L^2} = \int_{I_p} \langle v,w \rangle d\xi
\end{equation*}
for $v,w\in W_p^0$.
Also define $L^2_l(I_p)$\label{glo:sobltwo} to be the Sobolev space of functions on $I_p$ with $l$ derivatives in $L^2$. 
Often we will just write $\langle,\rangle$ where we mean the $L^2$ inner product or the pairing between elements in dual Sobolev spaces. 

One has to be slightly careful when defining spaces of distributions on manifolds with boundary due to the different choices that can be made when taking completions. 
It will be more apparent why this concerns us when we consider dual spaces in Section~\ref{sec:formaladjoint}. 
These subtleties are dealt with in H\"ormander's book \cite[Appendix B2]{hor85}, and we will adopt his notation. 
Let\label{pag:sobs}\label{glo:Lbar}
\begin{equation*}
\Lbar{l}[a,b] = L^2_l(\R) / \sim
\end{equation*}
where \label{glo:supp}
\begin{equation*}
f \sim g \Leftrightarrow \langle f-g,v\rangle =0
\quad \forall v\in L^2_{-l}(\R)\ 
\textrm{such that\ }\supp~v \subset [a,b]
\end{equation*}
and\label{glo:Ldot} 
\begin{equation*}
\Ldot{l}[a,b] = \{ v\in L^2_l(\R):\supp~v\subset [a,b] \}.
\end{equation*}
It follows immediately that $\Lbar{l}[a,b]$ is the dual of $\Ldot{-l}[a,b]$, where the pairing is given by the $L^2$ inner product. 
We define $W^l_p$ to consist of $\Lbar{l}(I_p)$ sections of $Y_p$, and in what follows we will often just write $L^2_l$ to mean $\Lbar{l}$ on a manifold with boundary. 
Note that differentiation is a well-defined map
\begin{equation*}
\frac{d}{dt}:\Lbar{l}[a,b]\rightarrow \Lbar{l-1}[a,b]
\end{equation*}
since if $f\sim g$ then
\begin{equation*}
\langle \frac{df}{dt} - \frac{dg}{dt},v\rangle = \langle f-g,\frac{dv}{dt} \rangle =0
\end{equation*}
for all $v\in L^2_{1-l}(\R)$ supported on $[a,b]$. 
Also note that if $f\in\Lbar{1}[a,b]$ then the values $f(a)$ and $f(b)$ are well defined: by the Sobolev embedding theorem each representative of $f$ is continuous, and any two representatives must agree on $[a,b]$. 
This ends the technical aside on the definition of the Sobolev spaces, and we return to the definition of the Nahm operator. 

We adopt the following terminology from \cite{hur89}.
Suppose $m_p\geq m_{p-1}$.
At the boundary point $\mu_p$ of $[\mu_{p+1},\mu_p]$, $Y_p$ decomposes as a direct sum $Y_p(\mu_p)=Y_{p-1}(\mu_p)\oplus Y_{p-1}(\mu_p)^{\perp}$ using the inclusion of $X_{p-1}(\mu_p)$ into $X_{p}(\mu_p)$. 
We call vectors in the first component \label{glo:contcmpt} `continuing', and vectors in the second component \label{glo:termcmpt}\label{pag:termcpt} `terminating'.
If $m_p\leq m_{p-1}$ then all the vectors in $Y_p(\mu_p)$ are continuing.
We adopt similar terminology at the other end of the interval. 
See Figure~\ref{fi:nahmbcsa} for an illustration. 

Let $\Woone{p}\subset W_p^1$\label{glo:Wop} be the subset of sections of $Y_p$ whose terminating components vanish at both ends of the interval (this definition makes sense following the remark above about $\Lbar{1}$ functions).
Define\label{glo:Dp}
\begin{gather*}
D_p(x):\Woone{p} \rightarrow W_p^0 \\ D_p(x) = i\nabla_p +iT_p +x
\end{gather*}
where \label{glo:Tp}
\begin{equation*}
T_p=\sum_{j=1}^{3} \gamma_j\otimes T^j_p, 
\end{equation*}
and 
\begin{equation*}
x=x_0+\sum_{j=1}^{3} \gamma_j\otimes x_j.
\end{equation*}
Then $D_p$ is well defined since the component of the section acted on by the singular part of $T$ is zero: using the Cauchy Schwartz inequality one obtains 
\begin{equation*}
\| (\xi-\mu_p)^{-1}f \|_{L^2}\leq C\| f \|_{L^2_1}
\end{equation*}
for any function $f\in \Lbar{1}(I_p)$ vanishing at $\mu_p$.

The zero jumps need special consideration.
Let $\xi=\mu_q$ be a zero jump \label{glo:zerojump3}(\ie suppose $m_q=m_{q-1}$).
The boundary condition on the Nahm data at a zero jump fixes a $1$ (complex-)dimensional subspace of $Y_q(\mu_q)$, which we denote $J_q$\label{glo:Jq} (the `jumping space'\label{glo:jumpingspace} at $\mu_q$). 
For each zero jump we fix an element $\zeta_q$\label{glo:zetaq} of $J_q$ with norm $1$, and let\label{glo:piq}
\begin{equation}\label{eq:defcmptspi}
\pi_q(w)=\langle w(\mu_q),\zeta_q \rangle
\end{equation}
for any continuous section $w$  of $Y_q$.
Let $\Zeros$\label{glo:Zeros} be the set of zero jumps 
\begin{equation*}
{\mathcal{J}}=\{ q : m_q = m_{q-1} \}
\end{equation*}
and let $\Nzer=| \Zeros |$\label{glo:Nzer}.

We are now in a position to define the Nahm operator $\Delta(x):W\rightarrow V$.
Let
\begin{equation}\label{eq:defW}
W = \{ (w_1,\ldots,w_{n-1})\in 
\Woone{1}
\oplus\cdots\oplus
\Woone{n-1}: 
w_p(\mu_p)=w_{p-1}(\mu_p)~\textrm{for}~p=2,\ldots,n-1\}
\end{equation}
and 
\begin{equation}\label{eq:defV}
V = W_1^0\oplus\cdots\oplus W_{n-1}^0\oplus\C^{\Nzer}.
\end{equation}
The Nahm operator $\Delta$ is the direct sum of the $D_p$ operators, together with the projection at each zero jump:
\begin{gather*}
\Delta(x) : W\rightarrow V, \\
\Delta(x)w = [ D_{1}(x)w_{1},\ldots,D_{n-1}(x)w_{n-1} ] \oplus[\pi w].
\end{gather*}
The map $\pi:W\rightarrow \C^{\Nzer}$ has components $\pi_q$ for $q\in\Zeros$.
Note that we deal with the zero jumps slightly differently from Hurtubise and Murray. Instead of $\C^{\Nzer}$ their projection $\pi$ maps into $\bigoplus J_q$---we fix a basis for this space (the $\zeta_q$), and work in this basis. At some stages we will need to deform the projections $\pi_q$; with our set-up the deformed operator will still be a map $W\rightarrow V$, whereas with Hurtubise and Murray's version, the spaces would change with the deformation, which would cause problems. 
We have to check that $\COKER~\Delta$ is independent of the choice of $\zeta_q$: making a different choice for the $\zeta_q$ is just equivalent to a unitary change of basis in $\C^{\Nzer}$. 
Hence the choice of $\zeta_q$ does not affect $\COKER~\Delta$ up to isomorphism.

\subsection{Results from Hurtubise and Murray}\label{sec:HurandMur}

Given a Nahm operator $\Delta(x):W\rightarrow V$ constructed from a set of $U(n)$ monopole Nahm data, Hurtubise and Murray prove the following results in \cite{hur89}:
\begin{itemize}
\item $\Delta(x)$ is injective and has index $-n$ for all $x$ \ie $\COKER~\Delta$ is well-defined (recall Definition~\ref{def:COKER}),
\item Nahm's equation implies that $\COKER~\Delta$ is anti-self-dual,
\item $\COKER~\Delta$ is a translation invariant $U(n)$ bundle and connection, and
\item $\COKER~\Delta$ satisfies the monopole boundary conditions.
\end{itemize}
In other words, they show that the Nahm transform takes a set of monopole Nahm data and produces a monopole.
Since Hurtubise and Murray assume weaker boundary conditions for monopoles than we do, we cannot assume that the Nahm transform on an element of $\monSD{N}{\monbdarydata}$ gives a monopole configuration as defined by Definition~\ref{def:monopole}. 
Thus we must state precisely which results from~\cite{hur89} we are free to assume. 
We assume:

\begin{lemma}\label{lem:HM1}
If $\Delta(x)$ is the Nahm operator constructed from some set of $U(n)$ monopole Nahm data in $\monSD{N}{\monbdarydata}$ then
$\Delta(x)$ is injective and Fredholm with index $-n$ for all $x\in\rfour$. 
\end{lemma}

\begin{lemma}\label{lem:HM2}
Under the same assumptions as Lemma~\ref{lem:HM1}, 
\begin{equation*}
\int_{[0,\perflat]\times\rthree}\chtwo~\COKER~\Delta =-\frac{1}{\muo}(\mu_1 k_1+\cdots+\mu_n k_n).
\end{equation*}
(This is just equation~$\eqref{eq:caloroncharge}$ except with $k_0=0$ since $\COKER~\Delta$ is a monopole configuration.)
\end{lemma}

Both Lemmas are implicit in~\cite{hur89}.

Later on it will be useful to consider Nahm data that does not satisfy Nahm's equation. 
In addition, it is useful to drop the requirement that the Nahm data is discontinuous accross zero jumps. 
We therefore make the following definition:

\begin{definition}\label{def:looseNahmmon}
The space $\mon{N}{\monbdarydata}$\label{glo:monopolenahmdata} consists of gauge equivalence classes of Nahm data with boundary data $\monbdarydata$, such that the data does not necessarily satisfy Nahm's equation. 
Moreover, at a zero jump $\xi=\mu_q$, either
\begin{enumerate}
\item the endomorphisms $T^1,T^2,T^3$ are discontinuous, as described previously, or
\item the endomorphisms are continuous, but there is still a projection operator $\pi_q$ and some $1$-dimensional subspace $J_q\subset Y_q(\mu_q)$ associated to the zero jump. 
\end{enumerate}
\end{definition}

Some of the results of Hurtubise and Murray continue to hold for this wider class of data. 
In particular, it is easy to check that the proof of Lemma~\ref{lem:HM1} still holds. 
We will also assume that Lemma~\ref{lem:HM2} holds for data in $\mon{N}{\monbdarydata}$: this does not follow from Hurtubise and Murray's results, but we will give a proof later. 
Since we will not require Nahm's equation to hold most of the time, we will refer to data from $\mon{N}{\monbdarydata}$ as `monopole Nahm data' and point out where we use Nahm's equation specifically. 

\section{The construction of calorons up to boundary conditions}\label{sec:calnahmdata}

\subsection{Definition of caloron Nahm data}\label{sec:nahmdataforcal}

A set of $U(n)$ caloron Nahm data with boundary data $\bdarydata$ consists of $n$ bundles $X_1,\ldots,X_n$ over the intervals $I_{p}\subset\Sdual=\R / \mu_0 \Z$ defined by~$\eqref{eq:bdlintervals}$.
The rank of $X_p$ is $m_p$, where $m_p$ is defined by~$\eqref{eq:constituentcharges}$.
The bundles are equipped with exactly the same structures as monopole Nahm data, and glued together in the same way at each $\mu_p+\mu_0 \Z$, $p=1,\ldots,n$. 
Note that $X_{1}$ is glued to $X_{n}$ at $\mu_{1}+\muo\Z$, and the data satisfy exactly the same gluing conditions there. 
The gauge transformations are also defined in an entirely analogous way. 

Next we have to build the Nahm operator $\Delta (x):W\rightarrow V$ from this data.
This is entirely analogous to the monopole case in Section~\ref{sec:mononahmop}.
Define $W_p^0$, $\Woone{p}$, and $D_p$ for $p=1,\ldots,n$ in the same way as in Section~\ref{sec:mononahmop}. Label the zero jumps by $q\in\Zeros$, and define $J_q,\zeta_q,\pi_q$ and $\Nzer$ just as previously.
Then
\begin{multline}\label{eq:defnW}
W = \{ (w_1,\ldots,w_{n})\in\Woone{1}\oplus\cdots\oplus\Woone{n}: 
w_p(\mu_p)=w_{p-1}(\mu_p)\ \textrm{for\ }p=2,\ldots,n\ \\
\textrm{and\ }w_1(\mu_1)=w_n(\mu_1-\muo) \}.
\end{multline}
In other words, $W$ consists of sections that are continuous around the circle.
Let
\begin{equation*}
V = [W_1^0\oplus\cdots\oplus W_{n}^0]\oplus[\C^{\Nzer}]
\end{equation*}
and
\begin{gather}
\Delta(x) : W\rightarrow V, \notag \\
\Delta(x)w = [ D_{1}(x)w_{1},\ldots,D_{n}(x)w_{n} ] \oplus[\pi w].
\label{eq:defdelta}
\end{gather}

In the monopole case every Nahm operator $\Delta(x)$ built from Nahm data was injective: Hurtubise-Murray prove this by showing
\begin{equation*}
\| \Delta(x) f \|^2 = \| \nabla_p f\|^2 +~\textrm{positive terms}.
\end{equation*}
Thus any element of the kernel of $\Delta(x)$ has to be constant, and the vanishing condition at the end points $\mu_1$ and $\mu_n$ ensures that any such section is trivial.
(In fact this is a slight over-simplification because zero jumps also have to be taken into account.)
This proof does not carry over to the caloron case: there could exist covariantly constant sections of the bundles that remain in the continuing component at each $\mu_p$. 
It is quite easy to construct examples of caloron Nahm data satisfying our definition that do not give rise to injective Nahm operators: for example take a set of $U(n)$ monopole Nahm data and glue on a rank $k_0$ bundle over $\Sdual$ equipped with the trivial connection and endomorphisms. 
This satifies the conditions to be a valid set of caloron Nahm data, but the Nahm operator is not injective for all $x$. 
Thus we restrict to sets of Nahm data that \emph{do} give rise to injective operators:

\begin{definition}
Let $\spcSD{N}{\bdarydata}$\label{glo:spcNSD} be the set of gauge equivalence classes of $U(n)$ caloron Nahm data satisfying Nahm's equation, and with boundary data $\bdarydata$, such that the Nahm operator $\Delta(x)$ is injective for all $x\in\rfour$.
\end{definition}

It is useful to also consider caloron Nahm data that does not satisfy Nahm's equation. 
We therefore define $\spc{N}{\bdarydata}$\label{glo:spcN} analogously to Definition~\ref{def:looseNahmmon}. 
In general, we will refer to data in $\spc{N}{\bdarydata}$ as `caloron Nahm data' and point out where we specifically require Nahm's equation and discontinuities at zero jumps. 

We will sometimes use the following terminology:

\begin{definition}
A set of caloron Nahm data is \emph{principal}\label{glo:prinNahm} if it has principal boundary data (in the sense of Definition~\ref{def:bdarydata}). 
\end{definition}

\subsection{The index of $\Delta$}\label{sec:indexdelta}

Given a set of caloron Nahm data and the corresponding Nahm operator $\Delta(x) :W\rightarrow V$, we want to show that $\COKER~\Delta$ is well-defined \ie we want to show that $\Delta(x)$ is Fredholm with index $-n$ for all $x$.
(Recall that $\Delta(x)$ is injective by definition.) 
The calculation of the index is an adaptation of the $SU(n)$ monopole version \cite[Section 4]{hur89}.

To calculate the index of $\Delta(x)$ we want to identify the kernel and cokernel, and count dimensions. 
It is clear that the kernel of $\Delta(x)$ consists of $(w_1,\ldots,w_n)\in W$ satisfying
\begin{itemize}
\item $D_p(x) w_p =0$, and
\item at zero jumps $w_q(\mu_q)$ is in $J_q^\perp$
\end{itemize}
in addition to the constraints on $W$ (\ie that terminating components vanish at $\mu_p$ where $k_p\neq 0$ and that continuing components are continuous at any $\mu_p$).
To find the cokernel we integrate by parts. 
Fix $(w_1,\ldots,w_n)\in W$ and 
\begin{equation} \label{eq:elementV}
v=(v_1,\ldots,v_n)\oplus s\in V = [\bigoplus_{p=1}^{n}W_p^0]\oplus[\C^{\Nzer}].
\end{equation}
Then 
\begin{equation} \label{eq:vDeltaw}
\langle v, \Delta(x) w \rangle  = \sum_{p=1}^{n} \langle v_p,D_p(x) w_p \rangle_{L^2} + 
\sum_{q\in\Zeros} \langle s_q\zeta_q,  w_q(\mu_q) \rangle
\end{equation}
and $v\in\coker\Delta(x)$ if and only if this vanishes for all $w$. 
If we assume that $w_p(\mu_p)=0$ and $w_p$ is smooth for all $p$, then integration by parts of the first sum makes sense, and we obtain
\begin{equation*}
\langle v,\Delta(x) w\rangle = \sum_{p=1}^{n} \langle D_p^\ast(x) v_p, w_p \rangle_{L^2}
\end{equation*} 
where
\begin{equation*} %\label{eq:defDpstar}
D_p^\ast(x) = i\nabla_p -iT_p +x^\ast. 
\end{equation*}
(If we do not assume $w_p(\mu_p)=0$ for all $p$ then problems arise because $v_p(\mu_p)$ may not be well-defined and the boundary contribution in the parts integration may not make sense.)
Hence $v_p$ must satisfy $D_p^\ast(x)v_p=0$ for all $p$. 
It follows that $v_p$ is smooth on the interior of $I_p$, and the continuing components of $v_p$ are continuous up to the ends of the interval. 
Under these conditions integrating~$\eqref{eq:vDeltaw}$ by parts makes sense for all $w$, and we obtain
\begin{multline*}
\langle v, \Delta(x) w \rangle  = \sum_{p=1}^n\langle D_p^\ast(x) v_p, w_p \rangle_{L^2} +
\sum_{p=1}^n i\langle v_{p-1}^{\textrm{cont}}(\mu_p)-v_{p}^{\textrm{cont}}(\mu_p), w_p(\mu_p) \rangle \\ +
\sum_{q\in\Zeros} \langle s_q \zeta_q,  w_q(\mu_q) \rangle
\end{multline*}
where $v^{\textrm{cont}}_p$\label{glo:supcont} is the continuing component of $v_p$. 
Then $v\in\coker\Delta(x)$ iff
\begin{itemize}
\item $D_p^\ast(x) v_p =0$ on each interval $I_p$ for $p=1,\ldots,n$, 
\item the continuing components are continuous at each $\mu_p$ where $k_p\neq 0$, and
\item at zero jumps,  $v_q(\mu_q)-v_{q-1}(\mu_q) = -is_q\zeta_q$.
%\footnote{Note that this condition differs slightly from Hurtubise and Murray: %they seem to lose a factor of $i$ when they integrate by parts.}
\end{itemize}

To count the number of solutions in the kernel and cokernel we need some analysis of $D_p(x)$ and $D_p^\ast(x)$ close to the singular points 
$\xi=\mu_p$. 
This is taken more-or-less directly from Hitchin's paper on the construction of monopoles \cite[Section 2]{hit83}. 
Start by considering the case $k_p > 0$, and recall the conditions imposed on the endomorphisms $T_p^j$ at such a point. 
Let $t=\xi-\mu_p$ be a coordinate in a neighbourhood of $\mu_p$. 
For $t\in(-\epsilon,0)$ we have in some parallel gauge the block decomposition
\begin{equation*}
\sum \gamma_j \otimes T_p^j  = 
\begin{pmatrix}
0 & 0 \\ 0 & R_p/t
\end{pmatrix}
+ B(t)
\end{equation*}
where $B(t)$ is analytic and bounded, and $R_p=\sum \gamma_j\otimes R_p^j$\label{glo:Rp}. 
In terms of the representation $\rho$ defined by~$\eqref{eq:irrepbc}$, 
\begin{equation*}
R_p=-\frac{1}{2}\sum \gamma_j \otimes \rho(\gamma_j) .
\end{equation*} 
It is possible to express $R_p$ in terms of Casimir operators\label{glo:Casimir}. 
If $\Srep$ is a representation of $\sutwo$ then the Casimir operator is defined by $C(\Srep)=\sum_{j} \rho(\gamma_j)^2$.
Now 
\begin{equation*}
C(\Srep^1\otimes\Srep^{k_p-1}) = 1\otimes C(\Srep^{k_p-1}) + 2\sum  \gamma_j\otimes\rho(\gamma_j) + C(\Srep^1)\otimes 1
\end{equation*}
so
\begin{equation*}
R_p = -\frac{1}{4}[ C(\Srep^1\otimes\Srep^{k_p-1}) - 1\otimes C(\Srep^{k_p-1})
- C(\Srep^1) \otimes 1].
\end{equation*}
The Casimir operator on $\Srep^{k_p}$ is $-k_p(k_p+2)\cdot{\textrm{id}}$,
and decomposing into irreducibles we have $\Srep^1\otimes\Srep^{k_p-1}\cong \Srep^{k_p}\oplus\Srep^{k_p-2}$.
Hence
\begin{equation}\label{eq:repthy}
R_p=\frac{1}{2}(k_p-1)\ \textrm{on}\ \Srep^{k_p}\subset\Srep^1\otimes\Srep^{k_p-1}
\end{equation}
and
\begin{equation*}
R_p=-\frac{1}{2}(k_p+1)\ \textrm{on}\ \Srep^{k_p-2}\subset\Srep^1\otimes\Srep^{k_p-1}.
\end{equation*}
Now let $U_p$\label{glo:Up} be the $2m_p$ dimensional space of solutions to $D_p$ on $I_p$. 
Using the block decomposition of $D_p$ and the calculation of $R_p$, we see that $U_p$ decomposes as $U_p = B_p\oplus G_p \oplus C_p$ where
\begin{itemize}
\item $B_p$ is the $k_p +1$ dimensional space of solutions that are $O(t^{-(k_p -1) / 2})$ corresponding to elements of $\Srep^{k_p}$, 
\item $G_p$ is the $k_p -1$ dimensional space of solutions that are $O(t^{(k_p +1) / 2})$ corresponding to elements of $\Srep^{k_p -2}$, and 
\item $C_p$ is the $2m_{p -1}$ dimensional space of solutions that are $O(t^0)$  corresponding to elements of the other diagonal block.
\end{itemize}
The solutions in $B_p$ do not have vanishing terminating component and are not $L^2$ so cannot be elements of $\Wo_p$.
(Here `B' is for `bad', `G' is for `good' and `C' is for `continuing'. The `bad' solutions are the ones that cannot be in $\ker\Delta(x)$.)

Next consider the cases $k_p < 0$ and $k_p =0$:
the same decomposition of $U_p$ exists, but $B_p$ and $G_p$ are trivial because all solutions are in the continuing block.
Alternatively, one can decompose $U_p$ depending on the behaviour of solutions at the other end of the interval, and obtain $U_p = \hat{B}_p\oplus \hat{G}_p \oplus \hat{C}_p$.
If $k_{p+1} \geq 0$ then $\hat{B}_p$ and $\hat{G}_p$ are trivial, but when $k_{p+1}<0$ we have 
\begin{itemize}
\item $\hat{B_p}$ is the $1-k_{p+1}$ dimensional space of solutions that are $O(t^{(1+ k_{p+1}) / 2})$ corresponding to elements of $\Srep^{| k_{p+1} |}$, 
\item $\hat{G_p}$ is the $-1-k_{p+1}$ dimensional space of solutions that are $O(t^{(1-k_{p +1}) / 2})$ corresponding to elements of $\Srep^{| k_{p+1}| -2}$, and 
\item $\hat{C}_p$ is the $2m_{p +1}$ dimensional space of solutions that are $O(t^0)$  corresponding to elements of the other diagonal block.
\end{itemize}
For these estimates we have taken $t=\xi-\mu_{p+1}$.
For $u\in U_p$ let $B_p u, G_p u, C_p u$ \etc denote the projections with respect to the decomposition.

We can perform exactly the same kind of analysis for $D_p^\ast$.
Let $U_p^\ast$\label{glo:Upast} be the $2m_p$ dimensional space of solutions to $D_p^\ast$ (at the moment the `$\ast$' is just a label---but in fact $U_p^\ast$ will turn out to be the dual of $U_p$).
We obtain decompositions $U_p^\ast = B_p^\ast\oplus G_p^\ast \oplus C_p^\ast$ and $U_p^\ast = \hat{B}_p^\ast\oplus \hat{G}_p^\ast \oplus \hat{C}_p^\ast$.
In particular when $k_p>0$ we choose $B_p^\ast$ to correspond to elements of $\Srep^{k_p}$ so that  it contains solutions that are $O(t^{(k_p -1) /2})$, 
and when $k_{p+1}<0$ we choose $\hat{B}_p^\ast$ to correspond to elements of $\Srep^{| k_{p+1} |}$ so that it contains solutions that are $O(t^{-(1+k_{p+1}) /2})$. 
When $k_p\leq 0$ $B_p^\ast$ is trivial, and similarly for $\hat{B}_p^\ast$ when $k_{p+1}\geq 0$. 
Now suppose $u\in U_p, w\in U_p^\ast$ for some fixed $p$.
Then $\langle u,w \rangle (\xi)$ is well defined for any $\xi$ in the interior of $I_p$ because the elements of $U_p$ and $U^\ast_p$ are necessarily smooth on the interior.
But because $u$ and $w$ are solutions to $D_p$ and $D_p^\ast$, $\frac{d}{d\xi}\langle u,w \rangle (\xi) =0$ so $U_p^\ast$ really is the vector space dual of $U_p$, with the pairing
\begin{equation*}
\langle u,w \rangle_{\textrm{dual}}= (\mu_p-\mu_{p+1})^{-1}\langle u,w \rangle_{L^2}= \langle u,w \rangle (\xi)\ \textrm{for any $\xi$ in the interior of $I_p$}.
\end{equation*}
Moreover, because the solutions in $B_p$ and $B_p^\ast$ correspond to elements of $\Srep^{k_p}$, $B_p^\ast$ is the dual of $B_p$, and similarly for $G_p^\ast$ and $C_p^\ast$.
Hence
\begin{equation}\label{eq:decompcokerdelta}
\langle u,w \rangle_{\textrm{dual}} =
\langle B_p u,B_p^\ast w \rangle_{\textrm{dual}} +
\langle G_p u,G_p^\ast w \rangle_{\textrm{dual}} +
\langle C_p u,C_p^\ast w \rangle_{\textrm{dual}}
\end{equation}
where 
\begin{equation*}
\langle B_p u,B_p^\ast w \rangle_{\textrm{dual}}=
\langle B_p u,B_p^\ast w \rangle(\xi)
\end{equation*}
for any $\xi$ in the interior of $I_p$ etc.  
Note that 
\begin{equation*}
\langle C_p u,C_p^\ast w \rangle_{\textrm{dual}}=
\langle C_p u,C_p^\ast w \rangle(\mu_p)=
\langle C_p u,C_p^\ast w \rangle(\mu_{p+1})
\end{equation*}
because the values at the end-points are well defined. 
We obtain a similar expression to~$\eqref{eq:decompcokerdelta}$ using the decomposition of $U_p$ at the other end of the interval, in terms of $\hat{B}_p$, $\hat{G}_p$, and $\hat{C}_p$. 

The boundary conditions for the kernel and cokernel can be stated in terms of these decompositions. 
Suppose $u_p\in U_p$ and $u_{p-1}\in U_{p-1}$ satisfy the conditions to be in the kernel at $\mu_p$. 
This is equivalent to saying $B_p u_p =0 =\hat{B}_{p-1}u_{p-1}$ (so that the terminating components vanish), that $C_p u_p (\mu_p) = \hat{C}_{p-1}u_{p-1}(\mu_p)$ (the continuing components are continuous), and at zero jumps $C_q u_q (\mu_q)\in J_q^\perp$. 
Similarly, suppose $w_p\in U_p^\ast$ and $w_{p-1}\in U_{p-1}^\ast$ satisfy the conditions to be in the cokernel at $\mu_p$. \label{pag:cokerconds}
This is equivalent to saying $G_p^\ast w_p =0 =\hat{G}_{p-1}^\ast w_{p-1}$ (so that $w_p$ and $w_{p-1}$ are $L^2$), that at points $\mu_p$ where $k_p\neq 0$ we have $C_p^\ast w_p (\mu_p) =\hat{C}_{p-1}^\ast w_{p-1} (\mu_p)$ (the continuing components are continuous), and at zero jumps $\mu_q$, 
$C_q^\ast w_q (\mu_q) -\hat{C}_{q-1}^\ast w_{q-1} (\mu_q) \in J_q$.

To find the index of $\Delta(x)$ we introduce a map $\Theta$ between finite dimensional vector spaces whose index is easy to compute, but constructed so that $\ker\Delta(x) = \ker\Theta$ and $\coker\Delta(x) = \coker\Theta$.
Let $U_p^{\textrm{pairs}}\subset U_p \times U_{p-1}$ be the set of pairs $(u_p,\hat{u}_{p-1})$ satisfying the boundary conditions for the kernel of $\Delta(x)$ at $\mu_p$ for $p=2,\ldots,n$, and let $U_1^{\textrm{pairs}}\subset U_1 \times U_n$ in the same way.
Consider the map
\begin{gather*}
\Theta : \bigoplus_{p=1}^n U_p^{\textrm{pairs}} \rightarrow \bigoplus_{p=1}^n U_p \\ \Theta :
(u_{1},\hat{u}_{n}),(u_{2},\hat{u}_{1}),\ldots,(u_{n},\hat{u}_{n-1})\mapsto
(u_{1}-\hat{u}_{1}),(u_{2}-\hat{u}_{2}),\ldots,(u_{n}-\hat{u}_{n}).
\end{gather*}
By construction, the kernel of $\Theta$ is the kernel of $\Delta(x)$.
Also 
\begin{equation*}
\ind\Theta = \sum \dim U_p^{\textrm{pairs}} - \sum \dim U_p.
\end{equation*}
Checking each case ($k_p>0,k_p=0,$ and $k_p<0$) we have $\dim U_p^{\textrm{pairs}} = m_p + m_{p-1} -1$ while $\dim U_p = 2m_p$. 
Hence $\ind\Theta =-n$. 
If we can show the annihilator of $\im \Theta$ is $\coker \Delta(x)$ then we can conclude that $\ind\Delta(x) = -n$

Fix
\begin{equation*}
u=((u_{1},\hat{u}_{n}),(u_{2},\hat{u}_{1}),\ldots,(u_{n},\hat{u}_{n-1}))
\in \bigoplus_{p=1}^n U_p^{\textrm{pairs}}
\end{equation*}
and 
\begin{equation*}
w=(w_1,\ldots,w_n)\in\bigoplus_{p=1}^n U_p^\ast
\end{equation*}
Then
\begin{equation*}
\begin{split}
\langle \Theta u,w \rangle_{\textrm{dual}} &=
\sum_p \langle u_p - \hat{u}_p,w_p \rangle_{\textrm{dual}} \\ &=
\sum_p [ \langle C_p u_p,C_p^\ast w_p \rangle_{\textrm{dual}} +
\langle G_p u_p,G_p^\ast w_p \rangle_{\textrm{dual}} \\ & \qquad \qquad \qquad-
\langle \hat{C}_p \hat{u}_p,\hat{C}_p^\ast w_p \rangle_{\textrm{dual}} -
\langle \hat{G}_p \hat{u}_p,\hat{G}_p^\ast w_p \rangle_{\textrm{dual}}  ]
\end{split}
\end{equation*}
using~$\eqref{eq:decompcokerdelta}$ (plus its analogous version in terms of $\hat{B}_p,\hat{G}_p,\hat{C}_p$) and the fact that $B_p u_p = 0 = \hat{B}_p \hat{u}_p$.
Now
\begin{equation*}
\langle C_p u_p,C^\ast_p w_p \rangle_{\textrm{dual}}
=\langle C_p u_p,C^\ast_p w_p \rangle (\mu_p), \quad\textrm{and~}
\langle \hat{C}_p \hat{u}_p,\hat{C}^\ast_p w_p \rangle_{\textrm{dual}}
=\langle \hat{C}_p \hat{u}_p,\hat{C}^\ast_p w_p \rangle (\mu_{p+1}).
\end{equation*}
This implies that
\begin{multline*}
\langle \Theta u,w \rangle_{\textrm{dual}} =
\sum_p [ \langle C_p u_p,C_p^\ast w_p \rangle (\mu_p) 
-\langle \hat{C}_{p-1} \hat{u}_{p-1},\hat{C}_{p-1}^\ast w_{p-1} \rangle(\mu_p) 
\\ + \langle G_p u_p,G_p^\ast w_p \rangle (\xi^+) -
\langle \hat{G}_{p-1} \hat{u}_{p-1},\hat{G}_{p-1}^\ast w_{p-1} \rangle (\xi^-)  ]
\end{multline*}
for any $\xi^+\in I^{\circ}_p$ and any $\xi^-\in I^{\circ}_{p-1}$. 
But $C_p u_p (\mu_p) = \hat{C}_{p-1}\hat{u}_{p-1}(\mu_p)$ because $(u_p,\hat{u}_{p-1})$ satisfies the boundary conditions for $\ker\Delta(x)$ at $\mu_p$, so
\begin{multline}\label{eq:thetauw}
\langle \Theta u,w \rangle_{\textrm{dual}} =
\sum_p [ \langle C_p u_p,C_p^\ast w_p -\hat{C}_{p-1}^\ast w_{p-1}\rangle (\mu_p) \\ + \langle G_p u_p,G_p^\ast w_p \rangle (\xi^+) -
\langle \hat{G}_{p-1} \hat{u}_{p-1},\hat{G}_{p-1}^\ast w_{p-1} \rangle (\xi^-)  ].
\end{multline}
The annihilator of $\im\Theta$ consists of $w\in\bigoplus U_p^\ast$ for which this vanishes for all $u$.
Such $w$ satisfy the following conditions.
\begin{enumerate}
\item At points $\mu_p$ that are not zero jumps 
$C_p^\ast w_p (\mu_p) -\hat{C}_{p-1}^\ast w_{p-1} (\mu_p) =0$.
\item At zero jumps $\mu_q$, 
$C_q^\ast w_q (\mu_q) -\hat{C}_{q-1}^\ast w_{q-1} (\mu_q) \in J_q$, because $C_q w_q (\mu_q)\in J_q^\perp$. 
\item We have seen that at any $\mu_p$ either $G_p=0=G^\ast_p$ (when $k_p\leq 0$) or $\hat{G}_{p-1}=0=\hat{G}^\ast_{p-1}$ (when $k_p\geq 0$). 
In the latter case $\eqref{eq:thetauw}$ gives $\langle {G}_{p} {u}_{p},{G}_{p}^\ast w_{p} \rangle (\xi)=0$ for all $u$ and for all $\xi\in I^{\circ}_{p-1}$ so ${G}_{p}^\ast w_{p}$ has to vanish. 
Similarly, by considering the other case, 
$w$ satisfies $G_p^\ast w_p =0 = \hat{G}_p^\ast w_p$ for each $p$.
\end{enumerate}
These are precisely the conditions that $w\in\coker\Delta(x)$ as stated on page~\pageref{pag:cokerconds}, and this completes the proof of the following lemma.

\begin{lemma}\label{lem:indexdelta}
If $\Delta(x): W\rightarrow V$ is the Nahm operator corresponding to some set of $U(n)$ caloron Nahm data then $\Delta(x)$ is Fredholm with index $-n$ for all $x$. 
\end{lemma}

We also need to consider the index of deformations of $\Delta$, for which the following definitions will be useful. 

\begin{definition}\label{def:multiplicativeop}\label{glo:multiplicativeop}
A \emph{multiplicative operator} $W\rightarrow V$ is a map of the form
\begin{equation*}
(w_1,\ldots,w_n)\in W\mapsto (A_1 w_1,\ldots,A_n w_n)\oplus(0)\in V
\end{equation*}
where $A_p$ is a smooth uniformly-bounded matrix-valued function on $Y_p=\C^2\otimes X_p$ over each interval $I_p\subset\Sdual$. 
The second component is the zero vector in $\C^{\Nzer}$. 
The continuing components of the $A_p$ matrices are continuous at each $\xi =\mu_p$, except at zero jumps where we allow $A_p(\mu_p)$ and $A_{p-1}(\mu_p)$ to be different.
\end{definition}

\begin{definition}\label{def:frameddeformn}\label{glo:frameddeformn}
A deformation $\tilde{\Delta}(x):W\rightarrow V$ of $\Delta(x)$ is \emph{framed} if, for sufficiently large $r$, $\tilde{\Delta}(x) - \Delta(x)$ is a multiplicative operator that is independent of $x$. 
(Here $r$ is the polar coordinate on $\rthree$.)
\end{definition}

\begin{definition}\label{def:controlleddeformn}
Suppose $\Delta(x):W\rightarrow V$ is the Nahm operator corresponding to a set of $U(n)$ caloron Nahm data, and $\Delta$ is defined by~$\eqref{eq:defdelta}$.
A deformation $\tilde{\Delta}(x):W\rightarrow V$ of $\Delta(x)$ is a \emph{controlled deformation}\label{glo:controlleddef} if it is of the form
\begin{equation*}
\tilde{\Delta}(x)w = [ D_{1}(x)w_{1}+A_{1}w_{1},\ldots,D_{n}(x)w_{n}+A_{n}w_{n} ] \oplus[\tilde{\pi} w]+B(x)w
\end{equation*}
where the following conditions hold:
\begin{itemize}
\item For each $p=1,\ldots,n$, $A_p$ is a smooth uniformly-bounded matrix-valued function on the interval $I_p\subset\Sdual$ that is independent of $x$, and with the same continuity conditions as in Definition~\ref{def:multiplicativeop}. 
\item The projection $\tilde{\pi}:W\rightarrow\C^{\Nzer}$ has components
\begin{equation*}
\tilde{\pi}_q w = \langle w(\mu_q),\tilde{\zeta}_q\rangle
\end{equation*}
for $q\in\Zeros$. 
Here $\tilde{\zeta}_q$ is some (possibly $x$-dependent) vector in $Y_q(\mu_q)$, such that  $\tilde{\zeta}_q(x)=\zeta_q$ for all $x$ outside some compact set.
\item For all $x$, $B(x)$ is a compact operator $W\rightarrow V$ and $B$ has compact support \ie $B(x)=0$ for sufficiently large $r$. 
\end{itemize} 
\end{definition}

Note that any controlled deformation is also framed. 
Roughly speaking, the solutions in the cokernel of a framed deformation $\tilde{\Delta}$ of $\Delta$ will be asymptotically close to solutions in the cokernel of $\Delta$. 
The condition of being a controlled deformation is stronger, with implications on the interior as well as asymptotically: 

\begin{corollary}\label{cor:indexdefdelta}
Suppose $\tilde{\Delta}$ is a controlled deformation of $\Delta$. The proof of Lemma \ref{lem:indexdelta} applies to $\tilde{\Delta}$, and so $\tilde{\Delta}(x)$ is Fredholm with index $-n$ for all $x$. 
\end{corollary}

\subsection{The adjoint of $\Delta(x)$}\label{sec:formaladjoint}

Given the Nahm operator $\Delta(x):W\rightarrow V$, the `Hilbert space' adjoint $\Delta^\ast (x): V\rightarrow W$ is difficult to compute, involving the Sobolev space inner product on $L^2_1$. 
Instead it is easy to consider the adjoint as a map into the dual space of $W$, where linear functionals consist of pairing the $L^2_1$ sections in $W$ with sections in $L^2_{-1}$, because this can be computed by parts integration like that at the start of Section~\ref{sec:indexdelta}. 
Formally, therefore, the adjoint is the map\label{glo:Deltaast} $\Delta^\ast(x):V\rightarrow W^\ast$, where $W^\ast$\label{glo:Wdual} is the dual space of $W$, defined by 
\begin{equation*} %\label{eq:formaladjoint}
\langle v,\Delta(x) w\rangle = \langle \Delta^\ast(x) v, w\rangle_{\textrm{dual}}
\end{equation*}
for all $v\in V$ and $w\in W$. 
The bracket $\langle , \rangle_{\textrm{dual}}$\label{glo:dualpair} denotes the evaluation of an element in $W^\ast$ on an element of $W$ while the bracket on the LHS is just the inner product in $V$. 

It is useful to understand $W^\ast$ as a sum of Sobolev spaces on the intervals $I_p$.  
First we claim that
\begin{equation}\label{eq:dualLtwodot}
(\Lbar{1}[a,b])^\ast = \Ldot{-1}[a,b] = \Lbar{-1}[a,b]\oplus\{\textrm{Span~}\delta_a\}\oplus\{\textrm{Span~}\delta_b\}
\end{equation}
where $\delta_a,\delta_b$ are the evaluation functionals at $a$ and $b$, which are certainly contained in $\Ldot{-1}[a,b]$. 
The first equality is true from the definitions of $\Lbar{l}$ and $\Ldot{l}$ on page~\pageref{pag:sobs}. 
Consider the orthogonal complement $U$ of $\{\textrm{Span~}\delta_a\}\oplus\{\textrm{Span~}\delta_b\}$ in $\Ldot{-1}[a,b]$, and fix some element $f\in U$. 
Then the map taking $f$ (as an element of $L^2_{-1}(\R)$) to its equivalence class in $\Lbar{-1}[a,b]$ is an isomorphism $U\cong \Lbar{-1}[a,b]$, establishing~$\eqref{eq:dualLtwodot}$. 
It follows that 
\begin{equation*}
\big( W^1_p \big)^\ast=\big( W^{-1}_p \big)\oplus\big( Y_p(\mu_p) \big)^\ast\oplus\big( Y_p(\mu_{p+1}) \big)^\ast.
\end{equation*}
Recall the definition~$\eqref{eq:defnW}$ of $W$. 
Since $W\subset\bigoplus W_p^1$,
\begin{align*}
W^\ast &= \bigoplus_{p=1}^n (W^1_p)^\ast / \textrm{Ann~}W \\
& = \bigoplus_{p=1}^n
\big[ \big( W^{-1}_p \big)\oplus\big( Y_p(\mu_p) \big)^\ast\oplus\big( Y_p(\mu_{p+1}) \big)^\ast \big] / \textrm{Ann~}W
\end{align*}
where $\textrm{Ann~}W$ is the annihilator of $W$ in $\bigoplus W_p^1$. 
Now, since $W$ is the subset of $\bigoplus W_p^1$ consisting of sections with vanishing terminating component which are also continuous across the $\mu_p$, we obtain
\begin{equation*}
W^\ast = \big[\bigoplus_{p=1}^n W^{-1}_p\big]\oplus\big[ \bigoplus_{p=1}^n\big( Y_p^{\textrm{cont}}(\mu_p) \big)^\ast\big],
\end{equation*}
where $Y^{\textrm{cont}}_p(\mu_p)$ is the space of continuing vectors at $\mu_p$. 

We can now compute $\Delta^\ast(x)$ in terms of this decomposition, \ie as an operator
\begin{equation*}
V=\big[\bigoplus_{p=1}^{n}W_p^0\big]\oplus\big[\C^{\Nzer}\big]\rightarrow 
\big[\bigoplus_{p=1}^{n} W^{-1}_p\big]\oplus\big[ \bigoplus_{p=1}^n\big( Y_p^{\textrm{cont}}(\mu_p) \big)^\ast\big] = W^\ast. 
\end{equation*}
If we restrict $\Delta^\ast(x)$ to some domain $A\subset V$ consisting of elements $v\in V$ with $v_p$ smooth on the interior of $I_p$ and continuous up to the ends of the interval, then $A$ contains the cokernel, and integration by parts like that at the start of Section~\ref{sec:indexdelta} gives
\begin{multline} \label{eq:formaladjoint}
\Delta^\ast(x)v =
(D^\ast_1(x)v_1,\ldots,D^\ast_n(x)v_n)\oplus
\big[ \bigoplus_{p\notin\Zeros}\big( iv_{p-1}^{\textrm{cont}}(\mu_p)-iv_{p}^{\textrm{cont}}(\mu_p) \big)^\ast \big] \\
\oplus \big[ \bigoplus_{p\in\Zeros}\big( iv_{p-1}^{\textrm{cont}}(\mu_p)-iv_{p}^{\textrm{cont}}(\mu_p)+s_p\zeta_p \big)^\ast \big]
\end{multline}
where $y^\ast$ denotes the dual of a vector $y\in Y^{\textrm{cont}}_p(\mu_p)$, and $v$ is given by~$\eqref{eq:elementV}$. 
We will need this expansion in Section~\ref{sec:proofSD}. 

Finally, we make a few remarks about the projection $P=P_x$ onto $\coker\Delta(x)$. 
Since $\Delta(x)$ is Fredholm, it follows that its image is closed, so the projection $P$ exists and $\coker\Delta(x)=\ker\Delta^\ast(x)$. 
In addition, $\Delta(x)$ is injective, so $\Delta^\ast(x)$ is surjective, and $\Delta^\ast(x)\Delta(x)$ is invertible. 
The projection $P$ is given by
\begin{equation*}
P=1-\Delta(x)\big(\Delta^\ast(x)\Delta(x)\big)^{-1}\Delta^\ast(x). 
\end{equation*}

\subsection{Nahm's equation and anti-self-duality}\label{sec:proofSD}

We want to show that when the Nahm data satisfies Nahm's equation, $\COKER~\Delta(x)$ is anti-self-dual. 
The index calculation shows that the connection $\A$ defined by $\COKER~\Delta(x)$ is a well-defined $U(n)$ connection on $\rfour$. 
Mimicking Proposition~\ref{prop:NahmgivesASD}, the curvature is given by 
\begin{equation}\label{eq:justabove}
F_{\A} = Pdx(\Delta^{\ast}(x)\Delta(x))^{-1}d{x^\ast}P,
\end{equation}
so anti-self-duality follows if $\Delta^\ast(x)\Delta(x)$ commutes with the $\gamma_j$ matrices. 
In equation~$\eqref{eq:justabove}$ the domain of $(\Delta^{\ast}(x)\Delta(x))^{-1}$ is restricted to the image of $d{x^\ast}P$. 
Using the conditions for sections to be in the cokernel of $\Delta(x)$ given at the start of Section~\ref{sec:indexdelta}, it follows that $(\Delta^\ast(x)\Delta(x))^{-1}$ is only passed elements 
\begin{equation*}
u=(u_1,\ldots,u_n)\oplus(y_1,\ldots,y_n)\in W^\ast = \big[\bigoplus_{p=1}^n W^{-1}_p\big]\oplus\big[ \bigoplus_{p=1}^n\big( Y_p^{\textrm{cont}}(\mu_p) \big)^\ast\big]
\end{equation*}
that have $u_p$ smooth on the interior of $I_p$ for all $p$ and such that the continuing components of $u_p$ are continuous up to the ends of the interval. 
Let $A$ be the inverse image of this subspace of $W^\ast$. 
Since $(\Delta^{\ast}(x)\Delta(x))^{-1}$ is smoothing, sections in $A$ are even `nicer', and on the domain $\Delta(x)A$, $\Delta^\ast(x)$ is given by~$\eqref{eq:formaladjoint}$. 
It is therefore sufficient to prove that $\Delta^{\ast}(x)\Delta(x)$ commutes with the $\gamma_j$ matrices on the domain $A$, for which we take $\Delta^\ast(x)$ to be given by~$\eqref{eq:formaladjoint}$.

Fix $w=(w_1,\ldots,w_n)\in A\subset W$. 
Then 
\begin{multline}\label{eq:deltadagdelta}
\Delta^\ast(x)\Delta(x) w = (D_1^\ast(x) D_1(x) w_1,\ldots,D_n^\ast(x) D_n(x) w_n) \\
\oplus\big[ \bigoplus_{p\notin\Zeros}\big( iv_{p-1}^{\textrm{cont}}(\mu_p)-iv_{p}^{\textrm{cont}}(\mu_p) \big)^\ast \big] \\
\oplus \big[ \bigoplus_{p\in\Zeros}\big( iv_{p-1}^{\textrm{cont}}(\mu_p)-iv_{p}^{\textrm{cont}}(\mu_p)+s_p\zeta_p \big)^\ast \big]
\end{multline}
using equation~$\eqref{eq:formaladjoint}$, where $v_p=D_p(x)w_p$ and $s_p=\pi_p w_p(\mu_p)$. 
Expanding $D^\ast_p(x) D_p(x)$ on each interval gives
\begin{equation}\label{eq:DdagDOrsqu}
D_p^\ast(x) D_p(x)  = \nabla_p^\ast \nabla_p +\sum_j [x_j^2 +2ix_jT_p^j - (T_p^j)^2] 
- \sum_{i,j,k}[ \gamma_i\otimes\nabla_{p} T_p^i + \epsilon_{ijk}\gamma_i \otimes T_p^j T_p^k ].
\end{equation}
The $T_p^j$ satisfy Nahm's equation~$\eqref{eq:nahm}$ on the interior of each interval $I_p$, so the term involving the $\gamma_j$ matrices vanishes there. 
It remains to consider the components in $Y^{\textrm{cont}}_p(\mu_p)$. 
First consider a point $\xi=\mu_p$ with $k_p \neq 0$. 
At such a point the continuing components of $T_p$ and $w_p$ are continuous, so
\begin{equation}\label{eq:wantthisreal}
iv_{p-1}^{\textrm{cont}}(\mu_p)-iv_{p}^{\textrm{cont}}(\mu_p) = 
-\frac{d}{d\xi}w_{p-1}^{\textrm{cont}}(\mu_p)+\frac{d}{d\xi}w_{p}^{\textrm{cont}}(\mu_p). 
\end{equation}
(The RHS exists since $w\in A$.)
If we replace $w$ with $\gamma_j w$, the RHS of equation~$\eqref{eq:wantthisreal}$ is multiplied by $\gamma_j$, so this component of $(\Delta^{\ast}(x)\Delta(x))^{-1}$ commutes with the $\gamma_j$ matrices. 
At a zero jump the continuing component of $T_p$ is discontinuous, and
\begin{multline}\label{eq:zerocontrib}
iv_{p-1}^{\textrm{cont}}(\mu_p)-iv_{p}^{\textrm{cont}}(\mu_p) = 
-\frac{d}{d\xi}w_{p-1}(\mu_p)+\frac{d}{d\xi}w_{p}(\mu_p) \\
-T_{p-1}(\mu_p)w_p(\mu_p)+T_p(\mu_p)w_p(\mu_p).
\end{multline}
The first two terms on the RHS commute with the $\gamma_j$ matrices, just as for the $k_p\neq 0$ case. 
However, 
\begin{equation*}
\big(T_p(\mu_p)-T_{p-1}(\mu_p)\big)w_p(\mu_p) =-\zeta_p\times\pi_p w_p(\mu_p)=-s_p\zeta_p
\end{equation*}
and these terms cancel with the contribution from the $s_p$ in~$\eqref{eq:deltadagdelta}$. 
Substituting~$\eqref{eq:wantthisreal}$ and~$\eqref{eq:zerocontrib}$ into~$\eqref{eq:deltadagdelta}$ gives
\begin{multline*}
\Delta^\ast(x)\Delta(x) w = (D_1^\ast(x) D_1(x) w_1,\ldots,D_n^\ast(x) D_n(x) w_n) \\
\oplus\big[ \bigoplus_{p=1}^n\big( \frac{d}{d\xi}w_{p}^{\textrm{cont}}(\mu_p)-\frac{d}{d\xi}w_{p-1}^{\textrm{cont}}(\mu_p) \big)^\ast \big]
\end{multline*}
and this map commutes with the $\gamma_j$ matrices. 
It follows that $\COKER~\Delta$ is ASD. 

\subsection{Periodicity}\label{sec:calper}

Before considering how to frame $(\E,\A)=\COKER~\Delta$, it is useful to make some remarks about periodicity. 
In particular we will fix some notation required later, and explain how to quotient $(\E,\A)$ to obtain a connection on $\So\times\rthree$. 
As it stands, $\COKER~\Delta$ is a bundle and connection over $\rfour$.
Under translation in $\xo$, $\Delta$ satisfies \label{glo:UVWtau}
\begin{equation}\label{eq:transNahmop}
\Delta(\tau x)=U_{\tau,V}\Delta(x)U_{\tau,W}^{-1}
\end{equation}
where $\tau$\label{glo:tau} is the translation
\begin{equation}\label{eq:defdeltatau}
\tau:(x_0,x_1,x_2,x_3)\mapsto(\xo+\delta_\tau,x_1,x_2,x_3)
\end{equation}
and $U_{\tau,W},U_{\tau,V}$ are unitary maps, given by
\begin{gather*}
U_{\tau,W}(w)=(\exp i \delta_\tau \xi)w, \\
U_{\tau,V}((v_p)\oplus(s_q))=((\exp i \delta_\tau \xi)v_p)\oplus((\exp i\delta_\tau \mu_q)s_q).
\end{gather*} 
These maps are not well-defined for all $\delta_\tau$: in general $U_{\tau,W}(w)$ will not be periodic in $\xi$, so will fail to be an element of $W$. 
When $\delta_\tau$ is a multiple of $\perflat$, however, the maps are well-defined, so fix $\delta_\tau=\perflat$ 
\footnote{We can use $\Un$ Nahm data to construct a caloron with $m$ times the expected period for $m=1,2,3,\ldots$ by taking $\delta_\tau$ to be $m\times \perflat$. The point is, however, that while other periods are possible, the Nahm data does give rise to a caloron of the anticipated period. 
Changing the periodicity of the caloron in this way maps $\muo\mapsto\muo / m$ and $\ko\mapsto m\ko$ because in the quasi-periodic picture the clutching function is composed with itself $m$ times. 
This should correspond to some map between sets of $U(n)$ Nahm data. 
Similarly, given a set of $U(n)$ Nahm data, we can simply glue together $m$ copies of it to obtain a set of $U(nm)$ Nahm data with period $m\mu_0$. 
This should correspond to some map between caloron configurations, probably embedding the $U(n)$ caloron $m$ times in $U(nm)$.} 
and the corresponding maps $U_{\tau,W},U_{\tau,V}$.
Of course, $U_{\tau,W},U_{\tau,V}$ correspond to the gauge transformation $\eqref{eq:torusGT}$ for the transform on the $4$-torus. 
We will often just write $U_\tau$\label{glo:Utau} for $U_{\tau,V}$, so that
\begin{equation}\label{eq:translationmap}
U_{\tau}\big((v_p)\oplus(s_q)\big) = 
((\exp 2\pi i \xi / \muo)v_p)\oplus((\exp 2\pi i \mu_q / \muo)s_q).
\end{equation}
Now 
\begin{equation*}
\coker\Delta(\tau x) = U_{\tau}~\coker\Delta(x)
\end{equation*}
so $U_{\tau}$ defines an action of $\Z$ on $\E$.
It is easy to check that
\begin{equation}\label{eq:connectionperiodic}
\A(\tau x) = (U_{\tau}^{-1})^\ast\A(x)
\end{equation}
so the connection is compatible with this action.
Quotienting by this action gives an hermitian bundle on $\So\times\rthree$ with a compatible connection. 

Note that $U_{\tau,W}$ and $U_{\tau,V}$ are not determined uniquely: 
we can replace them with $(\exp i\lambda)U_{\tau,W}$ and $(\exp i\lambda)U_{\tau,V}$ for any $\lambda\in\R$ and $\eqref{eq:transNahmop}$ still holds. 
However, the choice~$\eqref{eq:translationmap}$ is the only choice that allows the caloron to be framed in the correct way. 

\subsection{Remarks on the rotation map}\label{sec:rotnahm}

The definition of the Nahm data for a caloron implicitly involves a choice of origin for the circle $\Sdual$.
Suppose $\Delta (x):W\rightarrow V$ is the Nahm operator constructed from some element of $\spc{N}{(B)}$ for some set of boundary data $B$.
The inner product on $V$ is independent of the choice of origin on $\Sdual$, and so, as a bundle and connection over $\rfour$, $\COKER~\Delta$ is also independent of this choice. 
Recall the rotation map $\rotbdary$ defined by~$\eqref{eq:defnrotbdary}$. 
Rotation of the Nahm data by $\muo / n$ defines a map\label{glo:rotnahm}
\begin{equation}\label{eq:defrotnahm}
\rotNahm:\spc{N}{(B)}\rightarrow \spc{N}{(\rotbdary B)}
\end{equation}
but, as it stands, $\COKER~\Delta$ is insensitive to this action. 
However, the way we frame $\COKER~\Delta$ does depend on the choice of origin: if the Nahm data has boundary data $B$ then we want the corresponding caloron to have boundary data $B$ too. 

In fact, we only give a construction of calorons from Nahm data with principal boundary data---for technical reasons the construction is much harder if the data is not principal, essentially because we can only recover $k_0$ when it is the smallest rank of the bundles $X_1,\ldots,X_n$. 
Given a set of Nahm data with boundary data $B$ (not necessarily principal), we can rotate it until its boundary data $B'$ is principal. 
Our construction will then give a framed caloron with boundary data $B'$, but by applying the rotation map $\rotcal$ defined in Section~\ref{sec:rotationbdary} to the caloron, we can transform it into a caloron with boundary data $B$ as desired. 
This method of defining the construction when $B$ is non-principal ensures the following diagram commutes:
\begin{equation*}
\begin{CD}
\spc{N}{(B)} @>\textrm{Nahm}>\textrm{construction}> \spc{C}{(B)}  \\
@V{\rotNahm}VV @VV{\rotcal}V \\
\spc{N}{(\rotbdary B)}@>\textrm{Nahm}>\textrm{construction}> \spc{C}{(\rotbdary B)}
\end{CD}
\end{equation*}
The diagram commutes by definition when there is only one principal rotation of the Nahm data. 
If there is more than one principal rotation, then without loss of generality we can assume $B$ and $\rotbdary B$ are both principal (by replacing the rotation maps $\rotbdary,\rotcal,\rotNahm$ with $\rotbdary^k,\rotcal^k,\rotNahm^k$ for some $k$)
The two calorons on the RHS of the diagram must be equivalent as quasi-periodic connections over $\rfour$---in other words they are related by a large gauge transformation---since $\COKER~\Delta$ is insensitive to the action of $\rotNahm$. 
It is easy to see the large gauge transformation must be $\rotcal$.

\section{The model operator $\tilde{\Delta}$}\label{sec:modelop}

\subsection{Strategy}\label{sec:deformstrategy}

The aim of the remainder of this Chapter is to show that the bundle and connection $\COKER~\Delta$ extends to $\sphinf$, and is framed there. 
This will complete the proof that the Nahm construction on an element of $\spcSD{N}{\bdarydata}$ produces an element of $\spcSD{C}{\bdarydata}$.
The method adopted is to consider a deformation $\modelop$\label{glo:Deltatilde} of $\Delta$, and prove that $\COKER~\modelop$ extends to $\sphinf$ and is a framed bundle $(\tilde{\E},f)$ with invariant $c_2(\tilde{\E},f)=\ko$. 
We then prove that the framing and the invariant $c_2$ are independent of the deformation, and so deduce that $\COKER~\Delta\in\spcSD{C}{\bdarydata}$.
The first task is to define the model operator\label{glo:modelop} $\modelop$ and prove that it is injective and Fredholm with index $-n$, which forms Section~\ref{sec:modelop}.
In Section~\ref{sec:modelframed} we prove that $\COKER~\modelop$ is a framed caloron configuration, and in Section~\ref{sec:modelko} calculate $c_2(\tilde{\E},f)$. 
This shows that $\COKER~\modelop$ is an element of $\spc{C}{\bdarydata}$. 
Finally, in Section~\ref{sec:deform}, we prove that $\tilde{\Delta}$ can be deformed into $\Delta$ in such a way that the framing is independent of the deformation, and deduce that $\COKER~\Delta$ is an element of $\spcSD{C}{\bdarydata}$.
This method is based on Hitchin's proof that a connection and Higgs field constructed from monopole Nahm data satisfy the monopole boundary conditions \cite[Section 2]{hit83}.

From this point on fix the following notation. 
Fix a set of $\Un$ caloron Nahm data with principal boundary data $\bdarydata$.
If the boundary data is not principal then we apply the rotation map $\rotNahm$ to the Nahm data to obtain a set of principal Nahm data; we then apply the corresponding rotation map $\rotcal^{-1}$ to the caloron constructed to obtain the correct framing, just as we explained in Section~\ref{sec:rotnahm}. 
Let $\Ainf,\Phiinf$ be the connection and Higgs field at infinity determined by $\monbdarydata$. 
Let $\Delta(x):W\rightarrow V$ be the Nahm operator.

Any generic choice of model operator $\modelop$ will be injective (since the space of non-injective operators is in some sense small). 
The problems encountered when defining $\modelop$ are therefore to ensure $\COKER~\modelop$ is framed and that $c_2(\tilde{\E},f)=k_0$. 
Much of the `engineering' we do is to make the proof that $c_2(\tilde{\E},f)=k_0$ straightforward. 
Note that throughout this Section we make no claims that the model operator $\modelop$ is a deformation of $\Delta$; often $\modelop$ will be defined in terms of a deformation, but we delay the proof of the existence of a path joining the two until later. 

When defining the model operator $\modelop$ it is easiest to consider two special cases first. 
The first is the case of vanishing monopole charges\label{glo:VMC2}---the case that $k_p=0$ for all $p=1,\ldots,n$ but $\ko\neq 0$. 
(Calorons of this type were introduced in Section~\ref{sec:kraan}.)
The second case is the opposite of this: the case that there are no zero jumps in the lowest rank block (we will make this precise later). 
We call this the case of `no zero jumps in the instanton block'. 
In the remaining cases, the model operator is very similar to that for no zero jumps in the instanton block, so the reader may prefer to concentrate on that case. 

\subsection{Defining $\modelop$: the case of vanishing monopole charges}
\label{sec:physmodelop}

Suppose that $k_p=0$ for all $p=1,\ldots,n$ but $\ko\neq 0$ 
\ie that every singularity $\mu_p$ in the Nahm data is a zero jump. 
By gluing together the bundles $X_p$, a single continuous vector bundle $\bigcup X_p$ over $\Sdual$ with rank $k_0$ is obtained. 
Similarly we can glue together the $Y_p = \C^2\otimes X_p$ to obtain a vector bundle $Y$ over $\Sdual$. 
Elements of $W$ are sections of $Y$ that are periodic, $L^2_1$, and continuous across the singularities $\mu_p$. 
Let $\bar{\eta}_1,\ldots,\bar{\eta}_{k_0}$ be an orthonormal basis of sections of $\bigcup X_p$ that are periodic, smooth on each interval $I_p \subset\Sdual$, and continuous across each $\xi=\mu_p$.
Let\label{glo:etas}
\begin{equation}\label{eq:definstblcksections}
\eta_l = \begin{pmatrix}1 \\ 0 \end{pmatrix}\otimes \bar{\eta}_l
~\textrm{and}~
\eta_l^\perp = \begin{pmatrix}0 \\ 1 \end{pmatrix}\otimes \bar{\eta}_l
\end{equation}
Working in this gauge fix
\begin{equation*}
\tilde{D}_p (x) = i\frac{d}{d\xi} +i(1\otimes\Lambda) +x
\end{equation*}
for each $p=1,\ldots,n$, where
\begin{equation*}
\Lambda = \diag(i\lambda_1,\ldots,i\lambda_{\ko})
\end{equation*}
for some pairwise distinct $\lambda_1,\ldots,\lambda_{\ko}\in(0,\perflat)$, 
and $x=\sum\gamma_a x_a$. 
Given $\Delta(x):W\rightarrow V$, consider the operator
\begin{gather*}
\modelop(x):W\rightarrow V \\
\modelop(x)w = [\tilde{D}_1 (x) w_1,\ldots,\tilde{D}_n (x) w_n]\oplus
[\pi w]
\end{gather*}
where the projection $\pi$ is that determined by $\Delta$, as in equation~$\eqref{eq:defdelta}$.

We want $\modelop(x)$ to be injective for all $x$, so consider the conditions for $w \in W$ to lie in the kernel of $\modelop(x)$. 
We have $\modelop(x) w=0$ if $\tilde{D}_p(x) w_p=0$ and $\pi_p w_p(\mu_p)=0$ for all $p$, and if $w$ is continuous and periodic.
Parallel translation by the $\tilde{D}_p$ round $\Sdual$ determines a holonomy 
\begin{equation*}
\textrm{Hol}(x) = \exp [(ix-\Lambda)\mu_0].
\end{equation*}
Consider finding solutions to this parallel transport problem that are continuous and periodic in $\xi$. 
Such solutions exist if and only if $\textrm{Hol}(x)$ has eigenvalue $1$ \ie iff $[1-\textrm{Hol}(x)]$ is singular.
This occurs if and only if
\begin{equation}\label{eq:resonatingpoints}
x=(\lambda_l+2\pi m / \muo,0,0,0)\textrm{\ for some\ }
l\in\{ 1,\ldots,\ko \}\textrm{\ and any\ }m\in\Z.
\end{equation}
We call such points \emph{resonating points}\label{glo:respts}, and label them $x_{lm}^{\textrm{res}}$\label{glo:xreslm} for $l=1,\ldots,\ko$ and $m\in\Z$. 
As it stands, $\modelop(x)$ is therefore injective away from the resonating points. 
When $x=x_{lm}^{\textrm{res}}$, there is a $2$-dimensional space of solutions to the parallel transport problem that is spanned by \label{glo:etaslm}
\begin{gather}\label{eq:resonatingsoln1}
\eta_{lm}:=[\exp (2\pi im\xi / \muo)]\eta_l, \\
\eta_{lm}^\perp:=[\exp (2\pi im\xi / \muo)]\eta_l^\perp.\label{eq:resonatingsoln2}
\end{gather}
To exclude the possibility that these could be solutions to $\modelop(x)$, we adjust the projection $\pi$ to be non-zero on $\eta_{lm},\eta_{lm}^\perp$ 
by deforming the vectors $\zeta_p$ used to define $\pi$ in a small neighbourhood of each resonating point. 
(Recall the definition of the components of $\pi$, equation~$\eqref{eq:defcmptspi}$.)
For $p=1,\ldots, n$ let \label{glo:zetatilde}
\begin{equation*}
\tilde{\zeta}_p : \rfour \rightarrow  Y_p(\mu_p) 
\end{equation*}
and let\label{glo:pitilde}
\begin{equation*}
\tilde{\pi}_p(x)(w)=\langle w(\mu_p),\tilde{\zeta}_p(x)\rangle.
\end{equation*}
Let $\tilde{\pi}(x):W\rightarrow \C^n$ have components $\tilde{\pi}_p(x)$. 
Fix an open $4$-ball $B_{lm}$\label{glo:Blm} around each resonating point, sufficiently small that the balls do not overlap. 
We deform $\zeta_p$ only inside the union $\bigcup B_{lm}$: define
\begin{equation*}
\tilde{\zeta}_p(x)=\zeta_p
\end{equation*}
if $x\notin\bigcup B_{lm}$.
Then, for each $l$, pick $q_l,q_l^\perp\in\{ 1,\ldots,n \}$ such that $q_l\neq q_l^\perp$. 
Inside $B_{lm}$ arrange the $\tilde{\zeta}_p$ so that near $x=x_{lm}^{\textrm{res}}$
\begin{align}
\tilde{\zeta}_{q_l}(x)&= \eta_{lm}(\mu_{q_l}), \label{eq:physcond1}\\
\tilde{\zeta}_{q_l^\perp}(x) &= \eta_{lm}^\perp(\mu_{q_l^\perp})\textrm{\  and}\label{eq:physcond2} \\
\tilde{\zeta}_{q}(x)&=0
\textrm{\ if\ }q\notin\{ q_l,q_l^\perp \}.\label{eq:physcond3}
\end{align}
The third condition is not required for injectivity, but simplifies the calculation of $\ko$ for $\COKER~\tilde{\Delta}$. 
Equations~$\eqref{eq:physcond1}$ and~$\eqref{eq:physcond2}$ imply that at a resonating point $x_{lm}^{\textrm{res}}$, $\tilde{\pi}_{q_l}$ and $\tilde{\pi}_{q_l^\perp}$ cannot both be zero on non-trivial linear combinations of $\eta_{lm}$ and $\eta_{lm}^\perp$. 
This ensures that
\begin{equation*}
\modelop(x)w = [\tilde{D}_1(x) w_1,\ldots,\tilde{D}_n(x) w_n]\oplus
[\tilde{\pi}w ]
\end{equation*}
is injective for all $x$. 
We can arrange the deformation of the vectors $\zeta_p$ so that $\tilde{\zeta}_p(\tau x)=\tilde{\zeta}_p(x)$ for all $x$ and $p$, which ensures that $\modelop$ satisfies~$\eqref{eq:transNahmop}$ under translation by $\perflat$. 
Note that $\modelop$ is a controlled deformation of $\Delta$ (recall Definition~\ref{def:controlleddeformn}).
Corollary~\ref{cor:indexdefdelta} therefore implies that $\modelop(x)$ is Fredholm with index $-n$ for all $x$. 
As a final remark, note that, unlike $\Delta$, $\modelop$ is continuous across the zero jumps: the discontinuity in $\Delta$ is required to ensure $\COKER~\Delta$ is anti-self-dual, but it is not required to ensure $\COKER~\Delta$ is framed. 

\subsection{Defining $\modelop$: the case of no zero jumps in the instanton block}\label{sec:modopnoinstzerojumps}

Moving on to the second special case, we first explain what is meant by `no zero jumps in the instanton block'. 
A zero jump in the instanton block is a point $\xi=\mu_p$ for which $m_p=\textrm{min}\{ m_1, \ldots, m_n \}$ which is also a zero jump (\ie $k_p=0$). 
The condition of no zero jumps in the instanton block is the opposite of the condition of vanishing monopole charges, for which every singularity was a zero jump in the instanton block. 
Figure~\ref{fi:classification} illustrates the situation, plotting the rank of the Nahm data versus $\xi$ for three examples. 

%%%%%%%%%%%%%%%%%%%%%%%%%%%%%%%%%%%%%%%%%%%%%%%%%%%%%%%%%%%%%%%%%%%%%%%%%%%%%%
%FIGURE: classification of Nahm data.
%%%%%%%%%%%%%%%%%%%%%%%%%%%%%%%%%%%%%%%%%%%%%%%%%%%%%%%%%%%%%%%%%%%%%%%%%%%%%%
\begin{figure}
\setlength{\unitlength}{0.75cm}
\begin{center}
\begin{picture}(14,21.5)(0,0)
\put(0,0){\begin{picture}(16,7)(-8,-1)
\put(-8,0){\vector(1,0){16}}
\put(8.4,-0.2){$\xi$}
\put(-7,1.5){\line(1,0){1}}
\put(-6,2.5){\line(1,0){2}}
\put(-4,3.5){\line(1,0){2}}
\put(-2,4.5){\line(1,0){2}}
\put(0,1.5){\line(1,0){4}}
\put(4,2.5){\line(1,0){2}}
\put(6,3.5){\line(1,0){1}}
\put(-6,0){\line(0,1){2.5}}
\put(-4,0){\line(0,1){3.5}}
\put(-2,0){\line(0,1){4.5}}
\put(0,0){\line(0,1){4.5}}
\put(2,0){\line(0,1){1.5}}
\put(4,0){\line(0,1){2.5}}
\put(6,0){\line(0,1){3.5}}
\put(-6.35,-1){\begin{picture}(13,1)(0,0)
              \put(0,0){$\mu_5$}
              \put(2,0){$\mu_4$}
              \put(4,0){$\mu_3$}
              \put(6,0){$\mu_2$}
              \put(8,0){$\mu_1$}
              \put(9.6,0){$\mu_0+\mu_5$}
              \put(11.6,0){$\mu_0+\mu_4$}
              \end{picture}}
\put(-8,6.5){$U(5)$ Nahm data with one zero jump in the instanton block}
%\put(-8,5.8){(Two adjacent rotations of the data are principal)}
\end{picture}}

% Separated principal rotations:-
\put(0,9.5){\begin{picture}(16,6)(-8,-1)
\put(-8,0){\vector(1,0){16}}
\put(8.4,-0.2){$\xi$}
\put(-7,1.5){\line(1,0){1}}
\put(-6,0){\line(0,1){3.5}}
\put(-6,3.5){\line(1,0){2}}
\put(-4,3.5){\line(0,-1){3.5}}
\put(-4,0){\line(0,1){1.5}}
\put(-4,1.5){\line(1,0){2}}
\put(-2,1.5){\line(0,-1){1.5}}
\put(-2,0){\line(0,1){2.5}}
\put(-2,2.5){\line(1,0){2}}
\put(0,2.5){\line(0,-1){2.5}}
\put(0,0){\line(0,1){2.5}}
\put(0,2.5){\line(1,0){2}}
\put(2,2.5){\line(0,-1){2.5}}
\put(2,1.5){\line(1,0){2}}
\put(4,0){\line(0,1){3.5}}
\put(4,3.5){\line(1,0){2}}
\put(6,3.5){\line(0,-1){3.5}}
\put(7,1.5){\line(-1,0){1}}
\put(-6.35,-1){\begin{picture}(13,1)(0,0)
              \put(0,0){$\mu_5$}
              \put(2,0){$\mu_4$}
              \put(4,0){$\mu_3$}
              \put(6,0){$\mu_2$}
              \put(8,0){$\mu_1$}
              \put(9.6,0){$\mu_0+\mu_5$}
              \put(11.6,0){$\mu_0+\mu_4$}
              \end{picture}}
\put(-8,5){$U(5)$ Nahm data with no zero jumps in the instanton block}
%\put(-8,4.3){(Two rotations of the data are principal)}
\end{picture}}

% Vanishing monopole charges:-
\put(0,17.5){\begin{picture}(16,4)(-8,-1)
\put(-8,0){\vector(1,0){16}}
\put(8.4,-0.2){$\xi$}
\put(-7,2){\line(1,0){14}}
\put(-6,0){\line(0,1){2}}
\put(-4,0){\line(0,1){2}}
\put(-2,0){\line(0,1){2}}
\put(0,0){\line(0,1){2}}
\put(2,0){\line(0,1){2}}
\put(4,0){\line(0,1){2}}
\put(6,0){\line(0,1){2}}
\put(-6.35,-1){\begin{picture}(13,1)(0,0)
              \put(0,0){$\mu_5$}
              \put(2,0){$\mu_4$}
              \put(4,0){$\mu_3$}
              \put(6,0){$\mu_2$}
              \put(8,0){$\mu_1$}
              \put(9.6,0){$\mu_0+\mu_5$}
              \put(11.6,0){$\mu_0+\mu_4$}
              \end{picture}}
\put(-8,3){$U(5)$ Nahm data with vanishing monopole charges}
\end{picture}}

\end{picture}
\end{center}
\caption{Examples of classification of caloron Nahm data.}
\label{fi:classification}
\end{figure}
%%%%%%%%%%%%%%%%%%%%%%%%%%%%%%%%%%%%%%%%%%%%%%%%%%%%%%%%%%%%%%%%%%%%%%%%%%%%%%

We assume, therefore, that the Nahm data has no zero jumps in the instanton block. 
We can split off a bundle over $\Sdual$ of constant rank, and thereby decompose each bundle $Y_p$ into two sub-bundles, in the following way. 
Recall that we are assuming that the Nahm data is principal so that
\begin{equation*}
k_0 = \textrm{min}\{ \textrm{rank\ }X_p : p=1,\ldots,n \}.
\end{equation*} 
There exists a set $\{ \bar{\eta}_1,\ldots,\bar{\eta}_{\ko} \}$ of sections of the bundles $X_1,\ldots, X_n$ that are defined over all of $\Sdual$, that are periodic, smooth on each interval $I_p\subset\Sdual$, continuous across each singularity $\xi=\mu_p$, orthogonal for each $\xi\in\R$, normalised with respect to the $L^2$ inner product, and linearly independent. 
In other words, we can find a $\ko$-dimensional sub-bundle $X_p^I$ of each $X_p$ such that the $X_p^I$ glue together to give a continuous bundle over $\Sdual$. Note that this decomposition is not unique: if rank~$X_p>k_0$ then there is some choice for the $\bar{\eta}_l$ over $I_p$. 
Define $\eta_l$ and $\eta_l^\perp$ by~$\eqref{eq:definstblcksections}$. 
We can choose the $\bar{\eta}_l$ so that for all $l=1,\ldots,\ko$ and $q\in\Zeros$
\begin{equation}\label{eq:zerojumpcond}
\pi_q \eta_l = 0 = \pi_q\eta_l^\perp.
\end{equation}
This is possible because each zero jump occurs in a block with rank greater than $\ko$.
Thus we have a decomposition of each block $Y_p$ of Nahm data into a subspace $Y_p^I$\label{glo:YpI} spanned by $\{ \eta_l(\xi),\eta_l^\perp(\xi) : l=1,\ldots,k_0, \xi\in I_p\}$ and the orthogonal complement $Y_p^M$:
\begin{equation*}
Y_p = Y_p^I\oplus Y_p^M.
\end{equation*}
The bundle formed by gluing together the $Y^I_p$ is called the \emph{instanton block}\label{glo:instblock}, and its orthogonal complement is called the \emph{monopole block}\label{glo:monoblock}. 
The spaces $W$ and $V$ decompose in a similar way:\label{glo:Wdecomp}
\begin{equation}\label{eq:decompWV}
W=W_I\oplus W_M,\quad V=V_I\oplus V_M, 
\end{equation}
where
\begin{equation*}
V_I = \bigoplus_{p=1}^n W^0(Y_p^I)
\end{equation*}
and 
\begin{equation*}
V_M = [\bigoplus_{p=1}^n W^0(Y_p^M)]\oplus[\C^{\Nzer}].
\end{equation*}
Here $W^0(Y_p^I)$ denotes the Sobolev space of $L^2$ sections of $Y_p^I$ \etc 
Note that we impose the condition~$\eqref{eq:zerojumpcond}$ to ensure that the jumping space $J_q$ spanned by $\zeta_q$ at a zero jump $\mu_q$ is contained in the monopole block $Y^M_q(\mu_q)$, so that there really are no zero jumps in the instanton block.

Next we want to specify the model operator $\modelop(x):W\rightarrow V$ which we construct using the decomposition into monopole and instanton blocks. 
Define\label{glo:DeltaI}
\begin{gather}
\Delta_I(x):W_I\rightarrow V_I \\
\Delta_I(x) = i\frac{d}{d\xi} +i(1\otimes\Lambda) +x \label{eq:defdeltaI}
\end{gather}
where $\Lambda = \diag(i\lambda_1,\ldots,i\lambda_{\ko})$ 
for some pairwise distinct $\lambda_1,\ldots,\lambda_{\ko}\in(0,\perflat)$. 
Just as we saw for the case of vanishing monopole charges, $\Delta_I(x)$ is injective, apart from at the resonating points defined by~$\eqref{eq:resonatingpoints}$. 

Restriction of the caloron Nahm data to the monopole block determines a set of $\Un$ monopole data,~\ie an element of $\mon{N}{\monbdarydata}$, and the  associated Nahm operator $\Delta_M(x):W_M\rightarrow V_M$. 
(Note that this data does not necessarily satisfy Nahm's equation.)
At a singularity $\xi=\mu_p$ with $k_p\neq 0$ the Nahm data determines a $| k_p|$-dimensional residue $R_p$. 
We deform $\Delta_M(x)$ so that near such singularities it is given by 
\begin{equation*}
i\frac{d}{d\xi}+\frac{iR_p}{\xi-\mu_p}+x
\end{equation*}
on the terminating component and
\begin{equation}\label{eq:contcptmodelop}
i\frac{d}{d\xi}+x
\end{equation}
on the continuing component. 
Near zero jumps we deform so that $\modelop(x)$ is also given by~$\eqref{eq:contcptmodelop}$. 
Thus $\Delta_M(x):W_M\rightarrow V_M$ is given by 
\begin{equation}\label{eq:defdeltaM}
\Delta_M(x)w=\big[\big( i\frac{d}{d\xi}+i\sum_{p\notin\Zeros}\frac{\psi_p(\xi) R_p}{\xi-\mu_p}+x \big)w \big]\oplus[\pi w]
\end{equation}
where $\psi_p$ is a bump function equal to $1$ on some neighbourhood of $\mu_p$ and zero elsewhere. 
We interpret $R_p$ as acting on each terminating component in the obvious way. 
Note that without condition~$\eqref{eq:zerojumpcond}$, $\Delta_M$ would not be a well-defined $U(n)$ monopole Nahm operator, since $\pi$ would be non-zero on the instanton block. 
Lemma~\ref{lem:HM1} implies that $\Delta_M(x):W_M\rightarrow V_M$ defined in this way is injective and Fredholm with index $-n$ for all $x$. 

Consider the operator
\begin{equation*}
\begin{pmatrix}
\Delta_I(x) &0 \\ 0& \Delta_M(x)
\end{pmatrix}:W\rightarrow V.
\end{equation*}
It is injective away from the resonating points~$\eqref{eq:resonatingpoints}$. 
We want to put something into the off-diagonal entries, supported near the resonating points, that will ensure the new operator is injective everywhere. 
We need the following: 

\begin{lemma}\label{lem:stupid}
Let $\Delta_M$ be the Nahm operator for some set of $\Un$ ($n\geq 2$) Nahm data in $\mon{N}{\monbdarydata}$, and let $x\in\rfour$.
Then we can find two non-trivial orthogonal continuous sections $u,u^\perp$ of the bundle $\COKER~\Delta_M$ defined on some open neighbourhood $R$ of $x$, such that, for all $y\in R$, $u(y)$ and $u^\perp(y)$ are continuous across each zero jump so have zero component in $\C^{\Nzer}$.
\end{lemma}

\proof
For any set of $\Un$ monopole Nahm data the singularities $\mu_1$ and $\mu_n$ cannot be zero jumps, so there are at most $n-2$ zero jumps in the data. 
The restriction that solutions be continuous across the zero jumps therefore rules out at most $n-2$ dimensions, and so $u(y)$ and $u^\perp(y)$ can be chosen from a $2$-dimensional subspace of $\coker\Delta_M(y)$. 
Moreover, this can be done smoothly on some neighbourhood $R$ of $x$. 
\eproof

Let $\varphi_{lm}$ be a set of bump functions on $\R\times\rthree$ that are 
equal to $1$ on some $4$-ball with centre $x=x_{lm}^{\textrm{res}}$ and zero outside some $4$-ball with centre $x=x_{lm}^{\textrm{res}}$. 
We can make the support of each bump function sufficiently small that they are disjoint, and arrange them so that $\varphi_{l,m+1}(\tau x)=\varphi_{l,m}(x)$. 
At $x=x_{lm}^{\textrm{res}}$, $\Delta_I(x)$ has two solutions 
$\eta_{lm},\eta_{lm}^\perp$ defined by~$\eqref{eq:resonatingsoln1}$ and~$\eqref{eq:resonatingsoln2}$.
Let $u_l$ and $u_l^\perp$ be the elements of $\coker\Delta_M(x)$ fixed by Lemma~\ref{lem:stupid} defined on some neighbourhood of $x_{l,m=0}^{\textrm{res}}$. 
Adjust the bump functions so they are supported within these neighbourhoods, and extend $u_l,u_l^\perp$ periodically by defining
\begin{equation*}
u_{lm}=[\exp (2\pi mi\xi / \muo)]u_l, \quad u_{lm}^\perp=[\exp (2\pi mi\xi / \muo)]u_l^\perp.
\end{equation*}
Define $B(x):W_M\rightarrow V_I$\label{glo:offdiag} by
\begin{equation}\label{eq:defnB} 
B(x)w_M=\sum_{l,m} \varphi_{lm}(x)\big[(\langle w_M,u_{lm}(x) \rangle_{L^2})\eta_{lm}+ (\langle w_M,u_{lm}^\perp(x) \rangle_{L^2})\eta_{lm}^\perp \big]
\end{equation}
and $B^\ast(x):W_I\rightarrow V_M$ by
\begin{equation}\label{eq:defnBstar} 
B^\ast(x) w_I=\sum_{l,m} \varphi_{lm}(x)\big[(\langle w_I,\eta_{lm} \rangle_{L^2})u_{lm}(x)+ (\langle w_I,\eta_{lm}^\perp \rangle_{L^2})u_{lm}^\perp (x)\big].
\end{equation}
We chose $u_l$ and $u_l^\perp$ to have no component in the jumping spaces so that all inner products can be made inside $L^2$---this simplifies the definition of $B$. 
Define the model operator $\modelop$ by
\begin{equation}\label{eq:defmodopA}
\modelop = \begin{pmatrix} \Delta_I & B \\ B^\ast & \Delta_M \end{pmatrix}.
\end{equation}

\begin{lemma}\label{lem:modelinj}
$\modelop(x)$ is injective for all $x\in\rfour$.
\end{lemma}

\proof
We have already shown $\modelop(x)$ is injective outside $\bigcup\supp~\varphi_{lm}$, so it remains to consider what happens near resonating points.

First we show that there are no non-trivial solutions to 
\begin{equation}\label{eq:nonontrivsoln}
\modelop(x)\begin{pmatrix}w_I \\ 0\end{pmatrix}=0.
\end{equation}
We know $\Delta_I(x) w_I(x)=0$ has periodic solutions only when  $x=x_{lm}^{\textrm{res}}$ for some $l,m$.
At such a point, the solutions have the form 
\begin{equation*}
w_I=C\eta_{lm} + C^{\perp}\eta_{lm}^{\perp}
\end{equation*}
for some constants $C,C^{\perp}$.
Then $B^\ast w_I = (Cu_l+C^{\perp}u_l^{\perp})$, which is zero if and only if $C=C^{\perp}=0$.

Next, note that the image of $B^\ast(x)$ is contained in $\coker\Delta_M(x)$ and so is orthogonal to the image of $\Delta_M(x)$. 
Hence, if
\begin{equation*}
\begin{pmatrix} \Delta_I & B \\ B^\ast & \Delta_M \end{pmatrix}
\begin{pmatrix} w_I \\ w_M \end{pmatrix} =0
\end{equation*}
then $\Delta_M(x) w_M(x) =0$, and so $w_M(x)=0$ because $\Delta_M(x)$ is injective. 
We conclude that $\modelop(x)$ is injective because there are no non-trivial solutions to~$\eqref{eq:nonontrivsoln}$.
\eproof

Consider the behaviour of $\modelop$ with respect to translation $\tau$ by one period in $\xo$. 
Suppose $x\in\supp~\varphi_{lm}$. 
Then
\begin{align*}
\begin{split}
B(\tau x)w &=
\varphi_{l,m+1}(\tau x)\big[\langle w, (\exp (2\pi i\xi / \muo))u_{lm}(x) \rangle_{L^2}\eta_{l,m+1}\\ &\qquad\qquad\qquad\qquad\qquad\qquad\qquad+ \langle w,(\exp (2\pi i\xi / \muo))u_{lm}^\perp(x) \rangle_{L^2}\eta_{l,m+1}^\perp \big] \end{split}\\
\begin{split}
&=\varphi_{l,m}(x)\big[\langle (\exp (-2\pi i\xi / \muo))w, u_{lm}(x) \rangle_{L^2}(\exp (2\pi i\xi / \muo))\eta_{l,m}\\ &\qquad\qquad\qquad\qquad+ \langle (\exp (-2\pi i\xi / \muo))w,u_{lm}^\perp(x) \rangle_{L^2}(\exp (2\pi i\xi / \muo))\eta_{l,m}^\perp \big]
\end{split}\\
&=U_{\tau,V}B(x)U_{\tau,W}^{-1}w.
\end{align*} 
$B^\ast$ satisfies a similar formula, and this ensures that $\modelop$ satisfies~$\eqref{eq:transNahmop}$ under translation. 

By construction, $\modelop$ satisfies the conditions of Definition~\ref{def:controlleddeformn} so is a controlled deformation of $\Delta$.
Corollary~\ref{cor:indexdefdelta} then implies that $\modelop$ is Fredholm with index $-n$. 
Note that the model operator for a set of monopole Nahm data is implicitly dealt with by the case of no zero jumps in the instanton block---we simply take the instanton block to be trivial.

\subsection{Defining $\modelop$: the remaining case}\label{sec:modelremaining}

We now generalize the model operator defined in Section~\ref{sec:modopnoinstzerojumps} to deal with the remaining case: that of zero jumps in the instanton block together with a non-trivial monopole block. 
The following classification of the singularities $\mu_p$ is required. 
Recall that we are assuming that the Nahm data is principal and that $\Zeros$ is the set of zero jumps. 
Define\label{glo:moreZeros}
\begin{align*}
\Zeros_I &:= \textrm{\ zero jumps in instanton block} \\
&=\{ p: m_p=m_{p-1}\textrm{\ and\ }m_p=\ko \}, \\
\Zeros_M &:= \textrm{\ zero jumps in monopole block} \\
&=\Zeros \setminus \Zeros_I, \\
\Zeros_O &:= \textrm{\ other singularities} \\
&=\{ 1,\ldots,n \} \setminus \Zeros.
\end{align*}
Table~\ref{tab:classification} lists these sets for the examples shown in Figure~\ref{fi:classification}. 
Let $N_I=| \Zeros_I |$ and $N_M=| \Zeros_M |$\label{glo:NINM}. 
We assume that $N_I\neq n$ \ie we are not dealing with the case of vanishing monopole charges so there is a non-trivial monopole block. 
It follows that $n-N_I\geq 2$, because we cannot have all but one $\mu_p$ being zero jumps. 

%%%%%%%%%%%%%%%%%%%%%%%%%%%%%%%%%%%%%%%%%%%%%%%%%%%%%%%%%%%%%%%%%%%%%%%%%%%%%
% TABLE: classification
%%%%%%%%%%%%%%%%%%%%%%%%%%%%%%%%%%%%%%%%%%%%%%%%%%%%%%%%%%%%%%%%%%%%%%%%%%%%%
\begin{table}
\begin{center}
\begin{tabular}{|l|ccc|}
\hline
Data set& $\Zeros_I$ & $\Zeros_M$ & $\Zeros_O$ \\
\hline
Vanishing monopole charges & $\{1,2,3,4,5\}$ & $\{\}$ & $\{\}$ \\
No zero jumps in instanton block & $\{\}$ & $\{2\}$ & $\{1,3,4,5\}$ \\
Zero jump in instanton block & $\{1\}$ & $\{\}$ & $\{2,3,4,5\}$ \\
\hline
\end{tabular}
\end{center}
\caption{Classification of the singularities for the example sets of Nahm data in Figure~\ref{fi:classification}.}
\label{tab:classification}
\end{table}
%%%%%%%%%%%%%%%%%%%%%%%%%%%%%%%%%%%%%%%%%%%%%%%%%%%%%%%%%%%%%%%%%%%%%%%%%%%%%

Next pick a basis $\{ \eta_l,\eta_l^\perp: l=1,\ldots,k_0 \}$ for the instanton block in the usual way, such that for each $l=1,\ldots,\ko$ condition~$\eqref{eq:zerojumpcond}$ holds for each $q\in\Zeros_M$. 
(We cannot make~$\eqref{eq:zerojumpcond}$ hold for every $q\in\Zeros$ when there are zero jumps in the instanton block.)
This splits the data into instanton and monopole blocks as in equation~$\eqref{eq:decompWV}$ so that
\begin{equation}\label{eq:defVI}
V_I = [\bigoplus_{p=1}^n W^0(Y_p^I)]\oplus[\C^{N_I}]
\end{equation}
and 
\begin{equation}\label{eq:defVM}
V_M = [\bigoplus_{p=1}^n W^0(Y_p^M)]\oplus[\C^{N_M}].
\end{equation}
The projection $\pi$ also decomposes:\label{glo:piIM} let $\pi_I : W\rightarrow\C^{N_I}$ have components $\pi_q$ for $q\in\Zeros_I$ and $\pi_M : W\rightarrow\C^{N_M}$ have components $\pi_q$ for $q\in\Zeros_M$.
We can then define
\begin{gather}
\Delta_I(x):W_I\rightarrow V_I \notag \\
\Delta_I(x)w = [(i\frac{d}{d\xi} +i(1\otimes\Lambda) +x)w]\oplus[\pi_I w]. \label{eq:defdeltaI2}
\end{gather}

Projection onto the monopole block determines a set of $U(n-N_I)$ monopole Nahm data (recall $n-N_I\geq 2$ since the case of vanishing monopole charges is excluded).
In particular, it determines a residue $R_p$ for each $p\in\Zeros_O$. 
Like the case of no zero jumps in the instanton block, define 
\begin{gather}
\Delta_M(x):W_M\rightarrow V_M \notag \\
\Delta_M(x)w=\big[\big( i\frac{d}{d\xi}+i\sum_{p\in\Zeros_O}\frac{\psi_p(\xi) R_p}{\xi-\mu_p}+x \big)w \big]\oplus[\pi_M w] \label{eq:defdeltaM2}
\end{gather}
where $R_p$ acts on the terminating component. 
Then Lemma~\ref{lem:HM1} implies that $\Delta_M$ is injective and Fredholm with index $N_I-n$. 
We impose condition $\eqref{eq:zerojumpcond}$ for every $q\in\Zeros_M$ to ensure that these zero jumps really do occur in the monopole block, and that $\Delta_M$ is the Nahm operator of some (deformed) $U(n-N_I)$ Nahm data. 

Consider the operator $\Delta_I(x)\oplus\Delta_M(x) : W\rightarrow V$;
it is injective away from the resonating points~$\eqref{eq:resonatingpoints}$ and we want to use the tricks from Sections~\ref{sec:physmodelop} and~\ref{sec:modopnoinstzerojumps} to make it injective at the resonating points. 
As in the case of no zero jumps in the instanton block, we add an off-diagonal term. 
Lemma~\ref{lem:stupid} proves the existence of sections $u,u^\perp$ of $\COKER~\Delta_M$ with zero component in $\C^{N_M}$. 
These can be used to define $B$ and $B^\ast$ as in equations~$\eqref{eq:defnB}$ and~$\eqref{eq:defnBstar}$. 
Defining $\modelop$ by equation~$\eqref{eq:defmodopA}$, we see that $\modelop$ is injective by applying Lemma~\ref{lem:modelinj}.

We added the off-diagonal term to prevent $\eta_{lm}$ and $\eta_{lm}^\perp$ being solutions at the resonating point $x^{\textrm{res}}_{lm}$; in fact the projection $\pi_I$ may prevent them from being solutions to $\Delta_I(x)$ at the resonating point, so adding the diagonal term may have been unnecessary.
However, when we come to recover $k_0$ for the model operator, it is convenient to assume that $\eta_{lm}$ and $\eta_{lm}^\perp$ \emph{are} solutions to $\Delta_I(x)$ at $x^{\textrm{res}}_{lm}$. 
We can deform $\pi_I$ to ensure this, rather like the deformation performed in the case of vanishing monopole charges.
For each $q\in\Zeros_I$ replace $\zeta_q$ with an $x$-dependent vector $\tilde{\zeta}_q$ such that 
\begin{equation}\label{eq:xitildecond}
\tilde{\zeta}(x)=0
\end{equation}
in some neighbourhood of $x_{lm}^{\textrm{res}}$ containing $\supp~\varphi_{lm}$, and $\tilde{\zeta}(x)=\zeta$ away from this neighbourhood. 
Changing the definition of $\modelop$ in this way does not affect the proof of injectivity. 

Finally, note that $\modelop$ is a controlled deformation of $\Delta$, and so Corollary~\ref{cor:indexdefdelta} proves $\modelop$ is Fredholm with index $-n$.  We also obtain a formula like~$\eqref{eq:transNahmop}$ for $\modelop$ in exactly the same way as previously. 

\subsection{The adjoint of the model operator}\label{sec:modeladjoint}

We need to be able to identify the cokernel of the model operator $\modelop(x)$. 
To do this it is easiest to write down the adjoint $\modelop^\ast(x)$ on some domain $A$ of suitably smooth elements of $V$ which contains the cokernel, just as we did in Section~\ref{sec:formaladjoint} for $\Delta^\ast(x)$. 
Since this is analogous to what we did in Section~\ref{sec:formaladjoint} we will not dwell on the details but just write down $\modelop^\ast(x)$ directly. 

The adjoint is given by\label{glo:modeladjt}
\begin{equation}\label{eq:modeladjt}
\modelop^\ast = \begin{pmatrix} \Delta_I^\ast & B \\ B^\ast & \Delta_M^\ast \end{pmatrix}
:V_I\oplus V_M\rightarrow W_I^\ast\oplus W_M^\ast.
\end{equation}
In the case of vanishing monopole charges the monopole block is trivial and this reduces to $\modelop^\ast=\Delta_I^\ast$. 
The spaces $V_I$ and $V_M$ are defined by~$\eqref{eq:defVI}$ and~$\eqref{eq:defVM}$, while
\begin{equation*}
W^\ast_I = \big[\bigoplus_{p=1}^n W^{-1}(Y^I_p)\big]\oplus\big[ \bigoplus_{p=1}^n\big( Y_p^{I,\textrm{cont}}(\mu_p) \big)^\ast\big]
\end{equation*}
and
\begin{equation*}
W^\ast_M = \big[\bigoplus_{p=1}^n W^{-1}(Y^M_p)\big]\oplus\big[ \bigoplus_{p=1}^n\big( Y_p^{M,\textrm{cont}}(\mu_p) \big)^\ast\big]
\end{equation*}
where $Y_p^{I,\textrm{cont}}(\mu_p)$ is the continuing component of $Y_p^I$ at $\mu_p$ \etc
The off-diagonal blocks $B,B^\ast$ are defined by~$\eqref{eq:defnB}$ and~$\eqref{eq:defnBstar}$ (extended to zero on the $\C^{N_M}$ and $\C^{N_I}$ component of $V_M$ and $V_I$ respectively). 
Given $v=(v_1,\ldots,v_n)\oplus s\in V_I$ we have\label{glo:DeltaIast}
\begin{multline} \label{eq:defDeltaIstar}
\Delta^\ast_I(x)v =
(D^\ast_\Lambda(x)v_1,\ldots,D^\ast_\Lambda(x)v_n)\oplus
\big[ \bigoplus_{p\notin{\Zeros_I}}\big( iv_{p-1}^{\textrm{cont}}(\mu_p)-iv_{p}^{\textrm{cont}}(\mu_p) \big)^\ast \big] \\
\oplus \big[ \bigoplus_{p\in{\Zeros_I}}\big( iv_{p-1}^{\textrm{cont}}(\mu_p)-iv_{p}^{\textrm{cont}}(\mu_p)+s_p\tilde{\zeta}_p(x) \big)^\ast \big]
\end{multline}
where
\begin{equation*}
D^\ast_\Lambda(x) = i\frac{d}{d\xi} +i(1\otimes\Lambda) +{x^\ast} : W^0(Y_p^I)\rightarrow W^{-1}(Y_p^I).
\end{equation*}
Finally, given $v=(v_1,\ldots,v_n)\oplus s\in V_M$ we have\label{glo:Dptildeast}
\begin{multline} \label{eq:defDeltaMstar}
\Delta^\ast_M(x)v =
(\tilde{D}^\ast_1(x)v_1,\ldots,\tilde{D}^\ast_n(x)v_n)\oplus
\big[ \bigoplus_{p\notin{\Zeros_M}}\big( iv_{p-1}^{\textrm{cont}}(\mu_p)-iv_{p}^{\textrm{cont}}(\mu_p) \big)^\ast \big] \\
\oplus \big[ \bigoplus_{p\in{\Zeros_M}}\big( iv_{p-1}^{\textrm{cont}}(\mu_p)-iv_{p}^{\textrm{cont}}(\mu_p)+s_p{\zeta}_p \big)^\ast \big]
\end{multline}
where 
\begin{equation*}
\tilde{D}_p^\ast(x) = i\frac{d}{d\xi} -\frac{iR_p}{\xi-\mu_p} +x^\ast
\end{equation*}
on the terminating component near $\xi=\mu_p$ and 
\begin{equation*}
\tilde{D}_p^\ast(x) = i\frac{d}{d\xi} +x^\ast
\end{equation*}
elsewhere. 

\section{Framing for $\tilde{\Delta}$}\label{sec:modelframed}

The aim of this Section is to show that the bundle and connection $\COKER~\modelop$ on $\R\times\rthree$ extend to $\R\times\threeball$ and determine a quasi-periodic connection framed by $\Ainf,\Phiinf$.
We do this by finding `approximate solutions'\label{glo:approxsolns} to $\coker\modelop$, by which we mean sections of the trivial bundle $V\times\rfour$ that are asymptotically close to elements of $\coker\modelop(x)$ as $r\rightarrow\infty$. 
Near infinity, the solutions of $\modelop^\ast(x)$ correspond in some sense with the singularities $\mu_p$ in the Nahm data. 
We construct an approximate solution for each singularity in the monopole block  in Section~\ref{sec:approxsolns}. 
For singularities in the instanton block, we can in fact write down an exact solution, which we do in Section~\ref{sec:exactsolns}. 
In Section~\ref{sec:goodapprox} we show the approximate solutions are exponentially close to exact solutions of $\modelop^\ast(x)$. 
In Section~\ref{sec:framingmodelop} we show that in the gauge determined by these exact solutions the matrices representing the connection $\tilde{\A}=\COKER~\modelop$ extend to $\sphinf$ and are framed there.
This is based on Hitchin's proof that $SU(2)$ monopoles constructed from Nahm data satisfy the BPS boundary conditions \cite[Section 2]{hit83}. 
In particular the representation theory in Section~\ref{sec:approxsolns} is taken directly from \cite[Section 2]{hit83}.

\subsection{Approximate solutions for $\modelop^\ast$ in the monopole block}\label{sec:approxsolns}

Fix a singularity $\mu_p$ in the monopole block of the Nahm data (\ie $p\in\Zeros_M\cup\Zeros_O$). 
For large $r$, we construct an approximate solution to $\modelop^\ast(x)$ supported near $\xi=\mu_p$ which is characterized by the sign of $k_p$---we will deal with each case in turn. 
Let $k=k_p$, and let $t=\xi-\mu_p$.

{\textbf{The case $k > 0$:}} 
We continue the representation theory started in Section \ref{sec:indexdelta}.
Fix a unit vector $\hat{x}\in\rthree$, and let $u=\sum_1^3 x_j\gamma_j$. 
Then $u$ generates a circle in $SU(2)$ and decomposes $\Srep^{k-1}\otimes\Srep^1\cong\Srep^k\oplus\Srep^{k-2}$ into weight spaces\label{glo:weightspc} with weight $k, k-2, \ldots, -(k-2),-k$ for $\Srep^k$ and $k-2, k-4, \ldots, -(k-4),-(k-2)$ for $\Srep^{k-2}$. 
Recall that $\Srep^k$ is the representation of $\sutwo$ on homogeneous polynomials of degree $k$. 
If $[\xi_0 : \xi_1]$ are homogeneous coordinates on $\C P_1=S^2$, it is easy to check that the polynomial
\begin{equation}\label{eq:polyhighest}
(\xi_0 z_0 + \xi_1 z_1)^k
\end{equation}
is the highest weight vector when $u=[\xi_0 : \xi_1]$. 
The action of $1\otimes u$ commutes with the action of $u\otimes u$ so $1\otimes u$ preserves the weight spaces with weights $\pm k$ since they occur with multiplicity one.
Let $v_+ , v_-$ be elements of $\Srep^k$ with unit norm and weights $+k,-k$ respectively.
Now $(1\otimes u)^2 =-1$, so $(1\otimes u)v_+=\pm iv_+$ and $(1\otimes u)v_- = \pm iv_-$. 
In fact using the explicit form~$\eqref{eq:polyhighest}$ of the highest weight vector, we have $(1\otimes u)v_+= iv_+$ and $(1\otimes u)v_- = - iv_-$.
As $u$ varies, $v_+\in\Srep^k$ spans out a line bundle $L$ over $S^2$, and using~$\eqref{eq:polyhighest}$ it follows that $L$ has Chern class $k$.
The action of $SO(3)$ on the direction vector $\hat{x}$ lifts to the adjoint action of $SU(2)$ on $u\in{\mathfrak{su}}(2)$. 
Hence, on a neighbourhood of $\hat{x}\in\rthree$, we can choose the highest weight vector $v_+$ according to
\begin{equation*}
v_+ (g\hat{x}) = gv_+(\hat{x})
\end{equation*}
where $g\in SO(3)$ acts by some unitary endomorphism on $\Srep^k$.

Next, we use $v_+$ to write down an approximate solution to $\modelop^\ast(x)$.
As previously, let $R_p$ be the residue of $\sum_j \gamma_j \otimes T_p^j$ at $\mu_p$.
Near $\xi=\mu_p$, $\modelop^\ast(x)$ is given by
\begin{equation*}
\tilde{D}_p^\ast = i\frac{d}{dt} -\frac{iR_p}{t} +\xo -r(1\otimes u)
\end{equation*}
on the terminating component,  
where $r$ is the usual polar coordinate on $\rthree$.
From~$\eqref{eq:repthy}$ we have that $R_p v_+ = \frac{1}{2}(k-1)v_+$ and $(1\otimes u)v_+ = iv_+$, so a solution is given on some interval $t\in[-2\delta,0]$ by
\begin{equation}\label{eq:defapproxa}
\tilde{u}_p = t^{(k-1) / 2}\big[ \exp \big(i\xo(t+\mu_p)\big) \big]\big[\exp (rt)\big]v_+.
\end{equation}
Note that $(t+\mu_p)$ in the expression above could be replaced by $(t+a)$ for any $a$ and $\tilde{u}_p$ would still be a solution. 
We chose $\tilde{u}_p$ as above so that the result of translating by one period $\tau$ is given by the action of $U_\tau$ (where $U_\tau$ is defined by~$\eqref{eq:translationmap}$).
This will ensure that the quasi-periodic connection we eventually obtain has a clutch map that is the identity on the boundary.
Also note that we could have used $v_-$ to construct a solution. 
However, such a solution would blow up in the limit $t\rightarrow 0$, and so fails to be $L^2$.

We want our approximate solutions to be smooth and have compact support, so we multiply by a bump function $\varphi_p(t)$ supported on $t\in [-2\delta,0]$ that takes value $1$ on $[-\delta,0]$.
We also want the approximate solutions to have unit norm, so define $\tilde{v}_p$, the approximate solution associated to $\mu_p$, to be
\begin{equation}\label{eq:defapproxb}
\tilde{v}_p = C_p^{-1}\varphi_p \tilde{u}_p
\end{equation}
where $C_p =\| \varphi_p \tilde{u}_p \|_{L^2}$.
Consider the element of $V$ which is $\tilde{v}_p$ on the interval $I_p$ and zero otherwise; abusing notation, let $\tilde{v}_p$\label{glo:tildevp1} denote this element.

It will be useful to obtain an estimate for $C_p=\| \varphi_p \tilde{u}_p \|_{L^2}$.
Let\label{glo:intIk}
\begin{equation}\label{eq:defIk}
{\mathfrak{I}}_k = \int_{-2\delta}^{0} \varphi_p^2\ t^k \exp (2rt)\ dt
\end{equation}
so
\begin{equation}\label{eq:normfromIk}
\| \varphi_p \tilde{u}_p \|^2_{L^2} = | {\mathfrak{I}}_{k-1} |.
\end{equation}
Now \label{pag:intest}
\begin{equation} \label{eq:estAxob}
| \int_{-\delta}^{0} t^k \exp (2rt)\ dt | \leq
|{\mathfrak{I}}_k| \leq
| \int_{-2\delta}^{0} t^k \exp (2rt)\ dt |.
\end{equation}
Integrating by parts gives
\begin{equation*}
\int_{-a}^{0} t^k \exp (2t / \chi)\ dt = \Big[ \chi^{k+1}\exp (2t / \chi) P_k(t / \chi) \Big]^0_{-a}
\end{equation*}
for some polynomial $P_k$ of degree $k$. Hence
\begin{equation*}
\int_{-a}^{0} t^k \exp (2t / \chi)\ dt = \chi^{k+1}[P_k(0) - \exp (-2a / \chi) P_k(-a / \chi)].
\end{equation*}
Then~$\eqref{eq:estAxob}$ implies that 
\begin{equation}\label{eq:Ikest}
{\mathfrak{I}}_k  = C\chi^{k+1} + \ \textrm{smooth exponentially decreasing term in\ }\chi
\end{equation}
where $C$ is some constant (used in the generic sense).
Thus, using~$\eqref{eq:normfromIk}$, we have an estimate $C_p=\| \varphi_p \tilde{u}_p \|_{L^2} = C\chi^{k/2}+$ exponentially decaying term.

{\textbf{The case $k < 0$:}} 
This is entirely analogous to the case $k>0$.
We can repeat the representation theory on $\Srep^{|k|}\otimes\Srep^1$ to obtain vectors $v_+,v_-$ in the same way so that $v_-$ has weight $k$, $(1\otimes u)v_- = -iv_-$, and $v_-$ determines a line bundle with Chern class $k$.
Define the approximate solution to be $\tilde{v}_p = C^{-1}_p\varphi_p\tilde{u}_p$ where $\varphi_p$ is a bump function supported on $I_{p-1}$ and 
\begin{equation*}
\tilde{u}_p = t^{( |k| -1)/2}\big[\exp i\big(\xo(t+\mu_p)\big)\big]\big[\exp (-rt)\big]v_- .
\end{equation*}
Just as for the case $k>0$, there is an analogous estimate on $C_p=\| \varphi_p \tilde{u}_p \|_{L^2}$.

{\textbf{The case $k = 0$:}} 
Fix a unit vector $\hat{x}\in\rthree$ and let $u=\sum \hat{x}_j\gamma_j$. 
Consider solutions to $\Delta_M^\ast(x)v=0$ where $\Delta_M^\ast$ is defined by~$\eqref{eq:defDeltaMstar}$. 
These must satisfy
\begin{equation*}
\big(i\frac{d}{dt} +\xo -r(1\otimes u)\big)v_j=0
\end{equation*}
on $I_j$ for $j=p,p-1$. 
The general solutions are
\begin{equation*}
\big[\exp i\big(\xo(t+\mu_p)\big)\big]\big[\exp -i(rtu)\big]s_-\textrm{\ on $I_p$}
\end{equation*}
and
\begin{equation*}
\big[\exp i\big(\xo(t+\mu_p)\big)\big]\big[\exp -i(rtu)\big]s_{+}\textrm{\ on $I_{p-1}$}
\end{equation*}
for some vectors $s_{-},s_{+}$. 
We want the solutions to match the conditions to be in $\ker\Delta_M^\ast(x)$ \ie we want a discontinuity at $t=0$ such that the jump is a multiple of $\zeta_p$. 
Now, $(1\otimes u)$ has eigenvalues $\pm i$ and let $\pi_\pm$ denote projection onto these eigenspaces. 
Let
\begin{equation}\label{eq:monzerojumpminus}
\tilde{u}_p^- = \big[\exp i\big(\xo(t+\mu_p)\big)\big]\big[\exp -i(rtu)\big]\pi_+\zeta_p = \big[\exp i\big(\xo(t+\mu_p)\big)\big]\big[\exp (rt)\big]\pi_+\zeta_p
\end{equation}
on $I_p$ and
\begin{equation}\label{eq:monzerojumpplus}
\tilde{u}_{p}^+ = -\big[\exp i\big(\xo(t+\mu_p)\big)\big]\big[\exp -i(rtu)\big]\pi_-\zeta_p = -\big[\exp i\big(\xo(t+\mu_p)\big)\big]\big[\exp (-rt)\big]\pi_-\zeta_p
\end{equation}
on $I_{p-1}$. 
Then
\begin{equation*}
\tilde{u}_p^+(t=0) - \tilde{u}_{p}^-(t=0) = -[\exp (i\xo\mu_p)][\pi_+s + \pi_-s] =-[\exp (i\xo\mu_p)]\zeta_p\in J_p.
\end{equation*}
Finally we smooth off by bump functions $\varphi_{p}^+, \varphi_p^-$ and normalize to define the approximate solution $\tilde{v}_p$ by
\begin{center}
\setlength{\unitlength}{1.0cm}
\begin{picture}(12,2.0)
\put(0,1.2){$\tilde{v}_p = C_p^{-1}
(0,\ldots,0,\varphi_{p}^+\tilde{u}_{p}^+,\varphi_{p}^-\tilde{u}_{p}^-,0,\ldots,0)  \oplus (0,\ldots,0,s_p,0,\ldots,0)\in V$}
\put(3.8,0.4){\vector(0,1){0.6}}
\put(5,0.4){\vector(0,1){0.6}}
\put(9.2,0.4){\vector(0,1){0.6}}
\put(1.5,0){on interval $I_{p-1}$}
\put(4.8,0){on interval $I_{p}$}
\put(8.2,0){corresponding to  $\zeta_{p}$}
\end{picture}
\end{center}
where
\begin{equation*}
C_p = 
{(\| \varphi_{p}^-\tilde{u}_{p}^- \|^2_{L^2} + \| \varphi_{p}^+\tilde{u}_{p}^+ \|^2_{L^2} +1)^{1/2}}
\end{equation*}
and 
\begin{equation*}
s_p = i\exp (i\xo \mu_p).
\end{equation*}

This completes the definition of the approximate solution $\tilde{v}_p$ for $p\in\Zeros_O\cup\Zeros_M$.

\subsection{Solutions for $\modelop$ in the instanton block}\label{sec:exactsolns}

For each $p\in\Zeros_I$ (\ie for each zero jump in the instanton block) we can write down an exact solution to $\modelop^\ast(x)$ that lies wholly in the instanton block and is defined away from the resonating points. 
A solution $v=(v_1,\ldots,v_n)\oplus s\in V_I$ to $\modelop^\ast(x)$ must satisfy $\Delta_I^\ast(x)v=0$, where $\Delta_I^\ast(x)$ is defined by~$\eqref{eq:defDeltaIstar}$. 
Hence
\begin{equation}\label{eq:instblockop}
(i\frac{d}{d\xi} +i(1\otimes\Lambda) +{x^\ast})v_j=0
\end{equation}
for all $j$, and $v$ is continuous across the $\mu_j$, except at any zero jump $\mu_j$ in the instanton block, at which $v$ can jump by a multiple of $\tilde{\zeta}_j(x)$. 
(Recall that the vectors $\zeta_j$ are deformed to give $x$-dependent vectors $\tilde{\zeta}_j(x)$ in the definition of $\modelop(x)$.)
Solving~$\eqref{eq:instblockop}$ round $\Sdual$ gives a holonomy\label{glo:holDeltaI}
\begin{equation}\label{eq:defHolinstblock}
\textrm{Hol}(x)=\exp [(ix^\ast-\Lambda)\mu_0].
\end{equation}
We know that $[1-\textrm{Hol}(x)]$ is invertible away from the resonating points~$\eqref{eq:resonatingpoints}$, so define
\begin{equation}\label{eq:nearlydefexactsoln}
u_p = -\big[\exp \big((ix^\ast-\Lambda)(\xi-\mu_p)\big]\big[1-\textrm{Hol}(x)\big]^{-1}\big[\exp (i\xo\mu_p)\big]\tilde{\zeta}_p
\end{equation}
for each $p\in\Zeros_I$, and for $\mu_p\leq \xi\leq \mu_p+\muo$. 
It is easy to check that this is a solution to $\modelop^\ast(x)$, jumping by $[\exp (i\xo\mu_p)]\tilde{\zeta}_p$ at $\mu_p$. 
Define the exact solution $v_p\in V$\label{glo:exactvp} by
\begin{equation}\label{eq:defexactsoln}
v_p = C_p^{-1}(u_p)\oplus(0,\ldots,0, i\exp (i\xo\mu_p),0,\ldots,0)
\end{equation}
where $i\exp (i\xo\mu_p)$ lies in the component of $\C^{N_I}$ corresponding to the jump $\mu_p$, and $C_p$ is a constant that normalizes $v_p$. 
Note that in the case of vanishing monopole charges, the $v_p$ are well-defined solutions to $\modelop^\ast(x)$ on the complement of the resonating points. 
In the other cases, when the model operator has an off-diagonal block $B$, the $v_p$ are well-defined solutions to $\modelop^\ast(x)$ on the complement of $\textrm{Supp~}B$. 
Also note that where some vector $\tilde{\zeta}_p$ has been deformed to zero, $v_p$ is just the corresponding vector in $\C^{\Nzer}$. 

Although we have written down an exact solution to $\modelop^\ast(x)$ corresponding to each singularity $\mu_p$ with $p\in\Zeros_I$, it is often more convenient to work with an approximate solution. 
Define
\begin{equation}\label{eq:instzerojumpplus}
\tilde{u}_p^+ = -\varphi_p^+\big[\exp (-\Lambda t)\big]\big[\exp \big(i\xo(t+\mu_p)\big)\big]\big[\exp (-rt)\big]\pi_-\zeta_p
\end{equation}
for some bump function $\varphi_p^+$ on $(-\epsilon,0]$, and
\begin{equation}\label{eq:instzerojumpminus}
\tilde{u}_p^- = \varphi_p^-\big[\exp (-\Lambda t)\big]\big[\exp \big(i\xo(t+\mu_p)\big)\big]\big[\exp (rt)\big]\pi_+\zeta_p
\end{equation}
for some bump function $\varphi_p^-$ on $[0,\epsilon)$.  
(Compare with~$\eqref{eq:monzerojumpminus}$ and~$\eqref{eq:monzerojumpplus}$.)
We can then define an approximate solution $\tilde{v}_p\in V$\label{glo:tildevp2} constructed from $\tilde{u}_p^-,\tilde{u}_p^+$ so that the $\C^{N_I}$ component of $\tilde{v}_p$ matches the jump at $\mu_p$. 
A short calculation shows that $\tilde{v}_p$ is exponentially close to the exact solution $v_p$ in the limit $r\rightarrow\infty$. 
This is because in the limit the exact solution $v_p$ becomes more and more peaked about the discontinuity at $\mu_p$. 

\subsection{How good is the approximation?}\label{sec:goodapprox}

We need to consider in what sense the set of approximate solutions $\tilde{v}_1,\ldots,\tilde{v}_n$ approximate the cokernel of $\modelop(x)$.
Let $\tilde{P}$\label{glo:approxproj} denote the orthogonal projection onto $\tilde{\E}=\COKER~\modelop$\label{glo:modelbdl} and consider $\tilde{w}_p(x):=\tilde{P}(x)\tilde{v}_p(x)$. 
Now
\begin{equation*}
\tilde{P} = 1 - \modelop (\modelop^\ast\modelop)^{-1}\modelop^\ast,
\end{equation*}
so we need to calculate $\modelop^\ast(x)\tilde{v}_p$ in each of the cases $k_p>0,k_p=0,k_p<0$. 
For example, in the case $k_p>0$ we obtain
\begin{equation*}
\modelop^\ast (x) \tilde{v}_p = C_p^{-1}(i\frac{d}{dt}\varphi_p)\tilde{u}_p
\end{equation*}
where $d\varphi_p / dt$ is supported on some interval $t\in [-2\delta,-\delta]$, and $\tilde{u}_p$ is defined by~$\eqref{eq:defapproxa}$.
Using integral estimates just like those on page~\pageref{pag:intest}, it follows that
\begin{equation*}
\| \modelop^\ast (x)\tilde{v}_p \|_{L^2} \leq C\exp (-\frac{1}{\chi})
\end{equation*}
for some constant $C$. 
The norm of $\modelop(x) (\modelop^\ast(x)\modelop(x))^{-1}$ is bounded as $r\rightarrow\infty$ (we prove this later: see equation~$\eqref{eq:estimateGreen}$) and so
\begin{equation*}
\tilde{w}_p = \tilde{P}\tilde{v}_p = \tilde{v}_p + (\textrm{exponentially decreasing term}).
\end{equation*}
The other cases $k_p<0$ and $k_p=0$ are entirely analogous, and so we obtain
\begin{equation}\label{eq:approxprojid}
\tilde{P}= 1 + (\textrm{exponentially small operator})
\end{equation}
as a map 
$\textrm{span}\{ \tilde{v}_1,\ldots,\tilde{v}_n \}\rightarrow\tilde{\E}$.

We would like, however, to replace $\tilde{P}$ with a unitary isomorphism between these spaces, so that the trivialisation $\tilde{w}_1,\ldots,\tilde{w}_n$ is orthonormal. 
We know that the set $\tilde{v}_1,\ldots,\tilde{v}_n$ is orthonormal by definition (by making the supports of the bump functions disjoint), so replacing $\tilde{P}$ with a unitary map will ensure we obtain a unitary basis of $\tilde{\E}$. 
Recall the `polar decomposition' of an invertible matrix $M$: 
let 
\begin{equation*}
A=(MM^\ast)^{1/2}\quad\textrm{and\ }U=A^{-1}M.
\end{equation*}
so that $A$ is a positive-definite self-adjoint matrix and $U$ is unitary, and these satisfy $M=AU$. 
Variational methods show that this decomposition minimizes $\| M-U \|$. 
Define 
\begin{equation*}\label{glo:approxunitaryproj}
\tilde{P}_U = (\tilde{P}\tilde{P}^\ast)^{-1/2}\tilde{P}
\end{equation*}
so that $\tilde{P}_U$ is a unitary map 
$\textrm{span}\{ \tilde{v}_1,\ldots,\tilde{v}_n \}\rightarrow\tilde{\E}$.
Equation~$\eqref{eq:approxprojid}$ implies that 
\begin{equation*}
\tilde{P}_U= 1 + (\textrm{exponentially small operator}),
\end{equation*}
so if we define $\tilde{w}_p = \tilde{P}_U \tilde{v}_p$\label{glo:exactwp}, then $\tilde{w}_p = \tilde{v}_p +$ exponentially decreasing term. 
Hence the approximate solutions $\tilde{v}_1,\ldots,\tilde{v}_n$ are exponentially close to a unitary basis $\tilde{w}_1,\ldots,\tilde{w}_n$ of $\tilde{\E}$ in the limit $r\rightarrow\infty$. 
Recall that the exact solutions corresponding to zero jumps in the instanton block are exponentially close to the corresponding approximate solutions. 
Hence for $p\in\Zeros_I$ we have $\tilde{w}_p = {v}_p +$ exponentially decreasing term. 

\subsection{Proof that $\modelop$ is framed}\label{sec:framingmodelop}

We want to show that $(\tilde{\E},\tilde{\A})=\COKER~\modelop$ extends to the boundary $\sphinf$ and is framed by $\Ainf,\Phiinf$.
The exact statement is as follows:

\begin{definition}\label{def:intrest}
Let $\F,\B$ be a $\Un$ bundle and connection on $\Ieps\times\rthree$ and let $\B^q$ be some $\Un$ framed quasi-periodic connection on $\Eq$. 
The pair $(\F,\B)$ is the \emph{interior restriction}\label{glo:intrest} of $(\E^q,\B^q)$ if there is a unitary isomorphism $F:\F\rightarrow \E^q |_{\Ieps\times\rthree}$ such that $\B=F^{\ast}(\B^q |_{\Ieps\times\rthree})$.
\end{definition}

\begin{proposition}\label{prop:approxframe}
$(\tilde{\E},\tilde{\A})$ is the interior restriction of $(\Eq,\Aq)$ where $\Aq$ is some $\Un$ quasi-periodic connection which is smooth up to the boundary and which is framed by $\Ainf,\Phiinf$. 
\end{proposition}

\proof
First fix local trivialisations of $\Einf$ in the following way. 
Fix a direction vector $\hat{x}\in\rthree$. 
Each approximate solution $\tilde{v}_p$ is associated to a vector $e_p$ that spans out a line bundle $L_{k_p}$ over $\sphinf$.
In the case $k_p>0$, $e_p$ is the highest weight vector $v_+$; for $k_p<0$, $e_p$ is the vector $v_-$; and for $k_p=0$, $e_p$ is the constant vector $\zeta_p\in J_p$. 
Since $\sum_1^n k_p=0$ the vectors $e_1,\ldots,e_n$\label{glo:ep} form a local trivialisation of the trivial bundle $\Einf$ over $\sphinf$ on some neighbourhood of $\hat{x}$.
By construction, there is a unitary action of $SO(3)$ on the $e_p$ such that
\begin{equation}\label{eq:etaso3invart}
e_p(g\hat{x}) = ge_p(\hat{x}) \quad\textrm{where\ }g\in SO(3).
\end{equation}

To prove the Proposition it is sufficient to show that in the local trivialisation $\tilde{w}_1,\ldots,\tilde{w}_n$ of $\tilde{\E}$, the matrices $\tilde{\A}_\chi,\tilde{\A}_{x_0},\tilde{\A}_{y_1},\tilde{\A}_{y_2}$ representing $\tilde{\A}$ extend smoothly to the boundary $\Ieps\times\sphinf$ and are appropriately framed there. 
In particular, we claim that
\begin{itemize}
\item $\tilde{\A}_\chi = 0$ on $\Ieps\times\sphinf$,
\item $\tilde{\A}_{x_0} = \diag(i\mu_1,\ldots,i\mu_n)$ on $\Ieps\times\sphinf$,
\item $\tilde{\A}_{y_j} = \diag(\langle \partial_{y_j} e_p,e_p \rangle)$ on $\Ieps\times\sphinf$ for $j=1,2$.
\end{itemize}
These conditions are sufficient to deduce that $\tilde{\A}$ is the interior restriction of some connection $\A^q$ framed by $\Ainf,\Phiinf$. 
In addition, we also have to verify that there is some clutching map 
\begin{equation*}
c(s):\tilde{\E}|_{x_0=s}\rightarrow \tilde{\E}|_{x_0=s+\perflat}
\end{equation*}
such that 
\begin{equation*}
\tilde{\A}(x_0=s+\perflat)=\big( c(s)^{-1} \big)^\ast\tilde{\A}(x_0=s)
\end{equation*}
and that in the gauge $\tilde{w}_1,\ldots,\tilde{w}_n$, $c\rightarrow 1$ as $r\rightarrow\infty$. 
It follows that $\tilde{\A}$ is the interior restriction of a quasi-periodic connection $\A^q$ framed by $\Ainf,\Phiinf$. 

To prove the claim about the framing we have to calculate the matrices $\tilde{\A}_\chi$, $\tilde{\A}_{x_0}$, $\tilde{\A}_{y_1}$, and $\tilde{\A}_{y_2}$ 
using a formula like~$\eqref{eq:cokerinlocalgauge}$:
\begin{equation*}
(\tilde{\A}_a)_{ij} = \langle \partial_a \tilde{w}_i, \tilde{w}_j \rangle.
\end{equation*}
Since $\tilde{w}_j = \tilde{v}_j +$ exponentially decaying term, we have that
\begin{equation*}
(\tilde{\A}_a)_{ij} = \langle \partial_a \tilde{v}_i, \tilde{v}_j \rangle+
(\textrm{smooth exponentially decaying term in\ }\chi).
\end{equation*}
Hence it is sufficient to consider the matrices 
$\langle \partial_a \tilde{v}_i, \tilde{v}_j \rangle$ for $a=\chi,\xo,y_1,y_2$.
By making the supports of the bump functions $\varphi_p$ used to define the approximate solutions sufficiently small, these matrices are diagonal, and since the approximate solutions are orthonormal, the diagonal entries in the matrices must be imaginary. 
Fix some $p$ and consider the $p$'th diagonal element of each matrix. 
Let $k=k_p$ and $t=\xi-\mu_p$. 

First consider $\tilde{\A}_\chi$. 
When $k >0$, $\tilde{v}_p$ is defined by~$\eqref{eq:defapproxb}$, and $\partial_\chi\tilde{v}_p$ is given by
\begin{equation*}
\partial_\chi\tilde{v}_p = -\Big(\frac{t}{\chi^2}+\frac{\partial_\chi C_p}{C_p}  \Big)\tilde{v}_p.
\end{equation*}
Thus $(\tilde{\A}_\chi)_{pp}= \langle\partial_\chi\tilde{v}_p,\tilde{v}_p\rangle_{L^2}$ is a real integral, and so must vanish since it is also imaginary. 
Hence $\langle\partial_\chi\tilde{w}_p,\tilde{w}_p\rangle_{L^2}$ is smooth up to the boundary and vanishes there. 
The cases $k=0$ and $k<0$ are entirely similar. 

Next consider $\tilde{\A}_{x_0}=\diag( \langle \partial_{x_0} \tilde{w}_p, \tilde{w}_p \rangle )$. 
We want to show that $\langle \partial_{\xo} \tilde{v}_p, \tilde{v}_p \rangle$ is smooth up to the boundary and has value $i\mu_p$ when $\chi=0$.
The proof depends on whether $k>0$, $k=0$, or $k<0$; 
start by assuming $k>0$ so that $\tilde{v}_p$ is given by~$\eqref{eq:defapproxa}$ and~$\eqref{eq:defapproxb}$.
Then $\partial_{\xo} \tilde{v}_p = i(t+\mu_p)\tilde{v}_p$ so
\begin{equation*}
\langle \partial_{\xo} \tilde{v}_p, \tilde{v}_p \rangle =  
\langle i(t+\mu_p) \tilde{v}_p, \tilde{v}_p \rangle_{L^2} =
i\mu_p + i\langle t \tilde{v}_p, \tilde{v}_p \rangle_{L^2}.
\end{equation*}
This last term is independent of $\xo$,  and because of the $SO(3)$ invariance~$\eqref{eq:etaso3invart}$, independent of $y_1,y_2$; we want to show it is a smooth function of $\chi$ and tends to zero as $\chi\rightarrow 0$.
Now 
\begin{equation*}
\langle t\tilde{v}_p,\tilde{v}_p \rangle_{L^2} = {\mathfrak{I}}_{k} / | {\mathfrak{I}}_{k-1} |
\end{equation*}
where ${\mathfrak{I}}_{k}$ is defined by~$\eqref{eq:defIk}$. 
So~$\eqref{eq:Ikest}$ implies that 
\begin{equation*}
\langle t\tilde{v}_p,\tilde{v}_p \rangle_{L^2} = C\chi + \ \textrm{smooth exponentially decreasing term},
\end{equation*}
which completes the case $k>0$.
The cases $k=0$ and $k<0$ are dealt with via very similar estimates.

Finally, consider $\tilde{\A}_{y_j}$ for $j=1,2$.
Fix $j$ and let $y=y_j$. 
It is sufficient to show that $\langle \partial_y \tilde{v}_p,\tilde{v}_p \rangle_{L^2} = \langle \partial_{y} e_p,e_p \rangle$ on $\sphinf$. 
As usual, we have to deal case-by-case with the sign of $k$. 
When $k>0$ or $k<0$ this can be seen immediately from the definition of $\tilde{v}_p$.
For example, when $k>0$:
\begin{equation*}
\langle \partial_y\tilde{v}_p,\tilde{v}_p \rangle_{L^2} =
\frac{1}{\| \varphi_p \tilde{u}_p \|^2_{L^2}}
\int_{I_p}[\varphi^2_p\ t^{k_p-1}\exp \frac{2t}{\chi}]\langle \partial_{y} e_p,e_p \rangle~dt=\langle \partial_{y} e_p,e_p \rangle.
\end{equation*}
For $k_p=0$ and assuming the zero jump is in the monopole block, we have
\begin{equation*}
\langle \partial_y \tilde{v}_p,\tilde{v}_p \rangle_{L^2} = 
C_p^{-1}\big[
\| \varphi_p \tilde{u}_p^- \|^2_{L^2}
\langle \partial_y (\pi_+ e_p),\pi_+e_p \rangle +
\| \varphi_p \tilde{u}_p^+ \|^2_{L^2}
\langle \partial_y (\pi_- e_p),\pi_-e_p \rangle 
\big].
\end{equation*}
But integral estimates show $\| \varphi_p^- \tilde{u}_p^- \|_{L^2} = O(\chi) =\| \varphi_p^+ \tilde{u}_p^+ \|_{L^2}$ as $\chi\rightarrow 0$.  
Hence $\langle \partial_y \tilde{v}_p,\tilde{v}_p \rangle_{L^2}=O(\chi)$ as $\chi\rightarrow 0$. 
However, $\langle \partial_{y} e_p,e_p \rangle=0$ since $k=0$, and so $\langle \partial_y\tilde{v}_p,\tilde{v}_p \rangle_{L^2}$ extends smoothly to $\partial X$ where it equals $\langle \partial_{y} e_p,e_p \rangle$. 
When the zero jump is in the instanton block the exact solution defined by~$\eqref{eq:instzerojumpplus}$ and~$\eqref{eq:instzerojumpminus}$ gives exactly the same result. 
This completes the proof of the claim about the framing.

It remains to prove that $\tilde{\A}$ clutches correctly. 
The map $U_\tau=U_{\tau,V}$ defined by~$\eqref{eq:translationmap}$ gives a map from $\tilde{\E} |_{\xo=s}$ to $\tilde{\E} |_{\xo=s+\perflat}$ since
$\modelop$ satisfies 
\begin{equation*}
\modelop(\tau x)=U_{\tau,V}\modelop(x)U_{\tau,W}^{-1}.
\end{equation*} 
This implies that $\tilde{\A}$ satisfies 
\begin{equation*}
\tilde{\A}(\tau x) = (U_\tau^{-1})^\ast\tilde{\A}(x).
\end{equation*}
In the gauge $\tilde{w}_1,\ldots,\tilde{w}_n$ of $\tilde{\E}$, $U_\tau$ is given by the matrix 
\begin{equation*}
\langle U_\tau \tilde{w}_i(x), \tilde{w}_j(\tau x) \rangle = 
\langle U_\tau \tilde{v}_i(x), \tilde{v}_j(\tau x) \rangle + 
\textrm{exponentially decreasing term}.
\end{equation*}
By construction, however, $U_\tau \tilde{v}_i (x) = \tilde{v}_i (\tau x)$, so  
\begin{equation*}
U_\tau= 1 + \textrm{\ smooth exponentially decaying term}.
\end{equation*}
This shows that $\tilde{\A}$ clutches correctly: $\tilde{\A}$ is the interior restriction of a framed quasi-periodic connection $\A^q$ with clutching function $c$, where $c$ extends smoothly to infinity and is the identity there. 
\eproof

\section{Calculating $\ko$ for $\tilde{\Delta}$}\label{sec:modelko}

Proposition~\ref{prop:approxframe} shows that $\COKER~\modelop$ determines an element of $\spc{C}{(\tilde{k}_0,\vec k,\muo,\vec\mu)}$ for some $\tilde{k}_0\in\Z$\label{glo:tildeko}. 
In this Section we prove that $\tilde{k}_0=\ko$, where $\bdarydata$ is the boundary data for the Nahm data fixed at the start of Section~\ref{sec:modelop}.
We do this by calculating $\int\chtwo(\tilde{\E},\tilde{\A})$ and using~$\eqref{eq:caloroncharge}$. 

The calculation is slightly different in each of the cases, but uses the following scheme. 
The basic idea is to compare $(\tilde{\E},\tilde{\A})=\COKER~\modelop$ with a caloron configuration $(\E_0,\A_0)$\label{glo:E0A0} which has the same framing $f$ as $(\tilde{\E},\tilde{\A})$, but which has $c_2(\E_0,f)=0$, and so is a deformation of a monopole. 
Regard $(\tilde{\E},\tilde{\A})$ as a bundle and connection over $\Ieps\times\rthree$. 
Then $(\tilde{\E},\tilde{\A})$ is the interior restriction of a framed quasi-periodic connection with boundary data $(\tilde{k}_0,\vec k,\muo,\vec\mu)$, so 
\begin{equation}\label{eq:kocharge1}
\int_{[0,\perflat]\times\rthree} \chtwo(\tilde{\E},\tilde{\A})
=-\tilde{k}_0- \frac{1}{\mu_0}(\mu_1 k_1 + \cdots +\mu_n k_n)
\end{equation}
using~$\eqref{eq:caloroncharge}$. 
Suppose $(\E_0,\A_0)$ is a bundle and connection on $\Ieps\times\rthree$ such that
\begin{equation}\label{eq:kocharge2}
\int_{[0,\perflat]\times\rthree} \chtwo(\E_0,\A_0)
=-\frac{1}{\mu_0}(\mu_1 k_1 + \cdots +\mu_n k_n).
\end{equation}
Moreover, suppose that on the complement $R^c$ of some closed region $R\subset (0,\perflat) \times\rthree$ there is a unitary isomorphism
\begin{equation*}
F:\tilde{\E}|_{R^c}\rightarrow\E_0 |_{R^c}
\end{equation*}
such that 
\begin{equation*}
\tilde{\A}|_{R^c} = F^\ast(\A_0 |_{R^c}). 
\end{equation*}
Using~$\eqref{eq:kocharge1}$ and~$\eqref{eq:kocharge2}$ we then have
\begin{equation*}
\tilde{k}_0=\int_{R}\{ \chtwo(\E_0,\A_0)-\chtwo(\tilde{\E},\tilde{\A})\}.
\end{equation*}
Suppose the isomorphism $F$ is such that $R$ is the  disjoint union of some small closed balls $B_l$ for $l=1,\ldots,\ko$, with each ball containing one resonating point. 
For each $l$ construct a bundle and connection $(\F_l,\B_l)$ over $S^4$ by gluing $\tilde{\E}|_{B_l}$ and $\E_0|_{B_l}$ at their boundaries via the isomorphism $F$. 
Then 
\begin{align}
\tilde{k}_0 &=\sum_{l=1}^{\ko}\int_{S^4}\chtwo(\F_l,\B_l) \notag \\
& = -\sum_{l=1}^{\ko} c_2(\F_l) \label{eq:chFl}
\end{align}
where $c_2(\F_l)$ is the second Chern class, 
so if we can show $c_2(\F_l)=-1$ then $\tilde{k}_0=\ko$. 
This is done by calculating the transition function $g_l$ from $\E_0|_{B_l}$ to $\tilde{\E}|_{B_l}$, and using the relation
\begin{equation}\label{eq:degreeclutch}
c_2(\F_l)=\frac{1}{24\pi^2}\int_{S^3}\trace(dg_l g_l^{-1})^3 = \deg g_l.
\end{equation}
The transition function $g_l$ is found by fixing gauges on $\E_0|_{B_l}$ and $\tilde{\E}|_{B_l}$. 
The precise nature of these gauges differs for the different types of model operator $\modelop$. 
In many ways, the case of no zero jumps in the instanton block is the most illustrative, and the reader may prefer to concentrate on that case first. 

The calculation of $\ko$ for the model operator $\modelop$ suggests another way of thinking about the topology of caloron configurations, 
which we call the `spotted dick' model. 
(The term `spotted dick' refers to a type of pudding consisting of a cylinder of sponge containing currants, often served with custard.)
The idea is that away from a small neighbourhood of each resonating point, $\COKER~\modelop$ is isomorphic to a monopole configuration. 
The obstruction to extending this to a global isomorphism comes at the  resonating points. 
Near each resonating point $\COKER~\modelop$ resembles a charge-$1$ instanton.  
Thus, up to deformation, we can think of a caloron as the pull-back of a monopole to $\cyl$ with $\ko$ charge-$1$ instantons embedded in it. 
This explains the name: the `sponge' is a monopole configuration while the `currants' are charge-$1$ instanton configurations.

\subsection{The case of no zero jumps in the instanton block}\label{sec:konoinstzerojumps}

The main idea here is that restricting to the monopole block gives a $U(n)$ monopole Nahm operator $\Delta_M$, and we take $(\E_0,\A_0)=\COKER~\Delta_M$. 
This monopole configuration is framed in the same way as $\tilde{\E}$, and so we can apply the scheme outlined above. 

Recall the definition of $\modelop$ from Section~\ref{sec:modopnoinstzerojumps}:
\begin{equation*}
\modelop = \begin{pmatrix} \Delta_I & B \\ B^\ast & \Delta_M \end{pmatrix}
\end{equation*}
and let $(\E_M,\A_M)=\COKER~\Delta_M$. 
We showed in Section~\ref{sec:modopnoinstzerojumps} that $\Delta_M$ is injective and Fredholm with index $-n$. 
Lemma~\ref{lem:HM2} gives
\begin{equation*}
\int_{[0,\perflat]\times\rthree}\chtwo(\E_M,\A_M) = -\frac{1}{\muo}(\mu_1 k_1+
\cdots+\mu_n k_n).
\end{equation*}
Next, recall the definition of $\modelop^\ast(x)$, equation~$\eqref{eq:modeladjt}$. 
Outside $\bigcup\supp~\varphi_{lm}$ (\ie away from the resonating points)
\begin{equation*}
\modelop^\ast(x) = \begin{pmatrix} \Delta_I^\ast(x) & 0 \\ 0 & \Delta_M^\ast(x) \end{pmatrix},
\end{equation*}
so $\coker\modelop(x)=\coker\Delta_M(x)$ for $x$ outside $\bigcup\supp~\varphi_{lm}$, 
because $\Delta_I^\ast(x)$ has no solutions. 
(From equation~$\eqref{eq:defDeltaIstar}$ we see that a solution to $\Delta^\ast_I(x)$ must be continuous across all the $\mu_p$ and be in the kernel of $D_\Lambda^\ast(x)$. There are no such solutions away from the resonating points.)
Setting $\E_0=\E_M$ and $\A_0=\A_M$ we can then use the scheme described at the start of Section~\ref{sec:modelko}: $\ko$ is given by~$\eqref{eq:chFl}$, and for each $l=1,\ldots,\ko$ we want to find the transition function $g_l$ from $\E_M |_{\supp~\varphi_l}$ to $\tilde{\E} |_{\supp~\varphi_l}$ where $\varphi_l=\varphi_{l,0}$, and $\varphi_{lm}$ are the bump functions used to define $B$. 

We have to identify $\coker\modelop(x)$ on $\supp~\varphi_l$. 
Let 
\begin{equation}\label{eq:basisA}
\{u_{l,1},\ldots,u_{l,n-2}\}\cup\{u_l,u_l^\perp\} 
\end{equation}
be an orthonormal basis of $\ker\Delta_M^\ast(x)$ over $\supp~\varphi_l$, where $u_l,u_l^\perp$ are the monopole solutions used to define $B$. 
Note that $u_{l,1},\ldots,u_{l,n-2}$ are all solutions of $\modelop^\ast(x)$ on $\supp~\varphi_l$ because $B u_{l,j}=0$ for all $j=1,\ldots,n-2$. 
We know $\coker\modelop(x)$ is $n$-dimensional, so there are still $2$ solutions to find. 
Suppose $\modelop^\ast(x)(v_I\oplus v_M)=0$ and decompose $v_M$ as
\begin{equation*}
v_M(x) = C_M(x) u_l(x) +C_M^\perp(x) u_l^\perp(x) + v_M^\circ(x)
\end{equation*}
where $v_M^\circ(x)$ is perpendicular to $u_l(x)$ and $u_l^\perp(x)$, and $C_M,C_M^\perp$ are functions of $x$. 
Then for each $x\in\supp~\varphi_l$
\begin{equation*}
B(x)v_M(x) = C_M(x) \varphi_l(x) \eta_l +C_M^\perp(x) \varphi_l(x) \eta_l^\perp,
\end{equation*}
so $\Delta_I^\ast(x) v_I(x) +B(x)v_M(x) =0$ has solution 
\begin{equation}\label{eq:spottedsoln1}
v_I(x) = C_I(x) \eta_l +C_I^\perp(x) \eta_l^\perp
\end{equation}
for $C_I(x),C_I^\perp(x)$ satisfying
\begin{equation}\label{eq:spotteddicktransfn}
(x^\ast-\lambda_l)\begin{pmatrix}C_I \\ C_I^\perp\end{pmatrix}
+\varphi_l\begin{pmatrix}C_M \\ C_M^\perp\end{pmatrix}=0.
\end{equation}
The condition $B^\ast(x) v_I(x) +\Delta_M^\ast(x) v_M(x)=0$ implies that
\begin{equation}\label{eq:spottedsoln2}
v_M = C_M u_l +C_M^\perp u_l^\perp -\varphi_lC_I(\Delta_M^\ast)^{-1}u_l 
-\varphi_lC_I^\perp(\Delta_M^\ast)^{-1}u_l^\perp.
\end{equation}

We want to extend the solutions $u_{l,1},\ldots,u_{l,n-2}$ by two further solutions $v,v^\perp$ to give a gauge for $\tilde{\E}|_{\supp~\varphi_l}$. 
Consider taking
\begin{equation}\label{eq:gaugeintE}
\begin{pmatrix}C_M \\ C_M^\perp\end{pmatrix}=
\psi(x)\frac{x^\ast-\lambda_l}{\rho}
\begin{pmatrix}A \\ A^\perp\end{pmatrix}
\end{equation}
where $A$ and $A^\perp$ are some constants, $\rho=\|x^\ast-\lambda_l \|$ and $\psi$ is a bump function:
\begin{equation}\label{eq:tempbump}
\psi(x) = \begin{cases}
          \rho & \textrm{when $x^\ast-\lambda_l$ is small, and}\\
          1 & \textrm{outside a small neighbourhood of $x^\ast-\lambda_l=0$.}
          \end{cases}
\end{equation}
Then $C_M(x)$ and $C_M^\perp(x)$ are well defined everywhere on $\supp~\varphi_l$ for any choice of $A,A^\perp$. 
We define $v$ and $v^\perp$ by taking
\begin{equation*}
\begin{pmatrix}A \\ A^\perp\end{pmatrix}=\begin{pmatrix}1 \\ 0\end{pmatrix}
\quad\textrm{and\ }
\begin{pmatrix}A \\ A^\perp\end{pmatrix}=\begin{pmatrix}0 \\ 1\end{pmatrix},
\end{equation*}
and using $\eqref{eq:spottedsoln1}$, $\eqref{eq:spotteddicktransfn}$, and $\eqref{eq:spottedsoln2}$. 
Then 
\begin{equation*}
\{u_{l,1},\ldots,u_{l,n-2}\}\cup\{v,v^\perp\}
\end{equation*}
is a gauge for $\tilde{\E}|_{\supp~\varphi_l}$ which is unitary on the boundary of $\supp~\varphi_l$, but not necessarily unitary on the interior. 
Equation~$\eqref{eq:gaugeintE}$ was constructed to ensure that the gauge was unitary on the boundary but well-defined on the interior of $\supp~\varphi_l$. 
There is no obstruction to taking the unitarisation over $\supp~\varphi_l$, without affecting the gauge near the boundary. 
Using~$\eqref{eq:gaugeintE}$ we can compare this trivialisation with the trivialisation~$\eqref{eq:basisA}$, and see that the transition function $g_l$ from $\E_M |_{\supp~\varphi_l}$ to $\tilde{\E} |_{\supp~\varphi_l}$ is
\begin{equation}\label{eq:transcanonicalform}
g_l=
\begin{pmatrix}{\textrm{id}}_{n-2} & 0 \\0 & \frac{x^\ast-\lambda_l}{\| x^\ast-\lambda_l\|}\end{pmatrix}.
\end{equation}
Substituting this into~$\eqref{eq:degreeclutch}$ gives $c_2(\F_l)=-1$, since with our definition of $\deg g$, the map
\begin{gather*}
g:S^3\subset\rfour\rightarrow U(2) \\
g(x) = \frac{x^\ast-\lambda_l}{\| x^\ast-\lambda_l\|}
\end{gather*}
has degree $-1$. 
Thus, using~$\eqref{eq:chFl}$, we have shown that $\tilde{k}_0=\ko$. 

\subsection{The case of vanishing monopole charges}\label{sec:kovanishingmoncharges}

Following the scheme outlined at the start of Section~\ref{sec:modelko}, we want to identify the bundle and connection $(\E_0,\A_0)$. 
Since there is no monopole block, this is rather different to the case of no zero jumps in the instanton block, where $\E_0$ was the cokernel of the restriction of $\modelop(x)$ to the monopole block. 
Essentially, we calculate the transition functions $g_l$ by taking a trivialisation consisting of the exact solutions $v_1,\ldots,v_n$ away from the resonating points, while at the resonating point $x^{\textrm{res}}_{lm}$ two of these solutions are replaced by $\eta_{lm},\eta^\perp_{lm}$ to give a local trivialisation.  
We take $\E_0$ to be the trivial bundle $\underline{\C}^n$ over $\Ieps\times\rthree$. 
Let $\{ B_{lm}: l=1,\ldots,k_0~\textrm{and}~m\in\Z \}$ be a collection of closed balls round the resonating points $x^{\textrm{res}}_{l,m}$, on which conditions~$\eqref{eq:physcond1}$--$\eqref{eq:physcond3}$ hold, and let $B_l=B_{l,0}$. 
Let $R=\bigcup_1^{k_0} B_l$ and let $R^c$ denote the complement. 
The exact solutions $v_1,\ldots,v_n$ defined in Section~\ref{sec:exactsolns} determine a bundle isomorphism 
\begin{equation*}
F:\tilde{\E} |_{R^c} \rightarrow \E_0 |_{R^c}.
\end{equation*}
After applying the Gram-Schmidt algorithm to $v_1,\ldots,v_n$ we obtain a unitary bundle isomorphism $F_U$ in the same way. 
Define a connection $\A_0$ on $\E_0$ by
\begin{equation*}
\A_0 |_{R^c} = (F_U^{-1})^\ast\tilde{\A}|_{R^c}
\end{equation*}
and continue $\A_0$ arbitrarily over $R$. 
It follows that $\A_0$ is the interior restriction of a framed quasi-periodic connection, and that $(\E_0,\A_0)$ extends to the boundary in the same way as $(\tilde{\E},\tilde{\A})$. 
If $f$ denotes the framing at infinity, then since $(\E_0,\A_0)$ has vanishing monopole charges, equation~$\eqref{eq:caloroncharge}$ implies that 
\begin{equation*}
\int_{[0,\perflat]\times\rthree}\chtwo(\E_0,{\A}_0)=-c_2(\E_0,f).
\end{equation*}
However, we claim that $\A_0$ has trivial clutching function, so that the RHS is zero. 
Now $\A_0 = (F_U^{-1})^\ast\tilde{\A}$ on $R^c$ and $\tilde{\A}(\tau x) =(U_\tau^{-1})^\ast\tilde{\A}(x)$ where $\tau$ is translation by $\perflat$ in $x_0$, so
\begin{align*}
\A_0(\tau x)&= 
\big(F^{-1}(\tau x)\big)^\ast\big(U_\tau^{-1}\big)^\ast\big(F(x)\big)^\ast\A_0(x) \\ & =
[F(x)U_\tau^{-1}F^{-1}(\tau x)]^\ast\A_0(x).
\end{align*}
However, the exact solutions $v_p$ satisfy
\begin{equation}\label{eq:exactclutching}
v_p(\tau x) = U_\tau v_p(x)
\end{equation}
so
\begin{equation}\label{eq:Ftriv}
F(x)U_\tau^{-1}F^{-1}(\tau x)\equiv 1.
\end{equation}
The Gram-Schmidt process gives some $GL(n,\C)$ function $\Theta(x)$ such that
\begin{equation*}
F_U(x)=\Theta(x)F(x)
\end{equation*}
and equation~$\eqref{eq:exactclutching}$ implies that $\Theta(\tau x)=\Theta(x)$. 
Substituting this into~$\eqref{eq:Ftriv}$ shows that $\A_0$ has trivial clutching function, so we have proved our claim, and shown that
\begin{equation*}
\int_{[0,\perflat]\times\rthree}\chtwo(\E_0,{\A}_0)=0.
\end{equation*}

Next we want to find a trivialisation of $\tilde{\E}|_{B_l}$ for each $l=1,\ldots,k_0$. 
Comparing this with the gauge $v_1,\ldots,v_n$ on $R^c$ will give the transition function $g_l$ on $\partial B_l$ between $\E_0|_{B_l}$ and $\tilde{\E}|_{B_l}$. 
Start by fixing some $l\in\{ 1,\ldots, k_0 \}$. 
From the definition of the model operator in Section~\ref{sec:physmodelop}, there exist $q_l,q_l^\perp\in\{ 1,\ldots,n \}$ satisfying conditions~$\eqref{eq:physcond1}$, $\eqref{eq:physcond2}$, and~$\eqref{eq:physcond3}$. 
Condition~$\eqref{eq:physcond3}$ implies that for $q\notin\{ q_l,q_l^\perp \}$ the solution $v_q$ is well-defined everywhere on $B_l$, since $v_q$ is just some vector in $\C^{\Nzer}$.  
The set $\{ v_q : q\notin \{ q_l,q_l^\perp \} \}$ can therefore be extended by two sections $\hat{v}_{q_l}, \hat{v}_{q_l^\perp}$ to give a trivialisation of $\tilde{\E}$ over $B_l$. 
Consider
\begin{equation}\label{eq:defvhat}
\begin{pmatrix}
\hat{v}_{q_l}\\ \hat{v}_{q_l^\perp}
\end{pmatrix}
=\frac{x^\ast-\lambda_l}{\| {x^\ast}-\lambda_l \|}
\begin{pmatrix}
v_{q_l} \\ v_{q_l^\perp}
\end{pmatrix}
\end{equation}
where $x=\sum x_a\gamma_a$. 
We want to show that $\hat{v}_{q_l},$ and $\hat{v}_{q_l^\perp}$ are well-defined at $x=x_{l}^{\textrm{res}}$, and that
\begin{equation}\label{eq:extbasis}
\{ v_q : q\notin \{ q_l,q_l^\perp \} \} \cup \{ \hat{v}_{q_l}, \hat{v}_{q_l^\perp} \}
\end{equation}
is a trivialisation of $\tilde{\E}$ on $B_l$. 
This is clearly a trivialisation away from $x_{l}^{\textrm{res}}$, so we only have to understand what happens at the resonating point. 

Recall the definition of the exact solution $v_p$ given by equations~$\eqref{eq:nearlydefexactsoln}$ and~$\eqref{eq:defexactsoln}$. 
Since $\tilde{\zeta}_{q_l}(x)=\eta_l(\mu_{q_l})$ near the resonating point (condition~$\eqref{eq:physcond1}$), and using~$\eqref{eq:definstblcksections}$, the exact solution $v_{q_l}$ is given by
\begin{multline}\label{eq:exactsolnnearres}
v_{q_l} = -C_{q_l}^{-1}\big[\exp \big(i(x^\ast-\lambda_l)(\xi-\mu_{q_l})\big)\big]\big[1-\exp \big(i(x^\ast-\lambda_l)\muo\big)\big]^{-1}
\\ \times\big[\exp (i\xo\mu_{q_l})\big]
\begin{pmatrix} 1 \\ 0\end{pmatrix}\otimes\bar{\eta}_l
\end{multline}
for $\mu_{q_l}\leq \xi \leq\mu_{q_l}+\muo$. 
Note that all the terms in this expression commute. 
Hence 
\begin{equation*}
v_{q_l} = \big[1-\exp \big(i(x^\ast-\lambda_l)\muo\big)\big]^{-1}C_{q_l}^{-1}f_{q_l}(x,\xi)\begin{pmatrix} 1 \\ 0\end{pmatrix}\otimes\bar{\eta}_l,
\end{equation*}
where
\begin{equation*}
f_{q_l}(x,\xi) = -\big[\exp \big(i(x^\ast-\lambda_l)(\xi-\mu_{q_l})\big)\big]\big[\exp (i\xo\mu_{q_l})\big]
\end{equation*}
is defined for all $x\in B_l$ and $\mu_{q_l}\leq \xi \leq\mu_{q_l}+\muo$. 
Similarly, for $v_{q_l^\perp}$, 
\begin{equation*}
v_{q_l^\perp} = \big[1-\exp \big(i(x^\ast-\lambda_l)\muo\big)\big]^{-1}C_{q_l^\perp}^{-1}f_{q_l^\perp}(x,\xi)\begin{pmatrix} 0 \\ 1\end{pmatrix}\otimes\bar{\eta}_l,
\end{equation*}
and so 
\begin{equation}\label{eq:clutchatzerojump}
\begin{pmatrix}
\hat{v}_{q_l}\\ \hat{v}_{q_l^\perp}
\end{pmatrix}
=\frac{x^\ast-\lambda_l}{\| x^\ast-\lambda_l \|}
\big[1-\exp \big(i(x^\ast-\lambda_l)\muo\big)\big]^{-1}
\begin{pmatrix}
f_{q_l} / C_{q_l} \\ f_{q_l^\perp} / C_{q_l^\perp}
\end{pmatrix}\otimes\bar{\eta}_l.
\end{equation}

We want to estimate $C_{q_l}(x)$ and $C_{q_l^\perp}(x)$ near the resonating point. 
Let $\rho=\| x^\ast-\lambda_l \|$. 
Then $\| 1-\exp i(x^\ast-\lambda_l)\muo \|=O(\rho)$ as $\rho\rightarrow 0$, and $\eqref{eq:exactsolnnearres}$ implies that 
\begin{equation}\label{eq:estonnormal}
C_{q_l}=O(\rho^{-1})=C_{q_l^\perp}
\end{equation}
so that $v_{q_l}$ and $v_{q_l^\perp}$ have unit norm. 
Next, consider~$\eqref{eq:clutchatzerojump}$ in the limit $\rho\rightarrow 0$. 
Equation $\eqref{eq:estonnormal}$ implies that
\begin{equation}\label{eq:reckoVMC1}
\lim_{\rho\rightarrow 0}~\frac{1}{\| x^\ast-\lambda_l \|}
\begin{pmatrix}
f_{q_l} / C_{q_l} \\ f_{q_l^\perp} / C_{q_l^\perp}
\end{pmatrix}
\end{equation}
exists. 
Moreover, 
\begin{align}
\lim_{\rho\rightarrow 0}~
(x^\ast-\lambda_l)\big[1-\exp \big(i(x^\ast-\lambda_l)\muo\big)\big]^{-1} &=
 \lim_{\rho\rightarrow 0}~
(x^\ast-\lambda_l)\big[-i(x^\ast-\lambda_l)\muo+O(\rho^2)\big]^{-1}\notag \\
&= -\frac{1}{i\muo}\label{eq:reckoVMC2}
\end{align}
Combining~$\eqref{eq:reckoVMC1}$ and~$\eqref{eq:reckoVMC2}$ shows that the right-hand side of~$\eqref{eq:clutchatzerojump}$ is well defined as $\rho\rightarrow 0$. 
Thus $\hat{v}_{q_l}$, and $\hat{v}_{q_l^\perp}$ are well-defined in the limit $\rho\rightarrow 0$, and at $x=x_{l}^{\textrm{res}}$ are given by some multiples of $\eta_l,\eta_l^\perp$ respectively. 
The set~$\eqref{eq:extbasis}$ is therefore a trivialisation of $\tilde{\E}$ over $B_l$. 

We can now calculate the transition function $g_l$. 
Comparing the trivialisation~$\eqref{eq:extbasis}$ with the exact solutions $\{v_p:p=1,\ldots,n\}$, we see that (up to some re-ordering)
\begin{equation*}
g_l=
\begin{pmatrix}{\textrm{id}}_{n-2} & 0 \\0 & \frac{x^\ast-\lambda_l}{\| x^\ast-\lambda_l \|}\end{pmatrix}.
\end{equation*}
The two trivialisations $v_1,\ldots,v_n$ and~$\eqref{eq:extbasis}$ are not unitary, but applying the Gram Schmidt process is equivalent to replacing $g_l$ with
\begin{equation*}
\Theta(x)g_l(x)\Theta^{-1}(x)\in U(n)
\end{equation*}
where $\Theta :\partial B_l\rightarrow GL(n,\C)$. 
This does not affect the degree. 
Comparing $g_l$ with the calculation of $\tilde{k}_0$ in the case of vanishing monopole charges, we have $\deg g_l=-1$ and so $c_2(\F_l)=-1$. 
This completes the proof that $\tilde{k}_0=\ko$ for vanishing monopole charges.

\subsection{The case $1\leq N_I< n$}\label{sec:kogeneral}

The calculation is very similar to that in Section~\ref{sec:konoinstzerojumps}.   
Recall the definition of the model operator in Section~\ref{sec:modelremaining}, the definition of the adjoint in Section~\ref{sec:modeladjoint}, and the definition of the exact solutions $v_p$, $p\in\Zeros_I$, given in Section~\ref{sec:exactsolns}. 
In Section~\ref{sec:exactsolns} we showed that the $v_p$ are well-defined solutions to $\modelop^\ast(x)$ on the complement of $\textrm{Supp~}B$, where $B$ is the off-diagonal block of the model operator. 
In fact the $v_p$ are well-defined on $\supp~B$, and are just vectors in $\C^{\Nzer}$, because $\tilde{\zeta}_p(x)=0$ on $\supp~B$. 
It follows that $B^\ast(x) v_p(x)=0$ for each $p\in\Zeros_I$, and since the $v_p$ solve $\Delta^\ast_I (x)v_p=0$ everywhere they are solutions of $\modelop^\ast(x)$ for all $x$. 
The $v_p$ therefore define a rank $N_I$ sub-bundle $\E_I$ of $\tilde{\E}$, which is equipped with a connection $\A_I$ given by projection from $V$ onto $\E_I$. The proof of Proposition~\ref{prop:approxframe} shows that $(\E_I,\A_I)$ is the interior restriction of a framed $U(N_I)$ quasi-periodic connection with vanishing monopole charges and clutching function $c_I$. 
The $v_p$ give a global trivialisation of $\E_I$ in which $\A_I$ is framed. 
Since $v_p(\tau x) =U_\tau v_p(x)$, using the definition of the clutching function in Proposition~\ref{prop:approxframe} we have $c_I\equiv 1$, and so $c_2(\E_I,f)=0$ where $f$ is the framing determined by the exact solutions~$v_p$. 
Moreover, since the $\zeta_p$ span trivial line bundles in $\Einf$, using~$\eqref{eq:upcaloroncharge}$ we have
\begin{equation}\label{eq:insttriv}
\int_{[0,\perflat]\times\rthree}\chtwo(\E_I,\A_I) =0.
\end{equation}
Now $\Delta_M(x)$ is the Nahm operator of some $U(n-N_I)$ monopole Nahm data. 
Let $(\E_M,\A_M)=\COKER~\Delta_M$. 
Then Lemma~\ref{lem:HM2} implies that 
\begin{align*}
\int_{[0,\perflat]\times\rthree}\chtwo(\E_M,\A_M) &=
-\frac{1}{\muo}\sum_{p\in\Zeros_O\cup\Zeros_M}\mu_p k_p
\\ &=-\frac{1}{\muo}(\mu_1 k_1+\cdots+\mu_n k_n)
\end{align*}
since $k_p=0$ for all $p\notin\Zeros_O\cup\Zeros_M$.
We can then define $(\E_0,\A_0)=(\E_M\oplus\E_I,\A_M\oplus\A_I)$. 
On the complement of the region $R=\supp~B$, 
\begin{equation*}
\modelop^\ast(x) = \begin{pmatrix} \Delta_I^\ast(x) & 0 \\ 0 & \Delta_M^\ast(x) \end{pmatrix}
\end{equation*}
so $\tilde{\E}|_{R^c}=\E_I\oplus\E_M$ and $\tilde{\A}|_{R^c}=\A_I\oplus\A_M$. 
Since $\ch(\E_I\oplus\E_M,\A_I\oplus\A_M)=\ch(\E_I,\A_I)+\ch(\E_M,\A_M)$, $(\E_0,\A_0)$ satisfies~$\eqref{eq:kocharge2}$. 

It remains only to specify gauges on $\E_0$ and $\tilde{\E}$ near each resonating point and calculate the degree of the transition function.  
Let $u_l(x)$ and $u_l^\perp(x)$ be the solutions of $\Delta_M^\ast(x)$ used to define the off-diagonal blocks of $\modelop$. 
Then, as in Section~\ref{sec:konoinstzerojumps}, let
\begin{equation*}
\{ u_{l,p} :p=1,\ldots,n-2-N_I \}\oplus\{u_l,u_l^\perp \}\oplus\{ v_p:p\in\Zeros_I \}
\end{equation*}
be a gauge for $\E_0|_{\supp~\varphi_l}$ where $u_{l,p}$ are solutions of $\Delta^\ast_M(x)$. 
Again, following Section~\ref{sec:konoinstzerojumps} and reproducing equations~$\eqref{eq:spottedsoln1}$ to $\eqref{eq:spottedsoln2}$, we construct $v,v^\perp$ in exactly the same way, so that 
\begin{equation*}
\{ u_{l,p} :p=1,\ldots,n-2-N_I \}\oplus\{v,v^\perp \}\oplus\{ v_p:p\in\Zeros_I \}
\end{equation*}
is a gauge for $\tilde{\E}|_{\supp~\varphi_l}$, and so that the transition function $g_l$ has $\deg g_l=-1$.
This completes the proof that $\tilde{k}_0=k_0$.

\section{Deforming $\Delta$ to $\tilde{\Delta}$}\label{sec:deform}

The last step of the construction of calorons is to prove that the Nahm operator $\Delta$ can be deformed into the model operator $\modelop$ in such a way that we can deduce that $\COKER~\Delta$ is a framed caloron configuration. 
In Section~\ref{sec:pathconn} we prove the existence of a path of injective Fredholm operators between $\Delta$ and $\modelop$, and in Section~\ref{sec:recoverframed} we use this to deduce that $\COKER~\Delta$ is a framed caloron configuration. 
Finally, in Section~\ref{sec:volumeforms}, we use the boundary conditions and the anti-self-duality equation to prove that $\COKER~\Delta$ can be equipped with a compatible volume form. 

\subsection{Existence of a path between $\Delta$ and $\modelop$}\label{sec:pathconn}

We want to show that we can deform $\Delta_0 = \Delta$ to $\Delta_1 = \modelop$ with a path $\Delta_s$ in the space of injective Fredholm operators with index $-n$. 
Let $\Fred$\label{glo:Fred} denote the space of Fredholm operators from $W$ to $V$ with index $-n$, and let $\Fredi$ denote the subset of injective operators. 
Thus $\Delta_0$ and $\Delta_1$ are maps from $\Ieps\times\rthree$ to $\Fredi$. 

\begin{proposition}\label{pro:deformation}
If $\Delta_1 = \modelop$ is a controlled deformation of $\Delta_0 = \Delta$ (in the sense of Definition~\ref{def:controlleddeformn}) then there is a path $\Delta_s$ in $\Map(\Ieps\times\rthree,\Fredi)$ between the two, such that for all $s\in[0,1]$,
\begin{gather}\label{eq:defcon1}
\Delta_s\textrm{\ is a framed deformation of $\Delta$, and}
\\ \label{eq:defcon2}
\Delta_s(\tau x)=U_{\tau,V}\Delta_s(x)U_{\tau,W}^{-1},
\end{gather}
where $U_{\tau,V}, U_{\tau,W}$ were defined in Section~\ref{sec:calper}, and Definition \ref{def:frameddeformn} gave the notion of a framed deformation. 
\end{proposition}
In Section~\ref{sec:recoverframed} we show that conditions~$\eqref{eq:defcon1}$ and~$\eqref{eq:defcon2}$ ensure that $\COKER~\Delta_s$ is a framed caloron configuration for each $s$. 

\proof
We first prove the existence of a path $\Delta_s$ satisfying~$\eqref{eq:defcon1}$ and~$\eqref{eq:defcon2}$ that lies in $\Map(\Ieps\times\rthree,\Fred)$ and then perturb it so that it lies in $\Map(\Ieps\times\rthree,\Fredi)$. 
Recall that
\begin{equation*}
\Delta(x)w = [ D_{1}(x)w_{1},\ldots,D_{n}(x)w_{n} ] \oplus[\pi w]
\end{equation*}
and
\begin{equation*}
\tilde{\Delta}(x)w = [ D_{1}(x)w_{1}+A_{1}w_{1},\ldots,D_{n}(x)w_{n}+A_{n}w_{n} ] \oplus[\tilde{\pi} w]+B(x)w
\end{equation*}
for some $A_p$, $\tilde{\pi}$, and $B(x)$ satisfying the conditions of Definition~\ref{def:controlleddeformn}. 
Let $\tilde{\pi}_s$ denote a deformation of $\pi$ to $\tilde{\pi}$ given by deforming the vectors $\zeta_q$, and let
\begin{equation*}
\Delta_s(x)w = [ D_{1}(x)w_{1}+sA_{1}w_{1},\ldots,D_{n}(x)w_{n}+sA_{n}w_{n} ] \oplus[\tilde{\pi}_s w]+sB(x)w.
\end{equation*}
Then, for each $s$, $\Delta_s$ is a controlled deformation of $\Delta$, so condition~$\eqref{eq:defcon1}$ is met (controlled implies framed) and Corollary~\ref{cor:indexdefdelta} implies that $\Delta_s$ lies in $\Map(\Ieps\times\rthree,\Fred)$ for each $s$. 
Since $\Delta$ and $\modelop$ satisfy~$\eqref{eq:defcon2}$, if we make the deformation $\tilde{\pi}_s$ strictly periodic in $\xo$, then $\Delta_s$ satisfies condition~$\eqref{eq:defcon2}$ for all $s$. 

Next we want to perturb this deformation so that it lies in $\Map(\Ieps\times\rthree,\Fredi)$. 
The deformation corresponds to a five (real) dimensional surface $\Sigma$ lying in $\Fred$ which is the image of a map $[0,1]\times\Ieps\times\rthree\rightarrow\Fred$. 
Note that $\Fred$ is not a smooth manifold, but is stratified, with the strata corresponding to the dimension of the kernel. 
Let $U\subset\Fred$ denote the subset of operators that are not injective. 
If the codimension of $U$ is sufficiently large, then the surface $\Sigma$ can be deformed into $\Fredi$. 
To perturb $\Sigma$ into $\Fredi$ we require one dimension orthogonal to $\Sigma$ and $U$ at each point on $\Sigma$, \ie we require the codimension of $U$ to be at least six (real) dimensions.  
The codimension of $U$ can be calculated in the following way. 
Fix $\alpha\in U$. 
Since $\alpha$ is not injective, it has a non-trivial kernel which is finite dimensional, so suppose $\{ w_1,\ldots,w_m \}$ is a basis for the kernel, where $m\geq 1$. 
Since $\ind\alpha=-n$, $\coker\alpha$ is $m+n$ dimensional, so fix a basis $\{ v_1,\ldots,v_{m+n} \}$ for the cokernel. 
Note that $m+n\geq 3$ because $n\geq 2$. 
The paths
\begin{equation*}
\alpha_i(t) = \begin{cases}
\alpha\textrm{\ on\ }(\ker\alpha)^\perp, \\
w_p\mapsto v_p\quad \textrm{for\ }p=1,\ldots,m-1, \\
w_m\mapsto tv_{m+i-1}
\end{cases}
\end{equation*}
for $i=1,2,3$ define a tangent plane to $U$ at $\alpha$, contained in $\Fredi$. 
The condition $m+n\geq 3$ implies that the three paths are well defined.
Hence the (real) codimension of $U$ is at least six, and so the deformation can be shifted into $\Fredi$. 

A problem might arise: we do not want to perturb $\Delta_s$ when $s=0$ or $1$, because the perturbed path would no longer join $\Delta$ to $\modelop$. 
Similarly, the other components of $\partial\Sigma$ might be affected. 
First consider what happens for large $r$. 
We want the path $\Delta_s$ to consist of framed deformations of $\Delta$, which is certainly true before we perturb $\Sigma$. 
For sufficiently large $r$
\begin{equation*}
\Delta_s(x) = \Delta(x) +sA
\end{equation*}
where $A$ is a multiplicative operator independent of $x$. 
Equation~$\eqref{eq:DdagDOrsqu}$ implies that 
\begin{equation*}
\| \Delta_s (x)w\|^2_{L^2} \geq Cr^2 \| w\|^2_{L^2}
\end{equation*}
for large $r$ and some constant $C$, because $A$ is uniformly bounded. 
Hence for sufficiently large $r$
\begin{equation*}
\| \Delta_s(x) w\| = 0\quad \Rightarrow\quad
\| w\|_{L^2} = 0\quad \Rightarrow\quad
\| w\|_{L^2_1} = 0
\end{equation*}
so $\Delta_s(x)$ is injective for sufficiently large $r$. 
Thus, combining the three boundary components, we only have to perturb $\Delta_s(x)$ on some compact subset of $\rthree$ to move $\Sigma$ into $\Fredi$. 
Finally consider $\partial \Ieps$. 
If we perturb $\Delta_s$ into $\Fredi$ for $\xo\in(-\epsilon,\epsilon)$ then the periodicity rule~$\eqref{eq:defcon2}$ fixes the perturbation for $\xo\in(\perflat-\epsilon,\perflat+\epsilon)$. 
Thus we have some fixed perturbation of $\Sigma$ in some neighbourhood of the boundary of $[0,1]\times\Ieps\times\rthree$, and we want to extend this to a perturbation of the whole of $\Sigma$. 
The Proposition follows from the Lemma below. 
\eproof

\begin{lemma}
Suppose $\Sigma$ is the image of some map $\sigma:M=[0,1]\times\Ieps\times\rthree\rightarrow\Fred$ and $A$ is a neighbourhood of the boundary of $M$. 
We know that while $M$ is $5$-dimensional, the complement of the space of non-injective Fredholm operators has codimension $6$.
Then any deformation of $\sigma(A)$ into $\Fredi$ extends to a deformation of $\Sigma$ into $\Fredi$. 
\end{lemma}

\proof
Milnor \cite[Theorem 1.35]{mil58} proves an equivalent result for smooth manifolds which extends to spaces of Fredholm operators readily.  
\eproof

\subsection{Recovering the boundary conditions}\label{sec:recoverframed}

We have a deformation $\modelop$ of $\Delta$ together with bundles and connections
\begin{equation*}
(\E,\A)=\COKER~\Delta\quad\textrm{and\ }(\tilde{\E},\tilde{\A})=\COKER~\modelop.
\end{equation*}
Given that $\tilde{\A}$ extends to $\sphinf$ in the sense of Proposition~\ref{prop:approxframe}, our aim is to prove that $\A$ extends to $\sphinf$ and is framed there, by comparing the two connections and showing that $\A$ is asymptotically close to $\tilde{\A}$. 
The precise statement is as follows:

\begin{proposition}\label{prop:framed}
The bundle and connection $(\E,\A)$ is the interior restriction of $(\Eq,\Aq)$ where $\Aq$ is some $\Un$ quasi-periodic connection which satisfies the smoothness conditions of Section~\ref{sec:smooth} and which is framed by $\Ainf,\Phiinf$.
\end{proposition}

\proof
The aim is to construct a unitary isomorphism $F$ and framed quasi-periodic connection $\Aq$ such that $F:\E\rightarrow \Eq |_{\Ieps\times\rthree}$ and $\A=F^\ast(\Aq |_{\Ieps\times\rthree})$. 
We start by comparing the projections $P(x)$ and $\tilde{P}(x)$ from $V$ onto $\coker\Delta(x)$ and $\coker\tilde{\Delta}(x)$ respectively, and show that
\begin{equation}\label{eq:pminustp}
 (\tilde{P}(x) - P(x)) |_{\tilde{\E}} = O(\chi)
\end{equation}
as $\chi\rightarrow 0$.
It follows that for sufficiently small $\chi$, $P$ is an isomorphism as a map from $\tilde{\E}|_{\{\chi\ll 1\}}$ to $\E$. 
We can then compare $\tilde{\A}$ with the pull-back of $\A$ under $P$.  

The proof of~$\eqref{eq:pminustp}$ is taken almost directly from \cite[Section 2]{hit83} and follows from estimates on the Green's function $G(x)=(\Delta^\ast(x)\Delta(x))^{-1}\Delta^\ast(x)$ of $\Delta(x)$. 
Equations~$\eqref{eq:deltadagdelta}$ and~$\eqref{eq:DdagDOrsqu}$ imply that 
\begin{equation}\label{eq:Ltwoest}
\langle \Delta^\ast(x)\Delta(x) w, w \rangle_{\textrm{dual}} \geq \frac{C^2}{\chi^2} \| w \|^2_{L^2}
\end{equation}
for all $w\in W$ and for sufficiently small $\chi$. 
(Throughout, $C$ is a constant used in the generic sense.)
Since $\Delta(x) G(x) =1-P(x)$ it follows that
\begin{equation*}
\| v \|^2_{L^2} - \| P(x)v \|^2_{L^2}\geq \frac{C^2}{\chi^2}\| G(x)v\|^2_{L^2}
\end{equation*} 
for all $v\in V$, so that
\begin{equation}\label{eq:estimateGreen}
\| G(x) \|_{L^2\rightarrow L^2}\leq\frac{\chi}{C},\quad\textrm{and\ }
\| G^\ast(x) \|_{L^2\rightarrow L^2}\leq\frac{\chi}{C}.
\end{equation}
Strictly speaking, $G$ is a map $L^2_0\rightarrow L^2_1$, so $\| G \|_{L^2\rightarrow L^2}$ is the norm of the composition of $G$ with inclusion into $L^2_0$. 
Similarly, $G^\ast(x)$ is a map $L^2_{-1}\rightarrow L^2_0$, so $\| G^\ast (x)\|_{L^2\rightarrow L^2}$ is the norm of the restriction of $G^\ast(x)$ to $L^2_0\hookrightarrow L^2_{-1}$. 
Fixing an element $\tilde{v}$ of $\tilde{\E}$, we have
\begin{align*}
(\tilde{P}(x)-P(x))\tilde{v} &= (1-P(x))\tilde{v} \\
&= G^\ast(x)\Delta^\ast(x)\tilde{v} \\
&= G^\ast(x) A^\ast\tilde{v}
\end{align*}
where 
\begin{equation*}
A=\Delta(x)-\modelop(x)
\end{equation*}
is a multiplicative operator that is independent of $x$ for sufficiently small $\chi$. 
Since $A$ is uniformly bounded and smooth, $A^\ast\tilde{v}$ is $L^2$, and together with estimate $\eqref{eq:estimateGreen}$ this proves equation~$\eqref{eq:pminustp}$.

The next step is to compare $\A$ and $\tilde{\A}$. 
Equation~$\eqref{eq:pminustp}$ implies that $P$ is an isomorphism as a map $P:\tilde{\E}\rightarrow\E$ for sufficiently small $\chi$. 
Let $R\subset \Ieps\times\rthree$ be some region $\chi<\epsilon$ on which $P$ is an isomorphism. 
Consider the pull-back of $\A |_{R}$ under $P$: the difference between the connections on $\tilde{\E}|_{R}$ is an endomorphism-valued $1$-form $\alpha$, given by
\begin{equation*}
\alpha(s) = \tilde{\A}(s) - P^{-1}_{\E}\A(Ps)
\end{equation*}
where $P^{-1}_{\E}$ is the inverse of $P$ as a map $\tilde{\E}\rightarrow\E$, so that $P^{-1}_{\E}P = \tilde{P}$. 
Expanding $\alpha$ using $\A=P\cdot d$ and $\tilde{\A}=\tilde{P}\cdot d$ gives
\begin{equation*}
\alpha(s) = -\tilde{P}dPs
\end{equation*}
where $s$ is a section of $\tilde{\E}$. 
Using the identity $\tilde{P}d\tilde{P}\tilde{P}=0$ and the fact that $\tilde{P}s=s$, this gives
\begin{equation*}
\alpha = \tilde{P}(d\tilde{P}-dP)\tilde{P}.
\end{equation*}

Our aim is to prove that $\alpha$ is $C_\chi^{0,1}$ and that the $dx_0, dy_1, dy_2$ components of $\alpha$ vanish at the boundary---this will imply that $\A$ and $\tilde{\A}$ are framed in the same way on $\sphinf$. 
Now, 
\begin{equation*}
\tilde{P}-P = \Delta\rho\Delta^\ast - \tilde{\Delta}\tilde{\rho}\tilde{\Delta}^\ast
\end{equation*}
where $\rho(x)=(\Delta^\ast(x)\Delta(x))^{-1}$ and similarly for $\tilde{\rho}(x)$. 
Since $\Delta(x)=\modelop(x) +A$ on $R$, we have
\begin{equation}\label{eq:expandpminustp}
\tilde{P} -P = (\modelop +A)\rho(\modelop^\ast +A^\ast)-\modelop\tilde{\rho}\modelop^\ast.
\end{equation}
Hence
\begin{equation}\label{eq:estimatealpha}
\tilde{P}(d\tilde{P} - dP)\tilde{P} = 
\tilde{P}\big( (dx+A)\rho A^\ast + A\rho (dx^\ast +A^\ast) +A (d\rho) A^\ast \big)\tilde{P}.
\end{equation}
By making the region $R$ sufficiently small, $A$ is multiplicative and therefore smooth on each interval. 
Thus $\rho$ is only passed $L^2$ sections in equation~$\eqref{eq:estimatealpha}$, so we regard $\rho$ as being restricted to $L^2$ sections. 
A similar comment applies to the image of $\rho$, so $\rho$ can be regarded as a map from $L^2$ sections to $L^2$ sections rather than $L^2_{-1}\rightarrow L^2_{1}$.  
As such, equations~$\eqref{eq:deltadagdelta}$ and~$\eqref{eq:DdagDOrsqu}$ give
\begin{equation}\label{eq:estimaterho}
\rho_{L^2\rightarrow L^2}(x) = \chi^2 +O(\chi^3).
\end{equation}
It follows from~$\eqref{eq:estimatealpha}$ that the $dx_0$, $dy_1$, and $dy_2$ components of $\alpha$ are $O(\chi)$, and similarly any derivative of these components of the form $\partial_{x_0}^i \partial_{y_1}^j \partial_{y_2}^k$ is $O(\chi)$. 
Hence the $dx_0$, $dy_1$, and $dy_2$ components of $\alpha$ are $C_\chi^0$ and vanish on the boundary.

Next consider the $d\chi$ component of $\alpha$: using equation~$\eqref{eq:estimaterho}$, it is given by 
\begin{multline*}
\tilde{P}\big(
(\hat{x}\chi^{-2}+A)\rho A^\ast d\chi + A\rho (\hat{x}^\ast\chi^{-2} +A^\ast) d\chi +A(\partial_\chi \rho) A^\ast d\chi
\big)\tilde{P} \\ =
\tilde{P}\big(
\hat{x}A^\ast d\chi +A\hat{x}^\ast d\chi
\big)\tilde{P} +O(\chi)
\end{multline*}
where $\hat{x} = \chi x$ is the unit vector in direction $x$. 
In the basis $\tilde{w}_1,\ldots,\tilde{w}_n$ of $\tilde{\E}$, $\alpha$ is given by the matrix 
\begin{equation*}
\langle \alpha\tilde{w}_i,\tilde{w}_j \rangle, \quad i,j=1,\ldots,n.
\end{equation*}
So, up to $O(\chi)$, the $d\chi$ component is given by 
\begin{equation}\label{eq:dchicomp}
\langle (\hat{x}A^\ast +A\hat{x}^\ast)\tilde{v}_i,\tilde{v}_j \rangle.
\end{equation}
In the limit $\chi\rightarrow 0$ the $\tilde{v}_i$ become `square roots of $\delta$-functions': integral estimates like those in Section~\ref{sec:approxsolns} show that
\begin{equation}\label{eq:sqrtdelta}
\lim_{\chi\rightarrow 0} \langle A\tilde{v}_i,\tilde{v}_j \rangle = 
\begin{cases}
0 & \textrm{if\ }i\neq j \\
\langle A(\mu_p)e_p,e_p \rangle & \textrm{if\ }i=j=p
\end{cases}
\end{equation}
for any multiplicative operator $A$ that is independent of $\chi$, where  $\{e_p\}$ is the gauge on $\Einf$ fixed in Proposition~\ref{prop:approxframe}. 
Hence the limit of~$\eqref{eq:dchicomp}$ exists as $\chi\rightarrow 0$ which shows that the $d\chi$ component of $\alpha$ is continuous up to the boundary. 
In fact it shows the $d\chi$ component is diagonal on the boundary and independent of $\xo$---so the $d\chi$ component satisfies the conditions given in Section~\ref{sec:smooth}. 
Similarly, by considering derivatives in $\xo$, $y_1$, and $y_2$, it can be shown that the $d\chi$ component of $\alpha$ is $C_\chi^0$ up to the boundary.
The same method shows that the $dx_0$, $dy_1$, and $dy_2$ components of $\alpha$ are $C^1_\chi$: 
differentiating~$\eqref{eq:estimatealpha}$ with respect to $\chi$, the coefficients of the $dx_0$, $dy_1$, and $dy_2$ components are of the form 
\begin{equation*}
\textrm{multiplicative operator\ }+O(\chi)
\end{equation*}
and so extend to the boundary. 
Thus $\alpha$ is $C^{0,1}_\chi$. 

Let $\tilde{F}$ be the isomorphism $\tilde{F}:\tilde{\E}\rightarrow \E^q |_{\Ieps\times\rthree}$ whose existence was proved in Proposition~\ref{prop:approxframe}. 
Let $F=\tilde{F} P^{-1}_{\E}$, so that $F$ is an isomorphism $\E \rightarrow\Eq$ defined on $R$.
We have shown that 
\begin{equation}\label{eq:defaq}
\A = F^\ast(\Aq |_{R})
\end{equation}
where $\Aq$ is a $C_\chi^{0,1}$ connection on $\Eq$ framed by $\Ainf,\Phiinf$. 
The isomorphism $F$ can be extended to the interior of $\rthree$ so that $\eqref{eq:defaq}$ holds everywhere on $\Ieps\times\rthree$. 
(Given $\A$, this determines $\A^q$ on the interior.)
To complete the proof of the Proposition, it remains to be shown that $\Aq$ is quasi-periodic, and that $F$ can be replaced with some unitary isomorphism. 

Define
\begin{equation*}
c(s) = F(\xo=s+\perflat) U_\tau F^{-1}(\xo=s).
\end{equation*}
Equation~$\eqref{eq:connectionperiodic}$ implies that
\begin{equation}\label{eq:Aqclutch}
\Aq(s+\perflat) = (c^{-1}(s))^\ast\Aq(s) = c\Aq(s)c^{-1} - dcc^{-1}
\end{equation}
on $\Ieps\times\rthree$.
We want to show that $c\rightarrow 1$ as $\chi\rightarrow 0 $ and that $c$ is $C_\chi^1$. 
In the gauge $FP\tilde{w}_1,\ldots,FP\tilde{w}_n$ on $\Eq |_{R}$, $c$ is given by the matrix 
\begin{multline}\label{eq:estimateclutch}
\langle F(s+\perflat)U_\tau P\tilde{w}_i(s),
 F(s+\perflat) P\tilde{w}_j(s+\perflat) \rangle_{L^2} = \\
\langle P^{-1}_{\E}U_\tau P\tilde{w}_j(s), \tilde{w}_j(s+\perflat) \rangle_{L^2}
\end{multline}
because $F=\tilde{F}P^{-1}_{\E}$ and $\tilde{F}$ is unitary. 
Now $\eqref{eq:pminustp}$ implies that the RHS is given by
\begin{align} 
\langle U_\tau \tilde{w}_i, \tilde{w}_j\rangle +O(\chi) &=
 \langle U_\tau \tilde{v}_i, \tilde{v}_j\rangle +O(\chi) \notag \\
& = \delta_{ij} +O(\chi), \notag
\end{align}
where $\tilde{v}_j$ is the approximate solution associated to the solution $\tilde{w}_j$ of $\modelop(x)$. 
Hence $c\rightarrow 1 $ as $\chi\rightarrow 0$. 
Equation~$\eqref{eq:Aqclutch}$ implies that
\begin{equation*}
\lim_{\chi\rightarrow 0} dc = 
\lim_{\chi\rightarrow 0} \{c\Aq(s) -\Aq(s+\perflat)c\}
\end{equation*}
and since $\Aq$ is $C_\chi^{0,1}$ it follows that $c$ is $C_\chi^1$. 

The final step is to replace $F$ with a unitary isomorphism and show that this does not affect the framing or clutching adversely. 
As in Section~\ref{sec:goodapprox} we replace $P$ with the unitary approximation $P_U$\label{glo:PU} defined by 
\begin{equation*}
P_U = (PP^\ast)^{-1/2}P,
\end{equation*}
and compare $\tilde{\A}$ with the pull-back of $\A$ under $P_U$. 
A calculation similar to the comparison above shows that the difference of the two connections is given by
\begin{align*}
\alpha_U & = \tilde{\A} - P_U^\ast(\A) \\
& = \tilde{P}(d\tilde{P} - P_U^{-1}dP_U)\tilde{P}.
\end{align*}
We can calculate $P_U$ quite explicitly as follows. 
Equation~$\eqref{eq:expandpminustp}$ implies that 
\begin{equation*}
(\tilde{P}(x)-P(x)) |_{\tilde{\E}} = (\tilde{\Delta}(x)+A)\rho A^\ast = \tilde{\Delta}(x)\rho A^\ast +O(\chi^2).
\end{equation*}
Hence 
\begin{equation*}
P(x)|_{\tilde{\E}} = 1 - \tilde{\Delta}(x)\rho A^\ast +O(\chi^2)
\end{equation*}
where $\tilde{\Delta}(x)\rho A^\ast = O(\chi)$, and 
\begin{equation*}
PP^\ast(x) = 1 - (\tilde{\Delta}(x)\rho A^\ast + A\rho\modelop^\ast(x)) + O(\chi^2).
\end{equation*}
We can use a power series expansion for $(PP^\ast)^{-1/2}$:
\begin{equation*}
(PP^\ast)^{-1/2}(x) = 1 +\frac{1}{2}(\modelop(x)\rho A^\ast + A\rho\modelop^\ast(x))+O(\chi^2).
\end{equation*}
Hence 
\begin{equation*}
P_U(x) = \big( 1 + \frac{1}{2}(\modelop(x)\rho A^\ast + A\rho\modelop^\ast(x)) \big) P(x) +O(\chi^2)
\end{equation*}
and we have 
\begin{equation}\label{eq:projnearid}
P_U(x) = 1+O(\chi), \quad P_U^{-1}(x) = 1 +O(\chi).
\end{equation}

We can then calculate $\alpha_U$:
\begin{equation*}
P_U^{-1} dP_U = dP +\frac{1}{2}d(\modelop(x)\rho A^\ast + A\rho\modelop^\ast(x))P+O(\chi)
\end{equation*}
so 
\begin{equation*}
\tilde{P}(P_U^{-1} dP_U)\tilde{P} =
\tilde{P}dP\tilde{P} +\frac{1}{2}\tilde{P}d(\modelop(x)\rho A^\ast + A\rho\modelop^\ast(x))P \tilde{P}+O(\chi).
\end{equation*}
The second term in this equation simplifies:
\begin{align*}
\tilde{P}d(\modelop(x)\rho A^\ast + A\rho\modelop^\ast(x))P\tilde{P} &=
\tilde{P}d(\modelop(x)\rho A^\ast + A\rho\modelop^\ast(x)) \tilde{P} +O(\chi) 
\\& = \tilde{P}(dx\rho A^\ast + A\rho dx^\ast) \tilde{P} +O(\chi)
\\&=  \tilde{P}(dx\chi^2 A^\ast + A\chi^2 dx^\ast) \tilde{P} +O(\chi).
\end{align*}
This implies that 
\begin{equation*}
\alpha_U = \alpha -\frac{1}{2}\tilde{P}(dx\chi^2 A^\ast + A\chi^2 dx^\ast) \tilde{P}+O(\chi)
\end{equation*}
and so 
\begin{equation*}
\alpha-\alpha_U = (\textrm{bounded multiplicative operator})d\chi +O(\chi).
\end{equation*}
It is then easy to apply the arguments used to prove that $\alpha$ is $C_\chi^{0,1}$ to $\alpha_U$, and conclude that $\alpha_U$ is also $C_\chi^{0,1}$. 
This shows that $\A$ is a $C_\chi^{0,1}$ connection. 
Moreover, it also follows that the $d\chi$ component of $\alpha_U$ is diagonal and independent of $\xo$ on the boundary, so $\A_\chi$ satisfies the conditions defined in Section~\ref{sec:smooth}. 

The final check is to ensure that the clutching behaviour of $\Aq$ has not been disturbed by the change to a unitary isomorphism. 
The equation for the matrix of $c$, equation~$\eqref{eq:estimateclutch}$, becomes
\begin{equation*}
\langle P_U^{-1}U_\tau P_U \tilde{w}_i, P_U \tilde{w}_j \rangle = 
\langle U_\tau \tilde{w}_i, \tilde{w}_j \rangle +O(\chi)
\end{equation*}
because of equation~$\eqref{eq:projnearid}$. 
But the RHS is $\delta_{ij}+O(\chi)$ so $c\rightarrow 1$ as $r\rightarrow\infty$, 
and $c$ is still $C_\chi^1$.
\eproof

It seems that the Proposition could be extended quite readily to show that $\A$ is the interior restriction of a framed quasi-periodic connection $\A^q$ that is \emph{smooth} up to the boundary, rather than $C^{0,1}_\chi$. 
First we should indicate why this seems difficult initially. 
The difficulty comes when one considers $\chi$ derivatives of
\begin{equation}\label{eq:notsmooth} 
\langle \alpha \tilde{w}_i, \tilde{w}_j \rangle = 
\langle \alpha \tilde{v}_i, \tilde{v}_j \rangle +\textrm{smooth exponentially small term}. 
\end{equation}
To prove smoothness, all the $\chi$ derivatives of~$\eqref{eq:notsmooth}$ must extend continuously to $\chi=0$. 
However, the $\chi$ derivative of $\tilde{v}_p$ includes terms like $1 / \chi$, and it is unclear how to make these terms cancel to obtain the desired smoothness result. 
However, the method used in the Proposition to show that the $d\chi$ component of $\alpha$ is $C^0_\chi$ could be used to obtain smoothness in the following way. 
The proof of the Proposition included the calculation of the first few terms in the $\chi$ power series expansion of $\alpha$: the proof relied on the fact that the leading coefficients for $dx_0,dy_1,dy_2$ vanished, and that the leading coefficient in $d\chi$ was multiplicative.  
Equation~$\eqref{eq:sqrtdelta}$ showed how this multiplicative term extended to the boundary---a more general operator would not have a limit like~$\eqref{eq:sqrtdelta}$. 
However, all the coefficients in the power series expansion of $\alpha$ will be `differential operators of negative degree', so should have limits like~$\eqref{eq:sqrtdelta}$. 
Some careful analysis of the smoothing properties of the Green's function $\rho$ could therefore lead to the stronger result. 

We can use the Proposition to prove Lemma~\ref{lem:HM2} for Nahm data in $\mon{N}{\monbdarydata}$. 
(Recall that we have not yet proved Lemma~\ref{lem:HM2} for monopole Nahm data that does not satisfy Nahm's equation and that might be continuous accross zero jumps.) 
Suppose that $\Delta$ is the Nahm operator associated to an element of $\mon{N}{\monbdarydata}$. 
Then Proposition~\ref{prop:framed} implies that $\COKER~\Delta$ is the interior restriction of a $U(n)$ quasi-periodic connection framed by $\Ainf,\Phiinf$. 
However, $\COKER~\Delta$ is translation invariant, and so must have $k_0=0$. 
Applying~$\eqref{eq:caloroncharge}$ therefore gives
\begin{equation*}
\int_{[0,\perflat]\times\rthree}\chtwo~\COKER~\Delta =-\frac{1}{\muo}(\mu_1 k_1+\cdots+\mu_n k_n).
\end{equation*}
and we have proved Lemma~\ref{lem:HM2}. 

Next we apply the Proposition to our deformation $\Delta_s$:

\begin{corollary}
Consider the path $\Delta_s$ defined by Proposition~\ref{pro:deformation}, and fix some $s\in[0,1]$. 
The proof of Proposition~\ref{prop:framed} goes through if we replace $(\E,\A)$ by $\COKER~\Delta_s$ since it only relies on properties~$\eqref{eq:defcon1}$  and~$\eqref{eq:defcon2}$. 
Hence for each $s$ there exists a quasi-periodic connection $\A^q_s$ on $\E^q$ which is framed by $\Ainf,\Phiinf$ and which is $C_\chi^{0,1}$, such that $\COKER~\Delta_s$ is the interior restriction of $(\E^q,\A^q_s)$. 
\end{corollary}

Recall that $\Delta$ was constructed from a set of Nahm data with boundary data \newline
% hbox problems
$\bdarydata$. 
Since $\A^q_s$ is framed by $\Ainf,\Phiinf$ for each $s$, it must have boundary data $(K(s),{\vec{k}},\muo,{\vec{\mu}})$ where $K(s)\in\Z$ for each $s$, and so equation~$\eqref{eq:caloroncharge}$ gives
\begin{equation}\label{eq:deformcharge}
\int_{[0,\perflat]\times\rthree}\chtwo(\COKER~\Delta_s)=-K(s)
-\frac{1}{\muo}(\mu_1 k_1 +\cdots+\mu_n k_n).
\end{equation}
But the LHS of this equation is continuous in $s$, and since $\muo,{\vec{\mu}},{\vec{k}}$ are constant, $K$ is constant. 
From Section~\ref{sec:modelko} we know $K(s=1)=\ko$, so $K(s=0)=\ko$, and we have proved the following:

\begin{quotedtheorem}[Nahm data $\rightarrow$ caloron, $U(n)$ version]
Suppose $\bdarydata$ is a set of principal $\Un$ caloron boundary data. 
Given an element of $\spc{N}{\bdarydata}$ let $\Delta(x):W\rightarrow V$ be the corresponding Nahm operator. 
Then, up to gauge transformation, $\COKER~\Delta$ is the interior restriction of a $\Un$ framed quasi-periodic connection $\Aq$ on $\Eq$ with boundary data $\bdarydata$, \ie an element of $\spc{C}{\bdarydata}$. 
This construction takes elements of $\spcSD{N}{\bdarydata}$ to anti-self-dual connections \ie elements of $\spcSD{C}{\bdarydata}$. 
Moreover, using the rotation maps $\rotNahm$ and $\rotcal$, the construction extends to non-principal boundary data as explained in Section~\ref{sec:rotnahm}.
\end{quotedtheorem}

\subsection{Volume forms}\label{sec:volumeforms}

A periodic volume form on $\E$, parallel with respect to $\A$, corresponds to a volume form $\nu$\label{glo:nu} on $\Eq$, parallel with respect to $\Aq$, and clutching according to the rule
\begin{multline*}
\nu( w_1(\xo=s+\perflat),\ldots,w_n(\xo=s+\perflat) )= \\ \nu( c(s)w_1(\xo=s),\ldots,c(s)w_n(\xo=s) ).
\end{multline*}
where $\Aq$ and $c$ are defined by Proposition~\ref{prop:framed}.
Our aim is to prove that such a volume form exists when the boundary data $\bdarydata$ is $\SUn$ and the caloron configuration $\COKER~\Delta$ is anti-self-dual.

Let $F_{\Aq}$ denote the curvature of $\Aq$; a parallel volume form exists on $\Eq$ if $\trace F_{\Aq}=0$. 
If $\Aq$ is anti-self-dual, then $\trace F_{\Aq}$ is the curvature of an anti-self-dual finite action abelian field on $\Ieps\times\rthree$, and so vanishes. 
Hence $\Eq$ can indeed be equipped with a parallel volume form $\nu$. 
Since $\nu$ is parallel and $\Aq$ is compatible with the clutching map $c$, it follows that there exists a constant $\lambda$ with $| \lambda |=1$ such that
\begin{equation}\label{eq:volforma}
\nu( c(s)w_1(\xo=s),\ldots,c(s)w_n(\xo=s) ) =
\lambda \nu( w_1(\xo=s),\ldots,w_n(\xo=s) )
\end{equation}
for any sections $w_1,\ldots,w_n$ of $\Eq$.
We want to show that $\lambda=1$, so that $\nu$ corresponds to a periodic object on $\E$. 
Since $\lambda$ is independent of $x$ and all the objects are continuous up to the boundary, we can evaluate $\lambda$ by working on $\sphinf$. 

Suppose that the sections $w_1,\ldots,w_n$ are parallel in the $x_0$ direction \ie satisfy 
\begin{equation*}
\iota_{\partial_{x_{0}}}(\nabla_{\A_q}(w_j))=0
\end{equation*}
\label{glo:iota}for each $j$. 
Then 
\begin{equation}\label{eq:volformb}
\nu( w_1(\xo=s+\perflat),\ldots,w_n(\xo=s+\perflat) )=\nu( w_1(\xo=s),\ldots,w_n(\xo=s) ).
\end{equation}
It is easy to construct parallel solutions on $\sphinf$: in the gauge $e_1,\ldots,e_n$ on $\sphinf$, the $d\xo$ component of $\A^q$ is $\A_{x_0}^q=\diag(i\mu_1,\ldots,i\mu_n)$. 
Hence $w_j=(\exp -i\xo\mu_j)e_j$ defines a linearly independent set of sections that are parallel in the $x_0$ direction. 
Substituting these sections into equation~$\eqref{eq:volformb}$ gives
\begin{align}
\nu_{(\xo=0)}(e_1,\ldots,e_n) &= 
\nu_{(\xo=\perflat)}((\exp -2\pi i\mu_1 / \mu_0)e_1, \ldots, (\exp -2\pi i\mu_n / \mu_0)e_n) \notag \\
& = \big( \prod_1^n \exp -2\pi i\mu_p / \mu_0 \big)
\nu_{(\xo=\perflat)}(e_1,\ldots,e_n)\notag \\
& = \nu_{(\xo=\perflat)}(e_1,\ldots,e_n),\label{eq:volformc}
\end{align}
since $\sum_1^n \mu_p=0$.
However, the clutching map $c$ is the identity on the boundary, so equation~$\eqref{eq:volforma}$ becomes
\begin{equation}\label{eq:volformd}
\nu_{(\xo=\perflat)}(e_1,\ldots,e_n) = \lambda\nu_{(\xo=0)}(e_1,\ldots,e_n).
\end{equation}
Together, equations~$\eqref{eq:volformc}$ and~$\eqref{eq:volformd}$ imply that $\lambda=1$, and so we have shown that the volume form corresponds to a periodic object on $\E$. 
Hence we have established:

\begin{quotedtheorem}[Nahm data $\rightarrow$ caloron, $SU(n)$ version]
Suppose $\bdarydata$ is a set of $\SUn$ caloron boundary data. 
Given an element of $\spcSD{N}{\bdarydata}$ let $\Delta(x):W\rightarrow V$ be the corresponding Nahm operator. 
Then $\COKER~\Delta$ is the interior restriction of an $\SUn$ framed quasi-periodic connection $\Aq$ on $\Eq$ corresponding to some element of $\spcSD{C}{\bdarydata}$. 
\end{quotedtheorem}

\chapter{From Calorons to Nahm Data}

We present the Nahm transform from calorons to Nahm data: our aim is to prove that the Nahm transform is a well-defined map from $\spcSD{C}{\bdarydata}$ to $\spcSD{N}{\bdarydata}$. 
Throughout this Chapter we work with a fixed $SU(n)$ anti-self-dual caloron configuration, \ie a bundle and connection $(\E,\A)$ in $\spcSD{C}{\bdarydata}$, framed by the pair $\Ainf,\Phiinf$.

\section{Generalizing the $4$-torus Nahm transform}

Many of the more formal aspects of the Nahm transform carry over directly from the $4$-torus case described in~\ref{sec:nahmtorus} to the caloron case. 
We cover these in this Section, delaying the real difficulties---the calculation of the index of the Dirac operators involved, and recovering the behaviour of the Nahm data at singularities---till later. 

\subsection{Defining the transform}\label{sec:definecaltonahm}

Given the caloron $(\E,\A)$, recall the definitions~$\eqref{eq:defDAplus}$ and~$\eqref{eq:defDAminus}$ of the Dirac operators $D_\A^\pm$ on $\cylo$. 
Following the ideas of Section~\ref{sec:nahmtorus}, the transform from the caloron to its Nahm data involves the kernel of a family of Dirac operators parameterized by the dual torus $\Sdual$.
For each $\xi\in\R$, let $D_{\A,\xi}^\pm$ denote the Dirac operators coupled to $\E$ via the connection 
\begin{equation*}
\A-i\xi~dx_0,
\end{equation*}
so that 
\begin{equation*}
\DAxiplus = D_\A^+ - i\xi 
\end{equation*}
and 
\begin{equation*}
\DAximinus = D_\A^- + i\xi.
\end{equation*}
Note that, since $\gamma_j^\ast=-\gamma_j$ for $j=1,2,3$, the two Dirac operators can be written as 
\begin{equation}\label{eq:decompdirac}
D_{\A,\xi}^{\pm}=\pm(\nabla_0 -i\xi) +D_A
\end{equation}
where $D_A$ is defined by
\begin{equation}\label{eq:defDA}
D_A = \sum_{j=1}^3 \gamma_j \nabla_j
\end{equation}
and $\nabla_0,\nabla_1,\nabla_2,\nabla_3$ are the components of $\A$ in the frame $(\partial_0,\partial_1,\partial_2,\partial_3)$. 
We use the same symbols $\DAxiplus,\DAximinus$ \etc to denote the extension of these operators to Sobolev spaces of sections: 
\begin{equation*}
D_{\A,\xi}^{\pm} : W^1(\cylo,S^{\pm}\otimes\E)
\rightarrow
W^0(\cylo,S^{\mp}\otimes\E)
\end{equation*}
where $W^l$ is the space of sections with $l$ derivatives in $L^2$.

\begin{definition}\label{def:Xising}
The set of singular values, $\Xising$\label{glo:xising}, for $\DAxiplus$ is defined by
\begin{equation*}
\Xising = \{ \mu_j + N\muo : j=1,\ldots,n\textrm{\ and\ }N\in\Z \}.
\end{equation*}
\end{definition}

\begin{lemma}\label{lem:diracinj}
When $\A$ is anti-self-dual, $\DAxiplus$ is injective provided $\xi\notin\Xising$.
\end{lemma}

\proof
Applying the Weitzenb\"ock formula~\ref{lem:weitzenbock}, we obtain
\begin{equation}\label{eq:DAxiDAxireal}
\DAximinus\DAxiplus = -(\nabla_0 -i\xi)^2 -\sum_{j=1}^3 \nabla_j\nabla_j.
\end{equation}
Then
\begin{equation*}
\| \DAxiplus s \|^2_{L^2} = 
\| (\nabla_0 -i\xi)s \|^2_{L^2} + \sum_{j=1}^3 \| \nabla_j s \|^2_{L^2},
\end{equation*}
and so $\DAxiplus s=0$ implies that 
$\| \nabla_j s \|_{L^2}=0$ for $j=1,2,3$ and $\| (\nabla_0 -i\xi)s \|_{L^2}=0$. 
By definition, $(\E,\A)$ is framed on $\partial X$ and we can extend the framing to a neighbourhood $U$ of $\partial X$, to give an identification of $\E |_U$ with $p^\ast E$. 
In this identification we can write
\begin{equation}\label{eq:3plus1A}
\nabla_\A -i\xi dx_0= \nabla_A+dx_0(\partial_{x_0} + \Phi-i\xi)
\end{equation}
where $\Phi(x)\in\End(E)$ for each $x$, and $\Phi\rightarrow\Phiinf$ as $r\rightarrow\infty$. 
When $\xi\notin\Xising$ there exists a compact subset of $\rthree$ outside which $\Phi-i\xi$ has distinct eigenvalues, none of which equals an integer multiple of $i\muo$. 
Thus, on the complement of this compact set, $(\partial_{\xo} + \Phi -i\xi)$ has no non-trivial periodic solutions, and so any solution to $\DAxiplus s=0$ must have compact support. 
However, any solution $s$ must also satisfy $\nabla_{\partial_r}s=0$. 
This is a first order ODE, so if $s$ has compact support, it must be identically zero. 
It follows that, provided $\xi\notin\Xising$, $\DAxiplus$ is injective. 
\eproof

\begin{quotedproposition}
$\DAxiplus$ is Fredholm iff $\xi\notin\Xising$.
\end{quotedproposition}

We prove this result in Section~\ref{sec:fredcond}.

From these two results it follows that $\coker\DAxiplus=\ker\DAximinus$ is finite dimensional and has rank independent of $\xi$ on each of the component intervals of $\R\setminus\Xising$. 
As $\xi$ varies, the vector spaces $\coker\DAxiplus$ define a vector bundle on each interval in $\R\setminus\Xising$. 
Let $\IopN\subset\R\setminus\Xising$\label{glo:IopN} be the interval $(\mu_{p+1}+N\muo,\mu_{p}+N\muo)$ for $p=1,\ldots,n-1$ and $N\in\Z$, and let $I^\circ_{n,N}=(\mu_1+(N-1)\muo, \mu_n+N\muo)$. 
Let $X_{p,N}$\label{glo:XpN} denote the bundle $\coker\DAxiplus$ over the interval $\IopN$. 
The bundles inherit an hermitian inner product from $W^0(\Sminus\otimes\E)$. 
We define a connection and three endomorphisms on each bundle in the following way. 
Let \label{glo:NahmpN}
\begin{equation}\label{eq:defnnablaxi}
\nabla_{p,N}s=\Ph(\partial_\xi s)
\end{equation}
and
\begin{equation}\label{eq:defnTj}
T_{p,N}^j s=\Ph(ix_j s)\textrm{\ for\ }j=1,2,3
\end{equation}
where $s$ is a section of $X_{p,N}$ and $\Ph$ is the projection $\Ph:W^0(\Sminus\otimes\E)\rightarrow W^0(\Sminus\otimes\E)$ onto $\coker\DAxiplus$. 
Here $x_j$ denotes the $j$-th coordinate function of $\rthree$. 
Note that if $s$ is $L^2$ then $x_j s$ is not necessarily $L^2$ so these endomorphisms may not be well-defined. 
However, in Section~\ref{sec:zeromodedecay} we prove that sections in $\coker\DAxiplus$ are necessarily exponentially decaying, so this problem does not arise, and the endomorphisms $T^j_{p,N}$ are in fact well-defined. 
The connection and endomorphisms are skew-hermitian by definition. 
In Section~\ref{sec:Nahmsequn} we check that they satisfy Nahm's equation on the intervals $\IopN$, and we check that the transform gives periodic Nahm data in Section~\ref{sec:Nahmdataperiodic}. 

\subsection{Nahm's equation}\label{sec:Nahmsequn}

We want to show that the connection $\nabla_{p,N}$ and endomorphisms $T^j_{p,N}$ satisfy Nahm's equation on the interval $\IopN$ for each $p,N$. 
This is an adaptation of Proposition~\ref{prop:NahmgivesASD}, and relies on the fact that $\DAximinus\DAxiplus$ commutes with the Clifford matrices $\gamma_a$, $a=0,1,2,3$. 
For the present we assume that sections in $\coker\DAxiplus$ are exponentially decaying so that the endomorphisms $T^j_{p,N}$ are well-defined. 

For brevity we fix the notation $D=\DAxiplus$ and $D^\ast=\DAximinus$ in this Section. 
Note that if $s$ is a section of $X_{p,N}$ (\ie if $D^\ast s=0$) then 
\begin{equation*}
D^\ast (x_j s) = -\gamma_j^\ast s.
\end{equation*}
Since
\begin{equation*}
\Ph = 1 - D(D^\ast D)^{-1}D^\ast
\end{equation*}
it follows that
\begin{equation*}
(\Ph x_j \Ph)s:=\Ph (x_j\Ph(s))=[x_j+D(D^\ast D)^{-1}\gamma_j^\ast]\Ph s.
\end{equation*}
Taking the adjoint of this expression gives
\begin{equation*}
(\Ph x_j \Ph)s=\Ph [x_j+\gamma_j (D^\ast D)^{-1}D^\ast]s.
\end{equation*}
Using these formulae and definition~$\eqref{eq:defnTj}$, we can calculate the commutator $[T_{p,N}^j,T_{p,N}^k]$:
\begin{align*}
[T_{p,N}^j,T_{p,N}^k] &=
\Ph (\Ph x_k\Ph)(\Ph x_j\Ph)\Ph -\Ph (\Ph x_j\Ph)(\Ph x_k\Ph)\Ph \\
&=\Ph\big[ (x_k+\gamma_k(D^\ast D)^{-1}D^\ast)(x_j+D(D^\ast D)^{-1}\gamma_j^\ast) \\ & \qquad\qquad
-(x_j+\gamma_j(D^\ast D)^{-1}D^\ast)(x_k+D(D^\ast D)^{-1}\gamma_k^\ast)\big]\Ph
\\ &= \Ph\big[ \gamma_k(D^\ast D)^{-1}D^\ast x_j 
+x_k D(D^\ast D)^{-1}\gamma_j^\ast +\gamma_k(D^\ast D)^{-1}\gamma_j^\ast
\\ & \qquad\qquad -\gamma_j(D^\ast D)^{-1}D^\ast x_k 
-x_j D(D^\ast D)^{-1}\gamma_k^\ast -\gamma_j(D^\ast D)^{-1}\gamma_k^\ast\big]\Ph
\\&=
\Ph\big[ -\gamma_k(D^\ast D)^{-1}\gamma^\ast_j 
-\gamma_k (D^\ast D)^{-1}\gamma_j^\ast +\gamma_k(D^\ast D)^{-1}\gamma_j^\ast
\\ & \qquad\qquad +\gamma_j(D^\ast D)^{-1}\gamma^\ast_k 
+\gamma_j (D^\ast D)^{-1}\gamma_k^\ast -\gamma_j(D^\ast D)^{-1}\gamma_k^\ast\big]\Ph
\\&=
\Ph\big[ \gamma_j(D^\ast D)^{-1}\gamma_k^\ast- \gamma_k(D^\ast D)^{-1}\gamma_j^\ast\big]\Ph.
\end{align*}
From~$\eqref{eq:DAxiDAxireal}$, $D^\ast D$ commutes with $\gamma_j$ for $j=1,2,3$, and since $\gamma_j\gamma_k^\ast - \gamma_k\gamma_j^\ast = 2\sum\epsilon_{ijk}\gamma_i$ we have
\begin{equation}\label{eq:RHSNahm}
\frac{1}{2}\sum_{j,k}\epsilon_{ijk}
[T_{p,N}^j,T_{p,N}^k] = 2\Ph\big[ \gamma_i(D^\ast D)^{-1} \big]\Ph. 
\end{equation}

Next, consider the left-hand side of Nahm's equation:
\begin{equation}\label{eq:LHSNahm}
\nabla_{p,N}T^i_{p,N} = i\Ph (\partial_\xi \Ph)x_i\Ph
+i\Ph x_i(\partial_\xi \Ph)\Ph.
\end{equation}
Now
\begin{align*}
\Ph\partial_\xi \Ph & = -\Ph \partial_\xi \big[ D(D^\ast D)^{-1}D^\ast \big] 
\\ & = -\Ph (\partial_\xi D)(D^\ast D)^{-1}D^\ast
\end{align*}
since $\Ph D=0$. 
But $\partial_\xi D =-i$ so 
\begin{equation*}
\Ph \partial_\xi \Ph = i\Ph (D^\ast D)^{-1}D^\ast
\end{equation*}
and 
\begin{equation*}
(\partial_\xi \Ph)\Ph = -iD (D^\ast D)^{-1}\Ph
\end{equation*}
Substituting this back into~$\eqref{eq:LHSNahm}$ gives 
\begin{align*}
\nabla_{p,N}T^i_{p,N} &= 
\Ph x_i D(D^\ast D)^{-1}\Ph-\Ph (D^\ast D)^{-1}D^\ast x_i \Ph \\
& = \Ph (D^\ast D)^{-1}\gamma^\ast_i \Ph -\Ph\gamma_i(D^\ast D)^{-1}\Ph\\
& = -2\Ph \gamma_i (D^\ast D)^{-1}\Ph.
\end{align*}
Comparing this with equation~$\eqref{eq:RHSNahm}$, we have Nahm's equation:
\begin{equation*}
\nabla_{p,N}T^i_{p,N} + \frac{1}{2}\sum_{j,k}\epsilon_{ijk}
[T_{p,N}^j,T_{p,N}^k] =0.
\end{equation*}

\subsection{Periodicity}\label{sec:Nahmdataperiodic}

We have seen how to construct the Nahm data over the intervals $\IopN\subset\R$. In this Section, we explain how to identify the fibre $\coker\DAxiplus$ with $\coker D^+_{\A,\xi+\mu_0}$, so that the Nahm data is defined on $\Sdual = \R / \muo\Z$. 
This is entirely analogous to Section~\ref{sec:calper}, where we showed that $\COKER~\Delta$ is periodic for a given Nahm operator $\Delta$. 

Define the translation\label{glo:tauhat}
\begin{equation*}
\hat{\tau}:\xi\mapsto \xi+\muo
\end{equation*}
and let \label{glo:Utauhat}
\begin{equation*}
\hat{U}_{\hat{\tau}}=\exp i\mu_0 x_0.
\end{equation*}
(Compare this with equation~$\eqref{eq:translationmap}$.) 
Then $\hat{U}_{\hat{\tau}}$ is a unitary periodic bundle isomorphism of $\E$, and 
\begin{equation*}
D^+_{\A,\hat{\tau}\xi}=\hat{U}_{\hat{\tau}}\DAxiplus\hat{U}_{\hat{\tau}}^{-1}.
\end{equation*}
Note that just as in Section~\ref{sec:calper} there is some choice for the map $\hat{U}_{\hat{\tau}}$: it can be replaced with any map of the form $\exp i\muo(\xo + a)$. 
It follows that 
\begin{equation*}
\coker D^+_{\A,\hat{\tau}\xi}=\hat{U}_{\hat{\tau}}\coker\DAxiplus
\end{equation*}
for $\xi\in\R\setminus\Xising$, and 
\begin{equation*}
\hat{P}_{\hat{\tau}\xi}=\hat{U}_{\hat{\tau}}\hat{P}_{\xi}\hat{U}_{\hat{\tau}}^{-1}
\end{equation*}
where $\hat{P}_\xi:W^0(\Sminus\otimes\E)\rightarrow W^0(\Sminus\otimes\E)$ is the projection onto $\coker\DAxiplus$ for each $\xi$. 
Substituting this back into definitions~$\eqref{eq:defnnablaxi}$ and~$\eqref{eq:defnTj}$ gives
\begin{equation*}
\nabla(\xi+\muo)=\hat{U}_{\hat{\tau}}\nabla(\xi)\hat{U}_{\hat{\tau}}^{-1}
\end{equation*}
and 
\begin{equation*}
T^j(\xi+\muo)=\hat{U}_{\hat{\tau}}T^j(\xi)\hat{U}_{\hat{\tau}}^{-1}\quad j=1,2,3
\end{equation*}
where we have dropped the subscript $p,N$ on the Nahm data. 
Thus $\hat{U}_{\hat{\tau}}$ defines an action of $\Z$ on the collection of bundles $X_{p,N}$, and the connection and endomorphisms defined by~$\eqref{eq:defnnablaxi}$ and~$\eqref{eq:defnTj}$ are compatible with this action. 
Quotienting by the action, the data reduces to a collection of hermitian bundles $X_p$ defined on the intervals $\Iop\subset\Sdual$\label{glo:Iop}, where
$\Iop = (\mu_{p+1},\mu_p)+\muo\Z$ for $p=1,\ldots,n-1$, and $I^\circ_{n}=(\mu_1-\muo,\mu_n)+\muo\Z$. 
Since the connection and endomorphisms are compatible with this action, under the quotient they map to a connection $\nabla_p$ and endomorphisms $T_p^j$, $j=1,2,3$, on each bundle $X_p$. 

\subsection{Further remarks on the rotation map}

In analogy with Section~\ref{sec:rotnahm}, we want the following diagram to commute:
\begin{equation*}
\begin{CD}
\spc{C}{(B)} @>\textrm{Nahm}>\textrm{transform}> \spc{N}{(B)}  \\
@V{\rotcal}VV @VV{\rotNahm}V \\
\spc{C}{(\rotbdary B)}@>\textrm{Nahm}>\textrm{transform}> \spc{N}{(\rotbdary B)}
\end{CD}
\end{equation*}
where $\rotcal$ and $\rotbdary$ were defined in Section~\ref{sec:rotationbdary}, and $\rotNahm$ is defined by rotating the Nahm data by $\muo / n$ as in equation~$\eqref{eq:defrotnahm}$. 
Given our caloron $(\E,\A)$, traversing the diagram round the top right (\ie performing the Nahm transform followed by a rotation) is equivalent to performing the Nahm transform with the Dirac operator
\begin{equation*}
D^+_{\B,\xi}=D^+_{\A,(\xi - \mu_0 / n)}
\end{equation*}
rather than the Dirac operator $\DAxiplus$, where $\B=\A+i(\mu_0 / n )dx_0$. 
By `equivalent' we mean the two sets of Nahm data are isomorphic over $\Sdual$ with a fixed origin. 
In the quasi-periodic picture suppose that $\A^q,\B^q$ are framed quasi-periodic connections on $\E^q$ corresponding to $\A$ and $\B$. 
Then
\begin{equation*}
\B^q = g(\A^q)
\end{equation*}
where $g(s)=\exp -i(\mu_0 s/ n)$. 
If $\A^q$ has clutching function $c$, then so does $\B^q$ since
\begin{equation*}
\B^q (s+\perflat) = \omega c\big( \A^q(s) \big) = \omega c\big( \B^q(s) \big),
\end{equation*}
and $\omega = g(\perflat)=\exp -2\pi i / n$ acts trivially as a gauge transformation. 
(Recall that clutching functions must be the identity at spatial infinity.) 

Conversely, traversing around the bottom left of the diagram is just the Nahm transform on $\rotcal(\A)$. 
The quasi-periodic pull-back of $\rotcal(\A)$ is $\rho(\A^q)$ where $\rho$ is the bundle automorphism of $\E^q$ defined in Section~\ref{sec:rotationbdary}. 
We know that $\Theta := g\rho^{-1}$ is a bundle automorphism of $\E^q$ taking $\rho(\A^q)$ to $\B^q$. 
If $\Theta$ descends to give a strictly periodic isomorphism identifying $\rotcal(\A)$ with $\B$ then the diagram commutes. 
A section $\psi$ of $\E^q$ descends to a periodic section under the quotient by a clutching function $c$, if and only if it satisfies
\begin{equation*}
\psi(s+\perflat) = c\psi(s).
\end{equation*}
Now, $\rho(\A^q)$ has clutching function $c_\rho$ given by~$\eqref{eq:rotclutch}$, so consider a section of $\E^q$ satisfying
\begin{equation*}
\psi(s+\perflat) = c_\rho \psi(s).
\end{equation*}
Then under the action of $\Theta$ we have
\begin{align*}
\Theta\psi(s+\perflat) &= g(s+\perflat)\rho^{-1}(s+\perflat)c_\rho(s)\psi(s) \\
&= \omega g(s) \rho^{-1}(s+\perflat)\omega^{-1}\rho(s+\perflat)c(s)\rho^{-1}(s)\psi(s) \\
& = c\Theta\psi(s)
\end{align*}
where $c$ is the clutching function of $\B^q$, and so $\Theta\psi$ descends to a periodic section under the quotient by $c$. 
Thus $\Theta$ descends to a strictly periodic bundle isomorphism taking $\rotcal(\A)$ to $\B$ and we have shown that the diagram commutes. 

\section{The Fredholm condition}\label{sec:fredcond}

We need to prove the Fredholm condition stated in Section~\ref{sec:definecaltonahm}:

\begin{proposition}\label{prop:fredcond}
$\DAxiplus$ is Fredholm iff $\xi\notin\Xising$.
\end{proposition}

This ensures that the bundles $X_{p,N}$ are well-defined. 
We also need to calculate the rank of these bundles. 
Since $\DAxiplus$ is injective, the rank of $X_{p,N}$ is given by minus the index of $\DAxiplus$ for any $\xi\in\IopN$, and so the problem is equivalent to finding the index of $\DAxiplus$, which we calculate in Section~\ref{sec:indexthm}. 
The proof of the Fredholm condition and the index calculation have been published jointly with my supervisor in~\cite{nye00}---the material in this Section and Section~\ref{sec:indexthm} is taken more-or-less directly from that paper. 

We give two proofs of Proposition~\ref{prop:fredcond}: the first uses Anghel's criterion~\cite{ang93}, while the second uses the machinery of pseudo-differential operators (\PhiDO 's)\label{glo:PhiDO} on manifolds with fibred boundary~\cite{maz99}. 
%Together with the index theorem, the `if' part of the Proposition is sufficient to deduce that $\DAxiplus$ is not Fredholm when $\xi=\mu_p$ and $k_p\neq 0$ (since the index is locally constant); thus the `only if' part of the Proposition is only required to prove $\DAxiplus$ is not Fredholm at zero jumps.  
Using this machinery in Section~\ref{sec:zeromodedecay} we prove that solutions in $\coker\DAxiplus$ are exponentially decaying in $r$, so that~$\eqref{eq:defnTj}$ makes sense. 

\subsection{Proof of the Fredholm condition using Anghel's criterion}

Theorem $2.1$ of \cite{ang93} gives conditions for $D_{\A,\xi}:=\DAxiplus\oplus\DAximinus$ to be 
Fredholm:
$D_{\A,\xi}$ is Fredholm if and only if there is a compact set $K\subset \Xo$ and a
constant $C >0$ such that 
\begin{equation*}
\| D_{\A,\xi} \psi \|_{L^2} \geq C \|\psi\|_{L^2} ,\ \textrm{when}\ 
\psi\in W^{1} (S\otimes\E )\ \textrm{and}\ \supp (\psi )\subset 
\Xo\setminus K.
\end{equation*}
Note that $D_{\A,\xi}$ is Fredholm if and only if $\DAxiplus$ is Fredholm.
Now for $\psi\in W^2(S\otimes\E)$,
\begin{equation*}
\| D_{\A,\xi}\psi \|_{L^2} = \langle (\DAxiplus\DAximinus)\oplus(\DAximinus\DAxiplus)\psi,\psi \rangle_{L^2},
\end{equation*}
since $\DAximinus$ is the adjoint of $\DAxiplus$ and vice versa.
Using~\eqref{eq:decompdirac}, we have
\begin{equation*}
\DAximinus\DAxiplus = -(\nabla_0-i\xi)^2+ [D_A,\nabla_0] + D_A^2.
\end{equation*}
The third term here is clearly positive because $D_A$ is self-adjoint, and
the boundary conditions allow us to  estimate the other two as 
follows.

{\textit{The first term.}}
As in the proof of Lemma~\ref{lem:diracinj}, extend the framing $f$ to a neighbourhood of $\partial X$. 
In this identification we can define $A,\Phi$ using~$\eqref{eq:3plus1A}$.  
As the boundary $\partial X$ is approached the eigenvalues of $\Phi$ converge 
to the eigenvalues of $\Phiinf$.
Using spherical polar coordinates on $\rthree$, 
let $i\lambda_{j}(r,y_1,y_2,x_0)$ be the eigenvalues of $\Phi$, and 
$i\mu_{j}$ be the eigenvalues of $\Phiinf$ ($j=1,\ldots,n$) such that 
$\lambda_{j}\rightarrow\mu_{j}$ as $r\rightarrow\infty$.
Let $\lambda(r,y_1,y_2,x_0)$ be the smallest element in 
$\{ |\lambda_{j}+N\muo -\xi | : j=1,\ldots,n;~N\in\Z \}$ and $\mu$ be the 
smallest element of the set 
$\{ |\mu_{j}+N\muo -\xi| : j=1,\ldots,n;~N\in\Z \}$.
The condition $\xi\notin\Xising$ implies that $\mu > 0$,
so there exists a compact set $K_{1}\subset \Xo$ such that $\lambda > 
\mu / 2$ on $\Xo\setminus K_{1}$.

Suppose $\psi\in W^{2}(S^{+}\otimes\E)$ and $\supp~\psi\subset 
\Xo\setminus K_{1}$.
Using the isomorphism $\Splus\cong p^{\ast}\spinthree$,
$\psi$ can be written as a Fourier series
\begin{equation*}
\psi = \sum_{N}\exp(iN\muo \xo) \phi_N
\end{equation*}
where $\phi_{N}$ is a section of 
$\spinthree\otimes E$. Let
\begin{equation*}
\psi^{(N)} = \exp(iN\muo \xo) \phi_N.
\end{equation*}
Then
\begin{equation*}
  (\nabla_{0}-i\xi)\psi^{(N)} = (iN\muo  +\Phi-i\xi)\psi^{(N)}
\end{equation*}
so
\begin{equation*}
\langle -(\nabla_0-i\xi)^2\psi^{(N)},\psi^{(N)}\rangle
\geq\frac{1}{4}{\mu}^{2}{\|\psi^{(N)}\|}^{2}, \qquad
\textrm{on\ }\Xo\setminus K_{1}
\end{equation*}
as a pointwise estimate.
(Since $\psi^{(N)}\in W^{2}(S^{+}\otimes \E)$, $\psi^{(N)}$ is 
actually continuous so both sides of the inequality exist.)
Since the inequality is independent of $N$ it holds for general $\psi$ 
and we obtain
\begin{equation} \label{fredhbd_1}
\supp (\psi)\subset \Xo\setminus K_{1} \Rightarrow
{\langle -(\nabla_0-i\xi)^2\psi,\psi\rangle}_{L^2}
\geq\frac{1}{4}{\mu}^{2}{\|\psi\|}^{2}_{L^2}.
\end{equation}

{\textit{The second term.}}
We have
\begin{equation*}
[D_A,\nabla_0] = \sum_j \gamma_j[\nabla_j,\nabla_0]
= \sum_j \gamma_j \{
\iota(\partial_j)(\nabla_A 
\Phi -\partial_{\xo} A)\}
\end{equation*}
where $\iota(\theta)$ denotes the interior product with a tangent vector $\theta$, $D_A$ is defined by~$\eqref{eq:defDA}$, and $A$ is defined by~$\eqref{eq:3plus1A}$. 
But, using~$\eqref{eq:decayforfred}$, $\|\nabla_{A}\Phi-\partial_{\xo} A\| \rightarrow 0$ as $r\rightarrow\infty$, 
so there exists a compact set $K_{2}\subset \Xo$ such that
\begin{equation} \label{fredhbd_2}
\supp (\psi)\subset \Xo\setminus K_{2} \Rightarrow
| {\langle [D_A,\nabla_0]\psi ,\psi \rangle}_{L^2} | \leq
\frac{1}{8}\mu^{2}{\|\psi\|}^2_{L^2}.
\end{equation}

Now let $K$ be a compact set containing $K_{1}$ and $K_{2}$. 
Combining~$\eqref{fredhbd_1}$ and~$\eqref{fredhbd_2}$ we obtain 
\begin{equation*}
  \supp (\psi)\subset \Xo\setminus K\Rightarrow
  \langle\DAximinus\DAxiplus\psi,\psi\rangle_{L^2}\geq 
\frac{1}{8}\mu^{2}{\|\psi\|}^{2}_{L^2}.
\end{equation*}
A similar bound is obtained for 
$\DAxiplus\DAximinus$, and so we obtain the 
following bound for $D_{\A,\xi}$:
\begin{equation*}
\psi\in W^{2}(S\otimes \E),~\supp (\psi)\subset \Xo\setminus 
K\Rightarrow
\|D_{\A,\xi}\psi\|_{L^2}\geq\frac{1}{\sqrt{8}}\mu\|\psi\|_{L^2}.
\end{equation*}
By density, the inequality in fact holds for $\psi\in W^{1}(S\otimes
\E)$. This completes the verification of Anghel's criterion and 
gives a proof of the `if' part of the Proposition.  
When $\xi\in\Xising$ it is possible to use Anghel's criterion and some analysis similar to that above to prove that $\DAxiplus$ is not Fredholm, but we choose to omit this. 
In fact the converse statement follows much more easily if we use \PhiDO 's, as we will see below.
\eproof

\subsection{Pseudo-differential operators on manifolds \\ with fibred boundaries}\label{sec:PhiDOs}

Mazzeo and Melrose \cite{maz99} study pseudo-differential operators (\PhiDO 's) on manifolds with fibred boundaries. 
These operators are particularly suitable for problems like our Fredholm condition and index theorem: \cite[Proposition 9]{maz99} contains a necessary and sufficient condition for a \PhiDO~on a manifold with boundary to be Fredholm, which we apply to our operator $\DAxiplus$. 
I am indebted to my supervisor for explaining \PhiDO 's on manifolds with boundary to me; the proof of the Fredholm condition using \PhiDO 's is due to him. 

The general situation considered in \cite{maz99} is a fibration of the boundary $\partial X$ of $X$: 
\begin{equation*}
U\longrightarrow \partial X \xrightarrow{~p~} Y
\end{equation*}
where $U$ is the fibre, and $p$ is projection onto the base $Y$. 
In our example, 
\begin{equation*}
\partial X = \So\times\sphinf,\quad U=\So,\quad \textrm{and\ } Y=\sphinf, 
\end{equation*}
so the fibration is trivial. 
Mazzeo and Melrose assume $X$ has a boundary defining function $\chi$ \ie a function $\chi\in C^\infty(X)$ such that $\chi\geq 0$, $\partial X=\{ \chi=0 \}$, and $d\chi\neq 0$ on $\partial X$. 
They consider differential operators of the form
\begin{equation}\label{eq:phidiff}
P(\chi,y,u;\chi^2\partial_\chi,\chi\partial_y,\partial_u),
\end{equation}
near $\partial X$, where $P$ is
smooth in the first three variables and polynomial in the last three
variables. 
Here $y$ and $u$ are coordinates on $Y$ and $U$ respectively. 
These operators form the algebra of $\Phi$-differential operators. 
(Note that this $\Phi$ has nothing to do with the $d\xo$ component of $\A$, but stands for `fibred cusp' in \cite{maz99}.)
In \cite[Proposition 9]{maz99} it is shown that such an operator is 
Fredholm in $L^2$ if and only if it is {\em fully elliptic} in the following sense. 
First, \eqref{eq:phidiff} must be elliptic in the usual sense over $\Xo$. 
This will always be the case for Dirac operators. 
Secondly, the associated {\em indicial family} must be 
invertible on every fibre $p^{-1}(y)\subset \partial X$. Given such a fibre, the
indicial family on $p^{-1}(y)$ is defined by picking a real number 
$\zeta$ and a real cotangent vector 
$\eta\in T_y^*Y$, and defining\label{glo:indicial}
\begin{equation*}
I_\Phi(P)_{(y,\eta,\zeta)} = P(0,y,u;i\zeta,i\eta,\partial_u)
\end{equation*}
as a differential operator on $p^{-1}(y)$. To say that the indicial 
family is invertible is to say that
$I_\Phi(P)_{(y,\eta,\zeta)}$ is invertible (in any reasonable 
space of sections over $p^{-1}(y)$), for each choice of
$(y,\eta,\zeta)$ as above.

For our example, we work with the boundary-adapted coordinates $\chi,y_1,y_2,\xo$ and denote the components of $\nabla_\A$ in these coordinates by
\begin{equation*}
\nabla_\chi =\partial_\chi + A_\chi,\,\nabla_{y_j}= \partial_{y_j} + A_{y_j},\, \nabla_{\xo}=\partial_{\xo} + A_{\xo}.
\end{equation*}
Relative to a suitable choice of basis for the spin-bundles, we have
then
\begin{equation}\label{eq:diracnearbdary}
\DAxiplus = \nabla_{\xo} + \gamma_1 \chi\nabla_{y_1} + \gamma_2 \chi\nabla_{y_2} +\gamma_3 \chi^2 \nabla_\chi -i\xi.
\end{equation}
Strictly speaking, we are making a choice of normal coordinates here; 
otherwise there will be additional zero-order terms coming from 
connection coefficients. 
Hence $\DAxiplus$ is a $\Phi$-differential operator as defined by \cite{maz99}. 
Following the recipe for the indicial family for $\DAxiplus$, we obtain
\begin{equation*}
I_\Phi(P)_{(y,\eta,\zeta)} =
(\nabla_{\xo}-i\xi) + i(\eta_1\gamma_1 + \eta_2\gamma_2 + \zeta \gamma_3)
\end{equation*}
where $\eta_1,\eta_2$ are real numbers. 
This operator in $C^\infty(\So, p^*S_{(3)}\otimes E_\infty)$ is a sum of 
two terms $B+A$, where $A=i(\eta_1\gamma_1 + \eta_2\gamma_2 + \zeta \gamma_3)$
is self-adjoint, $B= \nabla_{\xo}-i\xi$ is skew-adjoint and 
$[A,B] =0$. It follows by considering $(A+B)^*(A+B)$
that $(A+B)u=0$ if and only 
if $Au=0$ and $Bu=0$. Now $B$ has a non-trivial null-space  only if $\xi\in\Xising$. 
Hence under the assumption of the Proposition, $A+B$ is 
injective. 
Similarly the adjoint $(A+B)^*=A-B$ is injective, so that 
$\xi\notin\Xising$ implies that the 
indicial family is invertible, and so $\DAxiplus$ is Fredholm in
$L^2$. 
Conversely, if $\xi\in\Xising$, then $B$ is not
invertible, and nor is $B+A$ when $\eta_j=0=\zeta$. 
So in this case
$\DAxiplus$ is not fully elliptic and hence cannot be Fredholm in
$L^2$. 
This completes the proof of Proposition \ref{prop:fredcond} using material from \cite{maz99}.

The result that a \PhiDO\ $P$ is Fredholm if and only if fully elliptic holds in a very strong sense. 
If $P$ has degree $m$ then $P$ makes sense as an operator between Sobolev spaces of degree $l$ and $l-m$ for all $l\in\Z$. 
Mazzeo and Melrose prove that if $P$ is fully elliptic it is Fredholm between any such spaces, and the index is independent of this choice. 
In fact, they show that if $P$ is fully elliptic,
if $P \psi =0$,  and if for some real  $m$,
$\chi^m \psi \in L^2(X)$,  then $\psi \in C^\infty(X)$ and $\psi$ 
vanishes to all orders in $\chi$ at $\partial X$. 
There is a similar statement for the cokernel.
In particular, it follows that $\DAxiplus$ is Fredholm as an operator
\begin{equation*}
\DAxiplus: W^1(\cylo,\Splus\otimes\E)\rightarrow W^0(\cylo,\Sminus\otimes\E)
\end{equation*}
(\ie restricted to $\Xo$) if and only if it is Fredholm as a \PhiDO\ on $X$. 
These strong decay conditions also imply the following:

\begin{lemma}\label{lem:defdirac}
Let $\A,\B$ be two caloron configurations on $(\E,f)$,  
framed by $(A_\infty,\Phi_\infty)$. Then $D_{\A}^+$ is Fredholm 
if and only if $D_{\B}^+$ is Fredholm. 
When the two operators are Fredholm their $L^2$-indices coincide.
\end{lemma}

\proof
That $D_{\A}^+$ is Fredholm if and only if $D_{\B}^+$ is Fredholm follows directly from Proposition~\ref{prop:fredcond}. 
It remains to prove that the operators have the same $L^2$-index. 
Consider the linear deformation of $\A$ into $\B$. 
This clearly gives rise to a norm-continuous path of Dirac operators between Sobolev spaces
\begin{equation}\label{eq:spcA}
W^1(X,\E\otimes\Splus, d\mu)\rightarrow W^0(X,\E\otimes\Sminus, d\mu)
\end{equation}
where $d\mu$ is the volume form $d\mu = h_1 h_2 dx_0 dy_1 dy_2$ in terms of the usual boundary-adapted coordinates. 
However, $X$ is equipped with the volume form $dx_0 dx_1 dx_2 dx_3 = \chi^{-4}d\mu$ near $\chi=0$, so it does not necessarily follow that the deformation is norm-continuous between
\begin{equation}\label{eq:spcB}
W^1(X,\E\otimes\Splus, \chi^{-4}d\mu)\rightarrow W^0(X,\E\otimes\Sminus, \chi^{-4}d\mu).
\end{equation}
However, the decay properties stated above imply that $D_{\A}^+$ is Fredholm as an operator between spaces \eqref{eq:spcA} if and only if it is Fredholm as an operator between spaces \eqref{eq:spcB}, and the index is the same. 
Thus the deformation from $\A$ to $\B$ preserves the $L^2$-index.
\eproof

We can prove the Lemma without using \cite[Proposition 9]{maz99} in fact. 
Let $\A(s)=(1-s)\A +s\B$ be the linear path joining $\A$ and $\B$. 
Then
\begin{multline*}
\| D^+_{\A(s_1)}- D^+_{\A(s_2)}\| \leq
| s_1 -s_2 |\Big[ 
\| \A_{x_0}-\B_{x_0} \|_{L^2} +
\|\chi(\A_{y_1}-\B_{y_1})  \|_{L^2} + \\
\| \chi(\A_{y_2}-\B_{y_2}) \|_{L^2} +
\| \chi^2(\A_{\chi}-\B_{\chi}) \|_{L^2} 
\Big]
\end{multline*}
using an expansion like~$\eqref{eq:diracnearbdary}$ and working in some fixed gauge. 
It follows that the path of operators $D^+_{\A(s)}$ is continuous provided $\| \A_{x_0}-\B_{x_0} \|_{L^2}$, $\|\chi(\A_{y_j}-\B_{y_j})  \|_{L^2}$, and $\| \chi^2(\A_{\chi}-\B_{\chi}) \|_{L^2}$ are bounded. 
Since the volume form near the boundary is $\chi^{-4}d\mu$, this follows provided we have pointwise estimates $| \A_{x_0}-\B_{x_0} |$, $|\chi(\A_{y_j}-\B_{y_j})  |$, $| \chi^2(\A_{\chi}-\B_{\chi}) |=O(\chi^2)$ as $\chi\rightarrow 0$. 
However, this is true because $\A$ and $\B$ are framed in the same way (using the smoothness assumptions of Section~\ref{sec:smooth}). 

\subsection{Decay properties of zero modes}\label{sec:zeromodedecay}

We want to show that the solutions to $\DAximinus\psi=0$ are exponentially decaying as $r\rightarrow\infty$ so that the Nahm matrices $T^j_p$ $j=1,2,3$ are well-defined by~$\eqref{eq:defnTj}$. 
Define $w_\lambda$ by
\begin{equation}\label{eq:weightingfn}
w_\lambda(r)=\begin{cases} \exp -\lambda r~\textrm{when}~r\geq 1, \\
                           \textrm{some smooth non-zero continuation on~}r<1. \end{cases}
\end{equation}

\begin{lemma}\label{lem:gooddecay}
Fix some $\xi\notin\Xising$, and define 
\begin{equation*}
M_\xi:=\textrm{min}\{ | \mu_j+N\mu_0-\xi | :j=1,\ldots,n~\textrm{and}~N\in\Z \}=\textrm{distance}~(\xi,\Xising).
\end{equation*}
If $\psi\in \Lker\DAximinus$ then $w_\lambda^{-1}\psi$ is $L^2$ for all $\lambda\in\R$ such that $0\leq\lambda<M_\xi$. 
\end{lemma}

This Lemma is certainly sufficient to deduce that the Nahm matrices are well-defined. 
To prove the Lemma we introduce the `weighted' operators\label{glo:weighted}
\begin{align*}
L^+_{\xi,\lambda} &= w_\lambda \DAxiplus w_\lambda^{-1},\quad\textrm{and}\\
L^-_{\xi,\lambda} &= w_\lambda^{-1} \DAximinus w_\lambda.
\end{align*}
It follows that 
\begin{equation}\label{eq:weightedker}
\psi\in\Lker L^-_{\xi,\lambda}\quad\Rightarrow\quad w_\lambda\psi\in\Lker\DAximinus
\end{equation}
provided $\lambda\geq 0$. 
Next we prove that the weighted operators are Fredholm provided $|\lambda |<M_\xi$:

\begin{lemma}
$L^+_{\xi,\lambda}$ is Fredholm iff $|\lambda |<M_\xi$.
\end{lemma}

\proof
Using the ideas in Section~\ref{sec:PhiDOs} we prove that the indicial family is invertible iff $|\lambda |<M$, which is sufficient for the claim. 
Take $P$ to be the $\Phi$-differential operator $L^+_{\xi,\lambda}$. 
Constructing the indicial family $I_\Phi(P)$ as described in Section~\ref{sec:PhiDOs} gives
\begin{equation*}
I_\Phi(P)=(\nabla_{\xo}-i\xi) + i\eta_1\gamma_1 + i\eta_2\gamma_2 + i\zeta \gamma_3+\lambda\gamma_3
\end{equation*}
as an operator on $C^\infty(\So, p^*S_{(3)}\otimes E_\infty)$, in some suitable choice of basis for the spin-bundles. 
We can perform a Fourier decomposition in $x_0$: $I_\Phi(P)$ maps each Fourier mode to itself, and on the $N$'th mode is given by
\begin{equation*}
I_\Phi(P)_N=\Phiinf+iN\mu_0-i\xi + i\eta_1\gamma_1 + i\eta_2\gamma_2 + i\zeta \gamma_3+\lambda\gamma_3 .
\end{equation*}
Moreover, $I_\Phi(P)$ is invertible if and only if $I_\Phi(P)_N$ is invertible for all $N\in\Z$. 
Working on the eigenspace of $\Phiinf$ with eigenvalue $i\mu_j$ for some $j\in\{ 1,\ldots,n \}$, $I_\Phi(P)_N$ is given by
\begin{multline*}
i\mu_j+iN\mu_0-i\xi + i\eta_1\gamma_1 + i\eta_2\gamma_2 + i\zeta \gamma_3+\lambda\gamma_3 = \\
\begin{pmatrix}
i\mu_j+iN\mu_0-i\xi-\eta_1 & -i\eta_2-\zeta+i\lambda \\ 
i\eta_2-\zeta+i\lambda & i\mu_j+iN\mu_0-i\xi+\eta_1
\end{pmatrix}.
\end{multline*}
This matrix has determinant
\begin{multline}\label{eq:detindicialfam}
[i(\mu_j+N\mu_0-\xi)-\eta_1][i(\mu_j+N\mu_0-\xi)+\eta_1]-[(i\lambda-\zeta)-i\eta_2][(i\lambda-\zeta)+i\eta_2]=\\
-(\mu_j+N\mu_0-\xi)^2-\eta_1^2-\eta_2^2 +\lambda^2+2i\lambda\zeta-\zeta^2.
\end{multline}
Hence $L^+_{\xi,\lambda}$ is Fredholm if and only if~$\eqref{eq:detindicialfam}$ is non-zero for all $j=1,\ldots,n$, $N\in\Z$, and for all $\eta_1,\eta_2,\zeta$. 
However, the real part of the terms~$\eqref{eq:detindicialfam}$ is strictly negative whenever
\begin{equation*}
\lambda^2<(\mu_j+N\mu_0-\xi)^2
\end{equation*}
for all $j$ and $N$,~\ie whenever $|\lambda |<M_\xi$. 
Conversely, when $|\lambda |\geq M_\xi$ there exist $\eta_1,\eta_2,\zeta$ that make the determinant~$\eqref{eq:detindicialfam}$ vanish for some values of $j$ and $N$. 
Hence $I_\Phi(P)$ is invertible if and only if $|\lambda |<M_\xi$, completing the proof of the Lemma. 
\eproof

Fixing some $\lambda$ with $0\leq\lambda<M_\xi$ and $\xi\notin\Xising$, $L^+_{\xi,\lambda}$ is Fredholm and has the same index as $\DAxiplus$, since $\{L^+_{\xi,s\lambda} : 0\leq s\leq 1\}$ is a continuous path of Fredholm operators. 
It follows that $\dim\Lker L^-_{\xi,\lambda}\geq\dim\Lker\DAximinus$. 
Combining this with~$\eqref{eq:weightedker}$ completes the proof of Lemma~\ref{lem:gooddecay}.

\section{The index theorem}\label{sec:indexthm}
 
We want to prove the following:

\begin{theorem}\label{thm:indexgen}
Suppose $(\E,\A)$ is a $U(n)$ caloron configuration framed by $\Ainf,\Phiinf$.
If $D_{\A}^+$ is Fredholm then the $L^2$-index is given by
\begin{equation}\label{eq:indexgeneral}
\ind D_{\A}^+ = -c_2(\E,f)[X]-\sum_N c_1(E^+_{(N)})[\sphinf]
\end{equation}
where for each $N\in\Z$, $E^+_{(N)}$ is the sub-bundle of $\Einf$ on which $N\muo-i\Phiinf$ is positive definite. 
\end{theorem}

The Theorem implies the following:

\begin{corollary}\label{cor:index}
Suppose $(\E,\A)$ is a framed $U(n)$ caloron configuration with boundary data $\bdarydata$.  
When $\xi\in\IopN$ the index of $\DAxiplus$ is given by
\begin{equation}\label{eq:indexspecific}
\ind\DAxiplus = -m_p
\end{equation}
for $p=1,\ldots,n$ and $N\in\Z$, where $m_p$ is defined by~$\eqref{eq:constituentcharges}$. 
Thus the bundles $X_{p}$ with fibre $\coker\DAxiplus$ have the correct dimension to correspond to Nahm data in $\spc{N}{\bdarydata}$. 
\end{corollary}

To see that Theorem~\ref{thm:indexgen} implies the Corollary, replace $\A$ in the Theorem with $\A-i\xi dx_0$ and $\Phiinf$ with $\Phiinf-i\xi$. 
Then, when $\xi\in\IopN$ it follows that $E^+_{(N)}$ is the direct sum of the eigenbundles with eigenvalues $i\mu_1,\ldots,i\mu_p$, while $E^+_{(k)}$ is trivial for all $k\neq N$. 
Hence $c_1(E^+_{(N)})[\sphinf]=k_1+\cdots+k_p$, and~$\eqref{eq:indexgeneral}$ becomes
\begin{align*}
\ind\DAxiplus & = -(k_0+k_1+\cdots+k_p) \\ & = -m_p
\end{align*}
which is~$\eqref{eq:indexspecific}$. 

The proof of the Theorem involves two main steps. 
The first is a calculation of the index in the case that there is a trivialisation of $\E$ in which $\A$ is independent of $\xo$. 
By Fourier analysis in $\xo$, the index problem reduces to a problem on $\rthree$ which can be dealt with using Callias' index theorem, which we introduced on page~\pageref{pag:callias}. 
By a deformation argument, this calculation gives the index for any caloron configuration when $c_2(\E,f)=0$, completing the first step.  
The precise statement of Callias' theorem we will use is:

\begin{quotedtheorem}
Let $(A,\Phi)$ be a $U(n)$ monopole configuration on $\threeball$ framed by $(\Ainf,\Phiinf)$ in the sense of Definition~\ref{def:monopole}, and let
\begin{equation}\label{eq:caltype}
D_{A,\Phi}=D_A+1\otimes\Phi:W^1 (\rthree,\spinthree\otimes E)\rightarrow W^0 (\rthree,\spinthree\otimes E)
\end{equation}
where $\spinthree$ is the spin bundle on $\rthree$. 
Then $D_{A,\Phi}$ is Fredholm iff $\Phiinf$ is invertible, and the index is given by
\begin{equation}\label{eq:calind}
\ind D_{A,\Phi} = -c_1(E_{\infty}^{+})[\sphinf]
\end{equation}
where $E^+_\infty$ is the sub-bundle of $\Einf$ on which $-i\Phiinf$ is positive-definite. 
\end{quotedtheorem}

This follows immediately from R\aa de's version of Callias' theorem \cite{rad94}. 

The second step in the proof of Theorem~\ref{thm:indexgen} invokes an excision theorem for
operators of Dirac type due to Anghel \cite{ang93} and Gromov--Lawson 
\cite{gro83}. 
In our case, this result gives
$\ind(D^+_{\A}) -
\ind(D^+_{\B})= -c_2(\E,f)[X]$
if $\B$ is any caloron configuration agreeing with $\A$ near $\partial X$ 
but living on a new framed bundle $(\F,f)$, with $c_2(\F,f)[X]=0$. 
Since we calculated $\ind(D^+_{\B})$ in the first step, that completes the 
proof of Theorem \ref{thm:indexgen}. 

\subsection{Proof when $c_2(\E,f)[X]=0$}\label{sec:indexkozero}

In this case, by Lemmas~\ref{lem:caltomon} and \ref{lem:defdirac}
it is enough to 
compute the index when $\E = p^\ast E$ and $\A = p^*A + p^*\Phi dx_0$ is 
the pull-back of a monopole. 
Then the coefficients of $D_{\A}^+$ are independent of $\xo$ and we can use
Fourier analysis in $\xo$ to reduce the calculation of the index to that of a 
collection of operators of the form \eqref{eq:caltype}.

Let 
\begin{equation}\label{eq:fourierA}
  Z_{N}=\{\psi=\exp (iN\muo \xo)\phi:\phi\in W^0(\R^3,S_{(3)}\otimes E) \}
\end{equation}
so that 
\begin{equation*}
  W^{0}(S^{+}\otimes \E) = \{ 
  \sum_{N\in\Z}\psi^{(N)} : \psi^{(N)}\in Z_{N}\ \textrm{and\ } 
  \sum_{N\in\Z} {\| \psi^{(N)} \|}^{2} < \infty \}
\end{equation*}
using identification~$\eqref{eq:idspinthreefour}$. 
Since by assumption the coefficients are independent of 
$\xo$, $D_{\A}^+$ maps $Z_N\cap W^1$ into $Z_N$ and, using~$\eqref{eq:decompdirac}$, its restriction to 
this subspace is equal to
\begin{gather*}
D_N : W^{1}(\spinthree\otimes E)\longrightarrow 
W^{0}(\spinthree\otimes E)
\\ D_N =  D_{A}+iN\muo  + 1\otimes\Phi
\end{gather*}
where $D_A$ is the Dirac operator on $\rthree$ coupled to $E$ via $A$. 
Using the statement of Callias' index theorem on the previous page, and using Proposition~\ref{prop:fredcond}, 
$D_N$ is Fredholm for every $N\in\Z$ iff $D_{\A}^+$ is Fredholm.  
Equation~$\eqref{eq:calind}$ shows that:
\begin{equation*}
\ind D_N = -c_{1}(E^{+}_{(N)})[\sphinf]
\end{equation*}
where $E^{+}_{(N)}$ is the subbundle of $\Einf$ on which 
$(N\muo-i\Phiinf)$ has positive eigenvalues. 
Since $Z_j\cap Z_k = 0$ if 
$j\not=k$, the index of $D_{\A}^+$ is the sum of the indices of the 
$D_N$, \ie
\begin{equation*}
\ind D_{\A}^+=\sum_{N}\ind D_N=-\sum_N c_{1}(E^{+}_{(N)})[\sphinf].
\end{equation*}
This sum is finite because $E^+_{(N)}$ is trivial when $\| N \|$ is sufficiently large. 
That completes the proof of Theorem~\ref{thm:indexgen} when $c_2(\E,f)[X]=0$.

\subsection{Proof when $c_2(\E,f)[X]\neq 0$}

Anghel \cite{ang93}, generalizing work of Gromov and Lawson \cite{gro83}, 
has given an excision theorem which compares the $L^2$-indices of a 
pair of Dirac operators over a complete manifold that agree near 
infinity. In our case this result yields the following statement. 
Let $\E$ and $\F$ be a pair of bundles over $\Xo$ and let $\A$ and 
$\B$ be unitary connections on $\E$ and $\F$ respectively. 
Suppose that there is a bundle isometry $\theta:\E|_{\Xo\backslash K} \to
\F|_{\Xo\backslash K}$ which carries $\A$ to $\B$ outside some compact set $K\subset\Xo$. 
Then 
\begin{equation}\label{eq:Anghelsformula}
\ind D_{\A}^+ -\ind D_{\B}^+=\int_{\Xo} \chtwo(\E)-\int_{\Xo} \chtwo(\F).
\end{equation}

We will deduce Theorem~\ref{thm:indexgen} by taking for $\B$ a 
connection which agrees with $\A$ near $\infty$, but which 
lives on a framed bundle $(\F,f)$ with $c_2(\F,f)=0$. This will 
complete the proof in view of the results of Section~\ref{sec:indexkozero}.

Let $(\E,f)$ be a framed bundle and $\A$ a framed 
$U(n)$ caloron configuration on $\E$. 
Let $\Aq$ be a quasi-periodic pull-back of $\A$ (in the sense of Section~\ref{sec:quasi-per}) with clutching function $c$. 
Let $U$ be a neighbourhood of $\sphinf$ so that $U=\threeball\setminus K$ where $K$ is a compact set $K\subset\rthree$. 
Let $c_{\textrm{ext}}$ be a bundle isomorphism of $\E^q$ that satisfies 
\begin{equation*}
c_{\textrm{ext}}=
\begin{cases}
c & \textrm{on\ }(\per-\epsilon,\per+\epsilon)\times U, \\
1 & \textrm{on\ }\Ieps\times\sphinf,\\
1 & \textrm{on\ }(-\epsilon,\epsilon)\times\threeball
\end{cases}
\end{equation*}
and which is arbitrary elsewhere (compare with the proof of Lemma~\ref{lem:caltomon}). 
By gauge transforming by $c_{\textrm{ext}}$ we can assume that $c\equiv 1$ on $U$. 
Define $\B^q$ on $\Eq$ to agree with $\Aq$ on $U$, but extended over $\Ieps\times K$ to define a smooth quasi-periodic connection on $\Eq$ with trivial clutching function, $c_{\B}$. 
Note that $\Aq$ and $\B^q$ are framed in the same way. 
Let $(\B,\F)$ be the quotient of $\B^q,\E^q$ by $c_\B$, so that $\B$ is a framed caloron configuration with $c_2(\F,f)=0$.  

Applying \eqref{eq:Anghelsformula},
\begin{equation*}
\ind D_{\A}^{+} - \ind D_{\B}^{+} = \int_{\Xo}\chtwo (\E) - \int_{\Xo}\chtwo (\F).
\end{equation*}
But
\begin{equation*}
\int_{\Xo}\chtwo (\E)  =  -c_2(\E,f)[X] - \frac{1}{\muo}\sum_{1}^{N}\mu_j
c_{1}(E_{\mu_j})[\sphinf]
\end{equation*}
from~$\eqref{eq:upcaloroncharge}$, and
\begin{equation*}
\int_{\Xo}\chtwo (\F)  =  -\frac{1}{\muo}\sum_{1}^{n}\mu_j
c_{1}(E_{\mu_j})[\sphinf].
\end{equation*}
So
\begin{equation*}
\ind D_{\A}^{+} = \ind D_{\B}^{+} - c_2(\E,f)[X]
\end{equation*}
From Section~\ref{sec:indexkozero} we know that $\ind D_{\B}^{+} =-\sum_N
c_{1}(E^{+}_{(N)})[\sphinf]$ 
so we have proved that
\begin{equation*}
\ind D_{\A}^{+} = -c_2(\E,f)[X] -\sum_{N} c_{1}(E_{(N)}^{+})[\sphinf].
\end{equation*}
This completes the proof of Theorem \ref{thm:indexgen}.

\section{Extension to the singular points}\label{sec:extNahm}

The final problem we consider is how to prove the Nahm data constructed from a caloron satisfies the correct gluing and singularity conditions at the points $\xi=\mu_1,\ldots,\mu_n$. 
These conditions were specified in Sections~\ref{sec:nahmdataformon} and~\ref{sec:nahmdataforcal}. 
Nakajima \cite[Section 2]{nak90} showed how to obtain the singularity conditions for Nahm data constructed from an $SU(2)$ monopole, and our approach will follow his quite closely. 
Nahm data corresponding to an $SU(2)$ monopole cannot contain zero jumps, and at the two singularities the continuing component is trivial. 
(Recall the definitions of the terminating and continuing components on page~\pageref{pag:termcpt}.) 
While Nakajima's method therefore helps us to recover the behaviour for the terminating component, it does not provide much insight into the continuing component. 
Hurtubise and Murray's proof \cite{hur89} that the Nahm data constructed from on $SU(n)$ monopole satisfies the gluing and singularity conditions uses the spectral curve of the monopole. 
No `direct proof' via analysis of Dirac operators exists to date. 
Indeed, the proof for calorons would follow quite readily from a `direct' proof for monopoles. 
We start, in Sections~\ref{sec:monmodel} and~\ref{sec:monterm}, by considering the terminating component for $U(n)$ monopoles, using Nakajima's method and filling in some of the details he misses. 
In Section~\ref{sec:termcal} we show how to extend these results for calorons, while in Section~\ref{sec:decompkpneq0} we consider the continuing component, giving only sketch results and conjectures. 

The main idea is that solutions to $\DAximinus$ are characterized by their asymptotic behaviour on $\cylo$ close to each $\xi\in\Xising$. 
Suppose that $\xi$ is close to $\mu_p$ and let $t=\xi-\mu_p$ (we will use this definition of $t$ for the remainder of the Chapter). 
We saw in Section~\ref{sec:zeromodedecay} that $\psi\in\coker\DAximinus$ decays at least as fast as $\exp (-r |t|)$ as $r\rightarrow 0$. 
We will show that solutions in the terminating component at $\mu_p$ are of the form $\exp (-r |t|)~\times$ (non-$L^2$ function), while the motivation behind our results for the continuing component is that the corresponding solutions decay like $\exp (-r |t+\alpha|)$ for some $\alpha\neq 0$, and so continue across $\xi=\mu_p$ as $L^2$ sections. 

\subsection{The model operator for monopoles}\label{sec:monmodel}

We need to recall the boundary conditions for $U(n)$ monopoles from Chapter 2 and the definition of the Nahm transform for monopoles. 
Let $(E,A)$ be a $U(n)$ monopole framed by $\Ainf,\Phiinf$ with boundary data $\monbdarydata$. 
Just as in Section~\ref{sec:smooth}, we need to assume some additional smoothness conditions near the boundary. 
We assume that there are local gauges on $E$, defined over some region $0\leq\chi<1 / R$ in which 
\begin{itemize}
\item $\Phi = \diag(i\mu_1,\ldots,i\mu_n) -\frac{\chi}{2}\diag(ik_1,\ldots,ik_n)+O(\chi^{2})$,
\item $A_{{y_j}} = \diag( \langle \partial_{y_j} e_p, e_p \rangle )+O(\chi^2),\quad j=1,2$,
\item $A_\chi$ is diagonal on $\sphinf$, and
\item $\Phi$ is $C^1_\chi$ and $A$ is $C^{0,1}_\chi$. 
\end{itemize}
These are entirely analogous to the smoothness conditions we assume for calorons in Section~\ref{sec:smooth} (compare with equations~$\eqref{eq:smo2}$ and~$\eqref{eq:asympphi}$). 
Next recall the definition of the Nahm transform for monopoles given in Sections~\ref{sec:reviewNahm} and~\ref{sec:scformonopoles}. 
The transform is given by the cokernel of $D_\xi$\label{glo:monDirac} as defined by equation~$\eqref{eq:diracopformono}$, and the Nahm data is defined by equations~$\eqref{eq:nahmmoncon}$ and~$\eqref{eq:nahmmonmtx}$. 

Fix a singularity $\xi=\mu_p$ with $k_p>0$, and let $k=k_p$ and $t=\xi-\mu_p$. 
Nakajima's method is similar to the way we recovered the boundary conditions in Chapter 3. 
We define a model operator $\tilde{D}_\xi$\label{glo:modelmon} that approximates $D_\xi$ and find $k$ solutions to $\tilde{D}^\ast_{\xi}\psi=0$, defined on some neighbourhood $t\in (-\epsilon,0)$, for some $\epsilon>0$. 
We then show that these solutions are arbitrarily close to solutions to $D_\xi^\ast\psi=0$ in the limit $t\rightarrow 0_-$, thereby recovering the terminating component. 

The definition of the model operator uses the following facts about Dirac operators on $\sphinf$, taken from~\cite{nak90}. 
The spin bundle $S_{(2)}$\label{glo:spintwo} of $\sphinf$ decomposes into two line bundles $S_{(2)}=S^+_{(2)}\oplus S^-_{(2)}$ and there are two Dirac operators
\begin{equation*}
D^{\pm}:C^\infty(\sphinf,S^\pm_{(2)})\rightarrow C^\infty(\sphinf,S^\mp_{(2)}).
\end{equation*}
As previously, we can identify $\sphinf\cong\mathbf{P}_1(\C)$. 
Then
\begin{equation*}
S^+_{(2)}\cong \Lambda^{0,1}\otimes H^{-1}\cong H\quad\textrm{and}\quad
S^-_{(2)}\cong\Lambda^{0,0}\otimes H^{-1}\cong H^{-1}
\end{equation*}
where $H$\label{glo:hyperH} is the hyperplane bundle on $\mathbf{P}_1(\C)$ and $\Lambda^{p,q}$ is the space of $(p,q)$-forms. 
There is an identification $\Lambda^{0,1}\cong H^2$ so $\Lambda^{0,1}\otimes H^{-1}\cong H^2\otimes H^{-1}=H$. 
The Dirac operator $D^-$ is then a multiple of the Cauchy-Riemann operator:
\begin{equation}\label{eq:diracasCR}
D^-:S^-_{(2)}\cong\Lambda^{0,0}\otimes H^{-1}\xrightarrow{2\bar{\partial}}\Lambda^{0,1}\otimes H^{-1}\cong S^+_{(2)}.
\end{equation}
Similarly, we can consider the Dirac operators coupled to the line bundle $H^k$ via the homogeneous connection $a_k$ on $H^k$, which we denote \label{glo:Dak}
\begin{equation*}
D^{\pm}_{a_k}: C^\infty (\sphinf,S^\pm_{(2)}\otimes H^k)\rightarrow C^\infty(\sphinf,S^\mp_{(2)}\otimes H^k).
\end{equation*}
Equation~$\eqref{eq:diracasCR}$ becomes
\begin{equation*}
D^-_{a_k}:S^-_{(2)}\otimes H^k\cong\Lambda^{0,0}\otimes H^{k-1}\xrightarrow{2\bar{\partial}}\Lambda^{0,1}\otimes H^{k-1}\cong S^+_{(2)}\otimes H^k.
\end{equation*}
so $\ker D_{a_k}^- = H^0\big( \mathbf{P}_1(\C), \mathcal{O}(k-1) \big)$. 

Next, consider $M=(R,\infty)\times\sphinf$\label{glo:M} equipped with coordinates $r,y_1,y_2$ and the metric
\begin{equation*}
dr^2 + r^2(h_1 dy_1^2 + h_2 dy_2^2)
\end{equation*} 
(compare with the notation at the start of Chapter 2), so that $M$ is isometric to $\rthree \setminus \threeball_R$, where $\threeball_R$ is the closed $3$-ball with radius $R$. 
Let $\varrho$\label{glo:varrho} be the projection $\varrho:M\rightarrow\sphinf$. 
The spin-bundle of $M$ is isomorphic to $\varrho^\ast S_{(2)}$. 
Under this identification the Dirac operator on $M$ is given by
\begin{equation}\label{eq:DM}
D_M = \begin{pmatrix}
i(\partial_r + \frac{1}{r}) & \frac{1}{r}D^- \\
 \frac{1}{r}D^+ & -i(\partial_r + \frac{1}{r})
\end{pmatrix}
\end{equation}
in some suitable local gauges on $S^\pm_{(2)}$, where $D^\pm:S^\pm_{(2)}\rightarrow S^\mp_{(2)}$ are the Dirac operators on $\sphinf$. 
When we couple the Dirac operator on $M$ to $\varrho^\ast H^k$ via the connection $\varrho^\ast a_k$, equation~$\eqref{eq:DM}$ becomes\label{glo:DMHk}
\begin{equation}\label{eq:DMHk}
D_{M,H^k} = \begin{pmatrix}
i(\partial_r + \frac{1}{r}) & \frac{1}{r}D^-_{a_k} \\
 \frac{1}{r}D^+_{a_k} & -i(\partial_r + \frac{1}{r})
\end{pmatrix}.
\end{equation}

With this background material established we are in a position to define the model operator. 
Given $E_\infty\rightarrow\sphinf$ we work on the pull-back $\varrho^\ast E_\infty$. 
Recall that $\Phiinf$ decomposes $E_\infty$ into eigenbundles, $E_\infty = L_{k_1}\oplus\cdots\oplus L_{k_n}$ where $L_{k_j}\cong H^{k_j}$ and $k_1,\ldots,k_n$ are the monopole charges. 
For each $p=1,\ldots,n$ define
\begin{align}
(\tilde{D}_\xi^p)^\ast & : \Gamma(M,\varrho^\ast S_{(2)}\otimes\varrho^\ast L_{k_p})\rightarrow\Gamma(M,\varrho^\ast S_{(2)}\otimes\varrho^\ast L_{k_p})\notag \\
(\tilde{D}_\xi^p)^\ast & = -i(\mu_p-\xi-\frac{k_p}{2r})+D_{M,H^{k_p}}\notag \\
 & = \begin{pmatrix}
i(\partial_r  -\mu_p+\xi+\frac{k_p+2}{2r}) & \frac{1}{r}D^-_{a_{k_p}} \\
 \frac{1}{r}D^+_{a_{k_p}} & i(-\partial_r-\mu_p+\xi + \frac{k_p-2}{2r})
\end{pmatrix} \label{eq:monmodelp}
\end{align}
where $D_{M,H^{k_p}}$ is defined by~$\eqref{eq:DMHk}$. 
Using the decomposition of $E_\infty$ into line bundles, we can define
\begin{gather}
\tilde{D}_\xi^\ast : \Gamma(M,\varrho^\ast S_{(2)}\otimes\varrho^\ast E_\infty)\rightarrow\Gamma(M,\varrho^\ast S_{(2)}\otimes\varrho^\ast E_\infty)\notag \\
\tilde{D}_\xi^\ast =  (\tilde{D}_\xi^1)^\ast\oplus\cdots\oplus(\tilde{D}_\xi^n)^\ast. \label{eq:defntildeDxi}
\end{gather}
Fixing some identification of $E|_M$ with $\varrho^\ast E_\infty$ and working in the local gauges on $E$ in which $A$ and $\Phi$ satisfy the asymptotic conditions stated at the start of this Section, we have
\begin{align}
D^\ast_\xi & = D_A-\Phi+i\xi \notag \\
&= \diag (D_{M,H^{k_1}},\ldots,D_{M,H^{k_n}} ) +i\diag (\xi-\mu_1+\frac{k_1}{2r},\ldots,\xi-\mu_n+\frac{k_n}{2r} ) +O(r^{-2}) \notag\\
&= \tilde{D}_\xi^\ast +O(r^{-2}) \label{eq:tina}. 
\end{align}

Since $\tilde{D}_\xi^\ast$ is given so explicitly, we can write down solutions and these will be our approximate solutions to $D^\ast_\xi$. 
Working near the singularity $\xi=\mu_p$ (with $k:=k_p>0$), $\tilde{D}_\xi^\ast$ is given by
\begin{equation}\label{eq:modelnearsing}
\begin{pmatrix}
i(\partial_r + t+\frac{k+2}{2r}) & \frac{1}{r}D^-_{a_{k}} \\
 \frac{1}{r}D^+_{a_{k}} & i(-\partial_r+t + \frac{k-2}{2r})
\end{pmatrix}
\end{equation}
on $\varrho^\ast L_k$. 
Now $\ker D_{a_k}^-=H^0\big(\mathbf{P}_1(\C),\mathcal{O}(k-1)\big)$, and every element $f=f(y_1,y_2)\in\ker D_{a_k}^-$ determines a solution of~$\eqref{eq:modelnearsing}$ of the form
\begin{equation}\label{eq:defapproxsoln}
\tilde{\psi} = \begin{pmatrix}
0 \\ (\exp rt) r^{(k-2)/2}f(y_1,y_2)
\end{pmatrix}.
\end{equation}
Note that these solutions are $L^2$ when $t\in (-\epsilon,0)$ but fail to be $L^2$ when $t\in [0,\epsilon)$. 
Since $H^0\big(\mathbf{P}_1(\C),\mathcal{O}(k-1)\big)$ is $k$-dimensional (when $k>0$), taking an orthonormal basis gives $k$ linearly independent orthogonal solutions to $\tilde{D}_\xi^\ast\tilde{\psi}=0$ of the form~$\eqref{eq:defapproxsoln}$. 
Smoothing these off by a bump function
\begin{equation*}
\phi(r)=\begin{cases}
1\quad & r\geq R+\delta \\
0\quad & r\leq R \\
\end{cases}
\end{equation*}
and identifying $E=\varrho^\ast E_\infty$ over $r\geq R$ gives the approximate solutions $\tilde{\psi}_1,\ldots,\tilde{\psi}_k$\label{glo:psitildes} associated to the terminating component at $\xi=\mu_p$. 
Note that the approximate solutions are orthogonal (because the basis of $H^0\big(\mathbf{P}_1(\C),\mathcal{O}(k-1)\big)$ is orthogonal), but not normal. 

We need some estimates as to how closely $\tilde{\psi}_1,\ldots,\tilde{\psi}_k$ approximate solutions to $D^\ast_\xi$. 
Now
\begin{equation*}
C\int_{R+\delta}^{\infty}(\exp 2rt)r^{k-2}~r^2~dr \leq \| \tilde{\psi}_j \|^2_{L^2} \leq C\int_{R}^{\infty}(\exp 2rt)r^{k-2}~r^2~dr
\end{equation*}
for some constant $C$ (used throughout in the generic sense), where $\delta$ was used to define the bump function $\phi$. 
These integrals are the same (up to a change of variable) as the integral ${\mathfrak{I}}_k$ defined by~$\eqref{eq:defIk}$. 
We could estimate them using integration by parts, just as we did on page~\pageref{pag:intest}, but this time around it is easier to
change variable to $u=rt$, giving
\begin{equation}\label{eq:estapproxnorm}
C\int_{(R+\delta)t}^{-\infty}(\exp 2u)\frac{u^{k}}{t^{k+1}}~du \leq \| \tilde{\psi}_j \|^2_{L^2} \leq C\int_{Rt}^{-\infty}(\exp 2u)\frac{u^{k}}{t^{k+1}}~du.
\end{equation}
Similarly, since $D^\ast_{\xi}\tilde{\psi}_j = O(r^{-2})\times \tilde{\psi}_j$, we have
\begin{equation}\label{eq:estDapproxnorm}
C\int_{(R+\delta)t}^{-\infty}(\exp 2u)\frac{u^{k-4}}{t^{k-3}}~du \leq \| D^\ast_\xi\tilde{\psi}_j \|^2_{L^2} \leq C\int_{Rt}^{-\infty}(\exp 2u)\frac{u^{k-4}}{t^{k-3}}~du.
\end{equation}
Together, equations~$\eqref{eq:estapproxnorm}$ and~$\eqref{eq:estDapproxnorm}$ give
\begin{equation*}
\| D_\xi^-\tilde{\psi}_j \|_{L^2}\leq Ct^2 \| \tilde{\psi}_j \|_{L^2}
\end{equation*}
whenever $k>3$, because the limits
\begin{equation*}
\lim_{t\rightarrow 0_-}\int_{Rt}^{-\infty} (\exp 2u) u^{k-4}~du\quad\textrm{and}\quad 
\lim_{t\rightarrow 0_-}\int_{Rt}^{-\infty} (\exp 2u) u^{k}~du
\end{equation*}
both exist. 
However, this estimate does not hold in the cases $k=1,2,3$ because the limit on the left does not exist---note that Nakajima does not point this out. 
For the cases $k=1,2,3$ the estimates~$\eqref{eq:estapproxnorm}$ and~$\eqref{eq:estDapproxnorm}$ give
\begin{equation}\label{eq:poorest}
\| D^\ast_\xi\tilde{\psi}_j \|_{L^2} / \| \tilde{\psi}_j \|_{L^2} \sim 
t^2\big( A+B\int_{Rt}^{-1} u^{k-4}~du \big)^{1/2}
\end{equation}
for some constants $A,B$, where `$\sim$' means that for sufficiently small $t$ there exist $A,B$ such that LHS$\geq$RHS and there exist $A,B$ such that LHS$\leq$RHS. 
Evaluating the integral in~$\eqref{eq:poorest}$, the RHS becomes
\begin{equation*}
t^2(A+Bt^{k-3})^{1/2}
\end{equation*}
when $k=1,2$, and
\begin{equation*}
t^2(A+B\log |t|)^{1/2}
\end{equation*}
for $k=3$. 
Thus we obtain
\begin{equation}\label{eq:estratio}
\| D^\ast_\xi\tilde{\psi}_j \|_{L^2} / \| \tilde{\psi}_j \|_{L^2} \sim
\begin{cases}
C|t|\quad & \textrm{when}~k=1 \\
C|t|^{3/2}\quad & \textrm{when}~k=2 \\
Ct^2(\log |t|)^{1/2}\quad & \textrm{when}~k=3 \\
Ct^2\quad & \textrm{when}~k>3.
\end{cases}
\end{equation}
Note that, by very similar estimates,
\begin{equation}\label{eq:estrapprox}
\| r\tilde{\psi}_j \|_{L^2} = \| \partial_t\tilde{\psi}_j \|_{L^2} \sim C|t|^{-1}\| \tilde{\psi}_j \|_{L^2}
\end{equation}
as $t\rightarrow 0_-$. 
At this stage it is convenient to normalise the approximate solutions $\tilde{\psi}_p$, $p=1,\ldots,k$, so that they are of the form 
\begin{equation*}
C_p^{-1}\phi(r)\begin{pmatrix}
0 \\ (\exp rt) r^{(k-2)/2}f_p(y_1,y_2)
\end{pmatrix}
\end{equation*}
where $C_p=O(t^{-(k+1)/2})$. 

\subsection{The terminating component for monopoles}\label{sec:monterm}

We want to use the approximate solutions to recover the terminating component of the Nahm data. 
The method used is very similar to the proofs of Propositions~\ref{prop:approxframe} and~\ref{prop:framed}. 
First, however, we need the following results from \cite{nak90}.

\begin{lemma}[Nakajima]\label{lem:nak1}
For any $\omega\in L^2(\rthree,\spinthree\otimes E)$ we have
\begin{equation*}
\| D_\xi(D_\xi^\ast D_\xi)^{-1}\omega \|_{L^2}\leq C|t|^{-1}\| \omega \|_{L^2}
\end{equation*}
for some constant $C$ and all $t\in (-\epsilon,0)$, where $t=\xi-\mu_p$. 
\end{lemma}

\proof
Since $D_\xi$ is injective and Fredholm on $t\in (-\epsilon,0)$ for some sufficiently small $\epsilon$, $(D_\xi^\ast D_\xi)$ is invertible and
\begin{equation*}
\varphi:=(D_\xi^\ast D_\xi)^{-1}\omega 
\end{equation*}
exists. 
(Nakajima gives a rather more involved argument.)
For sufficiently large $R$ and small $|t|$, 
\begin{equation*}
|t|~|\varphi |\leq 2 |(\Phi-i\xi)\varphi |
\end{equation*}
pointwise on $B^c_{R/|t|} := \rthree\setminus\threeball_{R / |t|}$, so 
\begin{equation*}
|t|^2 \int_{B^c_{R/|t|}} |\varphi |^2~dV \leq C \| (\Phi-i\xi)\varphi \|^2_{L^2}.
\end{equation*}
Using H\"older's inequality and the Sobolev inequality, Nakajima obtains
\begin{equation*}
|t|^2 \int_{\threeball_{R / |t|}} |\varphi |^2~dV \leq C \| \nabla_A \varphi \|^2_{L^2}.
\end{equation*}
(Note that these inequalities hold for any $\varphi$.)
Combining the inequalities gives
\begin{equation}\label{eq:naklem1}
\| \varphi \|_{L^2}\leq C|t|^{-1}~\| D_\xi\varphi \|_{L^2}
\end{equation}
since
\begin{equation*}
\| D_\xi\varphi \|_{L^2}^2=\| \nabla_A \varphi \|^2_{L^2}+\| (\Phi-i\xi)\varphi \|^2_{L^2}.
\end{equation*}
Substituting~$\eqref{eq:naklem1}$ into
\begin{equation*}
\| D_\xi\varphi \|_{L^2}^2= \langle D_\xi^\ast D_\xi\varphi,\varphi \rangle \leq \| \varphi \|_{L^2}\times\| \omega \|_{L^2}
\end{equation*}
gives
\begin{equation*}
\| D_\xi\varphi \|_{L^2}\leq C|t|^{-1}\| \omega \|_{L^2}
\end{equation*}
proving the Lemma.
\eproof

\begin{corollary}\label{cor:nak1}
When $k>3$, the approximate solutions $\tilde{\psi}_1,\ldots,\tilde{\psi}_k$ satisfy
\begin{equation*}
\| (1-\hat{P})\tilde{\psi}_j \|_{L^2}  \leq C|t|~\| \tilde{\psi}_j \|_{L^2}
\end{equation*}
as $t\rightarrow 0_-$, where $\hat{P}=\hat{P}_\xi$ is the $L^2$ projection onto $\ker D^\ast_\xi$. 
We also have 
\begin{equation*}
\| (1-\hat{P})\tilde{\psi}_j \|_{L^2}  \leq \begin{cases}
C|t|\log |t|~\| \tilde{\psi}_j \|_{L^2}\quad &\textrm{when}~k=3 \\
C|t|^{1/2}~\| \tilde{\psi}_j \|_{L^2}\quad &\textrm{when}~k=2.
\end{cases}
\end{equation*}
\end{corollary}

\proof
Put $\omega=D_\xi^\ast\tilde{\psi}_j$ in Lemma~\ref{lem:nak1} and use~$\eqref{eq:estratio}$. 
\eproof

Note that (contrary to Nakajima) we do not obtain an estimate $\| (1-\hat{P})\tilde{\psi}_j \|_{L^2}\rightarrow 0$ as $t\rightarrow 0_-$ in the case $k=1$. 
However, we expect the case $k=1$ to be exceptional: when $k=1$ the irreducible representation of $\sutwo$ is trivial, so the Nahm data should be analytic (rather than meromorphic) in $t$ near $t=0$. 
We therefore have to deal with the case $k=1$ separately---see the remarks in Section~\ref{sec:decompkpneq0}. 

\begin{lemma}[Nakajima]\label{lem:nak2}
Let $R_j$ be the endomorphism of $H^0\big(\mathbf{P}_1(\C),\mathcal{O}(k-1)\big)$ defined by
\begin{equation*}
\langle R_j f_1, f_2\rangle = \int_{\mathbf{P}_1(\C)}\langle ix_j f_1,f_2 \rangle
\end{equation*}
for $j=1,2,3$ where $x_j$ is the standard cartesian coordinate on $\mathbf{P}_1(\C)=S^2\subset\rthree$. 
Then a non-zero constant multiple of the linear map
\begin{equation}\label{eq:multipleisrep}
\lambda_1\gamma_1 + \lambda_2\gamma_2 + \lambda_3\gamma_3 \mapsto \lambda_1 R_1 + \lambda_2 R_2 + \lambda_3 R_3 
\end{equation}
defines an irreducible $k$-dimensional representation of $\sutwo$. 
\end{lemma}

\proof
See the Appendix to \cite{nak90}. 
\eproof

We are now in a position to prove the following: 

\begin{proposition}\label{prop:montermcpt}
Given a Bogomolny monopole $(A,\Phi)$ with boundary data $\monbdarydata$, then for any singularity $\xi=\mu_p$ with $k:=k_p\geq 2$ there is a parallel gauge defined on some neighbourhood $t:=(\xi-\mu_p )\in (-\epsilon,0)$ of the singularity in which the matrices $T_p^j$, $j=1,2,3$, defined by~$\eqref{eq:nahmmonmtx}$ decompose as 
\begin{center}
\setlength{\unitlength}{1.2cm}
\begin{picture}(8,2.5)(-4,-1.0)

\put(-2,0){\line(1,0){4}}
\put(0,-0.6){\line(0,1){1.2}}

\put(-1.3,0.25){$\ast$}
\put(0.8,0.25){$\ast$}
\put(-1.3,-0.5){$\ast$}
\put(0.1,-0.5){$R^{j}_{p}/t+B_j(t)$}

\put(-3.15,-0.1){$T_{p}^{j} = $}
\put(-2.4,-0.1){$\Bigg($}
\put(2.1,-0.1){$\Bigg)$}

\put(-1.4,0.9){$m_{p-1}$}
\put(3,0.25){$m_{p-1}$}
\put(0.8,0.9){$k_{p}$}
\put(3,-0.5){$k_{p}$}

\put(2.75,0.05){\vector(0,1){0.55}}
\put(2.75,0.6){\vector(0,-1){0.55}}
\put(2.75,-0.05){\vector(0,-1){0.55}}
\put(2.75,-0.6){\vector(0,1){0.55}}

\put(1.4,1.0){\vector(1,0){0.6}}
\put(0.6,1.0){\vector(-1,0){0.6}}
\put(-1.5,1.0){\vector(-1,0){0.55}}
\put(-0.6,1.0){\vector(1,0){0.55}}

\end{picture}
\end{center}
such that
\begin{enumerate}
\item $R_p^1,R_p^2,R_p^3$ form an irreducible representation of $\sutwo$ following equation~$\eqref{eq:irrepbc}$, and
\item in the limit $t\rightarrow 0_-$ we have
\begin{equation*}
B_j(t)=
\begin{cases}
O(t^0)~&\textrm{when~}k>3~\textrm{(\ie $B_j$ is bounded)} \\
O(t^{-1/2})&\textrm{when~}k=2~\textrm{or}~3.
\end{cases}
\end{equation*}
\end{enumerate}
We make no claims about the entries marked $\ast$ at this stage. 
\end{proposition}

We deal with the top left block (the continuing component), and discuss the off-diagonal blocks, the case $k=1$, and analyticity in Section~\ref{sec:decompkpneq0}. 
Of course, to satisfy the conditions for Nahm data stated on page~\pageref{pag:decompTj}, $B_j(t)$ must be analytic---so in the cases $k=2,3$ it seems disturbing that we can only prove $B_j(t)$ is $O(t^{-1/2})$. 
However, Nahm's equations impose additional strong conditions on $B_j$, which we will discuss in Section~\ref{sec:decompkpneq0}. 
Note that an entirely analogous statement to~\ref{prop:montermcpt} holds for $k_p<0$, essentially by replacing $t$ with $-t$ in the proof of the Proposition. 

\proof
The proof has two main steps. 
First we work in the `approximate gauge' $\tilde{\psi}_1,\ldots,\tilde{\psi}_k$ and evaluate the matrices
\begin{equation}\label{eq:evaluatethese}
\langle \partial_\xi \tilde{\psi}_a,\tilde{\psi}_b\rangle_{L^2}\quad\textrm{and}\quad
\langle ix_j \tilde{\psi}_a,\tilde{\psi}_b\rangle_{L^2}~\textrm{for}~j=1,2,3,
\end{equation}
where $a,b\in\{ 1,\ldots,k \}$. 
Then, using Corollary~\ref{cor:nak1}, we use the approximate matrices~$\eqref{eq:evaluatethese}$ to deduce that the Nahm matrices decompose as described in the claim. 

First consider evaluating the matrix with entries $\langle \partial_\xi \tilde{\psi}_a,\tilde{\psi}_b\rangle_{L^2}$. 
Since $\partial_\xi \tilde{\psi}_a = r\tilde{\psi}_a$, $\partial_\xi \tilde{\psi}_a$ is orthogonal to $\tilde{\psi}_b$ when $a\neq b$ because $f_a$ is orthogonal to $f_b$, and the matrix is diagonal. 
Moreover, because $\tilde{\psi}_1,\ldots,\tilde{\psi}_k$ is an orthonormal set, the diagonal entries are imaginary. 
On the other hand, the integral $\langle r\tilde{\psi}_a,\tilde{\psi}_a\rangle_{L^2}$ is real, and so the matrix must be zero. 
Also
\begin{equation*}
\langle ix_j \tilde{\psi}_a,\tilde{\psi}_b\rangle_{L^2}=\frac{
\int_R^\infty dr~\phi^2(r)(\exp 2rt)r^{k+1}~\int_{\mathbf{P}_1(\C)}dA~\langle i\hat{x}_j f_a,f_b \rangle
}{
\int_R^\infty dr~\phi^2(r)(\exp 2rt)r^{k}
}
\end{equation*}
where $\hat{x}_j=x_j / r$ is the $j$'th unit coordinate function on $\mathbf{P}_1(\C)$, and $f_a,f_b\in H^0\big(\mathbf{P}_1(\C),$ $\mathcal{O}(k-1)\big)$ were used to define the approximate solutions. 
Substituting $u=rt$ into the integrals, we obtain
\begin{equation*}
\frac{
\int_R^\infty dr~\phi^2(r)(\exp 2rt)r^{k+1}
}{
\int_R^\infty dr~\phi^2(r)(\exp 2rt)r^{k}
}=\alpha t^{-1}
\end{equation*}
for some non-zero constant $\alpha$. 
Defining the matrix $R_p^j$ by
\begin{equation*}
(R_p^j)_{ab}=\alpha\int_{\mathbf{P}_1(\C)}dA~\langle i\hat{x}_j f_a,f_b \rangle
\end{equation*}
we have 
\begin{equation}\label{eq:theresidue}
\langle ix_j \tilde{\psi}_a,\tilde{\psi}_b\rangle_{L^2}= \frac{(R_p^j)_{ab}}{t}
\end{equation}
and Nakajima's Lemma~\ref{lem:nak2} implies that some non-zero constant multiple of the map~$\eqref{eq:multipleisrep}$ (with $R_j:=R_p^j$) is an irreducible $k$-dimensional representation of $\sutwo$. 

Corollary~\ref{cor:nak1} shows that the projection $\hat{P}$ on to the cokernel of $D_\xi$ satisfies $\hat{P} = (1 + \textrm{decaying term})$
on the span of $\tilde{\psi}_1,\ldots,\tilde{\psi}_k$. 
Just as we did in Proposition~\ref{prop:framed}, we can replace $\hat{P}$ with its unitarization $\hat{P}_U$, so that 
\begin{equation*}
\hat{P}_U = 1+Q(t)
\end{equation*}
on the span of $\tilde{\psi}_1,\ldots,\tilde{\psi}_k$,
where 
\begin{equation*}
Q(t) = 
\begin{cases}
O(t)&\textrm{when $k>3$, and}\\
O(t^{1/2})&\textrm{when $k=2$ or $3$.}
\end{cases}
\end{equation*}
We can in fact obtain a better estimate in the case $k=3$, but the estimate above is sufficient for our purposes. 
Next define
\begin{equation}\label{eq:defnpsis}
\psi_a:=\hat{P}_U\tilde{\psi}_a.
\end{equation}
It follows that $\psi_1,\ldots,\psi_k$ is an orthonormal set of sections of the bundle $X_p$ defined over some neighbourhood $t\in (-\epsilon,0)$. 
The set can be extended by $m_{p-1}$ further sections to give a local trivialisation of the bundle $X_p$. 

For the time being we will assume $k>3$ and return to the cases $k=2,3$ later. 
Now
\begin{align*}
\langle \partial_\xi {\psi}_a,{\psi}_b\rangle_{L^2} &=
\langle \partial_\xi \tilde{\psi}_a+ \partial_\xi({\psi}_a-\tilde{\psi}_a),\tilde{\psi}_b + (\psi_b-\tilde{\psi}_b)\rangle_{L^2} \\
&= \langle  \partial_\xi({\psi}_a-\tilde{\psi}_a),\tilde{\psi}_b \rangle_{L^2}
+\langle  \partial_\xi({\psi}_a-\tilde{\psi}_a), (\psi_b-\tilde{\psi}_b)\rangle_{L^2} 
+\langle \partial_\xi \tilde{\psi}_a,(\psi_b-\tilde{\psi}_b)\rangle_{L^2},
\end{align*}
because $\langle \partial_\xi \tilde{\psi}_a,\tilde{\psi}_b\rangle_{L^2}=0$. 
The first term is bounded as $t\rightarrow 0_-$ because $\psi_a-\tilde{\psi}_a=O(t)$. 
Similarly, the second term is $O(t)$, and the third term is bounded by 
\begin{equation*}
\| \partial_\xi \tilde{\psi}_a \|_{L^2} \times \| \psi_b-\tilde{\psi}_b \|_{L^2} \leq C|t|^{-1}\times O(t)
\end{equation*}
using~$\eqref{eq:estrapprox}$. 
Thus we obtain a bound
\begin{equation}\label{eq:boundT0}
| \langle \partial_\xi {\psi}_a,{\psi}_b\rangle_{L^2} |\leq C
\end{equation}
on $t\in (-\epsilon,0)$ for some fixed $C$. 
Note that this bound might fail to hold in the cases $k=2,3$ due to the weaker estimates. 
Let $T^0$ be the matrix with entries $\langle \partial_\xi {\psi}_a,{\psi}_b\rangle_{L^2}$. 
Then the gauge transformation 
\begin{equation*}
g(t) = \exp\int_t^0 T^0(s)~ds
\end{equation*}
satisfies 
\begin{equation}\label{eq:ggood}
g(t) = 1+O(t)
\end{equation}
because $T^0$ is bounded, and maps $\psi_1,\ldots,\psi_k$ to a unitary parallel set of sections. 
We will apply the gauge transformation later. 

Next we consider the endomorphisms $T^j_p$, $j=1,2,3$. 
Extend $\psi_1,\ldots,\psi_k$ by $m_{p-1}$ further sections to form a gauge for $X_p$ on $t\in (-\epsilon,0)$. 
In such a gauge the endomorphisms are given by
\begin{align}
(T^j_p)_{ab} &= \langle ix_j {\psi}_a,{\psi}_b\rangle_{L^2} \notag \\
&= \langle ix_j \tilde{\psi}_a+ ix_j({\psi}_a-\tilde{\psi}_a),\tilde{\psi}_b + (\psi_b-\tilde{\psi}_b)\rangle_{L^2} \notag \\
& = \frac{(R^j_p)_{ab}}{t}+
\langle ix_j \tilde{\psi}_a, (\psi_b-\tilde{\psi}_b)\rangle_{L^2}
-\langle ({\psi}_a-\tilde{\psi}_a),ix_j {\psi}_b \rangle_{L^2} \label{eq:wantbdd}
\end{align}
using equation~$\eqref{eq:theresidue}$. 
From~$\eqref{eq:estrapprox}$, $\| ix_j\tilde{\psi}_a \|_{L^2}=O(|t|^{-1})$ and so 
\begin{equation*}
\| ix_j{\psi}_a \|_{L^2} = \| ix_j\big( 1+O(t) \big)\tilde{\psi}_a \|_{L^2}=O(|t|^{-1}).
\end{equation*}
Thus, estimating the RHS of~\eqref{eq:wantbdd} gives
\begin{equation}\label{eq:Bjbdd}
(T^j_p)_{ab} = \frac{(R^j_p)_{ab}}{t}+B_j(t)
\end{equation}
where $B_j$ is bounded as $t\rightarrow 0_-$. 
Moreover, gauge transforming by $g(t)$ into a parallel gauge does not alter the form of~$\eqref{eq:Bjbdd}$ because $g(t)$ has the form~$\eqref{eq:ggood}$. 
Nahm's equation implies that multiplying the map~$\eqref{eq:multipleisrep}$ by $-2$ gives an irreducible $k$-dimensional representation of $\sutwo$, as in equation~$\eqref{eq:irrepbc}$. 
This completes the proof when $k>3$. 

For the case $k=2$ or $3$ the analysis is very similar. 
Equation~$\eqref{eq:boundT0}$ becomes a bound $|\langle \partial_\xi\psi_a,\psi_b \rangle_{L^2}|=O(t^{-1/2})$, and the gauge transformation to the parallel gauge, equation~$\eqref{eq:ggood}$, becomes $g(t)=1+O(t^{1/2})$. 
Estimates on $\eqref{eq:wantbdd}$ give $T^j_p$ of the form $\eqref{eq:Bjbdd}$ but with $B_j(t)=O(t^{-1/2})$. 
Gauge transforming by $g(t)$ does not alter the form of this expansion, and Nahm's equation fixes the constant for the irreducible representation. 
\eproof

We leave monopoles at this point to prove an analogue of Proposition~\ref{prop:montermcpt} for calorons. 

\subsection{The terminating component for calorons}\label{sec:termcal}

We prove the following analogue of Proposition~\ref{prop:montermcpt}:

\begin{proposition}\label{prop:caltermcpt}
Given an anti-self-dual caloron $(\E,\A)$ framed by $(\Ainf,\Phiinf)$ and with boundary data $\bdarydata$, then for any singularity $\xi=\mu_p$ with $k:=k_p>3$ there is a parallel gauge for $X_p$ defined on some neighbourhood $t:=(\xi-\mu_p )\in (-\epsilon,0)$ of the singularity in which the matrices $T_p^j$, $j=1,2,3$, defined by~$\eqref{eq:defnTj}$ decompose as 
\begin{center}
\setlength{\unitlength}{1.2cm}
\begin{picture}(8,2.5)(-4,-1.0)

\put(-2,0){\line(1,0){4}}
\put(0,-0.6){\line(0,1){1.2}}

\put(-1.3,0.25){$\ast$}
\put(0.8,0.25){$\ast$}
\put(-1.3,-0.5){$\ast$}
\put(0.1,-0.5){$R^{j}_{p}/t+B_j(t)$}

\put(-3.15,-0.1){$T_{p}^{j} = $}
\put(-2.4,-0.1){$\Bigg($}
\put(2.1,-0.1){$\Bigg)$}

\put(-1.4,0.9){$m_{p-1}$}
\put(3,0.25){$m_{p-1}$}
\put(0.8,0.9){$k_{p}$}
\put(3,-0.5){$k_{p}$}

\put(2.75,0.05){\vector(0,1){0.55}}
\put(2.75,0.6){\vector(0,-1){0.55}}
\put(2.75,-0.05){\vector(0,-1){0.55}}
\put(2.75,-0.6){\vector(0,1){0.55}}

\put(1.4,1.0){\vector(1,0){0.6}}
\put(0.6,1.0){\vector(-1,0){0.6}}
\put(-1.5,1.0){\vector(-1,0){0.55}}
\put(-0.6,1.0){\vector(1,0){0.55}}

\end{picture}
\end{center}
such that
\begin{enumerate}
\item $R_p^1,R_p^2,R_p^3$ form an irreducible representation of $\sutwo$ following equation~$\eqref{eq:irrepbc}$, and
\item in the limit $t\rightarrow 0_-$ we have
\begin{equation*}
B_j(t)=
\begin{cases}
O(t^0)~&\textrm{when~}k>3~\textrm{(\ie $B_j$ is bounded)} \\
O(t^{-1/2})&\textrm{when~}k=2~\textrm{or}~3.
\end{cases}
\end{equation*}
\end{enumerate}
\end{proposition}

This follows almost directly from Proposition~\ref{prop:montermcpt} applied to the Fourier modes of $\DAximinus$, because up to $O(r^{-2})$ a caloron is the pull-back of a monopole configuration. 
We can extend the framing from the boundary to the whole of $\So\times M$, and apply the ``$3+1$'' decomposition~$\eqref{eq:3plus1A}$ to define $A,\Phi$. 
However, up to $O(r^{-2})$, $A$ and $\Phi$ are exactly the same as a monopole configuration framed by $\Ainf,\Phiinf$. 
Using~$\eqref{eq:decompdirac}$ and~$\eqref{eq:tina}$ we have
\begin{align*}
\DAximinus &= -\partial_{x_0}+D_A-\Phi+i\xi \\
& =-\partial_{x_0}+\tilde{D}^\ast_\xi+O(r^{-2})
\end{align*}
where $\tilde{D}^\ast_\xi$ is the monopole model operator determined by $\Ainf,\Phiinf$. 
Identifying the spin bundles $S^\pm$ of $\So\times M$ with $S_{(3)}$ and using the Fourier decomposition~$\eqref{eq:fourierA}$ gives
\begin{equation*}
\DAximinus |_{Z_N} = -iN\mu_0 +\tilde{D}^\ast_\xi +O(r^{-2}).
\end{equation*}
We therefore take the model operator for $\DAximinus$ to be $-iN\mu_0+\tilde{D}^\ast_\xi$ on the $N$'th Fourier mode. 

Working near the singularity $\xi=\mu_p$ the approximate solutions $\tilde{\psi}_j$, $j=1,\ldots,k$, defined in Section~\ref{sec:monmodel} pull-back to $\So\times M$ and satisfy $\DAximinus\tilde{\psi}_j=O(r^{-2})\tilde{\psi}_j$ together with the analogue of the estimates~$\eqref{eq:estratio}$. 
The proof of Lemma~\ref{lem:nak1} goes through, replacing $D_\xi$ with $\DAxiplus$ and taking $\omega\in L^2(\cylo,S^+\otimes \E)$, as does the proof of Corollary~\ref{cor:nak1}.  
The proof of Proposition~\ref{prop:montermcpt} then gives~\ref{prop:caltermcpt} directly. 

\subsection{The continuing component and decomposition for $k_p\neq 0$}\label{sec:decompkpneq0}

We want to show that the continuing block of the Nahm data constructed from a caloron is continuous across singularities with $k_p\neq 0$, and obtain the full decomposition of the Nahm data at such a point, as described on page~\pageref{pag:decompTj}. 
While we also want to obtain the corresponding decomposition at zero jumps, this Section concentrates on the case $k_p\neq 0$ and we will only make some brief remarks about the zero jump case. 
At present, obtaining the decomposition of monopole Nahm data at singularities via analysis of the Dirac operator is an open problem (of course, Hurtubise-Murray obtained the decomposition via spectral curves). 
It should be clear that if we could obtain the decomposition of the Nahm data constructed from a $U(n)$ monopole---defining the Nahm data via the coupled Dirac operator rather than via spectral curves---then the caloron case would be very similar. 
For the remainder of this Section we therefore concentrate on the simpler case of $U(n)$ monopoles rather than calorons, although our conjectures and results will apply to calorons in an obvious way. 
Fix a $U(n)$ Bogomolny monopole $(A,\Phi)$ on a bundle $E$, framed by $\Ainf,\Phiinf$ and with boundary data $\monbdarydata$. 
Let $D_\xi$ be the coupled Dirac operator defined by~$\eqref{eq:diracopformono}$ and let $\{ X_p,\nabla_p,T^j_p:p=1,\ldots,n-1,~j=1,2,3 \}$ be the Nahm data defined in terms of $D_\xi$: 
$X_p\rightarrow(\mu_{p+1},\mu_p)$ is the bundle with fibre $\coker D_\xi$ while $\nabla_p$ and $T^1_p,T^2_p,T^3_p$ are defined by~$\eqref{eq:nahmmoncon}$ and~$\eqref{eq:nahmmonmtx}$. 
The full claim we want to prove is:

\begin{claim}\label{claim:}
Let $\mu_p$ be a singularity with $k_p>0$ and let $t=\xi-\mu_p$. 
Then there is a parallel gauge on $X_{p-1}$ for some neighbourhood $t\in(0,\epsilon)$ in which the limits
\begin{equation}\label{eq:conjlims}
T^{j,+}_{p-1}=\lim_{t\rightarrow 0_+} T^j_{p-1}
\end{equation}
exist for $j=1,2,3$, and $T^j_{p-1}(t)$ is analytic. 
Similarly, there is a parallel gauge on $X_p$ for some neighbourhood $t\in(-\epsilon,0)$ in which there is a decomposition
%%%%%%%%%%%%%%%%%%%%%%%%%%%%%%%%%%%%%%%%%%%%%%%%%%%%%%%%%%%%%%%%%%%%%%%%%%%%%%%
% Decomposition of Nahm data at singularities
%%%%%%%%%%%%%%%%%%%%%%%%%%%%%%%%%%%%%%%%%%%%%%%%%%%%%%%%%%%%%%%%%%%%%%%%%%%%%%%
\begin{center}
\setlength{\unitlength}{1.2cm}
\begin{picture}(8,2.5)(-4,-1.0)

\put(-2,0){\line(1,0){4}}
\put(0,-0.6){\line(0,1){1.2}}

\put(-2,0.25){$T^{j,+}_{p-1}+O(t)$}
\put(0.1,0.25){$O(t^{(k_{p}-1)/2})$}
\put(-2,-0.5){$O(t^{(k_{p}-1)/2})$}
\put(0.1,-0.5){$R^{j}_{p}/t+O(1)$}

\put(-3.15,-0.1){$T_{p}^{j} = $}
\put(-2.4,-0.1){$\Bigg($}
\put(2.1,-0.1){$\Bigg)$}

\put(-1.4,0.9){$m_{p-1}$}
\put(3,0.25){$m_{p-1}$}
\put(0.8,0.9){$k_{p}$}
\put(3,-0.5){$k_{p}$}

\put(2.75,0.05){\vector(0,1){0.55}}
\put(2.75,0.6){\vector(0,-1){0.55}}
\put(2.75,-0.05){\vector(0,-1){0.55}}
\put(2.75,-0.6){\vector(0,1){0.55}}

\put(1.4,1.0){\vector(1,0){0.6}}
\put(0.6,1.0){\vector(-1,0){0.6}}
\put(-1.5,1.0){\vector(-1,0){0.55}}
\put(-0.6,1.0){\vector(1,0){0.55}}

%\put(5.0,-0.5){$(\dag)$}

\end{picture}
\end{center}
%%%%%%%%%%%%%%%%%%%%%%%%%%%%%%%%%%%%%%%%%%%%%%%%%%%%%%%%%%%%%%%%%%%%%%%%%%%%%%%
The upper diagonal block is analytic in $t=\xi-\mu_{p}$; the lower diagonal block is meromorphic in $t$; and the off-diagonal blocks are of the form $t^{(k_{p}-1)/2}\times (\textrm{analytic in\ }t)$.
The residues $R^{j}_{p}$ define an irreducible representation of $\sutwo$.
\end{claim}

Working with the fixed monopole $(A,\Phi)$, fix some $\mu_p$ with $k:=k_p>0$ and let $m:=m_{p-1}$. 
It is easy to show that away from the singularities $\{ \mu_1,\ldots,\mu_n \}$ solutions to $D_p^\ast$ decay at least as fast as $\exp(-r|t|)$ as $r\rightarrow\infty$, by a calculation analogous to that in Section~\ref{sec:zeromodedecay}. 
On the other hand, the `Nakajima solutions' $\tilde{\psi}_1,\ldots,\tilde{\psi}_k$ that determine the terminating component at $\mu_p$ are of the form $\exp(-r|t|)\times (\textrm{non-$L^2$ function})$. 
The following conjecture is based on the idea that solutions in the continuing component decay like $\exp(-r|t+\alpha|)$ across $t=0$ for some $\alpha\neq 0$, and are in some sense small in the eigenbundle with eigenvalue $\mu_p$. 
Let $\psi_1,\ldots,\psi_k$ be the exact solutions defined by~\eqref{eq:defnpsis} that determine the terminating component. 

\begin{conjecture}\label{conj:contcmpt}
There is an orthonormal set $\{ \psi_{k+1},\ldots,\psi_{k+m} \}$ of maps
\begin{equation*}
\psi_j:(-\epsilon,\epsilon)\rightarrow L^2(\rthree,S_{(2)}\otimes E)
\end{equation*}
such that for all $j=k+1,\ldots,k+m$:
\begin{enumerate}
\item $\psi_j(t)\in\coker D_\xi$ when $t\neq 0$, 
\item $\psi_j(t)$ is continuous in $t$
\item for all $t\in(-\epsilon,\epsilon)$, $\psi_j(t)$ vanishes to all orders of $r$ in the limit $r\rightarrow\infty$, and
\item for all $i=1,\ldots,k$ and $t\in(-\epsilon,0)$, $\psi_i(t)$ is orthogonal to $\psi_j(t)$. 
\end{enumerate}
\end{conjecture}

If Conjecture~\ref{conj:contcmpt} holds, then using Proposition~\ref{prop:montermcpt} the following Conjecture immediately holds:

\begin{conjecture}\label{conj:neardecomp}
There is a parallel gauge on $X_{p-1}$ for some neighbourhood $t\in(0,\epsilon)$ in which the limits $T^{j,+}_{p-1}$ defined by~$\eqref{eq:conjlims}$ exist. 
There is also a parallel gauge on $X_p$ for some neighbourhood $t\in(-\epsilon,0)$ in which the $T^j$ decompose as in Proposition~\ref{prop:montermcpt}, except the top left-hand block has the form $T^{j,+}_{p-1} +O(t)$. 
\end{conjecture}

\proof
Condition $3$ of Conjecture~\ref{conj:contcmpt} ensures that $\langle ix_j\psi_a,\psi_b \rangle_{L^2}$ exists for all $t\in(-\epsilon,\epsilon)$ and $a,b\in\{ k+1,\ldots,k+m \}$, and is continuous in $t$. 
Moreover, transforming to a parallel gauge cannot introduce any discontinuities, because (fixing $T^0$ to be the matrix with entries $\langle \partial_\xi {\psi}_a,{\psi}_b\rangle_{L^2}$) the gauge transformation $\exp\int T^0 ds$ will always be continuous, even if $T^0$ is discontinuous. 
\eproof

We can provide some evidence to support Conjecture~\ref{conj:contcmpt}. 
Recall the model operator $\tilde{D}_\xi$ defined by~$\eqref{eq:defntildeDxi}$. 
If Conjecture~\ref{conj:contcmpt} does not hold for $\tilde{D}_\xi$ (in some sense) then we cannot reasonably expect it to hold for $D_\xi$, so we should try to understand the behaviour of the solutions to $\tilde{D}_\xi^\ast$ near the singularity $\mu_p$. 
First consider the component $(\tilde{D}_\xi^p)^\ast$ of $\tilde{D}_\xi^\ast$ defined by equation~$\eqref{eq:monmodelp}$. 
Consider a separable solution of the form
\begin{equation}\label{eq:sepsoln}
\psi(r,y_1,y_2)=\begin{pmatrix}
f(r)u(y_1,y_2) \\
g(r)v(y_1,y_2)
\end{pmatrix}.
\end{equation}
This is a solution if
\begin{equation*}
\begin{pmatrix}
(P^+f)u+g(D^-_{a_k}v) \\
f(D^+_{a_k}u)+(P^-g)v
\end{pmatrix}=0
\end{equation*}
where $P^+=i(r\partial_r+rt+(k+2)/2)$ and $P^-=i(-r\partial_r+rt+(k-2)/2)$. 
We therefore have 
\begin{equation*}
\begin{pmatrix}
0 & D^-_{a_k} \\ D^+_{a_k} & 0
\end{pmatrix}
\begin{pmatrix}
u \\ v
\end{pmatrix}=
\begin{pmatrix}
\lambda & 0 \\ 0 & \mu
\end{pmatrix}
\begin{pmatrix}
u \\ v
\end{pmatrix}
\end{equation*}
for some constants $\lambda,\mu$, and 
\begin{equation}\label{eq:seprequn}
\begin{pmatrix}
0 & P^- \\ P^+ & 0
\end{pmatrix}
\begin{pmatrix}
f \\ g
\end{pmatrix}=
\begin{pmatrix}
-\mu & 0 \\ 0 & -\lambda
\end{pmatrix}
\begin{pmatrix}
f \\ g
\end{pmatrix}.
\end{equation}
It follows that
\begin{align}
D^-_{a_k}D^+_{a_k}u &=\lambda\mu u,& D^+_{a_k}D^-_{a_k}v &=\lambda\mu v,\quad \textrm{and}\label{eq:SLy}\\
P^-P^+f &=\lambda\mu f,& P^+P^-g &=\lambda\mu g. \label{eq:PplusPminus}
\end{align}
The values $\alpha=\lambda\mu$ are fixed by the spectrum of the operator $D^-_{a_k}D^+_{a_k}$, which has a complete set of orthogonal eigenvectors. 
Given any non-zero $\lambda,\mu$, equation~$\eqref{eq:seprequn}$ can be solved explicitly using Maple: the solutions are Whittaker functions with argument $z=2rt$ (see \cite[Section 13]{abr92} for a description of Whittaker functions and their asymptotic expansions). 
In the limit $r\rightarrow\infty$ these have the form 
\begin{equation*}
\big( \exp(-rt) \big)\times \big( 1+\textrm{lower order terms} \big)
\end{equation*}
for $t<0$, and so are not normalizable. 
However, when $\lambda=\mu=0$ the equations decouple and normalizable solutions for $g$ exist---but these are just the Nakajima solutions. 
By completeness of the eigenfunctions in $\eqref{eq:SLy}$ and $\eqref{eq:PplusPminus}$ we can assume that any solution to $(\tilde{D}_\xi^p)^\ast$ is a sum of terms of the form~$\eqref{eq:sepsoln}$, and by orthogonality $(\tilde{D}_\xi^p)^\ast$ must kill each term in the sum. 
Thus we have shown that the only normalizable solutions to $(\tilde{D}_\xi^p)^\ast$ on $t\in(-\epsilon,0)$ are the Nakajima solutions. 
This analysis can be repeated for the other components $(\tilde{D}_\xi^q)^\ast$ ($q\neq p$) of $\tilde{D}^\ast_\xi$, and shows that their solutions decay like $\exp(-r| \xi-\mu_q |)$ as $r\rightarrow\infty$. 
These solutions therefore form continuous families across $t=0$ which satisfy the conditions of Conjecture~\ref{conj:contcmpt}. 

While this supports Conjecture~\ref{conj:contcmpt}, it certainly does not prove it. 
Given a family of solutions $\tilde{\psi}(t)$ to $\tilde{D}_\xi^\ast$ that is continuous across $t=0$, one might hope to use $\tilde{\psi}$ as an approximate solution, as we did for the terminating component, and show that $\psi=\hat{P}\tilde{\psi}$ is a continuous family of solutions to $D^\ast_\xi$. 
However, since $\tilde{\psi}(t)$ decays like $\exp(-r|t+\mu|)$ for some $\mu\neq 0$ the estimates~$\eqref{eq:estratio}$ do not hold, and we do not obtain $(1-\hat{P})\tilde{\psi}\rightarrow 0$ as $t\rightarrow 0$. 
Nakajima's analysis is therefore insufficient to prove that the family $\hat{P}\tilde{\psi}(t)$ is continuous across $t=0$. 
In any case, there may be the wrong number of solutions to $\tilde{D}^\ast_\xi$ to match the expected rank of the continuing component. 

An obvious approach to proving Conjecture~\ref{conj:contcmpt} is to use weighted operators, like those in Section~\ref{sec:zeromodedecay}, using the weighting to kill off the solutions corresponding to the terminating component. 
Consider the operators
\begin{equation*}
L_{\xi,\lambda}=w_\lambda D_\xi w_\lambda^{-1}\quad\textrm{and}\quad L_{\xi,\lambda}^\ast=w_\lambda^{-1} D_\xi^\ast w_\lambda
\end{equation*}
where $w_\lambda$ is defined by~$\eqref{eq:weightingfn}$. 
If we take $\lambda$ to be some small positive constant then, given the Nakajima solutions $\psi_1,\ldots,\psi_k$ of $D^\ast_\xi$, $w_\lambda^{-1}\psi_1,\ldots,w_\lambda^{-1}\psi_k$ are solutions to $L_{\xi,\lambda}^\ast$ but are not $L^2$. 
In other words, by weighting we have removed the Nakajima solutions from the $L^2$-kernel. 
If we could prove that $L_{\xi,\lambda}^\ast$ was Fredholm with $L^2$-index $m$ for $t\in(-\epsilon,\epsilon)$, then a trivialization $\psi_{k+1}(t),\ldots,\psi_{k+m}(t)$ of the $L^2$ kernel of $L_{\xi,\lambda}^\ast$ would give rise to a set of solutions to $D^\ast_\xi$ satisfying the properties in Conjecture~\ref{conj:contcmpt}. 
Unfortunately, a calculation of the indicial family like that in Section~\ref{sec:zeromodedecay} shows that $L_{\xi,\lambda}^\ast$ is not Fredholm when $t\in(-\lambda,\lambda)$, and so the strategy fails to work. 

Obtaining the continuing component at a zero jump presents further difficulties, and we will not be so bold as to make a formal conjecture as we did for $k_p\neq 0$. 
Given a zero jump $\mu_p$, we would not expect to find families $\psi_1(t),\ldots,\psi_m(t)$ defined on $t\in(-\epsilon,\epsilon)$ satisfying conditions $1$, $2$, and $3$ of the Conjecture, since the Nahm matrices would then be continuous across the zero jump. 
One possibility is that there are families $\psi_1(t),\ldots,\psi_m(t)$ defined on $t\in(-\epsilon,\epsilon)$ satisfying conditions $1$ and $2$ but not $3$. 
This would imply that the matrices $(T^j)_{ab}=\langle ix_j\psi_a,\psi_b \rangle_{L^2}$ do not exist at $t=0$ (because $x_j\psi_a$ might fail to be $L^2$), while the limits as $t\rightarrow 0$ from either side could exist but be different. 
Beyond this remark we will not consider zero jumps further. 

The final step is to go from Conjecture~\ref{conj:contcmpt} and Proposition~\ref{prop:montermcpt} to the full decomposition~\ref{claim:}. 
We make the following:

\begin{conjecture}\label{conj:fulldecomp}
Given that the data $\nabla_p$, $T_p^1$, $T_p^2$, $T_p^3$ satisfy Nahm's equation on the interior of each interval $I_p$, we obtain the full decomposition~\ref{claim:} from Conjecture~\ref{conj:neardecomp}. 
\end{conjecture}

Conjecture~\ref{conj:fulldecomp} follows immediately from:

\begin{conjecture}\label{conj:Nahmequn}
Suppose we have a rank $(m+k)$ solution $\nabla$, $T^1$, $T^2$, $T^3$ to Nahm's equation on a bundle over the interval $t\in(0,\epsilon)$, where $m,k\geq 0$. 
In addition, suppose there is a parallel gauge in which the $T^j$ decompose as 
\begin{center}
\setlength{\unitlength}{1.2cm}
\begin{picture}(8,2.5)(-4,-1.0)

\put(-2,0){\line(1,0){4}}
\put(0,-0.6){\line(0,1){1.2}}

\put(-1.7,0.25){$S^j+O(t)$}
\put(0.8,0.25){$\ast$}
\put(-1.3,-0.5){$\ast$}
\put(0.6,-0.5){$A^j(t)$}

\put(-3.6,-0.1){$T^{j}(t) = $}
\put(-2.4,-0.1){$\Bigg($}
\put(2.1,-0.1){$\Bigg)$}

\put(-1.2,0.9){$m$}
\put(3,0.25){$m$}
\put(0.95,0.9){$k$}
\put(3,-0.5){$k$}

\put(2.75,0.05){\vector(0,1){0.55}}
\put(2.75,0.6){\vector(0,-1){0.55}}
\put(2.75,-0.05){\vector(0,-1){0.55}}
\put(2.75,-0.6){\vector(0,1){0.55}}

\put(1.3,1.0){\vector(1,0){0.7}}
\put(0.7,1.0){\vector(-1,0){0.7}}
\put(-1.4,1.0){\vector(-1,0){0.65}}
\put(-0.7,1.0){\vector(1,0){0.65}}

\end{picture}
\end{center}
where $S^j$ is some fixed skew-hermitian matrix and $A^j$ satisfies
\begin{equation*}
A^j(t)=\begin{cases}
R^j/t+O(1) &\textrm{when~}k>3 \\
R^j/t+O(t^{-1/2})& \textrm{when~}k=2\textrm{~or~}3 \\
\textrm{could be unbounded~}&\textrm{when~}k=1
\end{cases}
\end{equation*}
as $t\rightarrow 0$. 
Here $R^1,R^2,R^3$ define an irreducible representation of $\sutwo$ in the usual way. 
Under these assumptions it necessarily follows that
\begin{enumerate}
\item the top left-hand block is analytic in $t$, even for $k=0$, 
\item the off-diagonal blocks are of the form $t^{(k-1)/2}\times\textrm{analytic function}$, and
\item $A^j(t)$ is meromorphic when $k>1$ but holomorphic when $k=1$. 
\end{enumerate}
\end{conjecture}

Conjecture~\ref{conj:Nahmequn} should be relatively easy to prove, and elements of a proof already exist in the literature: \cite[Section 2]{hur89b} contains related results.
The main assertion contained in the Conjecture is that when $k=2,3$ and $A^j(t)=R^j/t+O(t^{-1/2})$, Nahm's equation forces $A^j(t)$ to have the form $R^j /t +O(1)$.  
This is more straight-forward to prove when $m=0$, according to the following outline. 
Suppose that the matrices $T^1,T^2,T^3$ have rank $k=2$ or $k=3$, solve Nahm's equation (with $\nabla=\partial_t$) on $(0,\epsilon)$, and have the form 
\begin{equation*}
T^j=\frac{R^j}{t}+\frac{Q^j}{t^{1/2}}+\textrm{higher order terms}.
\end{equation*}
It follows that
\begin{equation}\label{eq:QR}
-\frac{1}{2}Q^1+R^2Q^3+Q^2R^3-R^3Q^2-Q^3R^2 =0
\end{equation}
and the two equations obtained from cyclic permutations of $\{ 1,2,3 \}$. 
In the case $k=2$ we can express each $Q^j$ as a sum $\sum_0^3 Q^{jl}\gamma_l$ and assume $R^j=-\frac{1}{2}\gamma_j$ for $j=1,2,3$. 
Substituting this into~$\eqref{eq:QR}$ and the other two equations, it is easy to show that $Q^{jl}=0$ for all $j,l$. 
A similar proof using more sophisticated representation theory should work for the case $m=0, k=3$. 
However, when $m>0$ the off-diagonal blocks make the Conjecture harder to prove. 

To conclude this Section we return to calorons to give a precise statement of our results concerning the transform from calorons to Nahm data. 
Our aim was to prove that the Nahm transform is a well-defined map from $\spcSD{C}{\bdarydata}$ to $\spcSD{N}{\bdarydata}$. 
We have proved that the transform of an element of $\spcSD{C}{\bdarydata}$ consists of a well-defined connection and endomorphisms $\{ \nabla_p,T_p^1,T_p^2,T_p^3 \}$ on bundles $X_p\rightarrow (\mu_{p+1},\mu_p)\subset \R / \muo\Z$ for each $p=1,\ldots,n$; that the data satisfy Nahm's equation; and that the data has the correct rank to be an element of $\spcSD{N}{\bdarydata}$. 
To complete the proof we must also obtain the decomposition of the Nahm data at each point $\xi=\mu_p$ and prove that the Nahm operator constructed from the Nahm data is injective (since $\spcSD{N}{\bdarydata}$ is by definition the set of caloron Nahm data that determine injective Nahm operators). 
We have not addressed the problem of injectivity, but made some remarks about this in Section~\ref{sec:further}. 
Although the Conjectures above are stated for $U(n)$ monopoles they also apply to calorons in an obvious way, and assuming they hold, we have obtained the correct decomposition of the caloron Nahm data at singularities with $k_p\neq 0$. 
We refer the reader back to Section~\ref{sec:further} for remarks about the invertibility of the transform and problems that could be tackled with the Nahm transform in place.

%%%%%%%%%%%%%%%%%%%%%%%%%%%%%%%%%%%%%%%%%%%%%%%%%%%%%%%%%%%%%%%%%%%%%%%%%%%%
%                                  GLOSSARY

\chapter*{Glossary of Notation}
\addcontentsline{toc}{chapter}{Glossary}

\newcommand{\glonotitem}[3]{
{#1} & {#2} & \pageref{#3} \\
}

\label{pag:glo}

%Final step when glossary complete: fiddle round to make the table nice.

\begin{tabular}{lp{9cm}l}
\glonotitem{$x_0,x_1,x_2,x_3$}{standard coordinates on $\rfour$}{glo:rfourcoords}
\glonotitem{$\ast$}{Hodge star operator}{glo:ast}
\glonotitem{$\ast_3,\ast_4$}{Hodge stars on $\rthree$ and $\rfour$}{glo:ast34}
\glonotitem{$(A,\Phi)$}{monopole configuration}{glo:monconfig}
\glonotitem{$S^+,S^-$}{spin bundles on $\rfour$, $S^1\times\rthree$ or $4$-torus}{glo:spin4}
\glonotitem{$\gamma_0,\gamma_1,\gamma_2,\gamma_3$}{spin matrices}{glo:gammaj}
\glonotitem{$\Gamma(V)$}{sections of a bundle $V$}{glo:Gamma}
\glonotitem{$D_\A^+,D_\A^-$}{Dirac operators coupled to connection $\A$}{glo:DA}
\glonotitem{$\spinthree$}{spin bundle on $\rthree$}{glo:spinthree}
\glonotitem{$T,T^\ast$}{the torus $\rfour / \Lambda$ and its dual}{glo:TTast}
\glonotitem{$\hat{S}^+,\hat{S}^-$}{spin bundles on the dual torus}{glo:spindual}
\glonotitem{$\xi$}{coordinate on the dual torus}{glo:xi}
\glonotitem{WFF}{without flat factors}{glo:WFF}
\glonotitem{$D^+_\xi,D^-_\xi$}{Dirac operators coupled to the flat line bundle parameterized by $\xi$}{glo:Dxi}
\glonotitem{$\hat{P}=\hat{P}_\xi$}{orthogonal projection on $\ker D^-_\xi$}{glo:Phat}
\glonotitem{$(\hat{\E},\hat{\A})$}{Nahm transform of $(\E,\A)$}{glo:nahmtrans}
\glonotitem{$D_x^+,D^-_x$}{Dirac operators coupled to the flat line bundle parameterized by $x$}{glo:Dx}
\glonotitem{$(\check{\E},\check{\A})$}{inverse Nahm transform of $(\E,\A)$}{glo:invnahmtrans}
\glonotitem{$T^n$}{the $n$-dimensional torus $S^1\times\cdots\times S^1$}{glo:Tn}
$X_p$ & vector bundle on which Nahm data is defined & \pageref{glo:Xp1}, \pageref{glo:Xp2}, \pageref{glo:Xp3}\\
$\nabla_p$ & connection on $X_p$ & \pageref{glo:nabp1}, \pageref{glo:nabp2}\\
$T^j_p$ & Nahm matrices on $X_p$ & \pageref{glo:Tjp1}, \pageref{glo:Tjp2}\\
\glonotitem{$LG$}{the group of smooth loops in a group $G$}{glo:LG}
\glonotitem{$L\curlyg$}{the Lie algebra of $LG$}{glo:Lalg}
\glonotitem{$\Lh G$}{semi-direct product of $LG$ and $U(1)$}{glo:LhG}
$\mu_0$ & $\perflat$ is the period of the caloron & \pageref{glo:mu01}, \pageref{glo:mu02} \\
$\So$ & $\So = \R / (\per\Z)$& \pageref{glo:So}, \pageref{glo:So2}\\
\glonotitem{$(\hat{A},\hat{\Phi})$}{a $\Lh SU(n)$ monopole}{glo:loopmon}
$I_p$ & The interval $[\mu_{p+1},\mu_p]$  & \pageref{glo:Ip1}\\
\glonotitem{$\threeball$}{the closed $3$-ball}{glo:threeball}
$c_2(\E,f),c_2(\E,f)[X]$ & obstruction to extending the framing $f$ to the interior of $\E$ (the instanton charge) & \pageref{glo:c21}, \pageref{glo:c22}\\
\glonotitem{$X$}{$X=\So\times\threeball$}{glo:X}
\glonotitem{$\Xo$}{interior of $X$}{glo:Xo}
\glonotitem{$\partial X$}{boundary $S^1\times\sphinf$ of $X$}{glo:partialX}
\glonotitem{$\sphinf$}{boundary of $\threeball$}{glo:sphinf}
\glonotitem{$p$}{projection $\So\times M\rightarrow M$ for some manifold $M$}{glo:projp}
%pagebreak
\end{tabular}
\newpage
\begin{tabular}{lp{9cm}l}
\glonotitem{$r,y_1,y_2$}{polar coordinates on $\rthree$}{glo:ry1y2}
\glonotitem{$\chi$}{boundary defining function on $\threeball$, $\chi=r^{-1}$}{glo:chi}
\glonotitem{$E$}{trivial $U(n)$ bundle on $\threeball$}{glo:E}
\glonotitem{$\Einf$}{$\Einf=E | \sphinf$}{glo:Einf}
\glonotitem{$(\Ainf,\Phiinf)$}{connection and Higgs field on $\Einf$}{glo:Ainf}
\glonotitem{$\mu_1,\ldots,\mu_n$}{eigenvalues of $\Phiinf$}{glo:muj}
\glonotitem{$k_1,\ldots,k_n$}{Chern classes of eigenbundles of $\Phiinf$ (monopole charges)}{glo:kj}
\glonotitem{$\monbdarydata$}{monopole boundary data}{glo:veckvecmu}
\glonotitem{$\Ieps$}{the interval $(-\epsilon,\perflat+\epsilon)$}{glo:Ieps}
\glonotitem{$\Auto E$}{unitary automorphisms of $E$ that are the identity at infinity}{glo:AutoE}
\glonotitem{$\deg c$}{degree of a map $c:S^3\rightarrow U(n)$}{glo:deg}
\glonotitem{$\SAuto E$}{trace-free elements of $\Auto E$}{glo:SAutoE}
\glonotitem{$k_0$}{the instanton charge $c_2(\E,f)$}{glo:k0}
\glonotitem{$\bdarydata$}{caloron boundary data}{glo:bdarydata}
$m_p$ & rank of $X_p$, $m_p=\sum_0^p k_j$ for calorons & \pageref{glo:mp}\\
\glonotitem{$q$}{the projection $q:\Ieps\times\threeball\rightarrow\threeball$}{glo:q}
\glonotitem{$\E^q$}{the trivial bundle $\E^q=q^\ast E$}{glo:Eq}
\glonotitem{$\A^q$}{a quasi-periodic connection on $\E^q$, often the pull-back of $\A$}{glo:Aq}
\glonotitem{$\monopoles$}{space of monopoles with boundary data $\monbdarydata$}{glo:spcmon}
\glonotitem{$\curlyL\bdarydata$}{space of loops of monopoles in $\monopoles$ with period $\mu_0$ and degree $k_0$}{glo:loopsofmon}
\glonotitem{$C^k_\chi(X)$}{functions with $k$ derivatives in $\chi$ that are smooth up to the boundary}{glo:Ckchi}
\glonotitem{$C_\chi^{0,1}$}{space of $1$-forms such that the $d\chi$ component is $C^0_\chi$, while the other components are $C^1_\chi$}{glo:C01chi}
\glonotitem{$\spc{C}{\bdarydata}$}{space of framed caloron configurations with boundary data $\bdarydata$}{glo:spccal}
\glonotitem{$\spcSD{C}{\bdarydata}$}{subspace of ASD calorons}{glo:spcSDcal}
\glonotitem{$\ch(\E,\A)$}{Chern character of the bundle and connection $(\E,\A)$}{glo:ch}
\glonotitem{$c_1(L)$}{first Chern class of a bundle $L$}{glo:c1}
\glonotitem{$\rotbdary$}{the rotation map on sets of boundary data}{glo:rotbdary}
\glonotitem{$\rotcal$}{the rotation map on caloron configurations}{glo:rotcal}
\glonotitem{$R^j_p$}{residue in the terminating component of the Nahm data at $\xi=\mu_p$}{glo:Rjp}
\glonotitem{$\Srep^k$}{irreducible $(k+1)$-dimensional representation of $\sutwo$}{glo:srep}
\glonotitem{$T_p^+,T_p^-$}{limits of the continuing component of the Nahm matrices either side of a singularity}{glo:Tpplus}
\glonotitem{$\monSD{N}{\monbdarydata}$}{the set of monopole Nahm data with boundary data $\monbdarydata$}{glo:nahmmonSD}
\glonotitem{$\Delta(x):W\rightarrow V$}{the Nahm operator}{glo:Deltax}
\glonotitem{$\COKER~\Delta$}{the cokernel of $\Delta(x)$ regarded as a bundle over $\rfour$}{glo:COKER}
\glonotitem{$P=P_x$}{projection onto $\coker\Delta(x)$}{glo:P}
\glonotitem{$Y_p$}{$Y_p=\C^2\otimes X_p$}{glo:Yp}
\glonotitem{$W^l_p$}{Sobolev space of sections of $Y_p$ with $l$ derivatives in $L^2$}{glo:Wlp}
\glonotitem{$L^2_l(I_p)$}{Sobolev space of functions on $I_p$ with $l$ derivatives in $L^2$}{glo:sobltwo}
%pagebreak
\end{tabular}
\newpage
\begin{tabular}{lp{9cm}l}
\glonotitem{$\Lbar{l}[a,b]$}{space of restrictions to $[a,b]$ of distributions in $L^2_l(\R)$}{glo:Lbar}
\glonotitem{$\Ldot{l}[a,b]$}{space of distributions in $L^2_l(\R)$ supported on $[a,b]$}{glo:Ldot}
\glonotitem{$\supp~f$}{support of a function $f$}{glo:supp}
\glonotitem{$\Woone{p}$}{elements of $W_p^1$ with vanishing terminating component}{glo:Wop}
\glonotitem{$D_p(x)$}{the operator $i\nabla_p +iT_p +x$ on $Y_p$}{glo:Dp}
\glonotitem{$T_p$}{the sum $\sum_{j=1}^{3} \gamma_j\otimes T^j_p$}{glo:Tp}
\glonotitem{$J_q$}{subspace of $Y_q(\mu_q)$ corresponding to the zero jump at $\mu_q$}{glo:Jq}
\glonotitem{$\zeta_q$}{a vector in $J_q$}{glo:zetaq}
\glonotitem{$\pi_q(w)$}{inner product of a vector $w$ with $\zeta_q$}{glo:piq}
\glonotitem{$\Zeros$}{the set of zero jumps $\{q : m_q = m_{q-1}\}$}{glo:Zeros}
\glonotitem{$\Nzer$}{the number of zero jumps, $\Nzer=| \Zeros |$}{glo:Nzer}
\glonotitem{$\mon{N}{\monbdarydata}$}{monopole Nahm data not necessarily satisfying Nahm's equation}{glo:monopolenahmdata}
\glonotitem{$\spcSD{N}{\bdarydata}$}{space of caloron Nahm data with boundary data $\bdarydata$}{glo:spcNSD}
\glonotitem{$\spc{N}{\bdarydata}$}{caloron Nahm data not necessarily satisfying Nahm's equation}{glo:spcN}
\glonotitem{$v^{\textrm{cont}},V^{\textrm{cont}}$}{continuing component of a vector or vector space}{glo:supcont}
\glonotitem{$R_p$}{the sum $\sum_1^3 \gamma_j\otimes R_p^j$}{glo:Rp}
\glonotitem{$U_p$}{space of solutions to $D_p(x)$}{glo:Up}
\glonotitem{$U_p^\ast$}{space of solutions to $D_p^\ast(x)$}{glo:Upast}
$\Delta^\ast(x)$ & the adjoint of $\Delta(x)$ & \pageref{glo:Deltaast}\\ 
\glonotitem{$W^\ast$}{the dual space of $W$}{glo:Wdual}
\glonotitem{$\langle , \rangle_{\textrm{dual}}$}{pairing of an element of $W$ and an element of $W^\ast$}{glo:dualpair}
\glonotitem{$U_{\tau,V},U_{\tau,W}$}{action of translation $x_0\mapsto x_0+\perflat$ on $V,W$}{glo:UVWtau}
\glonotitem{$\tau$}{translation by one period in the $x_0$ direction}{glo:tau}
\glonotitem{$U_\tau$}{$U_\tau=U_{\tau,V}$}{glo:Utau}
\glonotitem{$\rotNahm$}{rotation of Nahm data by $\mu_0 / n$}{glo:rotnahm}
\glonotitem{$\modelop(x)$}{the model operator}{glo:Deltatilde}
\glonotitem{$\eta_l,\eta_l^\perp$}{basis of sections of the instanton block}{glo:etas}
\glonotitem{$x_{lm}^{\textrm{res}}$}{the resonating point $(\lambda_l+2\pi m / \muo,0,0,0)$}{glo:xreslm}
\glonotitem{$\eta_{lm},\eta_{lm}^\perp$}{}{glo:etaslm}
\glonotitem{$\tilde{\zeta}_p$}{deformation of the vector $\zeta_p$}{glo:zetatilde}
\glonotitem{$\tilde{\pi}_p(w)$}{inner product of $w$ with $\tilde{\zeta}_p$}{glo:pitilde}
\glonotitem{$B_{lm}$}{$4$-ball round resonating point $x_{lm}^{\textrm{res}}$}{glo:Blm}
\glonotitem{$Y_p^I,Y_p^M$}{instanton and monopole blocks of $Y_p$}{glo:YpI}
\glonotitem{$W=W_I\oplus W_M$}{decomposition of $W$ into instanton and monopole blocks}{glo:Wdecomp}
\glonotitem{$\Delta_I(x),\Delta_M(x)$}{components of the model operator on the instanton and monopole blocks}{glo:DeltaI}
\glonotitem{$B(x)$}{the off-diagonal block of the model operator}{glo:offdiag}
\glonotitem{$\Zeros_I,\Zeros_M$}{zero jumps in instanton and monopole blocks}{glo:moreZeros}
\glonotitem{$\Zeros_O$}{singularities in the Nahm data that are not zero jumps}{glo:moreZeros}
\glonotitem{$N_I,N_M$}{$N_I=| \Zeros_I |$ and $N_M=| \Zeros_M |$}{glo:NINM}
\glonotitem{$\pi_I,\pi_M$}{decomposition of the projection $\pi$ into instanton and monopole blocks}{glo:piIM}
\glonotitem{$\modelop^\ast(x)$}{adjoint of the model operator}{glo:modeladjt}
\glonotitem{$\Delta_I^\ast(x),\Delta_M^\ast(x)$}{adjoints of $\Delta_I,\Delta_M$}{glo:DeltaIast}
%pagebreak
\end{tabular}
\newpage
\begin{tabular}{lp{9cm}l}
\glonotitem{$\tilde{D}_p^\ast(x)$}{the adjoint of $\tilde{D}_p(x)$}{glo:Dptildeast}
$\tilde{v}_1,\ldots,\tilde{v}_n$ & approximate solutions to the model operator & \pageref{glo:tildevp1}--\pageref{glo:tildevp2} \\
\glonotitem{${\mathfrak{I}}_k$}{}{glo:intIk}
\glonotitem{$\textrm{Hol}(x)$}{holonomy of $\Delta_I^\ast(x)$}{glo:holDeltaI}
\glonotitem{$v_p,~p\in\Zeros_I$}{exact solutions to $\modelop^\ast(x)$ corresponding to zero jumps in the instanton block}{glo:exactvp}
\glonotitem{$\tilde{P}$}{projection onto the cokernel of the model operator}{glo:approxproj}
\glonotitem{$\tilde{P}_U$}{unitary approximation of $\tilde{P}$}{glo:approxunitaryproj}
\glonotitem{$\tilde{w}_1,\ldots,\tilde{w}_n$}{solutions to $\modelop^\ast(x)$, $\tilde{w}_p = \tilde{P}_U\tilde{v}_p$}{glo:exactwp}
\glonotitem{$e_1,\ldots,e_n$}{local trivialization of $\Einf$}{glo:ep}
\glonotitem{$\tilde{k}_0$}{instanton charge of $\COKER~\modelop$}{glo:tildeko}
\glonotitem{$(\E_0,\A_0)$}{caloron configuration with zero instanton charge, used to calculate $\tilde{k}_0$}{glo:E0A0}
\glonotitem{$\Fred,~(\Fredi)$}{space of all (injective) Fredholm operators $W\rightarrow V$ with index $-n$}{glo:Fred}
\glonotitem{$P_U$}{unitary approximation to the projection $P$}{glo:PU}
\glonotitem{$\nu$}{volume form}{glo:nu}
\glonotitem{$\iota_\alpha(\beta)$}{inner derivative of a vector field $\alpha$ with a form $\beta$}{glo:iota}
\glonotitem{$\Xising$}{$\Xising = \{ \mu_j + N\muo : j=1,\ldots,n\textrm{\ and\ }N\in\Z \}$}{glo:xising}
\glonotitem{$\IopN$}{$\IopN=(\mu_{p+1}+N\muo,\mu_{p}+N\muo)$}{glo:IopN}
\glonotitem{$X_{p,N}$}{the bundle with fibre $\coker\DAxiplus$ over $\IopN$}{glo:XpN}
\glonotitem{$\nabla_{p,N},T_{p,N}^j$}{Nahm data on $X_{p,N}$}{glo:NahmpN}
\glonotitem{$\hat{\tau}$}{the translation $\hat{\tau}:\xi\mapsto \xi+\muo$}{glo:tauhat}
\glonotitem{$\hat{U}_{\hat{\tau}}$}{action of the translation $\hat{\tau}$ as a bundle isomorphism on $\E$}{glo:Utauhat}
\glonotitem{$\Iop$}{$\Iop=(\mu_{p+1},\mu_p)\subset \R / \mu_0\Z$}{glo:Iop}
\glonotitem{\PhiDO 's}{pseudo-differential operators}{glo:PhiDO}
\glonotitem{$I_\Phi(P)$}{the indicial family of a \PhiDO~$P$}{glo:indicial}
\glonotitem{$L^\pm_{\xi,\lambda}$}{`weighted' Dirac operators}{glo:weighted}
\glonotitem{$D_\xi$}{Dirac operator coupled to a monopole}{glo:monDirac}
\glonotitem{$\tilde{D}_\xi$}{model operator approximating $D_\xi$}{glo:modelmon}
\glonotitem{$S_{(2)},S_{(2)}^\pm$}{spin bundles $S_{(2)}=S_{(2)}^+\oplus S_{(2)}^-$ on $\sphinf$}{glo:spintwo}
\glonotitem{$H$}{hyperplane bundle on $\sphinf$}{glo:hyperH}
\glonotitem{$D^{\pm}_{a_k}$}{Dirac operators on $\sphinf$ coupled to $H^k$}{glo:Dak}
\glonotitem{$M$}{$M=\rthree \setminus \threeball_R$}{glo:M}
\glonotitem{$\varrho$}{projection $\varrho:M\rightarrow\sphinf$}{glo:varrho}
\glonotitem{$D_{M,H^k}$}{Dirac operator on $M$ coupled to the pull-back of $H^k$}{glo:DMHk}
\glonotitem{$\tilde{\psi}_1,\ldots,\tilde{\psi}_k$}{approximate solutions to $D^\ast_\xi$ that determine the terminating component}{glo:psitildes}
%\glonotitem{}{}{glo:}
\end{tabular}

%%%%%%%%%%%%%%%%%%%%%%%%%%%%%%%%%%%%%%%%%%%%%%%%%%%%%%%%%%%%%%%%%%%%%%%%%%%%%

%%%%%%%%%%%%%%%%%%%%%%%%%%%%%%%%%%%%%%%%%%%%%%%%%%%%%%%%%%%%%%%%%%%%%%
%                                BIBLIOGRAPHY

\bibliographystyle{amsplain}
\bibliography{general}

\providecommand{\bysame}{\leavevmode\hbox to3em{\hrulefill}\thinspace}
\begin{thebibliography}{10}

\bibitem{abr92}
Milton Abramowitz and Irene~A. Stegun (eds.), \emph{Handbook of mathematical
  functions with formulas, graphs, and mathematical tables}, Dover Publications
  Inc., New York, 1992, Reprint of the 1972 edition.

\bibitem{ang93}
Nicolae Anghel, \emph{An abstract index theorem on non-compact {R}iemannian
  manifolds}, Houston Journal of Mathematics \textbf{19} (1993), 223--237.

\bibitem{ang93b}
\bysame, \emph{On the index of {C}allias-type operators}, Geom. Funct. Anal.
  \textbf{3} (1993), no.~5, 431--438.

\bibitem{ati79}
M.~F. Atiyah, \emph{Geometry of {Y}ang-{M}ills fields}, Fermi Lectures, Scuola
  Normale Superior, Pisa, 1979, (also in Collected Works, Vol. 5).

\bibitem{adhm78}
M.~F. Atiyah, V.~G. Drinfeld, N.~J. Hitchin, and Yu.~I. Manin,
  \emph{Construction of instantons}, Phys. Lett. \textbf{65A} (1978), 185--187.

\bibitem{bra89}
Peter~J. Braam and Pierre van Baal, \emph{Nahm's transformation for
  instantons}, Comm. Math. Phys. \textbf{122} (1989), no.~2, 267--280.

\bibitem{cal78}
C.~Callias, \emph{Axial anomalies and index theorems on open spaces}, Comm.
  Math. Phys. \textbf{62} (1978), 213--234.

\bibitem{cha99}
A.~Chakrabarti, \emph{Implementation of an iterative map in the construction of
  (quasi)periodic instantons: chaotic aspects and discontinuous rotation
  numbers}, J. Math. Phys. \textbf{40} (1999), no.~2, 635--673.

\bibitem{cor84}
E.~Corrigan and P.~Goddard, \emph{Construction of instanton and monopole
  solutions and reciprocity}, Ann. Physics \textbf{154} (1984), no.~1,
  253--279.

\bibitem{gro81}
R.~D.~Pisarski D.~J.~Gross and L.~G. Yaffe, \emph{{QCD} and instantons at
  finite temperature}, Rev. Modern Phys. \textbf{53} (1981), no.~1, 43--80.

\bibitem{don84}
S.~K. Donaldson, \emph{Nahm's equations and the classification of monopoles},
  Comm. Math. Phys. \textbf{96} (1984), no.~3, 387--407.

\bibitem{don90}
S.~K. Donaldson and P.~B. Kronheimer, \emph{The geometry of four-manifolds},
  Oxford University Press, 1990.

\bibitem{gar88}
H.~Garland and M.~K. Murray, \emph{{K}ac-{M}oody monopoles and periodic
  instantons}, Comm. Math. Phys. \textbf{120} (1988), 335--351.

\bibitem{gro83}
M.~Gromov and H.~B. Lawson, \emph{Positive scalar curvature and the index of
  the {D}irac operator on complete {R}iemannian manifolds}, Inst. Hautes Etudes
  Scientifiques Publ. Math. \textbf{58} (1983), 295--408.

\bibitem{har78}
B.~J. Harrington and H.~K. Shepard, \emph{Periodic {E}uclidean solutions and
  the finite-temperature {Y}ang-{M}ills gas}, Phys. Rev. \textbf{D17} (1978),
  no.~8, 2122--2125.

\bibitem{hit82}
N.~J. Hitchin, \emph{Monopoles and geodesics}, Comm. Math. Phys. \textbf{83}
  (1982), 579--602.

\bibitem{hit83}
\bysame, \emph{On the construction of monopoles}, Comm. Math. Phys. \textbf{89}
  (1983), 145--190.

\bibitem{hor85}
Lars H{\"o}rmander, \emph{The analysis of linear partial differential
  operators}, vol. III, Springer-Verlag, 1985.

\bibitem{hur89b}
J.~Hurtubise, \emph{The classification of monopoles for the classical groups},
  Comm. Math. Phys. \textbf{120} (1989), 613--641.

\bibitem{hur89}
J.~Hurtubise and M.~K. Murray, \emph{On the construction of monopoles for the
  classical groups}, Comm. Math. Phys. \textbf{122} (1989), 35--89.

\bibitem{jar01}
Marcos Jardim, \emph{Construction of doubly-periodic instantons}, Comm. Math.
  Phys. \textbf{216} (2001), no.~1, 1--15.

\bibitem{kra00}
Thomas~C. Kraan, \emph{Instantons, monopoles and toric hyper{K}\"ahler
  manifolds}, Comm. Math. Phys. \textbf{212} (2000), no.~3, 503--533.

\bibitem{kra98a}
Thomas~C. Kraan and Pierre van Baal, \emph{Exact ${T}$-duality between calorons
  and {T}aub-{N}{U}{T} spaces}, Phys. Lett. B \textbf{428} (1998), no.~3-4,
  268--276.

\bibitem{kra98b}
\bysame, \emph{Periodic instantons with non-trivial holonomy}, Nuclear Phys. B
  \textbf{533} (1998), no.~1-3, 627--659.

\bibitem{lee98b}
Kimyeong Lee, \emph{Instantons and magnetic monopoles on {${\bf {R}}\sp 3\times
  {S}\sp 1$} with arbitrary simple gauge groups}, Phys. Lett. B \textbf{426}
  (1998), no.~3-4, 323--328.

\bibitem{lee98}
Kimyeong Lee and Changhai Lu, \emph{{$SU(2)$} calorons and magnetic monopoles},
  Phys. Rev. \textbf{D58} (1998), no.~2, 025011.

\bibitem{maz99}
Rafe Mazzeo and Richard~B. Melrose, \emph{Pseudodifferential operators on
  manifolds with fibred boundaries}, Asian J. Math. \textbf{2} (1998), no.~4,
  833--866, Mikio Sato: a great Japanese mathematician of the twentieth
  century.

\bibitem{mel95}
Richard~B. Melrose, \emph{Geometric scattering theory}, Cambridge University
  Press, Cambridge, 1995.

\bibitem{mil58}
J.~Milnor, \emph{Differential topology}, Unpublished lecture notes.

\bibitem{muk81}
Shigeru Mukai, \emph{Duality between ${D}({X})$\ and ${D}(\hat {X})$\ with its
  application to {P}icard sheaves}, Nagoya Math. J. \textbf{81} (1981),
  153--175.

\bibitem{mur84}
M.~K. Murray, \emph{Non-abelian magnetic monopoles}, Comm. Math. Phys.
  \textbf{96} (1984), 539--565.

\bibitem{nah82}
W.~Nahm, \emph{The construction of all self-dual multimonopoles by the {ADHM}
  method}, Monopoles in quantum field theory (N.~S. Craigie, P.~Goddard, and
  W.~Nahm, eds.), World Scientific, Singapore, 1982.

\bibitem{nah83}
W.~Nahm, \emph{Self-dual monopoles and calorons}, Group theoretical methods in
  physics (Trieste, 1983), Springer, Berlin, 1984, pp.~189--200.

\bibitem{nak90}
H.~Nakajima, \emph{Monopoles and {N}ahm's equations}, Lecture Notes in Pure and
  Appl. Math., vol. 145, Dekker, New York, 1993.

\bibitem{nor00}
Paul Norbury, \emph{Periodic instantons and the loop group}, Comm. Math. Phys.
  \textbf{212} (2000), no.~3, 557--569.

\bibitem{nye00}
T.~M.~W. Nye and M.~A. Singer, \emph{An {$L^2$}-index theorem on
  {$S^1\times{\mathbb{R}}^3$}}, J. of Funct. Anal. \textbf{177} (2000),
  203--218.

\bibitem{pre86}
A.~Pressley and G.~Segal, \emph{Loop groups}, Oxford University Press, 1986.

\bibitem{rad94}
J.~R{\aa}de, \emph{{C}allias' index theorem, elliptic boundary conditions, and
  cutting and gluing}, Comm. Math. Phys. \textbf{161} (1994), 51--61.

\bibitem{roe88}
John Roe, \emph{Elliptic operators, topology and asymptotic methods}, Longman,
  1988.

\bibitem{uhl82}
K.~K. Uhlenbeck, \emph{Removable singularities in {Y}ang-{M}ills fields}, Comm.
  Math. Phys. \textbf{83} (1982), 11--29.

\bibitem{baa96}
Pierre van Baal, \emph{Instanton moduli for {${T}\sp 3\times{\bf {R}}$}},
  Nuclear Phys. B Proc. Suppl. \textbf{49} (1996), 238--249, Theory of
  elementary particles (Buckow, 1995).

\bibitem{baa99}
\bysame, \emph{Nahm gauge fields for the torus}, Phys. Lett. B \textbf{448}
  (1999), no.~1-2, 26--32.

\bibitem{war77}
R.~S. Ward, \emph{On self-dual gauge fields}, Phys. Lett. A \textbf{61} (1977),
  no.~2, 81--82.

\end{thebibliography}

\end{document}